\documentclass{SciPostAnonymous}
\usepackage{graphicx,cite}
\begin{document}

\begin{center}{\Large \textbf{
Anderson localisation in two dimensions: insights from Localisation Landscape Theory, exact diagonalisation, and time-dependent simulations
}}\end{center}

\begin{center}
S.S. Shamailov\textsuperscript{1*}, 
D.J. Brown\textsuperscript{1,2},
T.A. Haase\textsuperscript{1},
M.D. Hoogerland\textsuperscript{1}
\end{center}

\begin{center}
{\bf 1} Dodd-Walls Centre for Photonic and Quantum Technologies, Department of Physics, University of Auckland, Private Bag 92019, Auckland 1142, New Zealand.
\\
{\bf 2} Present Address: Light-Matter Interactions for Quantum Technologies Unit, Okinawa Institute of Science and Technology Graduate University, Onna, Okinawa 904-0495, Japan.
\\
* sophie.s.s@hotmail.com
\end{center}

\begin{center}
\today
\end{center}


\section*{Abstract}
{\bf 
Motivated by rapid experimental progress in ultra-cold atomic systems, we aim to provide a simple, intuitive description of Anderson localisation that allows for a direct quantitative comparison to experimental data, as well as yielding novel insights. To this end, we advance, employ and validate a recently-developed theory -- Localisation Landscape Theory (LLT) -- which has unparalleled strengths and advantages, both computational and conceptual, over alternative methods. We focus on two-dimensional systems with point-like random scatterers, although an analogous study in other dimensions and with other types of continuous disordered potentials would proceed similarly. We begin by showing that exact eigenstates cannot be efficiently used to extract the localisation length. We then provide a comprehensive review of known LLT, and confirm that the Hamiltonian with the effective potential of LLT has very similar low energy eigenstates to that with the physical potential. Next, we use LLT to compute the localisation length for very low-energy, maximally localised eigenstates and (manually) test our method against exact diagonalisation. Furthermore, we propose a transmission experiment that optimally detects Anderson localisation, and demonstrate how one may extract a length scale which is correlated with (and in general smaller than) the localisation length. In addition, we study the dimensional crossover from one to two dimensions, providing a new explanation to the established trends. The prediction of a mobility edge coming from LLT is tested by direct Schr\"odinger time evolution and is found to be unphysical. Moreover, we investigate expanding wavepackets, to check if these can be useful in detecting and quantifying Anderson localisation in a transmission experiment. We find that this is indeed the case, and the only disadvantage of such probing waves is the inability to resolve the energy dependence of the localisation length. Then, we utilise LLT to uncover a connection between the Anderson model for discrete disordered lattices and continuous two-dimensional disordered systems, which provides powerful new insights. From here, we demonstrate that localisation can be distinguished from other effects by a comparison to dynamics in an ordered potential with all other properties unchanged. Finally, we thoroughly investigate the effect of acceleration and repulsive interparticle interactions, as relevant for current experiments.
}

\vspace{10pt}
\noindent\rule{\textwidth}{1pt}
\tableofcontents\thispagestyle{fancy}
\noindent\rule{\textwidth}{1pt}
\vspace{10pt}

\section{Introduction}
\label{Intro}
In this section, we provide a ``gentle'', global introduction, giving some general background and motivating the research undertaken in the rest of the paper. More specific introductions, including detailed literature reviews, are to be found in the subsequent sections, as the range of topics covered is quite broad.

Anderson localisation \cite{Greek_review, DelandeLectures} is a universal wave interference phenomenon, whereby transport (i.e. wave propagation) is suppressed in a disordered medium due to dephasing upon many scattering events from randomly-positioned obstacles. This can be understood from Feynman's interpretation of quantum mechanics, where one must sum over all possible paths from the initial to the final points of interest to obtain the total transmission probability. The random positions of the scatterers guarantee dephasing between the different paths, leading to an attenuation of the amplitude of the wavefunction. First discovered in the context of quantised electron conduction and spin diffusion \cite{DrPhil}, Anderson localisation of particles thus provides direct evidence for the quantum-mechanical nature of the universe at a small scale.

This phenomenon can occur if the the de-Broglie wavelength is larger than the correlation length of the disorder, so that the wave ``sees'' the potential as random -- this is one of the factors responsible for the profound dependence of localisation properties on the energy of the probing wave. Moreover, to ensure sufficiently strong dephasing for localisation to take place, the wave must scatter either frequently or strongly, or both. Therefore, the density and strength of the impurities, as well as the system size, determine whether the wave is Anderson-localised at all, and if so, to what degree. Under Anderson localisation, the wavefunction decays exponentially in the tails with a length scale known as the localisation length. If transport is measured across a system the size of which is less than the localisation length, one finds that transport is reduced but does not vanish \cite{Sheng}.

Anderson localisation has been observed in many physical systems, including electron conduction in crystals \cite{Ying2016} and quantum wells \cite{LLT2018}, light waves \cite{Sperling, Lahini2008, Wiersma, Scheffold, Storzer2006, Segev2007, Riboli2011, Segev3}, microwaves \cite{Weiland1999, Laurent2007, McCall, Genack1997}, electromagnetic waves \cite{Chabanov}, ultrasound \cite{Weaver1990, Hu} and photonic crystal waveguides \cite{Vollmer2007}. With the rapid advance of ultra-cold atomic physics, the possibility of observing Anderson localisation directly for a coherent matter-wave soon became a reality. A momentum-space analogy has been employed to demonstrate localisation in a kicked rotor system \cite{Chabe2008, Lopez2012, Delande2015}, complemented by real-space localisation observations in one and three dimensions (1D and 3D, respectively) \cite{Aspect2005, Aspect2008, Aspect2012, deMarco2011, deMarco2013, Inguscio2008, Inguscio2015}. Two dimensions (2D) has been more challenging: for several years, classical trapping has prevented the detection of Anderson localisation with cold atoms \cite{Aspect2010, Aspect2011, Esslinger2012, Inguscio2005} (however, other systems have proved more fruitful \cite{Delande2015, Riboli2011, Segev2007, Vollmer2007}). Extremely recently, an innovative experimental approach has led to claims of direct observation in 2D as well \cite{BS}.

Despite the undeniable tour-de-force achievements on the practical side of these ultra-cold atomic experiments, often many open questions remain regarding what exactly happened in the experiment, why, what it means, and the implications that follow. To some degree, this is due to the lack of a simple, accessible and transparent theory that experimentalists could use to understand their findings. For continuous systems, researchers commonly draw on the predictions of scaling theory \cite{ScalingTheory}, which, due to its elegance and universality, is indeed very appealing. However, its applicability is limited to infinite systems with white noise (and finite-range hopping), conditions that are never satisfied in real experiments, and its predictions are often too general to be of practical use. A classical diffusive picture, applicable in the weakly-localised\footnote{The terms `weak' and `strong' localisation refer to the degree of transport suppression across the system, which depends on its size. For a system much smaller than the localisation length, such that the exponential decay of the wavefunction amplitude is not noticeable, a diffusion picture can assist with the description. In contrast, strong localisation is said to take place when the system is sufficiently large to allow the density to decay almost fully within its boundaries. Notice that these are limiting cases, with a wide range of intermediate scenarios connecting them.} regime, is commonly employed (e.g.~\cite{Delande2015,Muller2005}) because it can be easily grasped, sometimes well outside the limit where it is relevant. An alternative approach favoured by many theorists is Green's functions \cite{Sheng} which is exact (as long as all the assumptions are satisfied) but extremely cumbersome and involved. Finally, brute-force time-dependent simulations with the Schr\"{o}dinger \cite{deMarco2015} or Gross-Pitaevskii (GP) \cite{BS} equations are employed to mimic experiments as closely as possible, but this approach is very time-consuming and yields little insight into the physics. (Note that other methods are additionally reviewed in section \ref{litrev}).

In a sense, all the information concerning localisation properties is contained in the Hamiltonian of the system and can be accessed through its eigenspectrum. Exact diagonalisation is indeed a useful tool, but it is certainly limited by system size from the computational point of view, and, as we shall see, it is not obvious how one can extract the relevant information from the eigenstates. If one poses questions about \textit{dynamics} specifically, then indeed solving the Schr\"{o}dinger equation may be the most efficient way to obtain answers, but system size and spatial resolution are again serious limiting factors. If the particles are weakly interacting (which would naturally be the case for cold atoms), the GP equation is the simplest way of accounting for the effect of the nonlinearity. However, its numerical solution is even more demanding than that of its linear counterpart. Nonetheless, both exact diagonalisation and time-dependent simulations are powerful methods and will play an important role in our study, as much for their own merits as for benchmarking purposes.

Meantime, a break-through new theory -- coined Localisation Landscape Theory (LLT) \cite{Marcel2012, FnM2014, FnM2016, FnM2016b, part1, part2, part3} -- was developed recently, completely revolutionising the field. It allows for intuitive and transparent new insights into the physics, as well as a practical, efficient way of performing calculations. To give a brief overview, this theory relies on the construction of a function, the localisation landscape, which governs all the low-energy, localised physics. One can treat finite problems so that boundary effects are accounted for, and yet push the algorithms to very large system sizes, where alternative methods are completely impractical. The validity of this theory is not restricted to a specific noise type, making it widely applicable to a range of problems. An effective potential can be constructed, such that quantum interference effects can be captured instead by quantum tunnelling through this effective potential (but this is restricted to low energies, as we shall show). One can predict the main regions of existence (referred to as ``domains'') of the low-energy localised eigenstates, reconstruct the eigenstates on these domains, as well as compute the associated energy eigenvalues. Thus, Anderson localisation can be fully reinterpreted in this picture, including the energy dependence of the localisation length (so far, qualitatively). Very recently, LLT has been used to support an experimental study of Anderson localisation \cite{LLT2018}. Localisation landscape theory is a very young theory; in this article, we will somewhat advance it, clarify its limitations, and help link its predictions to realistic experiments.

In this regard, to date, the vast majority of experiments on Anderson localisation with cold atoms have examined the density profiles of wavefunctions expanding into a disordered potential (usually speckle), using the variance to quantify the size of the cloud (e.g.~\cite{Inguscio2008, Inguscio2015, Aspect2005, Aspect2010, Aspect2012}). The exception is the recent study \cite{BS}, where the authors chose to allow their wavefunction to \textit{transmit} through a region filled with random scatterers. A dumbbell geometry was chosen, in line with earlier work \cite{Eckel2016, Edwards2016, Esslinger2012, Esslinger2012b}, and the atomtronic LCR model suggested in these papers was employed to analyse the data. With the appearance of new experimental approaches, there is a need for a better theoretical description of such scenarios. Here we will show that indeed much can be learned from a transmissive experiment, but we will advocate a different key observable, proportional to the quantum-mechanical transmission coefficient through the disordered potential.

Thus, at the outset, our goals in this work are several. First of all, we wish to find a simple, intuitive picture that allows one to understand Anderson localisation conceptually. Second, it is desirable to develop a framework that allows for the computation of key quantities easily and directly, such that the theory is transparent to all. Finally, we aim to propose an experimental scenario that cleanly exposes the essence of the physics, suggest what should be measured, and by employing several theoretical methods, demonstrate how the observations are to be interpreted, i.e.~how one can extract meaning from the data.
\subsection{Article overview}
We begin by introducing the system of interest in section \ref{System}, and proceed to demonstrate what can and cannot be learned from an exact diagonalisation of the Hamiltonian in section \ref{Diag}. From here, section \ref{LLTold} reviews known LLT, highlights its strengths and advantages, and presents a quick survey of the effect of the key parameters on localisation. In section \ref{Wmeaning}, we show that the effective potential of LLT can be used to access the exponential decay in the low-energy eigenstates of the physical potential by comparing the eigenstates of the Hamiltonian with the two potentials. Then, in section \ref{XiSaddles}, we extend known LLT to calculate the localisation length at very low energies, as defined by the length scale of exponential decay in the tails of the eigenstates of the Hamiltonian, and directly test the method by comparison to exact eigenstates. This method breaks down at higher energies, together with the tunnelling picture, as we describe in detail in section \ref{HigherEs}. Here, we discuss the mechanisms by which the eigenstates extend to cover larger areas at higher energies, and explain why our method cannot capture this behaviour, which can no longer be viewed as a simple tunnelling process in the effective potential. In the course of our work (relevant at low energies), we develop a simple and practical approximation to multidimensional tunnelling, discussed in section \ref{MultiDimTun}, which has many potential applications in other contexts.

Following a brief motivational discussion in section \ref{Transport}, section \ref{Trans} moves on to propose an experimental configuration that would allow to unambiguously observe Anderson localisation. An excellent observable is examined which is robust, readily accessible in experiments, and has a clear physical interpretation. We show that this observable allows for the extraction of a length-scale that is correlated with the localisation length, as obtained from the density profiles in large systems. Advantages and disadvantages of this approach are discussed.

Furthermore, LLT and time-dependent simulations enable us to naturally study finite-size effects and observe a dimensional crossover from 1D to 2D as the width of the system is increased, as explored in section \ref{WidthDep}. Then, in section \ref{ME}, we use LLT to compute the mobility edge, and test this prediction using time-dependent simulations. Section \ref{Expansion} demonstrates that expanding wavepackets can also be used to probe Anderson localisation in a useful, quantitative manner, except for the fact that the energy dependence cannot be truly probed.

Next, in section \ref{BHM}, we use LLT to demonstrate a connection between the Anderson model for discrete disordered lattices and continuous 2D disordered systems, which provides powerful new insights. Crucially, we complement our study by contrasting systems with randomly-positioned scatterers to ones with a regular lattice in section \ref{Ordered}. This allows to isolate the effect of disorder and provides a means of testing whether the observed effects arise from Anderson localisation or other mechanisms. Finally, in section \ref{Secondary}, we consider the effect of various secondary features that would be present in a realistic experiment. We study the effects of acceleration and interparticle interactions in some detail, both of which are believed to be detrimental to Anderson localisation, carrying out definitive tests and obtaining novel understanding.

Conclusions are presented in section \ref{Conc} and several ideas are discussed as directions for a potential forthcoming investigation. Four appendices give technical details that enable interested parties to fully reproduce our work. These focus on exact diagonalisation (appendix \ref{appDiag}), an implementation of known LLT (appendix \ref{appLLTold}), details on the numerical solution of time-dependent partial differential equations (PDEs) used in the main text (appendix \ref{apptdep}), and finally, the new LLT ``technology'' developed in our work here (appendix \ref{appLLTnew}).
\section{System of interest}
\label{System}
For the purpose of this article, we restrict our investigation to 2D; primarily, this is because our work was inspired by the experiment \cite{BS}, concerning Anderson localisation is 2D. Performing an analogous study in 1D would be absolutely straight-forward as the computational cost decreases significantly and all the numerical procedures, including LLT machinery, are simplified considerably. Conceptually, it is clear that 3D could be treated by an extension of our work here, but in practice, the computational cost will increase and the complexity of the LLT methodology will grow as well. So far, LLT has been used in 3D in a limited capacity (only to compute the localisation landscape and the density of states from the effective potential; see section \ref{LLTold}) -- a full development is a matter for a future endeavour.

Thus, consider a (non-interacting) particle of mass $m$ confined to a 2D plane, whose motion is restricted to a rectangular region defined by $x\in[0,L]$ and $y\in[0,W]$. At the boundaries of this rectangular region, we impose Dirichlet boundary conditions, requiring the wavefunction to vanish. The particle moves in an external potential $V(x,y)$, so that the Hamiltonian is simply
\begin{equation}
\label{Ham}
H = -\frac{\hbar^2}{2m}\nabla^2 + V(x,y).
\end{equation}
Because we are interested in studying Anderson localisation, the potential $V(x,y)$ is taken as a sum of $N_s$ randomly-placed Gaussian peaks of the form
\begin{equation}
V_0 \exp\left\{-\frac{(x-x_0)^2 + (y-y_0)^2}{2\sigma^2}\right\},
\end{equation}
constituting what is known as ``point-like'' disorder, chosen for its lower percolation threshold \cite{deMarco2015}.

This system could be experimentally realised with cold atoms as in \cite{BS}, where an attractive 2D trap is used to contain atoms in a planar geometry, a repulsive custom potential generated by a spatial light modulator (SLM) allows the atoms to be confined to, for example, a rectangular box, and Gaussian point-like scatterers are generated by imaging squares of light produced by the SLM.

Next, we must introduce a set of dimensionless units, to be used throughout the paper. Let $\ell$ be a typical physical length scale relevant for the problem (for example, $\ell\sim\sigma$). Lengths will be measured in units of $\ell$, energy is units of $E_0 = \hbar^2/(2m\ell^2)$, and time in $t_0=\hbar/E_0$. Typically, for a cold-atom experiment such as \cite{BS}, $\ell\sim1\ \mu$m, $E_0\sim1$ nK $\times k_B$, and $t_0\sim5$ ms. 

Note that the coordinates $(x_0,y_0)$ of the Gaussian scatterers are drawn from a uniform distribution of \textit{half-integers} between $[0,L/\ell]$ and $[0,W/\ell]$, respectively. In all the simulations to follow, $L/\ell$ and $W/\ell$ are further chosen as integers. This restriction is imposed to stay in line with the discrete nature of the pixelated SLM used in \cite{BS} to both set the geometry and produce the scatterers. In the case of this experiment, one could reasonably choose $\ell$ to be the length of the side of the squares imaged on the SLM to produce the disorder.

The density of the scatterers is a more meaningful quantity to quote than their number, especially when one wishes to examine the effect of system size. Therefore, we define a dimensionless density, referred to as the fill factor, $f$, as
\begin{equation}
f = \frac{N_s\ell^2}{LW}.
\end{equation}

In later sections, we will wish to examine time evolution. Let us introduce a transmissive scenario, to be studied in more detail in section \ref{Trans}. First we have to slightly modify the geometry of the system we are examining. The region occupied by the potential scatterers remains precisely the same, $x\in[0,L],\ y\in[0,W]$, but we add empty ``reservoirs'' on either side of the disorder where the potential is zero. These occupy $x\in[-R,0],\ y\in[0,W]$ (first reservoir, $R_1$) and $x\in[L,L+R],\ y\in[0,W]$ (second reservoir, $R_2$). Usually, we choose $R=30\ell$, just large enough to contain the initial condition that will be used. In the transmissive scenario, a wavefunction with centre of mass (CoM) translation starts out in $R_1$ and goes through the disorder, finally arriving in $R_2$.

The most common initial condition we will use in this set up is a 1D Gaussian wavepacket\footnote{The use of similar probing waves was independently suggested by \cite{Piraud2012} and used in the experiment \cite{Berthet}.} (Gaussian along $x$ and uniform along $y$), which is fairly wide in position space and therefore has a rather localised energy distribution. The functional form is simply
\begin{equation}
\label{1DG}
\psi = \exp(i k_0 x)\exp\left[-\frac{(x+R/2)^2}{4\bar{\sigma}^2}\right],
\end{equation}
where we leave out the normalisation constant. In this case, we have initialised the 1D Gaussian at the centre of $R_1$, but by changing the shift of $x$, we can place it in other locations as well. Typical parameters would be $\bar{\sigma}=5\ell$, $k_0=1/\ell$, so that the momentum distribution is quite localised and the mean energy is $E\approx 1.17E_0$.
\section{Exact diagonalisation}
\label{Diag}
We begin our investigation by directly diagonalising (\ref{Ham}) and inspecting the eigenstates and energies, with the goals of (a) gaining intuition for our system and (b) checking whether useful quantitative predictions may be readily obtained in this framework. Details on the numerical implementation are given in appendix \ref{appDiag}.

\begin{figure}[htbp]
\includegraphics[width=6in]{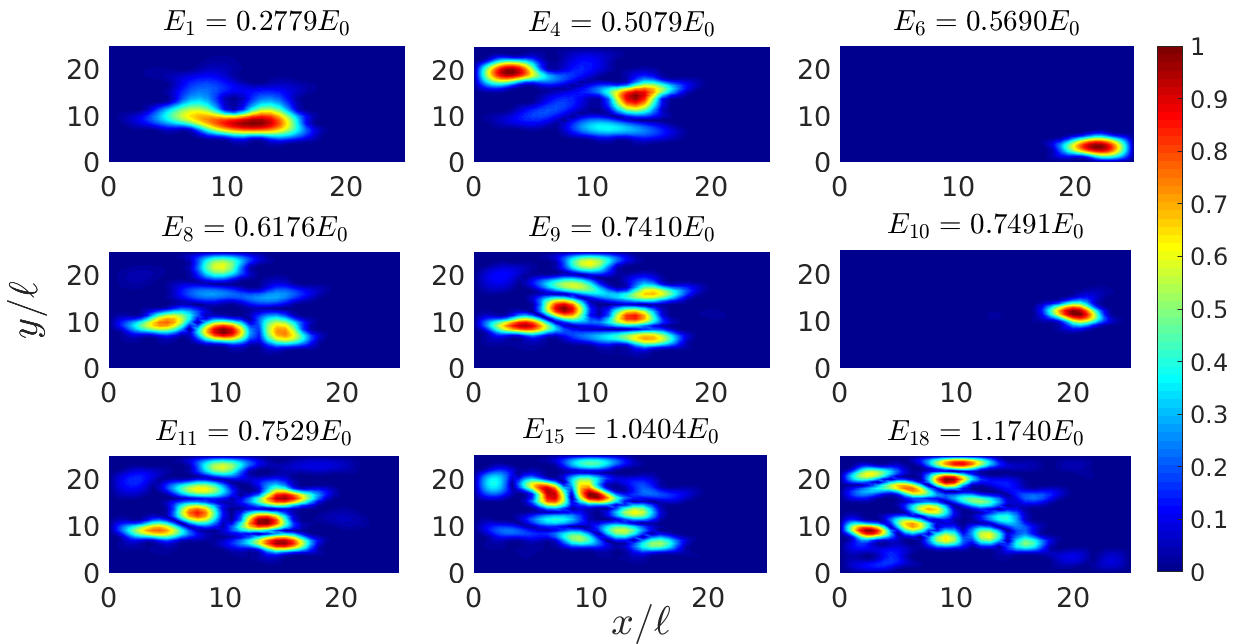}
\caption{\label{estEdep} Nine low-energy eigenstates of the Hamiltonian (\ref{Ham}) for a given noise realisation with $L=W=25\ell$, $f=0.1$, $V_0=20E_0$, $\sigma=\ell/2$, showing the absolute value of the eigenstates as a colour-map. Note that all eigenstates are normalised such that the maximum is one so that the values can be read on the same colour bar. We see that overall, the spatial extent of the eigenstates increases with energy, quoted above each panel. However, occasionally, very localised states are encountered at higher energies, on account of the considerable kinetic energy such eigenstates carry.}
\end{figure}

\begin{figure}[htbp]
\includegraphics[width=6in]{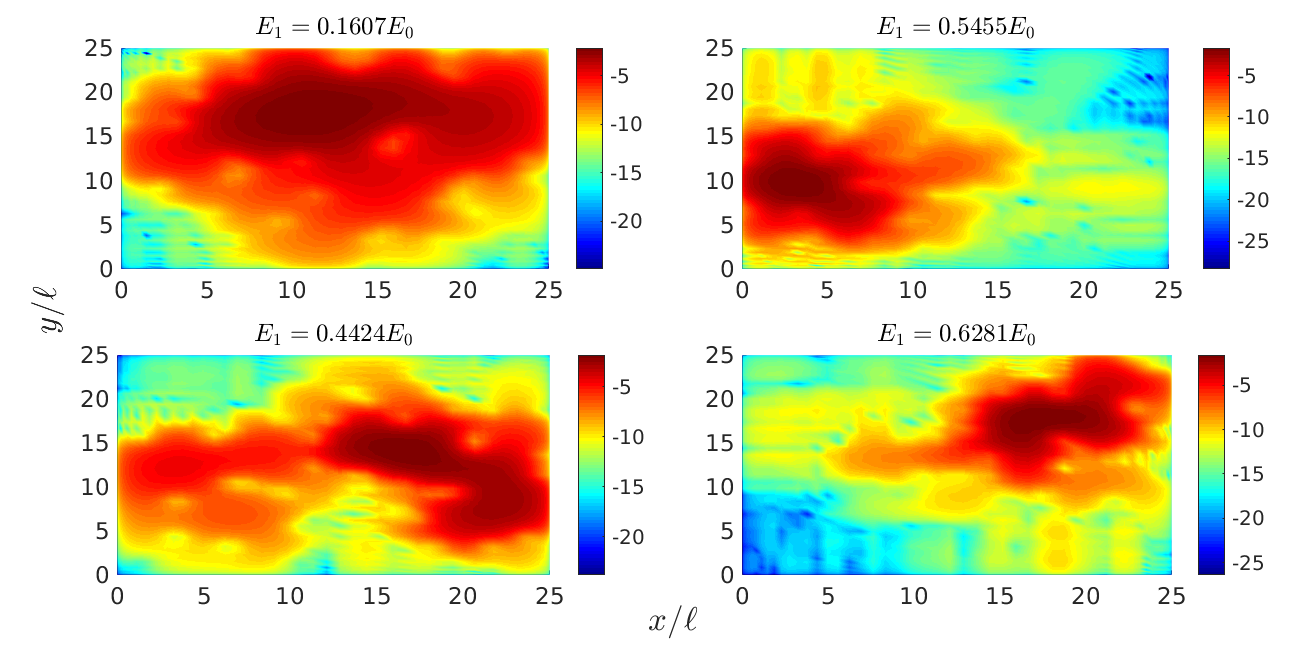}
\caption{\label{est_ffg} The lowest eigenstate of the Hamiltonian (\ref{Ham}) for some noise realisations with $L=W=25\ell$ and $\sigma=\ell/2$, showing the logarithm of the absolute value of the eigenstates as a colour-map. Top left: $f=0.1$, $V_0=10E_0$, top right: $f=0.2$, $V_0=10E_0$, bottom left: $f=0.1$, $V_0=20E_0$, bottom right: $f=0.2$, $V_0=20E_0$. We observe that the degree of localisation is controlled both by the density of the scatterers and their height. The energy eigenvalue is quoted above each panel: it increases as the area of the (node-free) localised mode decreases. Note that the horizontal and vertical stripes seen most prominently in the right-hand-side panels are an artefact of the diagonalisation algorithm (at the given spatial resolution used) and are non-physical.}
\end{figure}

As expected, the localised eigenstates lie at low energies, and the degree of localisation decreases as the energy increases. This can be easily seen by eye when inspecting the eigenstates, plotting their amplitude, $\left|\psi\right|$. An example is shown in Fig.~\ref{estEdep}, depicting nine low-energy eigenstates for a particular noise realisation. Overall, as energy increases, the weight of the eigenstates spreads out over a larger area (see Fig.~3 of \cite{Marcel2012} for another example). This process, however, is not monotonic: occasionally we encounter very localised states with a fairly high energy, where most of the energy comes from the rapidly changing wavefunction rather than the spatial extent and the associated potential energy. Also quite intuitively, if $f$ or $V_0$ are increased, the strength of localisation increases and the area within which the weight of the eigenstates is contained shrinks. Figure \ref{est_ffg} demonstrates this by visually comparing the lowest energy eigenvector for different combinations of $f$ and $V_0$ (different noise realisations are used for each panel). We see that both the fill factor and the scatterer height are equally important parameters, influencing localisation properties just as strongly.

Increasing the width of the scatterers $\sigma$ also leads to stronger localisation (not illustrated), because the area occupied by the Gaussian peaks increases, but the dependence on the scatterer width is not methodically explored here. Note, however, that when the width of the scatterers becomes sufficiently large, there is a decrease in the randomness of the potential as we approach the limit where the entire system is covered by overlapping potential bumps (the same of course happens as the fill factor is increased strongly). Once this regime is reached, localisation weakens with further increases of the fill-factor and the scatterer width.

The shape of the scatterers also naturally plays a role, but as long as the (``volume'') integral over a single scatterer is kept constant, the specific functional form is expected to have a much weaker effect on the physics than $f$ and $V_0$. The shape of the scatterers influences the spectral properties of the disordered potential, the relation of which to a (possible) mobility edge\footnote{The mobility edge is a cut-off energy as a function of disorder strength below which eigenstates are localised, and above which they are delocalised.} could be investigated in the future.
%

Next, let us consider how the localisation length may be extracted from the exact eigenstates of (\ref{Ham}). By definition, the localisation length is the length scale on which the localised states decay exponentially, far away from the region where their main weight is concentrated. This decay can be seen in Fig.~\ref{est_ffg} as a change of colour from dark red to red to orange to yellow to green to blue, as the wavefunction gradually drops by orders of magnitude. The localisation length increases with energy, depends on the strength of the disorder, and should only be discussed in a configuration-averaged context.

If we inspect any one given eigenstate, assuming the energy is sufficiently low or localisation is strong enough, there is usually only one peak -- one local maximum -- in $\left|\psi\right|$. If we temporarily place our origin there and vary the azimuthal angle $\theta$, then the curve $\left|\psi(r)\right|$ along different directions will certainly be different depending on $\theta$. Still, we could average these curves over $\theta$, and attempt fitting an exponential function to the tail of the resultant. If the peak is located in a corner of our rectangular system, for example, the average should only be taken over those angles along which one has reasonable extent along $r$.

However, as energy increases (or localisation decreases due to changes in parameters), the eigenstates develop a multi-peak structure: there are several ``bumps'' (see Fig.~\ref{estEdep}), and it is not clear where to place our origin. Furthermore, the energy eigenvalues are of course quantised, so any extracted localisation lengths from single-peak eigenstates need to be averaged over noise realisations, only using eigenstates of roughly the same energy (binning within a reasonable range). This makes such an approach very limited.

Now, a very common solution to this problem -- heavily used in the literature (e.g.~\cite{deVries, Inguscio2008, Inguscio2015, Aspect2005, Aspect2010, Aspect2012, Donsa}) -- is to compute the spatial variance of the localised states instead. Since we are working in 2D, we could tentatively examine the quantity
\begin{equation}
\left[\Delta x^2 \Delta y^2 \right]^{1/4},
\end{equation}
where the variance along $x$ is
\begin{equation}
\label{varR}
\Delta x^2 = \left\langle x^2 \right\rangle - \left\langle x \right\rangle^2 = \int\limits_0^L\ dx \int\limits_0^W\ dy\ x^2 \left|\psi\right|^2 - 
\left[ \int\limits_0^L\ dx \int\limits_0^W\ dy\ x \left|\psi\right|^2 \right]^2,
\end{equation}
assuming the wavefunction is normalised to one, and $\Delta y^2$ is defined similarly.

\begin{figure}[htbp]
{\includegraphics[width=3.1in]{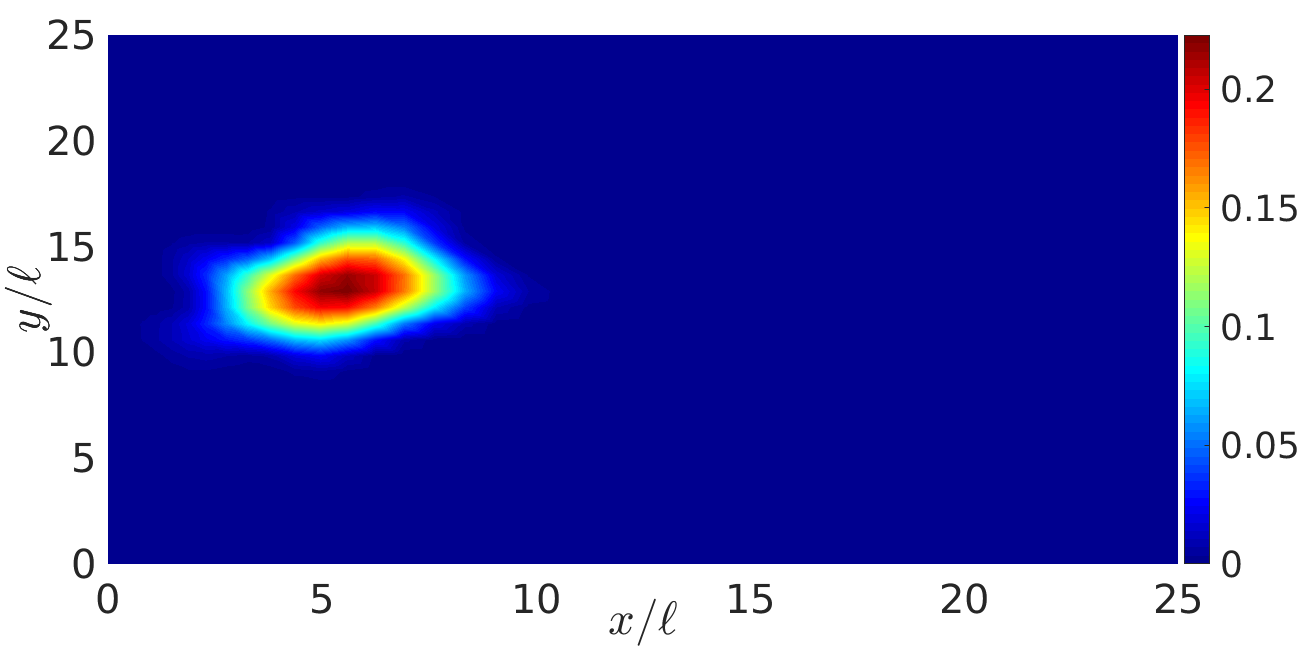}}
{\includegraphics[width=3.1in]{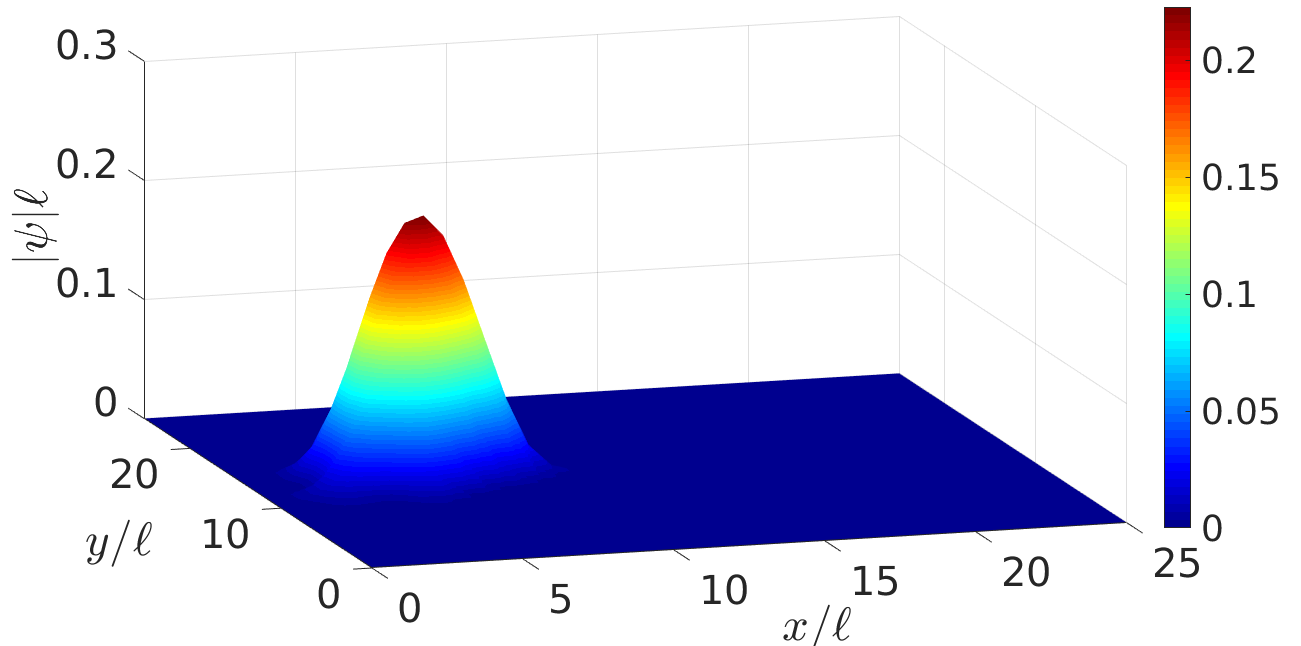}}
\caption{\label{eg_bump} The lowest eigenstate of the Hamiltonian (\ref{Ham}) for a given noise realisation with $L=W=25\ell$, $f=0.2$, $V_0=20E_0$, $\sigma=\ell/2$, plotting $\ell\left|\psi\right|$ as a colour-map. The exponential decay away from the main region of existence of the eigenstate is unresolvable on a linear scale.}
\end{figure}

Figure \ref{eg_bump} shows a typical low energy eigenstate, plotting $\left|\psi\right|$ on a linear scale. The small-amplitude yet large-scale structure seen on the logarithmic plots of Fig.~\ref{est_ffg}, capturing the exponential decay of the eigenstates away from their main region of existence, is completely invisible on such a plot. When there is a single main ``bump'' in the eigenstates, the variance-based length scale of (\ref{varR}) reports mostly on the width of the main peak (seen in Fig.~\ref{eg_bump}, for example) -- analogous to the full-width-at-half-maximum or the standard deviation of a Gaussian peak. It measures the size of the main bump, and carries only indirect information on the exponential decay in the tails. In cases when there are smaller, secondary bumps in the eigenstates, their presence increases the variance, even if their width and decay rate are identical to those of the main bump. Therefore, the variance does not report on the localisation length, as such. We thus advise caution when using the variance to quantify localisation properties, a common practice in the literature.
\section{LLT to date}
\label{LLTold}
A powerful new theory has recently been pioneered by Marcel Filoche and Svitlana Mayboroda \cite{Marcel2012}: LLT is a purely linear theory which describes localisation effects, whether due to Anderson localisation or other factors. It carries the information contained in the Hamiltonian, its eigenstates, and its spectrum in a different, more accessible form. In particular, LLT yields intuitive and transparent conceptual insights, as well as providing a framework where quantitative calculations can be performed directly and simply. In this section, we provide an overview of the main results and arguments of LLT known so far \cite{Marcel2012, FnM2014, FnM2016, FnM2016b, part1, part2, part3}. Technical details regarding our numerical implementation can be found in appendix \ref{appLLTold}.

\begin{figure}[htbp]
{\includegraphics[width=3.1in]{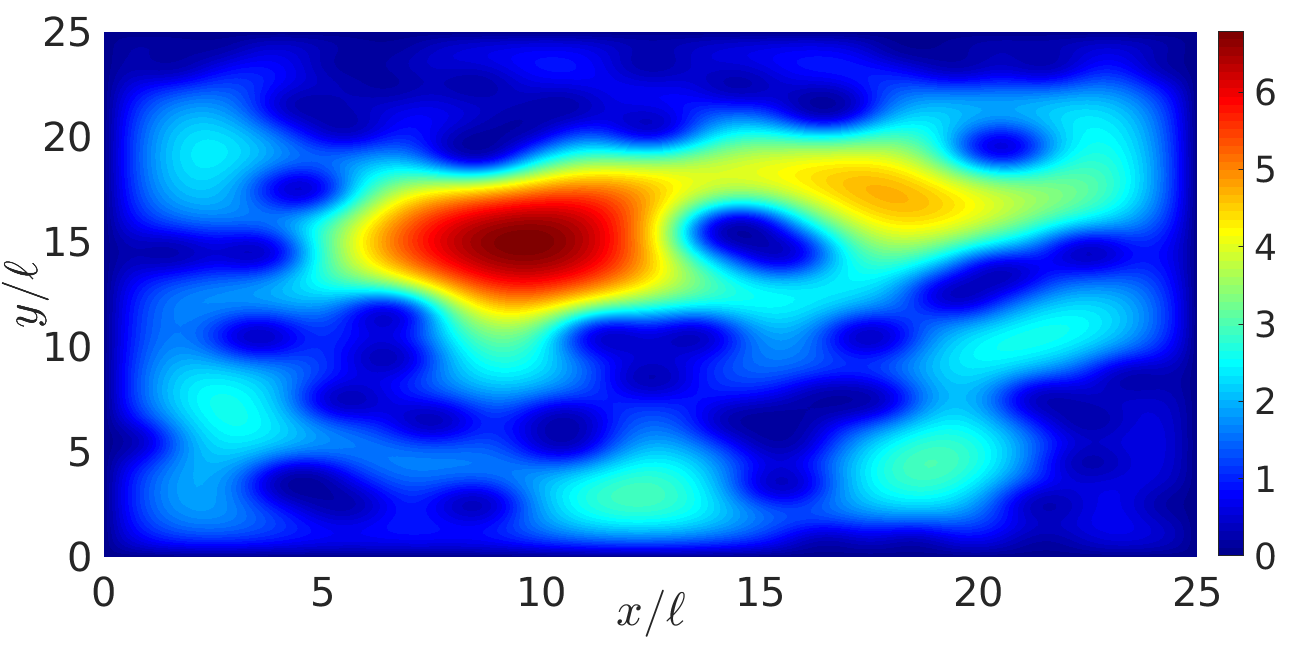}}
{\includegraphics[width=3.1in]{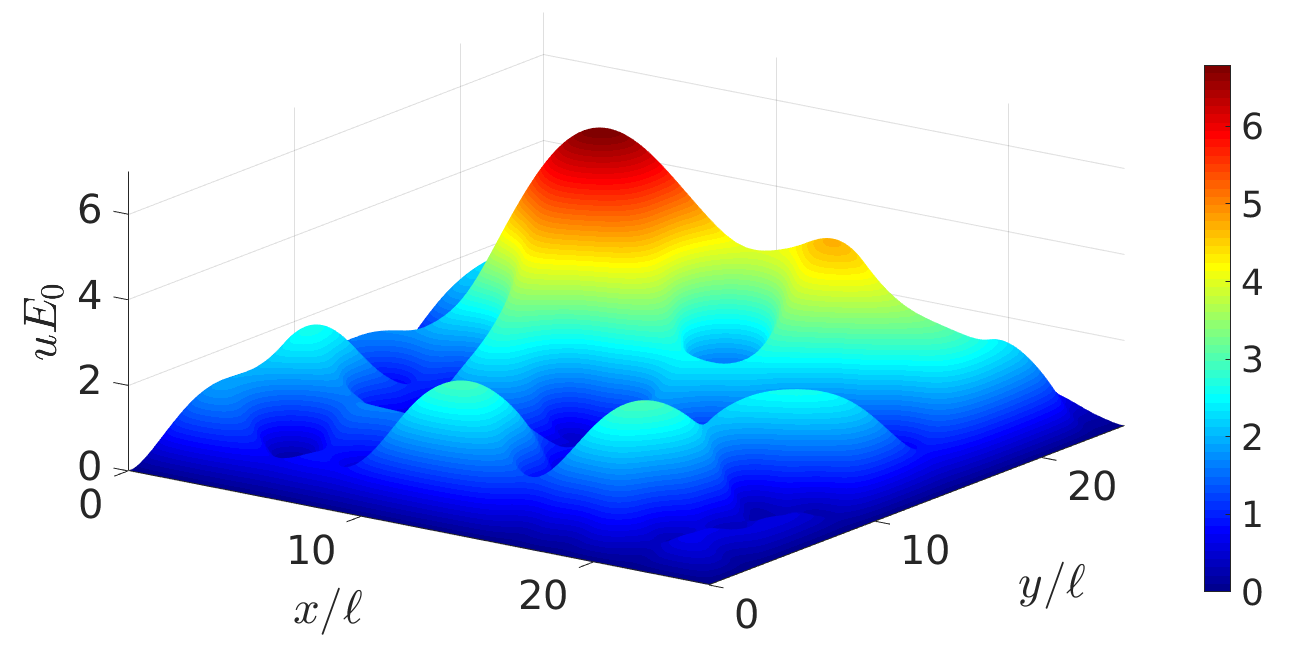}}
\caption{\label{Ueg} The localisation landscape $u$ for a given noise realisation with $L=W=25\ell$, $f=0.1$, $V_0=10E_0$, $\sigma=\ell/2$, viewed from the top and from the side.}
\end{figure}

\begin{figure}[htbp]
{\includegraphics[width=3.1in]{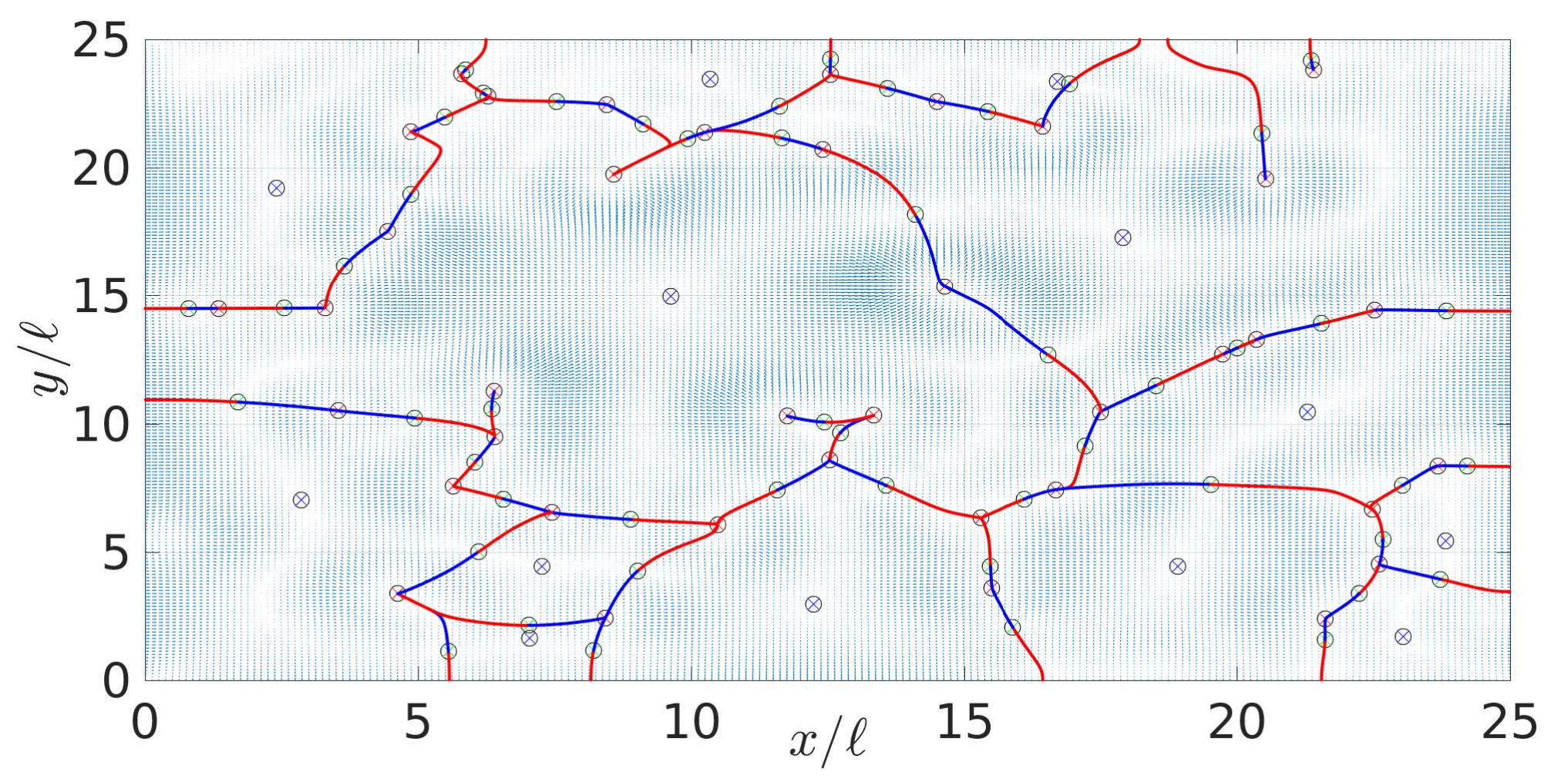}}
{\includegraphics[width=3.1in]{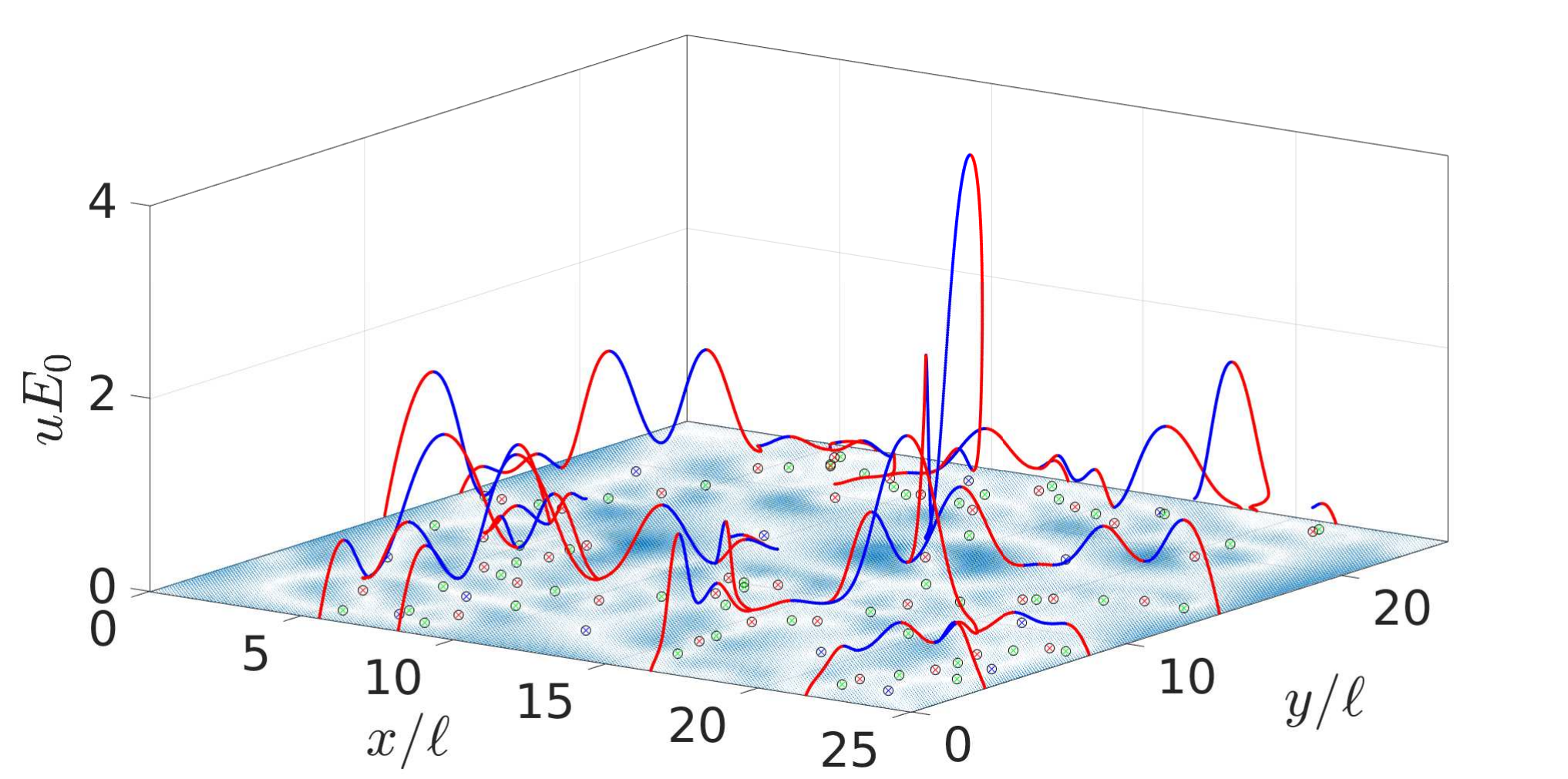}}
\caption{\label{Neteg} The valley network for the localisation landscape shown in Fig.~\ref{Ueg}. The blue and red lines show the valley lines; one blue and one red valley line emanates from each saddle point (the different colours are used to make the structure of the valley network clearer). The extrema of the landscape are also shown as symbols (maxima in blue, minima in red, saddles in green). The valley lines connect saddle points to minima, and each closed domain contains a maximum. The ``velocity field'' of $u$ is depicted by the small blue arrows (see appendix \ref{appLLTold} for the significance of the extrema and velocity field to the network construction process). The right panel shows a rotated view, highlighting the importance of the value of $u$ on the the valley lines: as energy increases, the network is cut down from the top, moving down, according to (\ref{effNet}).}
\end{figure}

The central object of LLT is the localisation landscape $u$, defined by
\begin{equation}
\label{uPDE}
Hu=1,
\end{equation}
where $H$ is the Hamiltonian and $u$ is required to vanish on the boundary of the system. It is simple to prove that $u$ is a real and positive function as long as $V\geq0$ everywhere. In 2D, it is a surface, and a typical example is shown in Fig.~\ref{Ueg}. We notice that the surface is ``pitted'': it has many local maxima and minima and a complicated shape, with features on an intermediate length scale between the system size, and the size and spacing of the random scatterers.
The significance of the localisation landscape arises from the inequality
\begin{equation}
\label{mainman}
\left|\psi(\mathbf{x})\right|\leq E u(\mathbf{x}),
\end{equation}
where $\mathbf{x}$ is the position vector (keeping the system dimensionality general), $\psi$ is an eigenstate of $H$ with eigenvalue $E$, normalised (without loss of generality) such that
\begin{equation}
\label{est_norm}
\max \left|\psi(\mathbf{x})\right| = 1.
\end{equation}
Since the function $u$ constrains the eigenstates from above (the effect of the energy will be discussed later), it is sensible that the valleys of this landscape should play an important role in confining the eigenstates: at its valleys, $u$ is small and the eigenstates are forced down. Thus, the so-called ``valley network'' is a collection of all the valley lines (anti-watersheds) of $u$ (see appendix \ref{appLLTold} for how the valley lines are formally defined). Note that since we are interested in a 2D system, the valleys are indeed lines: in 1D, they are points and in 3D, surfaces. The valley network divides the entire system into a collection of ``domains'', separated by valley lines, completed by the boundaries of the system itself. However, not all valley lines must necessarily form closed structures: when localisation is fairly weak, it is very common to have ``open'' valley lines that extend into the interior of some closed domain without constituting part of a domain wall themselves. An example valley network is presented in Fig.~\ref{Neteg}. The value of $u$ on the valley lines is extremely important and is discussed below. At this point, we simply remark that if the valley lines are plotted as trajectories in $x-y-u$ space, they appear as a collection of ``bridges'' (see Fig.~\ref{Neteg}), with the top of each bridge being a saddle point and the low ends located at minima of $u$.

\begin{figure}[htbp]
\includegraphics[width=6in]{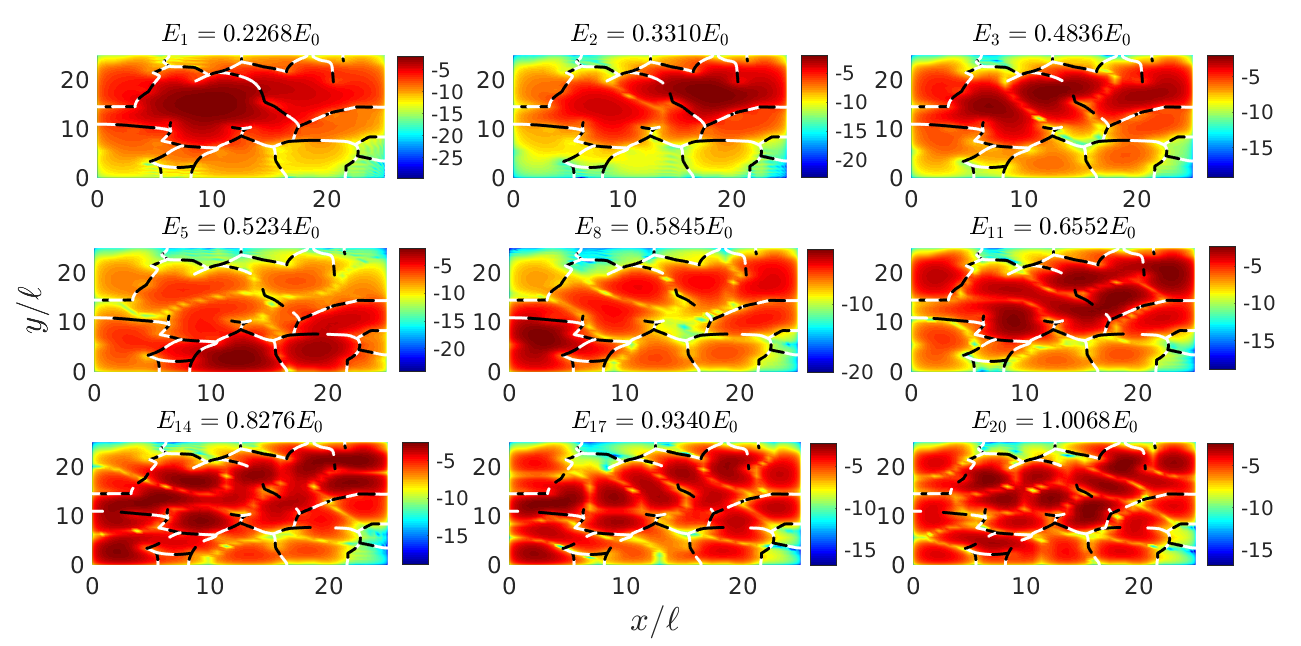}
\caption{\label{estNet} Some of the low energy eigenstates for the same noise realisation as used in Fig.~\ref{Ueg}, showing the logarithm of the absolute value of the eigenstates as a colour-map and the \text{effective} valley network as white and black lines (the different colours are used to make the structure of the valley network clearer). The first panel demonstrates the localisation of an eigenstate to a single domain of the network, with any occupation in the other domains resulting from decay across the valley lines from this single main peak. As energy increases, the network shrinks (compare the first and last panels), allowing the eigenstates to spread out over a larger area. In this example, all shown eigenstates apart from the first have independent occupation of several domains. In many cases, we see several distinct peaks inside a domain: this is an excitation of a mode higher than the fundamental one on that domain. An order of magnitude drop in the amplitude of the eigenstates can be seen when a peak decays across a valley line, as for example in the first panel.}
\end{figure}

In fact, the domains defined by the valley network are of key importance: the eigenstates of the Hamiltonian are localised such that their main weight lies precisely within these domains. Superimposing the valley network on top of several of the lower energy eigenstates clearly demonstrates this (see Fig.~\ref{estNet}). Recall that the governing inequality (\ref{mainman}) depends on energy. Combining this with the normalisation of the eigenstates (\ref{est_norm}), we see that the valley lines only effectively constrain the eigenstates where
\begin{equation}
\label{effNet}
u < 1/E.
\end{equation}
In other words, depending on the energy of the eigenstate, part of the valley network needs to be dropped. As the energy increases, the ``bridges'' formed by the valley lines in $x-y-u$ space are cut down from the top -- i.e.~``breaks'' in the domain walls start from the saddle points and grow as energy goes up. This allows the higher energy eigenstates to extend to neighbouring domains, leaking out through the openings in the domain walls. A detailed discussion of this phenomenon at higher energies is given in section \ref{HigherEs}. Several eigenstates are shown in Fig.~\ref{estNet}, with the effective network superimposed, demonstrating the eigenstates extending to occupy larger areas with increasing energy. In fact, this mechanism is one of the factors behind the dependence of the localisation length on energy.

Even when there are no breaks in the domain walls, however, $u$ is practically never zero on its valleys (merely small), which means that the eigenstates can leak out into neighbouring domains, but the amplitude drops by an order of magnitude in the process. In contrast, within a domain, the eigenstate amplitude remains a single order of magnitude. Both statements can be confirmed by noticing the change in colour in Fig.~\ref{estNet} as one crosses a domain wall, while within a domain, one colour dominates.

Now, let us discuss the importance of disorder for this picture. A prudent question is why the valley lines contain the eigenstates, rather than just forcing $\psi$ to be small at the valleys and allowing it to spill out into adjacent domains with a large amplitude. As the authors of \cite{Marcel2012} show, an eigenvector $\psi$ can be non-zero in a given domain if the corresponding energy eigenvalue $E$ is close to the energy of a mode of the original eigenvalue problem (with Dirichlet boundary conditions) restricted to that domain. If the global mode energy $E$ does not match a local mode energy, the (global) eigenstate weight is expelled from this domain. From this, it follows that the shapes of the global mode and the local domain mode must match quite closely. Furthermore, a global eigenmode can extend across neighbouring domains if the valley network has shrank sufficiently (due to a high energy value) to allow this ``spillage'', or if the two domains have nearly matching eigen-energies. This is why a noisy potential creates localisation: it mismatches the eigenspectra of domains. More will be said about this in sections \ref{BHM} and \ref{Ordered}.

\begin{figure}[htbp]
{\includegraphics[width=3.1in]{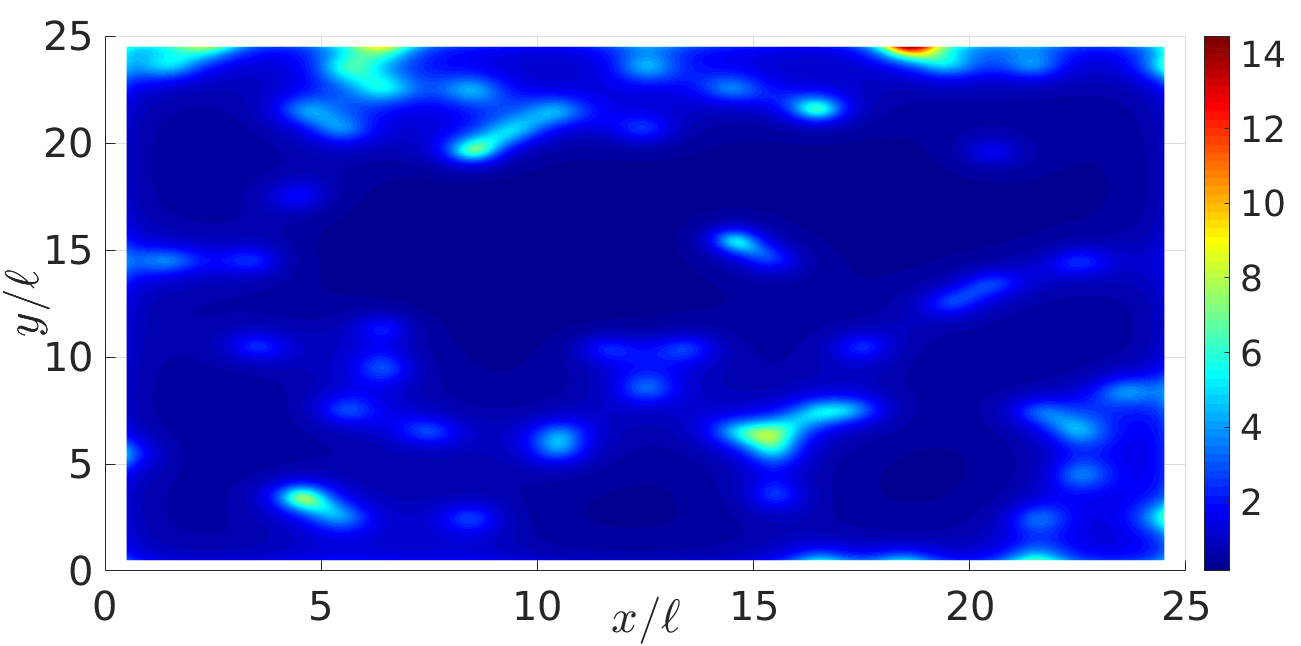}}
{\includegraphics[width=3.1in]{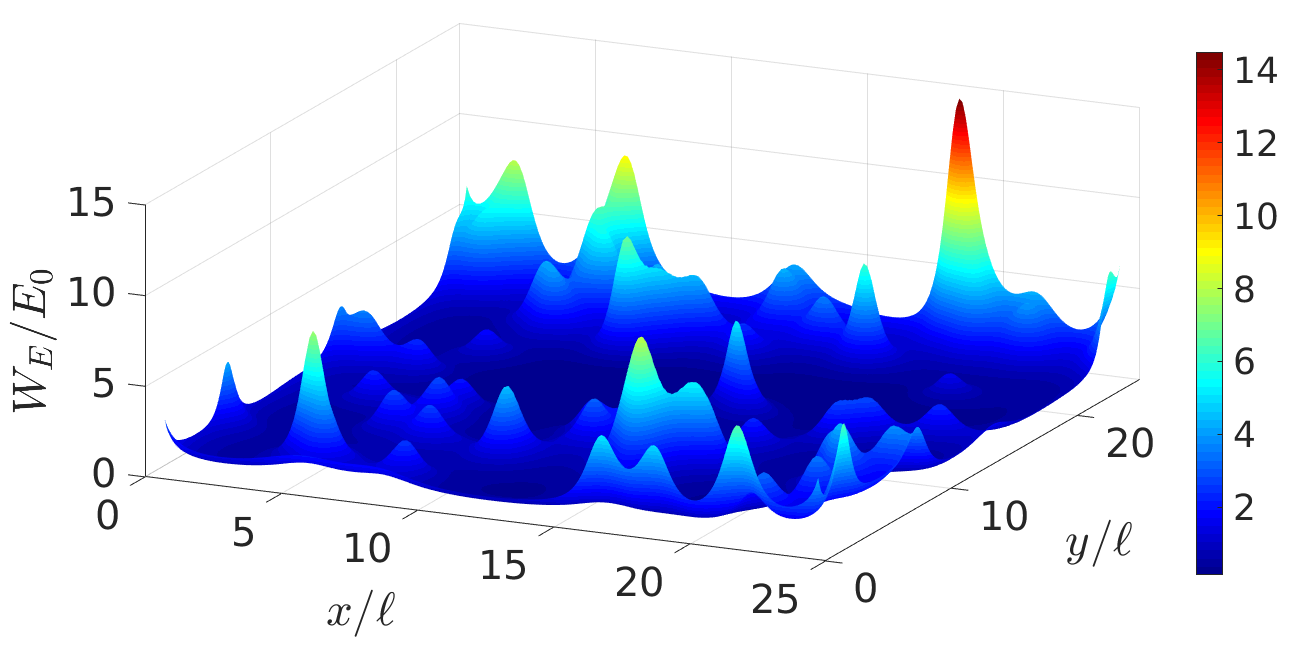}}
{\includegraphics[width=3.1in]{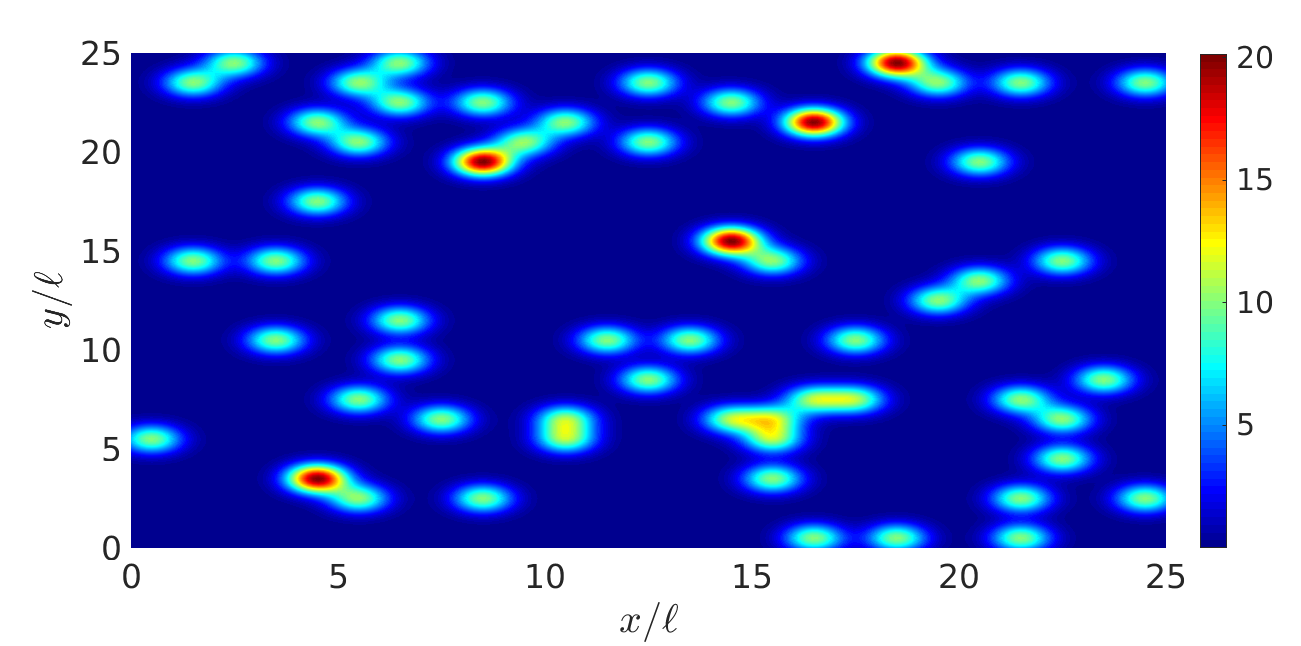}}
{\includegraphics[width=3.1in]{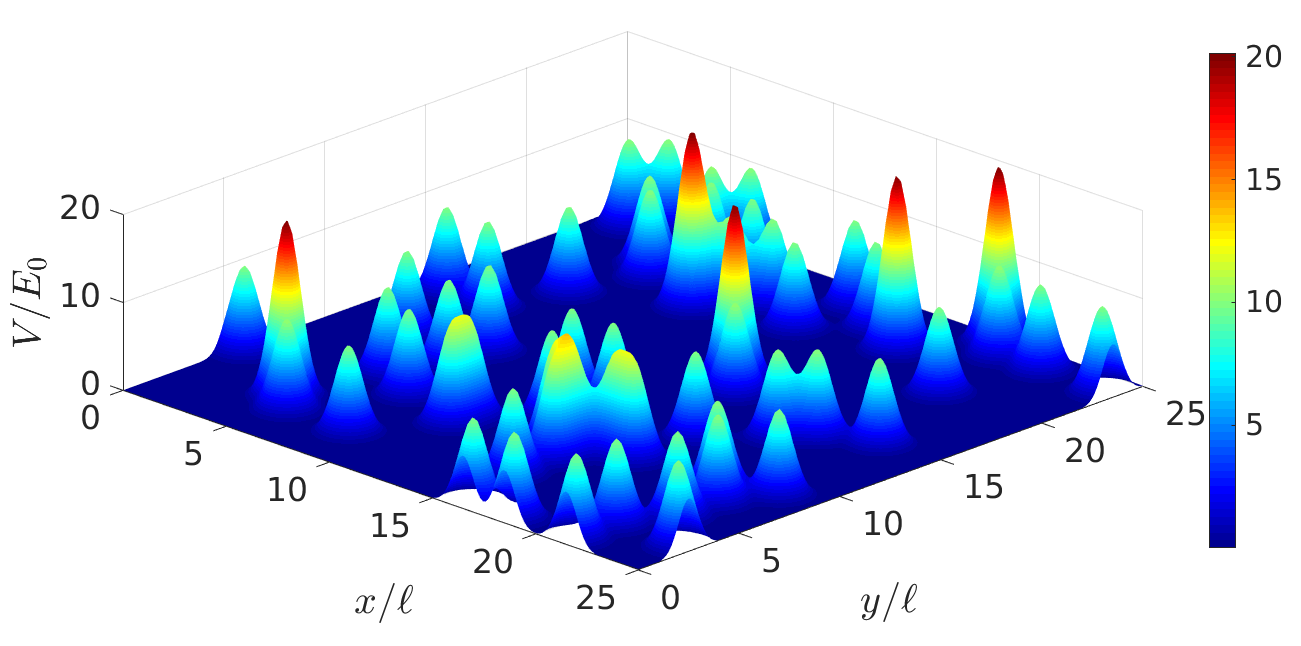}}
\caption{\label{Weg} The effective potential $W_E$ (top panels) and the physical potential $V$ (bottom panels) for the localisation landscape shown in Fig.~\ref{Ueg}, viewed from the top and from the side. The peak ranges of $W_E$ correspond to the valley lines of $u$ and govern both the localisation regions of the eigenstates and their decay outside of their main domains of existence. While the potential barriers of $W_E$ are located largely at the positions of the scatterers in $V$, $W_E$ can be thought of as a smoothed out version of $V$, so that the clear gaps between the scatterers in $V$ are annealed, leading to the formation of proper local wells that can support classical trapping. The smoothing operation also has the effect of creating lower peaks in $W_E$ compared to $V$, which is relatively more significant for weak disorder, at the expense of creating an almost-constant, non-zero background value to the potential away from the peaks. In addition, $W_E$ inherits the random nature of $V$, and is capable of supporting Anderson localisation.}
\end{figure}

The next major step forward in LLT came in Ref.~\cite{FnM2016b}, where it was realised that
\begin{equation}
W_E = 1/u
\end{equation}
can be thought of as an effective potential for our problem, to some degree capable of replacing $V$ at very low energies (see below and section \ref{Wmeaning}). The essence of the effective potential is that quantum interference effects in $V$ are translated to ordinary quantum tunnelling in $W_E$, which is much more familiar and easier to work with. The valleys of $u$ are the peak ranges of $W_E$, and it is not surprising that to cross them, the wavefunction must tunnel through the barriers and thus decays by an order of magnitude. An example of $W_E$ is shown in Fig.~\ref{Weg}, with ``mountain ranges'' (where $u$ has valleys) being the most prominent feature. One caveat of using $W_E$, however, is that since $u$ has Dirichlet boundary conditions, $W_E$ diverges on the edges of the system. This is not reporting on Anderson localisation (as opposed to the peaks of $W_E$ in the interior of the global domain), and as such, we must avoid including the section of $W_E$ in the immediate proximity of the system edges in any numerical calculations.

Now, one may wonder whether the low-energy, localised states seen in Figs.~\ref{estEdep} and \ref{est_ffg} are simply trapped in local minima of the potential $V$, formed by surrounding Gaussian scatterers. When examined, the effective potential $W_E$ resembles the physical potential $V$ quite closely, as demonstrated in Fig.~\ref{Weg}. Scatterer positions in $V$ largely coincide with peak positions in $W_E$, but the latter is a smoothed-out version of the former (on a length-scale which depends on $V(x,y)$), as discussed in \cite{FnM2016b}. In particular, while $V$ has clear gaps between scatterers (as long as the fill factor and scatterer width are not too great to cause significant scatterer overlap), $W_E$ has continuous potential ridges that encircle domains, allowing for classical trapping in these regions (this was also pointed out in \cite{FnM2016b}). Meantime, due to the smoothing, $W_E$ has lower peaks than $V$ (which is more noticeable at weaker disorder), and an almost constant, non-zero background value away from these peaks.

We note that since $W_E$ inherits so many of its features from $V$, it is also intrinsically a random potential, and will give rise to Anderson localisation (as was already realised in \cite{FnM2016b}). These quantum interference effects in $W_E$ will be similar to those in $V$ in as much as the two potentials are similar, but of course there will be differences in the localisation properties as well: for example, the lower peaks in $W_E$ would cause weaker localisation than one would have in $V$.

We now clarify in what sense $W_E$ is an effective potential for our system. If $\psi$ is an eigenstate of the Hamiltonian, the authors of \cite{FnM2016b} define $\phi=\psi/u$, and rewrite the eigenvalue problem for the Hamiltonian as
\begin{equation}
\label{auxWE}
-\frac{\hbar^2}{2m}\left[\frac{1}{u^2}\mathbf{\nabla}\cdot\left(u^2\mathbf{\nabla}\phi\right)\right] + W_E\phi = E\phi.
\end{equation}
This has a similar form to the stationary Schr\"{o}dinger equation, with $W_E$ replacing $V$ and a modified kinetic energy term. Two further potentially useful results are: for any state $\left|\psi\right\rangle$
\begin{equation}
\left\langle\psi\right|H\left|\psi\right\rangle = \frac{\hbar^2}{2m}\left\langle u \mathbf{\nabla}(\psi/u) \right|\left. u \mathbf{\nabla}(\psi/u) \right\rangle + \left\langle\psi\right|W_E\left|\psi\right\rangle,
\end{equation}
and
\begin{equation}
V-W_E = \frac{\hbar^2}{2m} \frac{\nabla^2u}{u}.
\end{equation}
Next, Ref.~\cite{FnM2016b} provides a simple and direct method of obtaining the number of states below a given energy $E$ -- the integrated density of states. Starting from Weyl's law, it is shown to be proportional to
\begin{equation}
\label{Weyl}
N\left(E\right)\propto \int\limits_{W_E(\mathbf{x})<E} \sqrt{E-W_E(\mathbf{x})}\ d\mathbf{x},
\end{equation}
with the proportionality constant dependent on the dimensionality of the system. This very simple formula reproduces the density of states very accurately \cite{FnM2016b,part1}. We will uncover other important aspects of the physical significance of $W_E$ in section \ref{Wmeaning}.

Another extraordinary feature of LLT is that it allows us to compute the fundamental eigen-mode and -energy of the Hamiltonian eigenvalue problem restricted to each domain of the valley network \cite{part1}. For the j$^{\mathrm{th}}$ domain, we have
\begin{eqnarray}
\label{local_mode}
\psi^{(j)} &=& \frac{u}{||u||},\\
\label{local_energy}
E_j &=& \frac{\langle 1|u\rangle}{||u||^2},
\end{eqnarray}
where $\langle 1|u\rangle$ and $||u||^2$ are the integrals of $u$ and $u^2$, respectively, over the area of the domain.

\begin{figure}[htbp]
{\includegraphics[width=6in]{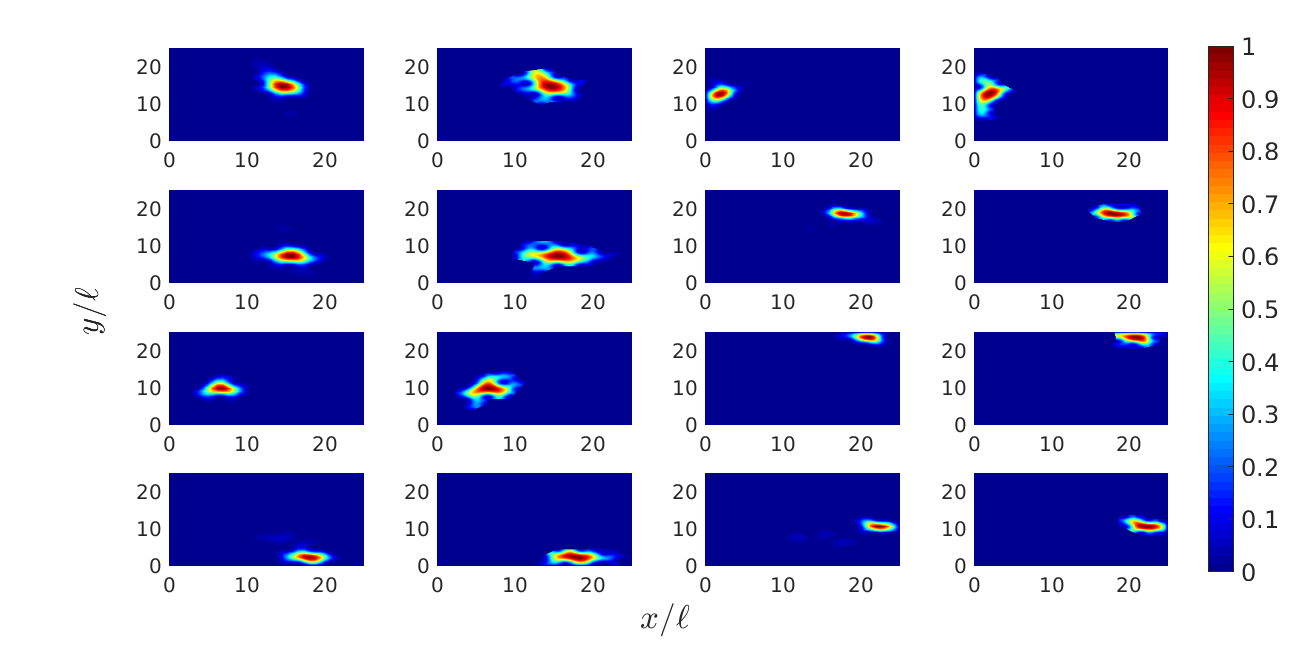}}
{\includegraphics[width=6in]{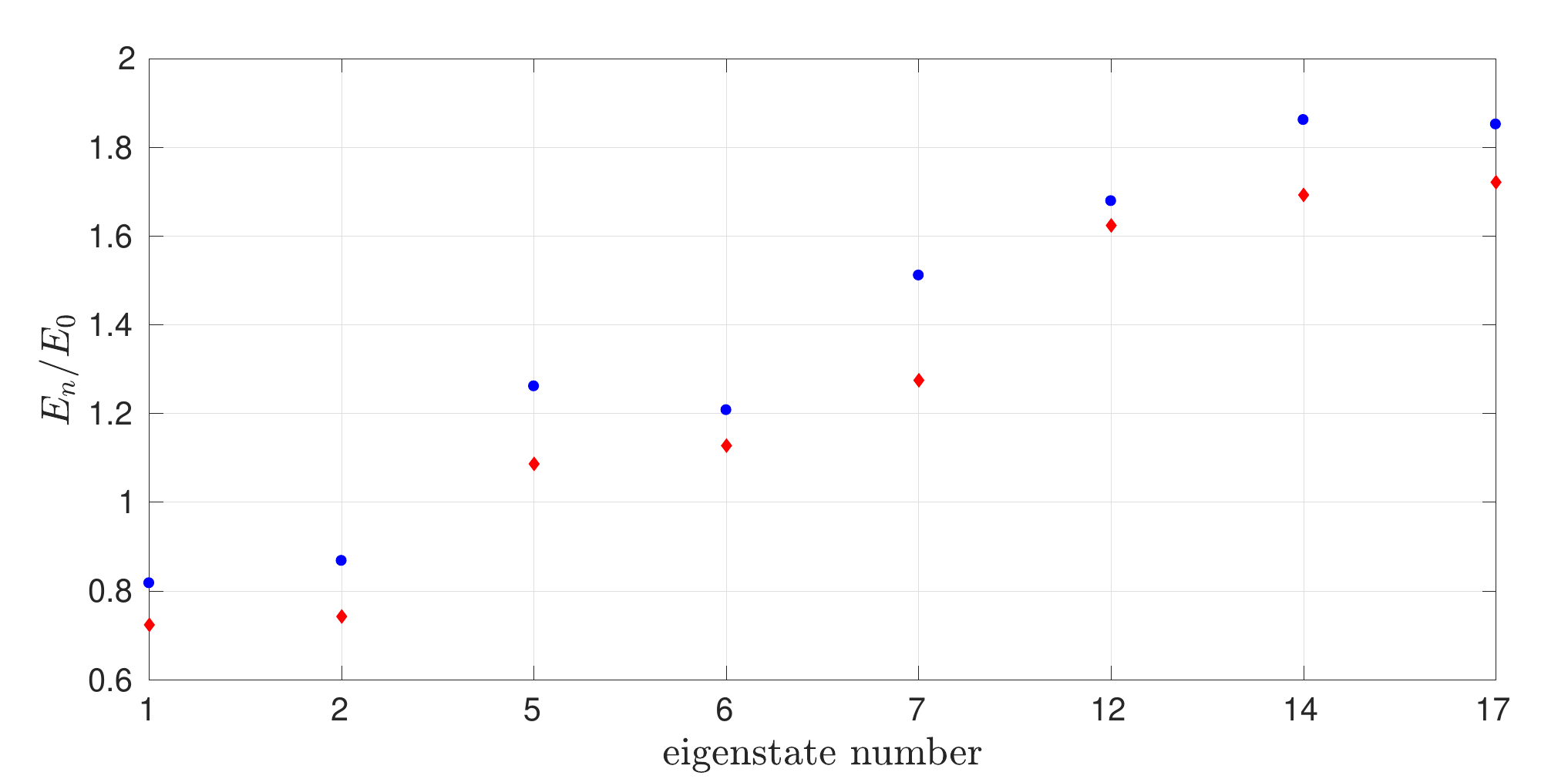}}
\caption{\label{TestLocal} Top: Eigenstate amplitude (normalised such that the maximum is one) for a given noise realisation with $L=W=25\ell$, $f=0.2$, $V_0=21.33E_0$, $\sigma=0.48\ell$. Columns 1 \& 3 show the exact eigenstates and 2 \& 4 the approximation from LLT, equation (\ref{local_mode}). Going down columns and then across, the shown eigenstates are numbers 1, 2, 5, 6, 7, 12, 14, 17 in the spectrum, ordered by increasing energy. Bottom: the corresponding eigen-energies, computed exactly (red diamonds) and calculated from LLT according to (\ref{local_energy}) (blue circles). It is clear that the local fundamental modes and energies are well approximated by LLT.}
\end{figure}

Moreover, as discussed above, the low-energy eigenmodes of the full Hamiltonian that only have strong occupation of a single domain with a single peak in the density are very similar to the fundamental local state on the relevant domain, and the eigen-energies are also in close agreement. This can be readily verified by direct comparison of the exact eigenstates and eigen-energies to the predictions of (\ref{local_mode}) and (\ref{local_energy}), as is done in Fig.~\ref{TestLocal}. The chief difference is that in the global eigenstates, some weight spills out into neighbouring domains. We can estimate the amplitude of the full eigenstates outside of the primary domain via the following method \cite{FnM2016b}. Define the energy-dependent quantity known as the Agmon distance:
\begin{equation}
\label{Agmon}
\rho_E(\mathbf{x_0},\mathbf{x}) = \min\limits_{\gamma}\left(\int\limits_{\gamma}\Re\sqrt{2m[W_E(\mathbf{x})-E]}/\hbar\ ds\right).
\end{equation}
Because only the real part of the square root is used, the integrand is zero if $E$ exceeds $W_E$ at position $\mathbf{x}$. The integral should be minimised over all possible paths $\gamma$ going from $\mathbf{x_0}$ to $\mathbf{x}$, and $ds$ is the differential arc length. If we have a local domain eigenstate peaked at position $\mathbf{x_0}$, then the full corresponding  eigenstate will have amplitude at position $\mathbf{x}$ outside of this main domain bounded by
\begin{equation}
\label{decay}
|\psi(\mathbf{x})|\lesssim |\psi(\mathbf{x_0})| \exp\left[-\rho_E(\mathbf{x_0},\mathbf{x})\right].
\end{equation}
In a way, this tells us how the wavefunction decays across the barriers of $W_E$, and constitutes another important aspect of its physical meaning. As the authors of \cite{FnM2016b} point out, the formula (\ref{Agmon}) is commonly encountered in the context of the Wentzel–Kramers–Brillouin (WKB) approximation in 1D (and higher dimensions), and constitutes a semiclassical approximation of multidimensional tunnelling. The inequality (\ref{decay}) assumes that the connection between the wavefunction at points $\mathbf{x_0}$ and $\mathbf{x}$ is quantum mechanical tunnelling through the potential barriers between them. In 1D, Ref.~\cite{FnM2016b} has shown that (\ref{decay}) can be used to predict the shape of the eigenstates very closely. Further discussion of this equation, physical insight, and practical computational considerations are given later in the article.
\subsection{Advantages of LLT}
Considering the fact that LLT reports on the information contained in the spectrum and eigenstates of the Hamiltonian, one might wonder if it actually presents any significant advantages over traditional methods such as exact diagonalisation and Schr\"{o}dinger evolution when it comes to describing Anderson localisation. For one, exact diagonalisation cannot be pushed to very large system sizes. The ``active area'' (filled with disordered scatterers) used in \cite{BS} was very large, and it is not simple to push the numerical algorithms to such extensive sizes. Parallelising such a problem is difficult and memory constraints are also an issue. Simulating time evolution (described later) suffers from the same limitations, with the additional problem that resolving high energy components requires a fine grid, which makes the computational cost scale up with system size \textit{and} energy. On the other hand, LLT relies on the one-off solution of a stationary PDE which can be done very efficiently even for extremely large systems (see appendix \ref{appLLTold}), and one immediately gets information about the behaviour of all energy components (at least in principle) through the effective potential $W_E$. Another key strength of LLT is the ability to learn about finite size effects (this will be illustrated later).
\subsection{Effect of parameters}
\begin{figure}[htbp]
{\includegraphics[width=3.1in]{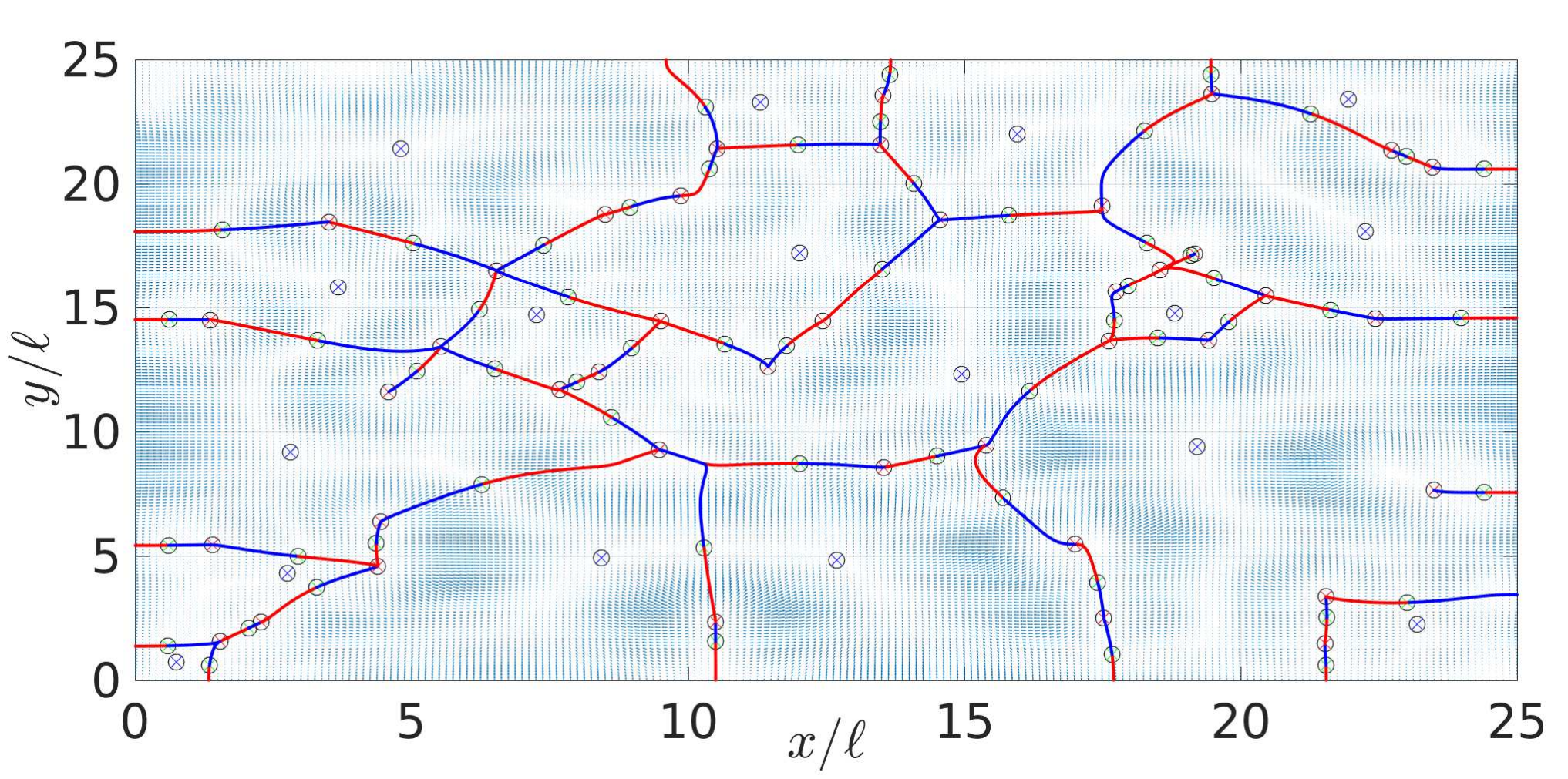}}
{\includegraphics[width=3.1in]{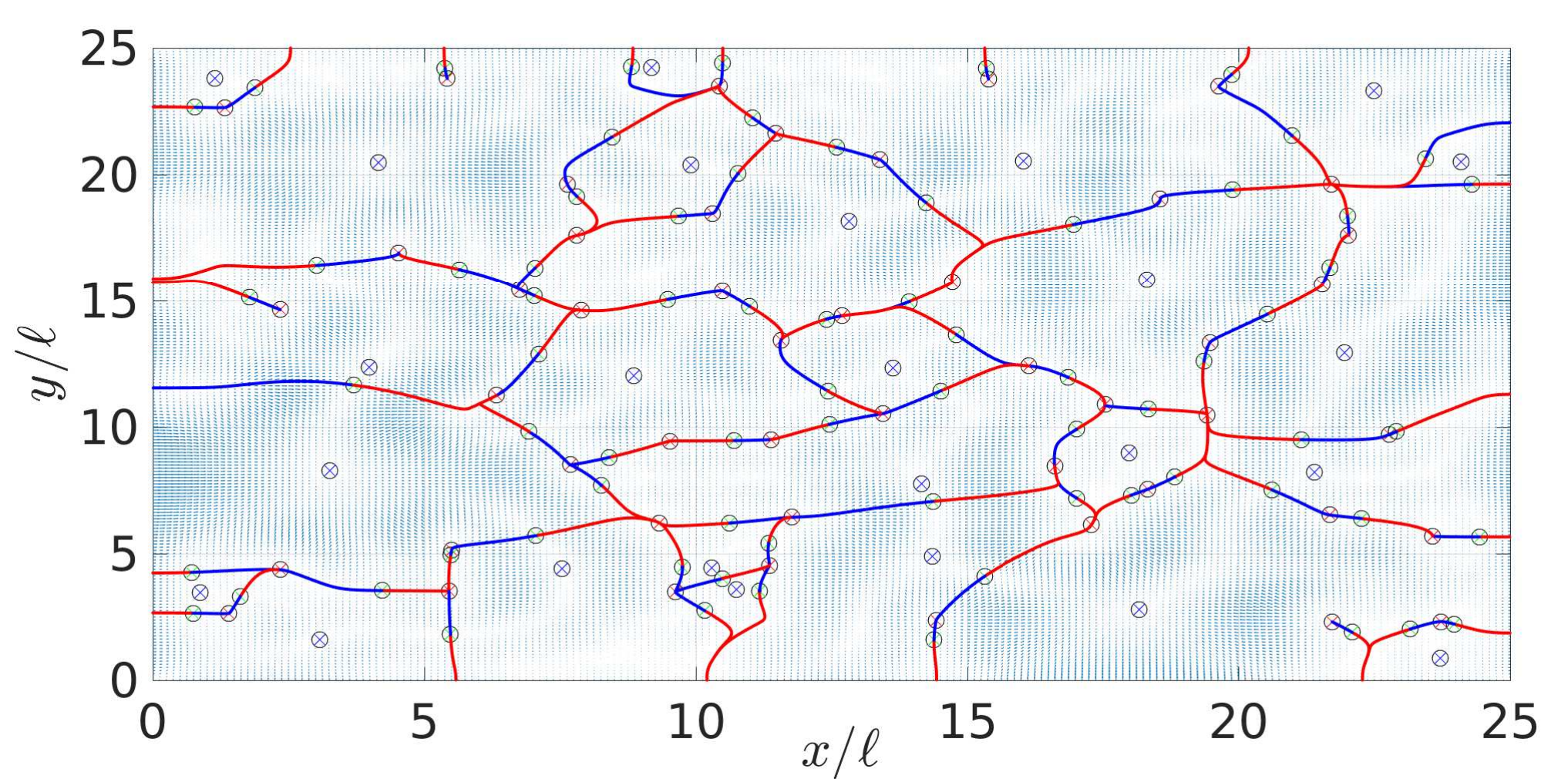}}
{\includegraphics[width=3.1in]{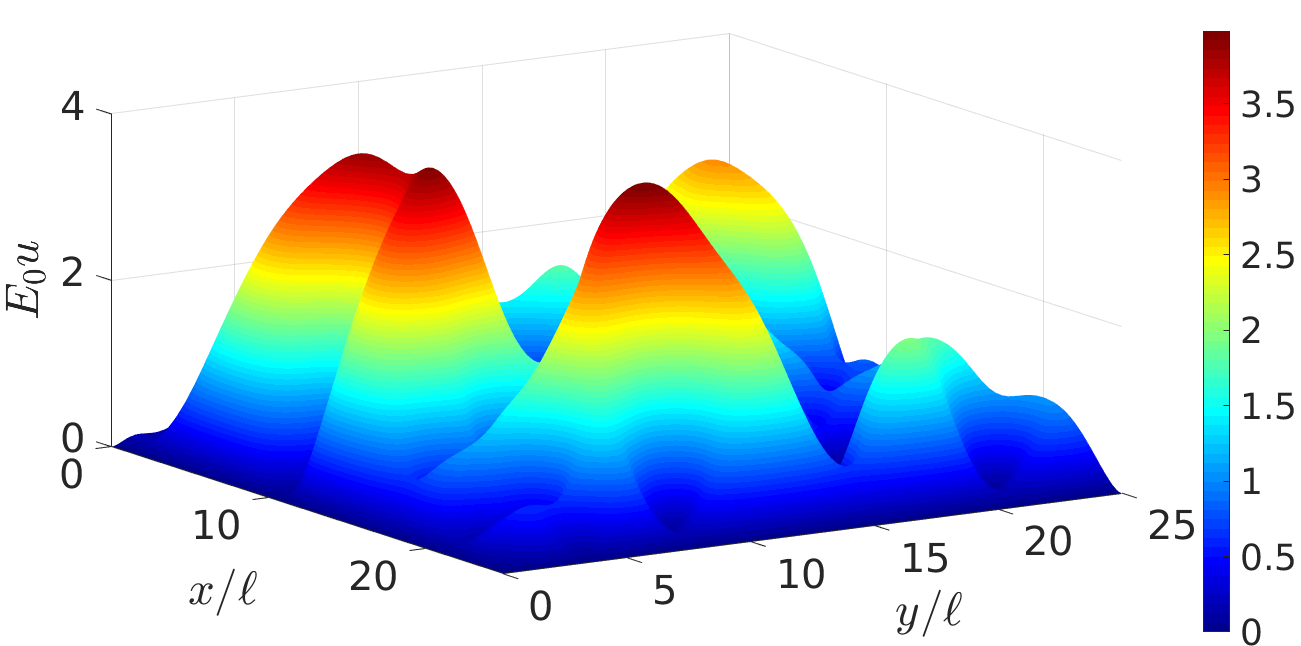}}
{\includegraphics[width=3.1in]{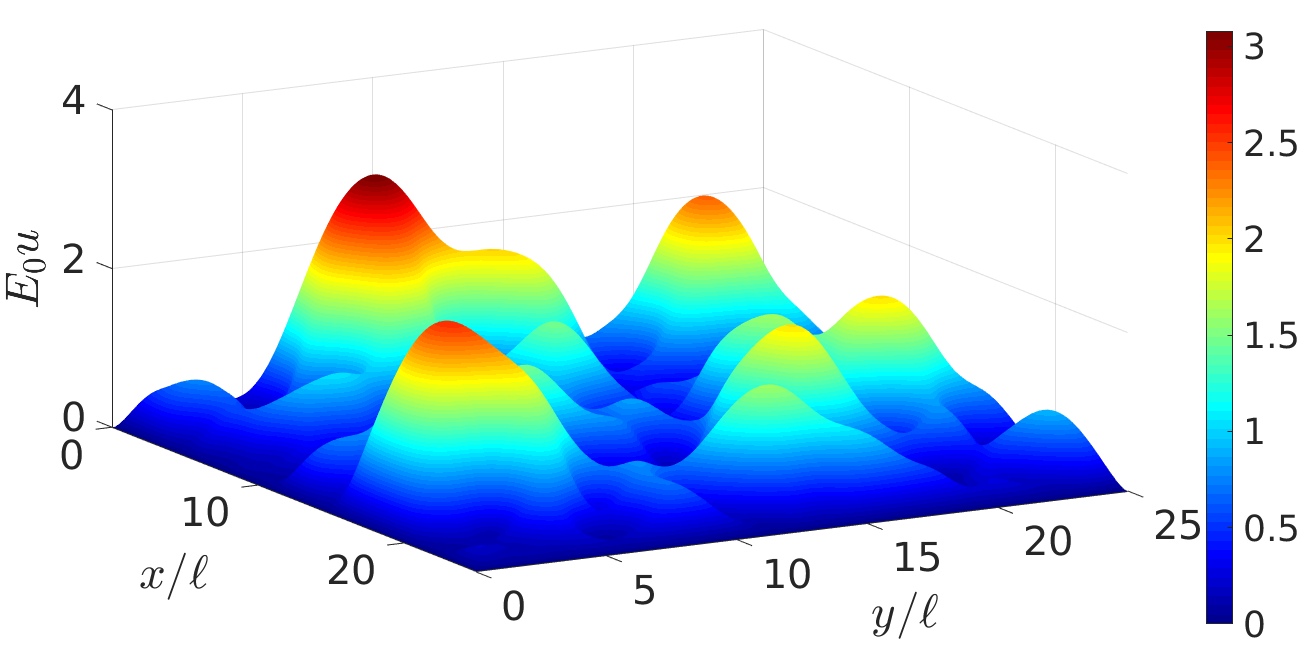}}
{\includegraphics[width=3.1in]{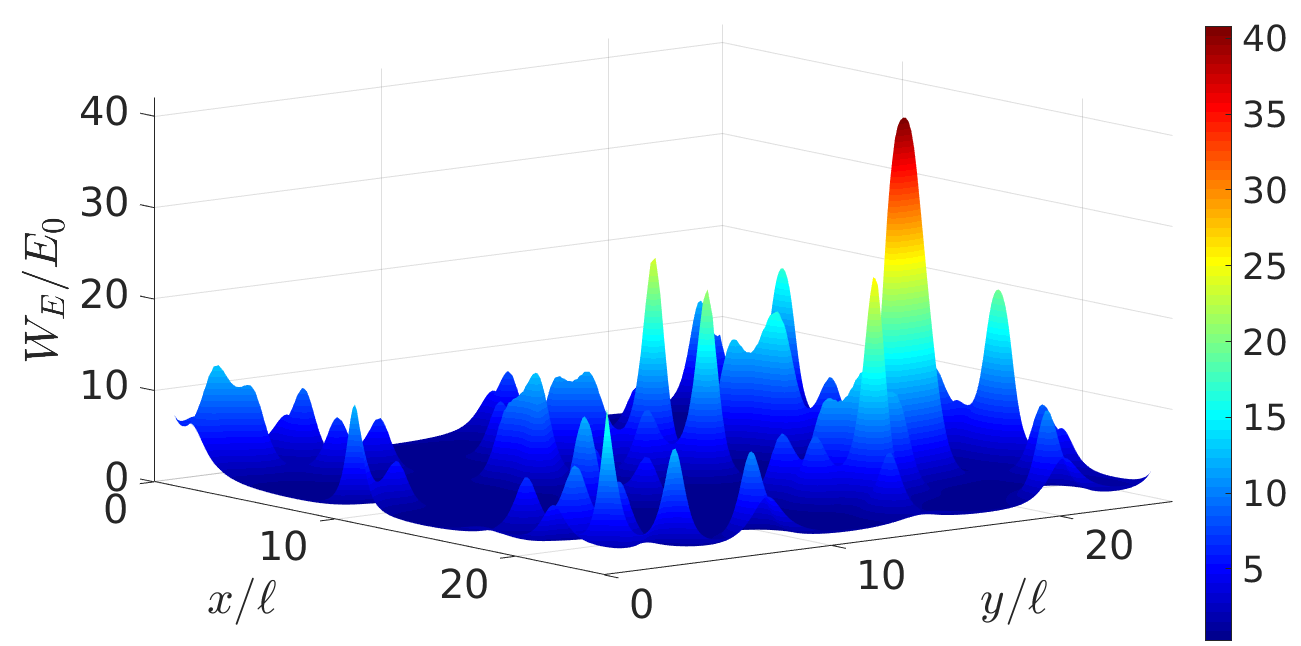}}
{\includegraphics[width=3.1in]{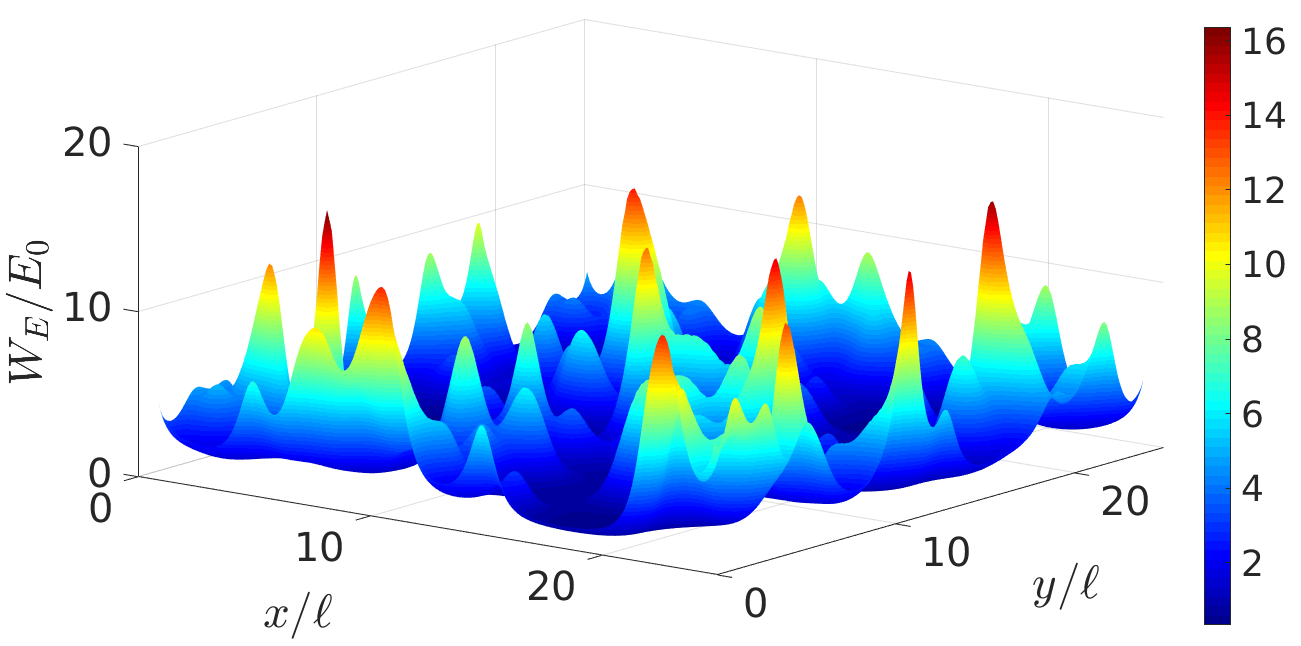}}
\caption{\label{ParEff} The valley network (top), localisation landscape (middle), and effective potential (bottom) are shown for the same parameters as in Figs.~\ref{Ueg}, \ref{Neteg}, \ref{Weg}, except that in the left column, we set $V_0=20E_0$ and in the right, $f=0.2$. The networks are denser, the entire surface of $u$ (including its valleys) is lower, and the peaks in $W_E$ are higher compared to the case of $V_0=10E_0, f=0.1$, indicating a stronger degree of localisation.}
\end{figure}

We can easily use LLT to investigate (at this stage, qualitatively) the effect of the different parameters in our system on localisation. Increasing either $f$ or $V_0$ unambiguously strengthens localisation (Fig.~\ref{ParEff}). This manifests as denser valley lines, forming smaller domains, with the value of $u$ on the valleys significantly reduced. The number of valley lines that are not part of closed domain walls reduces. Simultaneously, the peak ranges in $W_E$ become much taller. In fact, the entire localisation landscape $u$ drops to smaller values. All these factors are in agreement with one another and point to stronger Anderson localisation upon increasing the density or height of the scatterers, consistent with what we have learned by examining the exact eigenstates in section \ref{Diag}. The width of the scatterers $\sigma$ has a similar effect, but it is not studied here and therefore not illustrated.
%
%
\section{The effective potential}
\label{Wmeaning}
So far, LLT has given us several extremely useful results involving the effective potential $W_E$ which allow to make physical predictions for a system with real potential $V$ -- in our case, a disordered one. In particular, $W_E$ controls the regions of localisation of the eigenstates at different energies, the density of states according to Weyl's law (\ref{Weyl}), and the decay of the eigenstates through the valley lines according to the Agmon distance (\ref{Agmon}). While the authors of \cite{FnM2016b,part1} motivate this remarkable success of the effective potential by the auxiliary wave equation (\ref{auxWE}), it appears that $W_E$ may, to a good approximation, be able to replace $V$ in the real Schr\"{o}dinger equation, directly in the Hamiltonian (\ref{Ham}), simply based on its successful use in place of $V$ in so many different formulae.

Ultimately, the main advantage of using the effective potential for us will lie in applying a semiclassical approximation to describe tunnelling at low energies in this landscape, but the semiclassical theory is an approximation to the full quantum-mechanical problem, and so before we explore the additional complexity of this approximation, we should check whether the substitution is valid in the full quantum mechanical treatment. This can be achieved by comparing the eigenstates in the two potentials and checking for similarity, which will justify the application of semiclassical tunnelling theory based on the effective potential $W_E$ to predict the behaviour of the eigenstates in the physical potential $V$.

\begin{figure}[htbp]
{\includegraphics[width=6in]{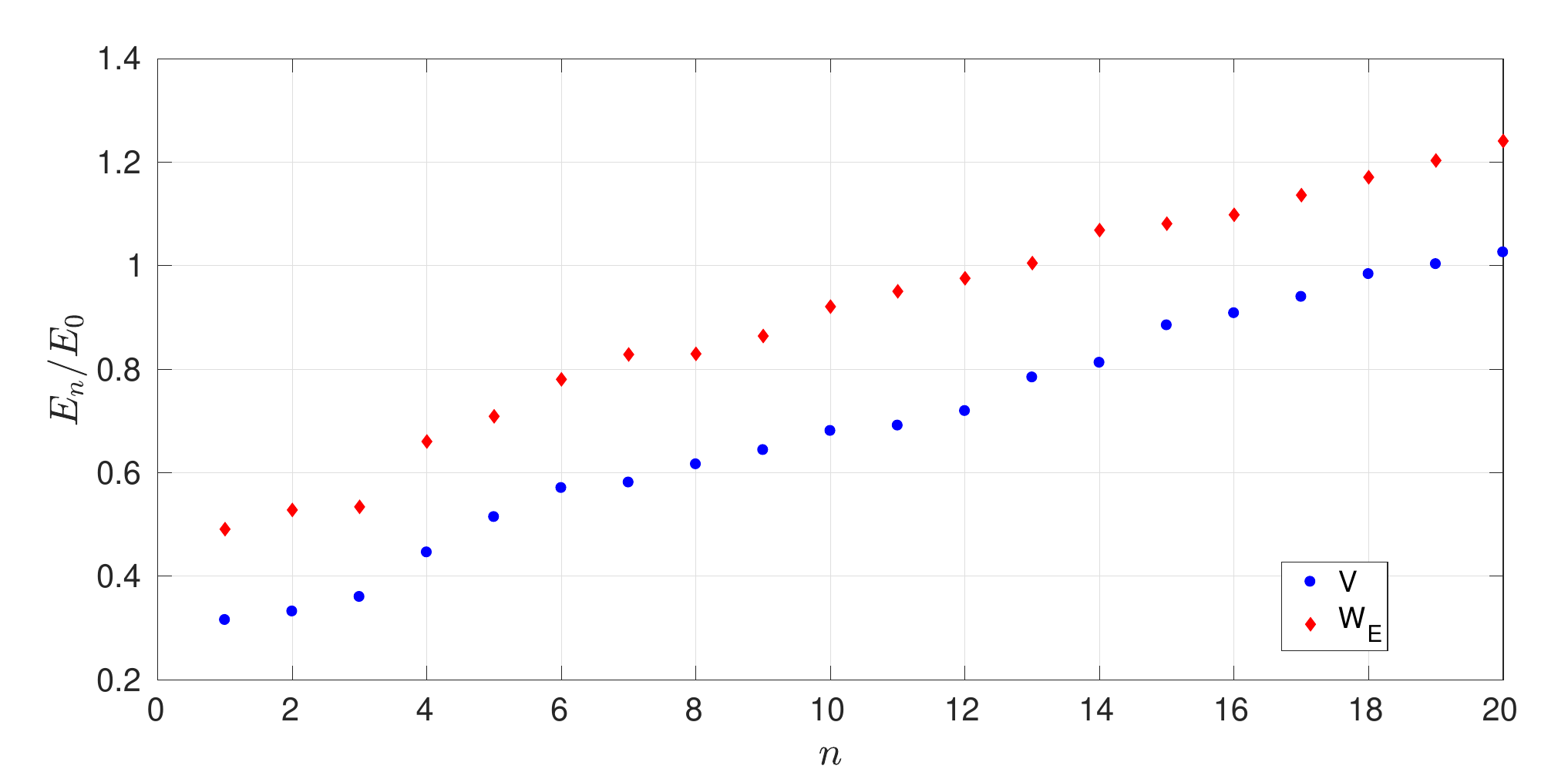}}
{\includegraphics[width=6in]{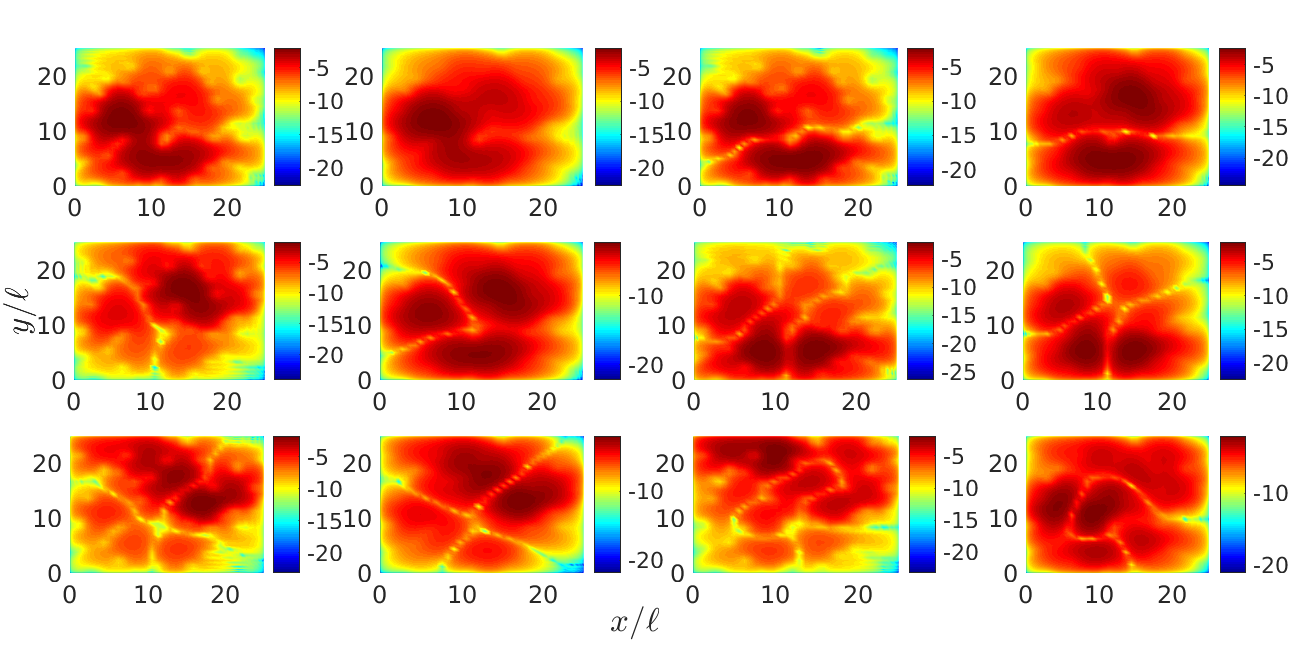}}
\caption{\label{CompSpec} Low-energy eigenspectrum (top) and six of the lowest eigenstates with $L=W=25\ell$, $f=0.1$, $V_0=10E_0$, $\sigma=\ell/2$, showing the logarithm of the absolute value of the eigenstates as a colour-map (bottom). A direct comparison is drawn between the spectrum of the Hamiltonian (\ref{Ham}) with potential $V$ and with $W_E$ for the same noise realisation. The eigenvalues seem very similar, up to a global energy shift. In the bottom panel, going across the rows, we plot consecutively the $n^{\mathrm{th}}$ eigenstate using $V$ and the $n^{\mathrm{th}}$ eigenstate using $W_E$, alternating between the potentials before increasing $n$. Thus the first and second panels can be directly compared, the third and fourth, etc. Up to the fifth eigenstate, the correlation between the mode shapes is clear. From the sixth eigenstate onward, there is no visible relation between the eigenmodes of the Hamiltonian with the two potentials.}
\end{figure}

We therefore check whether the eigen-states and -energies of $H$ with $W_E$ are similar to those of $H$ with $V$. To some extent, this is indeed the case, as demonstrated in Fig.~\ref{CompSpec}. The energy spectrum seems very similar up to a global energy shift, first proven to exist and derived in \cite{FnM2016b}, while the eigenstates themselves are closely correlated for sufficiently low energies. We find that for eigenstates that are localised to a handful of domains, involving fundamental local modes, the similarity is immediately obvious. Once localisation is weakened (due to an increase in energy) to allow the occupation of many domains (possibly in excited local states), the correlation is lost. If Anderson localisation is strengthened (by increasing either or all of $V_0$, $f$, $\sigma$), more low-energy eigenstates match between the spectra of $H$ with $V$ and $H$ with $W_E$, and the agreement between the eigenstates is improved. We will discuss this further in section \ref{HigherEs}.

%
As a final note, if one evolves the same initial wavepacket in $V$ compared to $W_E$, one finds that transmission in the effective potential always happens more readily than in the real. This may be explained by the observation that the eigenstates of $H$ with $W_E$ are somewhat more extended than the exact and have higher overlaps. Moreover, we would expect Anderson localisation in $W_E$ to be weaker due to the lower peaks, which would lead to the same effect.

To conclude, we have shown that the lowest energy eigenstates in $V$ are similar to those in $W_E$. This will later allow us to apply a semiclassical approximation to tunnelling in $W_E$ and use it to make quantitative predictions about the decay of eigenstates in $V$, thus granting access to the localisation length.
\section{Eigenstate localisation length}
\label{XiSaddles}
%
In this section we extend LLT to compute the localisation length for very low energy, maximally localised eigenstates, defined as the length scale of exponential decay in the tails of the eigenstates of the Hamiltonian. A combination of several LLT concepts allows for the development of a general methodology that can be applied to other systems, with other kinds of disorder, or in other dimensions.

In the regime where our calculation is applicable, we explicitly test our ideas by direct comparison to exact eigenstates and find good agreement. We highlight the unavailability of other reliable methods for the purpose of comparison to and validation of our new technique. For example, the transfer matrix method is commonly used for discrete systems, and may be extended to 1D continuous systems \cite{PiraudThesis}), but to the best of our knowledge, not to 2D. Previous papers that have used point-like disorder have faced a similar problem: Refs.~\cite{deMarco2015, BS} ran time-dependent simulations to extract the localisation length from the density decay rate, but were not able to compare their results to any other accurate or reliable computation.

In principle, we could compare our LLT calculation to time-dependent simulations, but in practice, in order to have a sufficient energy range over which the LLT results are valid so as to accommodate a translating Gaussian wavepacket in this narrow interval, localisation must be very strong indeed. In this regime, edge effects (described in more detail later) become important and cause the localisation lengths obtained from LLT and time-evolution to differ. Time-dependent simulations can, however, be used outside of the regime of applicability of the LLT calculation and be qualitatively tested for consistency with the indication provided by exact eigenstates regarding the question ``in which way does the LLT calculation fail at higher energies, and how does it deviate from the true result?''.

Before introducing our new method, however, we remind the reader of the alternative approaches available to date.
\subsection{Literature review}
\label{litrev}
The computation of the localisation length is by no means straight-forward. For continuous systems, a rough estimate can be obtained by setting the renormalised diffusion coefficient, derived in the limit of weak scattering where it is only slightly reduced from its classical value, to zero \cite{Sheng, Vollhardt, DelandeLectures}. While the resulting analytical formula is not expected to be accurate, it is of course convenient, and is thus used by many researchers \cite{Piraud2012, Muller2005, deMarco2015, Delande2015}. The diffusive picture is in general often employed to describe Anderson localisation, even though it is strictly inapplicable in this limit \cite{Muller2005, Delande2015}. A rigorous calculation can be performed using Green's functions \cite{Sheng, Vollhardt, russian_guys, DelandeLectures}, but it requires many assumptions regarding the nature of the disorder and is quite involved. On the other hand, Green's functions can be used to extend the classical diffusive picture into the weakly-localised regime by computing the correction to the diffusion coefficient \cite{Sheng, Vollhardt, Muller2005, DelandeLectures}, and even push this picture into the strongly localised limit by making the renormalised diffusion integral equation self-consistent \cite{Sheng, Vollhardt, Muller2005, Ono, DelandeLectures}.

Another approach to obtain the localisation length is the Born approximation, commonly utilised for weak scattering  \cite{deMarco2015, Piraud2012, Benoit}: here, one takes the total wave in the extended scattering body as the incident wave only, assuming that the scattered wave is negligibly small in comparison. Understandably, this method is inaccurate for strong disorder. Exact time-dependent simulations with the Schr\"{o}dinger \cite{Piraud2012, deMarco2015, Thouless, deVries} or Gross-Pitaevskii \cite{BS, Donsa} equations can be used instead, but this approach is somewhat of a ``brute force'' one, as discussed in the general introduction of section \ref{Intro}. Finally, access to the localisation length directly through the eigenstates of the Hamiltonian is hampered by practical considerations (as we have shown).

Other, more model-specific methods have also been employed in the literature: \cite{Peres} solved the Schr\"odinger equation via a random walk on a hyperboloid, \cite{Hilke} derived a non-linear wave equation to extract the Lyapunov exponents corresponding to the linear problem of interest, \cite{Kicked} solved the kicked-rotor model analytically, and \cite{Stefan} derived analytical expressions relevant for the weak disorder limit.

For discrete models, a plethora of methods to calculate the localisation length likewise exists. The most renowned is of course the transfer matrix method, allowing for the calculation of Lyapunov exponents and thus the localisation length \cite{Kunz, Sarma, Heinrichs, Fan, Romer, Romer2, finite_scaling, Su, 2D_corr}. Such calculations have commonly been used to confirm the predictions of finite scaling theory \cite{finite_scaling, Fan}. While often used together, transfer matrices and Lyapunov exponents have been combined with other elements to obtain the localisation length: the former with analytical continuation \cite{Kirkman} to compute moments of resistance and the density of states, and the latter in a perturbative expansion, with numerical simulations of a quantum walker \cite{Derevyanko}. The Kubo-Greenwood formalism has also proved highly successful \cite{Fan, LeeFisher, ScalingTheory}.

Green's functions have been as invaluable for discrete systems as for continuous \cite{HerbertJones, Thouless, Greek, Fan, russian_guys, Benoit}, allowing for renormalisation techniques to be applied \cite{Greek, Aoki}, or alternatively scattering matrices, treated with the Dyson equation \cite{russian_guys}. Out of these references, \cite{HerbertJones} examined the off-diagonal elements of the Green's matrix as a localisation order parameter, \cite{Thouless} the distribution of eigenstates which was related to the spatial extent of the eigenstates, \cite{russian_guys} the characteristic determinant related to the poles of the Green's function, and Ref.~\cite{Greek} developed a renormalised perturbation expansion for the self energy. Recursion formulae encoding the exact solution \cite{He, Mertsching} can also sometimes allow one to calculate the localisation length (and the density of states \cite{Mertsching}).

Out of the studies above, 1D \cite{Piraud2012, Kunz, HerbertJones, Thouless, Kirkman, Sarma, Peres, Heinrichs, Greek, Derevyanko, He, Hilke, Mertsching, Kicked} and 2D \cite{Piraud2012, deMarco2015, BS, Kunz, Sarma, russian_guys, Heinrichs, Fan, Romer, Romer2, deVries, finite_scaling, Benoit, Su, Aoki, Stefan, Muller2005} models have been numerically explored far more thoroughly than three-dimensional (3D) \cite{Piraud2012, HerbertJones, finite_scaling}, simply because of the increased computational requirements of higher-dimensional spaces. Possibly the most heavily studied model of localisation is the Anderson model, also known as a tight-binding Hamiltonian \cite{Sheng, HerbertJones, Thouless, Kirkman, Sarma, russian_guys, Fan, Heinrichs, Greek, Romer, Romer2, He, finite_scaling, Benoit, Aoki, Chinese, 2D_corr, LeeFisher, Malyshev, Weaire, Greek2, 1D_corr, Makarov}, but other examples include the kicked rotor \cite{Kicked} (formally equivalent to the Anderson model), the Lloyd model \cite{Kunz, Thouless}, the Peierls chain \cite{Mertsching}, a quantum walker \cite{Derevyanko}, and the continuous Schr\"odinger equation \cite{Thouless, Peres, deVries}, with either a speckle potential \cite{Donsa, Muller2005}, delta-function point scatterers \cite{Piraud2012, russian_guys}, or more realistic Gaussian scatterers \cite{deMarco2015, BS}.

It is worth noting that for 2D continuous potentials with arbitrary disorder, there is no numerically-exact or even a fairly accurate, approximate method to compute the localisation length, leaving the direct integration of the time-dependent Schr\"odinger equation as the only currently viable approach.

We now demonstrate how the localisation length can be obtained from LLT, a method that can be applied to continuous systems with any potential (as long as $V>0$, to satisfy the applicability requirements of LLT), for any strength of the disorder, and which will provide accurate results for a range of (reasonably low-lying) energies. Our description is in 2D, a 1D version is much simpler and can be implemented with no additional effort, while a 3D version can be eventually developed by a direct extension.
\subsection{Outline of the LLT method}
Recall that LLT has taught us that the low-energy eigenstates are localised inside domains of the valley network, and must tunnel through the peaks of the effective potential in order to spread to neighbouring domains (this is in contrast to the physical potential $V$, where there are gaps between scatterers, with the domains connected classically \footnote{This statement holds at reasonable fill-factors and scatterer widths. If either parameter is increased excessively such that the scatterers join and form closed regions in the plane, then classical trapping becomes possible.}). Within any given domain, there is nothing to induce exponential decay -- the decay does not happen continuously (as commonly believed), but in discrete steps, every time the wavefunction crosses a valley line. This was originally shown in Ref.~\cite{FnM2016b}, but is also visible in essentially all the figures depicting eigenstates in the sections above. Furthermore, valley lines which are not part of a closed domain are irrelevant, as the wavefunction simply goes around them without losing amplitude.

If we approximate the domains on average as circular in shape and denote the diameter $D$, then every distance $D$, the wavefunction undergoes a decay. The cost of crossing a valley line will be bounded below by the Agmon distance (motivated later), so we may safely use the symbol $\rho_E$ to denote the exponent, such that the amplitude of the wavefunction drops by at least a factor of $\exp(-\rho_E)$ on average every time. If we assume for the moment that $\rho_E$ faithfully captures the decay rate, combining these two quantities, we see that the localisation length is simply given by
\begin{equation}
\xi_E = D/\rho_E,
\end{equation}
where the subscript $E$ on $\xi$ stands for ``eigenstate''. Remarkably, the difference between $D$ and $\xi_E$ was already realised in \cite{Greek}.

Now, evaluating $\rho_E$ between any two arbitrary points in the $x-y$ plane is extremely difficult, as discussed in section \ref{MultiDimTun}. However, this is not strictly necessary for our purposes. With the understanding that the system is divided into network domains, we can estimate the Agmon distance between the minima of $W_E$ (equivalently, the maxima of $u$), considering only nearest neighbour domains. In other words, if we have two neighbouring domains (which share some common segment of domain walls), we aim to find the least-cost path, according to (\ref{Agmon}), that connects the two unique maxima of $u$ which reside in these domains. Evaluating $\rho_E$ along this path would then be straight-forward.

\begin{figure}[htbp]
{\includegraphics[width=6in]{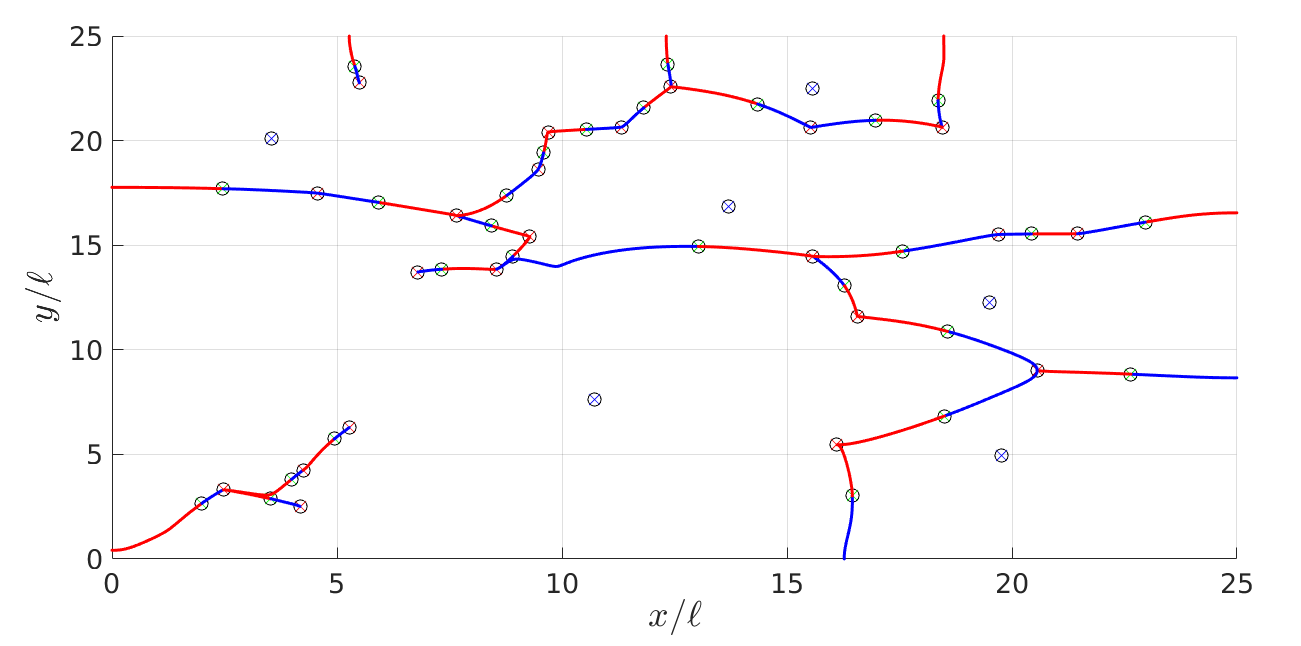}}
{\includegraphics[width=6in]{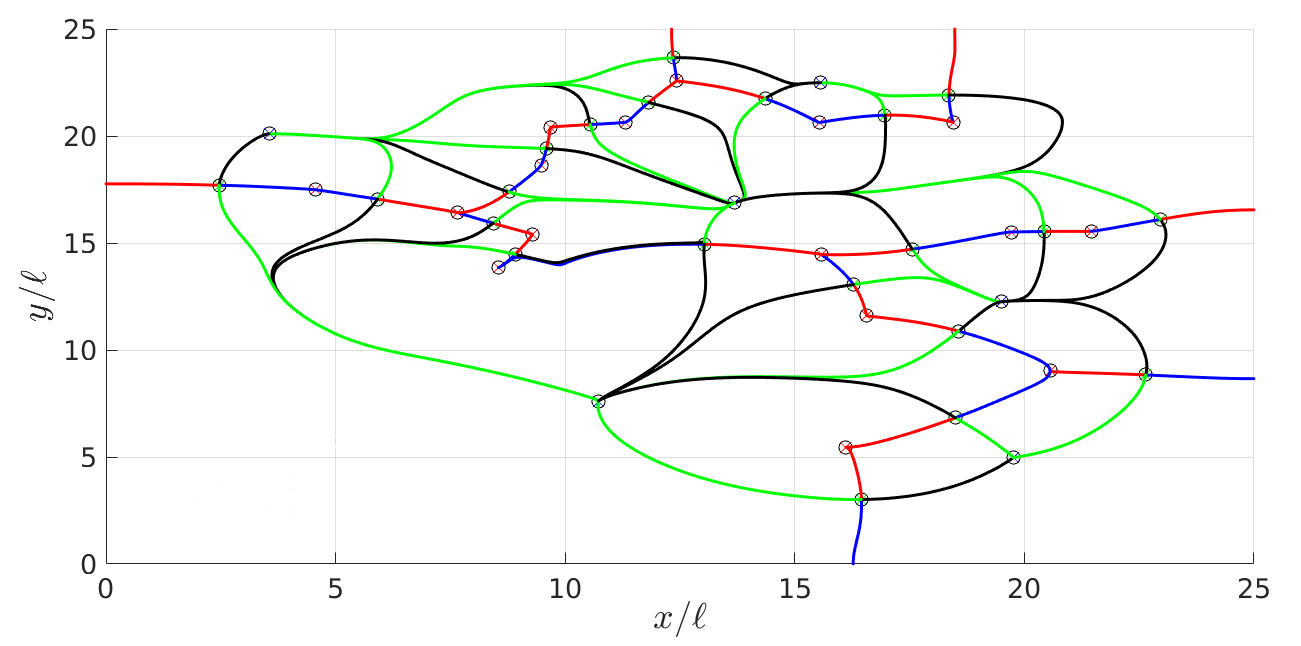}}
\caption{\label{C2demo} The original valley network (top) for some given noise realisation with $L=W=25\ell$, $f=0.06$, $V_0=5E_0$, $\sigma=\ell/2$, and the same network after all ``open'' valley lines have been removed (bottom). Both panels plot the valley lines in red and blue (different colours are used simply to make it easier to see the structure of the network). The extrema of $u$ are also shown, as usual (maxima in blue, minima in red, saddles in green). The bottom panel displays in addition all candidate approximate paths of least cost with respect to the Agmon metric as green and black lines (again, for more accessible visual interpretation), connecting neighbouring maxima of $u$ through the linking saddle points.}
\end{figure}

Again, formally, finding the true least-cost path is a difficult task. We have found an approximate solution to this problem that seems much simpler to implement compared to all currently known alternatives, while not sacrificing much in terms of accuracy at all (see section \ref{MultiDimTun} to gain perspective). As explained in appendix \ref{appLLTold}, the valley lines are the paths of steepest descent, starting from each saddle point and ending at minima of $u$ (valley lines may also terminate by exiting the system). Consider now curves that start from the saddle points and follow paths of steepest \textit{ascent}, ending at maxima of $u$. Each saddle point thus links two maxima of $u$, and the curve formed in this way is the lowest-lying path on the inverse landscape $W_E$ that connects the two minima of $W_E$ in question. Figure \ref{C2demo} first shows an example of the valley network as originally defined, and then with open valley lines removed (as they do not matter for eigenstate confinement and decay) and the candidate minimal paths connecting maxima of $u$ through the saddle points overlaid.

We will use these paths to compute $\rho_E$ between any two neighbouring maxima of $u$. First of all, we highlight that the Agmon distance is an energy-dependent quantity. Thus, along each path, the integral in (\ref{Agmon}) must be done separately at each energy of interest, $E$. Now, generally speaking, any two neighbouring domains have several common saddles on the shared section of their domain walls. At each energy, we must choose the minimal path which has the smallest Agmon integral out of the finite, discrete number of available options (which is computationally trivial). The path integral along that curve then becomes the Agmon distance $\rho_E$ between the domain maxima in question at the energy considered. This must be done for all neighbouring domains and at all energies in any given landscape $u$.

One may wonder, at this point, how well does our approximation capture the ``real'' Agmon distance, obtained by proper path minimisation, as described in section \ref{MultiDimTun}. We have tested this for several examples by solving the semiclassical equations and comparing the Agmon integral to that taken over the minimal lowest-lying path on the surface of $W_E$. We found that the true minimal path always lies very close to the minimal lowest-lying path and the integral along the latter is only slightly greater than the smallest possible cost obtained by proper path minimisation; the results are summarised in Table \ref{AgmonTest}.
\begin{table}
\begin{center}
\begin{tabular}{|c|c|c|}
\hline
  $E/E_0$ & LLT approx. & True semiclassical \\
  \hline\hline
   0 & 6.8349 & 6.2586 \\
 0.1 & 5.7521 & 5.3130 \\
 0.2 & 4.3178 & 4.0233 \\
 0.3 & 2.0687 & 1.9918 \\
 0.4 & 0.9004 & 0.8831 \\
 0.5 & 0.2545 & 0.2491 \\
 0.6 & 0 & 0 \\
  \hline
 0   & 6.6619 & 6.1395 \\
 0.1 & 5.0021 & 4.6693 \\
 0.2 & 2.4690 & 2.3319 \\
 0.3 & 0 & 0 \\
  \hline
 0   & 5.4327 & 5.1905 \\
 0.1 & 4.0763 & 3.9064 \\
 0.2 & 1.9386 & 1.8750 \\
 0.3 & 0.2674 & --- \\
 0.4 & 0 & 0 \\
  \hline
\end{tabular}
\caption{\label{AgmonTest}
A comparison of the Agmon distance found using the approximate LLT path (the path of minimal cost out of the finite set of lowest-lying paths on $W_E$ that connect neighbouring minima of $W_E$ through the saddle points), to the real semiclassical result, obtained by solving differential equations, as described in section \ref{MultiDimTun}. In this case, we used a single noise realisation with $L=W=25\ell$, $f=0.06$, $V_0=21.33E_0$, $\sigma=0.48\ell$, and three domain pairs, the results for which are separated by horizontal lines in the table. For each domain pair, we computed the Agmon distance at several energies, until the domains became classically connected and the cost vanished. In the one case where the table entry is missing (replaced by a hyphen), the true semiclassical path could not be found. It is immediately clear that the LLT approximation of the Agmon distance is a very close one, and that in all cases, our method only slightly overestimates the true minimal cost. This is a small price to pay for the remarkable computational advantages of our scheme compared to the real semiclassical solution (see section \ref{MultiDimTun}).
}
\end{center}
\end{table}

In the decay picture painted so far, restricting our consideration exclusively to neighbouring domains does not introduce an additional level of approximation: we only need to know the average cost of crossing from one domain into another, and decay over large distances can be simply composed of several such domain-to-domain tunnelling events. That is, our calculation only requires the computation of \textit{local} quantities, which makes it largely system-size independent. Indeed, up to finite size effects which change the spacing of the valley lines at small system sizes (as studied in section \ref{WidthDep}), averaging over a few large systems  will give the localisation length to the same precision as averaging over a bigger number of smaller systems: the only important factor is how many typical domains (for the area) and domain-pairs (for the tunnelling coefficient) are averaged over, not whether they are in one or several valley networks.

The next question is whether the Agmon distance $\rho_E$ faithfully captures the decay rate between neighbouring domains: after all, it is a lower bound on the decay coefficient, not an estimate thereof. We test this in the bottom panel of Fig.~\ref{rhoComp} (see the next subsection for details), finding that the Agmon distance itself systematically underestimates the true decay rate seen in the exact eigenstates. Therefore, rather than choosing the minimal-integral path, we take the average of the path integrals over \textit{all} candidate paths from LLT (lowest-lying paths going through the saddle points), to obtain what we will coin the ``mean'' Agmon distance, $\bar{\rho}_E$. As we demonstrate in the top panel of Fig.~\ref{rhoComp} below, this method of computation actually captures the true decay rate much better, so we proceed with the understanding that
\begin{equation}
\label{Eqn_xi_E}
\xi_E = D/\bar{\rho}_E.
\end{equation}
Note that this modified decay rate fully obeys the Agmon inequality (\ref{decay}), and that this is an advancement of semiclassical multidimensional tunnelling, as so far, it has only been possible to calculate the lower bound of the decay rate, but not an approximation of the real value (see section \ref{MultiDimTun} for a further discussion).

As pointed out, $\bar{\rho}_E$ between neighbouring domains is an intrinsically energy-dependent quantity. Once the energy is so high that the saddle points of the candidate paths on the effective potential $W_E$ are below $E$, the cost of crossing from one domain to the other vanishes: $\bar{\rho}_E$ becomes zero as breaks develop in the domain wall separating the two maxima of $u$. For our computation of $\xi_E$, we need the average of all non-zero $\bar{\rho}_E$ across the 2D system as a function of energy, but we also need to compute the domain area to extract the diameter, $D$. This requires integrating over the individual domain areas (at $E=0$), averaging over all domains, assuming the area is that of a circle, and computing the diameter. However, as energy goes up and domain walls break down, domains effectively \textit{merge}, so that the area increases with energy as well. Thus, in our calculation, domains are merged once $\bar{\rho}_E$ between them vanishes.

To summarise, the main steps of the calculation are as follows. Take a precomputed valley network, remove any open valley lines and calculate all the ``candidate minimal paths'' connecting saddles to maxima of $u$. Next, identify the valley lines (and potentially segments of the system boundary) that form the domain walls for each domain and perform local, on-domain integrals (for now we only need the area, so the integrand is one). From here, identify all saddles linking any two neighbouring domains, calculate the path integral in (\ref{Agmon}) over all linking paths between them, and finally obtain $\bar{\rho}_E$ by averaging over these integrals (including any paths that give a vanishing cost) at every energy. Then, for each noise configuration, the mean of $\bar{\rho}_E$ is computed over all neighbouring domain pairs, and the mean domain area yields the diameter $D$. Both of these quantities are energy dependent: zero-cost links are excluded from the average of $\bar{\rho}_E$ and domain areas are merged as the walls between them break down. Finally, many noise configurations need to be averaged over to get a reasonable estimate of the localisation length.

Note that an analogue of our LLT calculation cannot be usefully performed by using $V$ directly, instead of the effective potential $W_E$. This is because the exponential cost of crossing most domain walls would be zero (exceptions would be caused by scatterer overlap), as the scatterers in $V$ are separated by gaps. In other words, since classical trapping in $V$ is not possible (at reasonable fill factors and scatterer widths), a semiclassical tunnelling picture would predict no exponential decay. This is in addition to the fact that in order to find the domains, one needs the localisation landscape $u$ (1/$V$ would not yield closed domains in the valley network due to gaps between the scatterers). Thus, LLT is essential for our method and one could not avoid using it.

We remark that this calculation can be performed for any given localisation landscape as long as it has (appropriate) extrema. This includes, in particular, cases when the potential $V$ is regular and Anderson localisation is impossible. The resulting ``localisation length'' is then of course meaningless. It is up to the researcher performing the calculation to identify cases when one is dealing with localisation before attaching any significance to the result. This can be done by examining the fundamental on-domain eigen-energies, and ensuring that they are randomised, as explained in detail in sections \ref{LLTold} and \ref{BHM}.
\subsection{Test of decay constants}
We have just outlined a proposed method for computing the localisation length at very low energies. Let us assume for the moment that the decay model we have developed applies (i.e.~that the eigenstates take the form of one or a handful of strongly occupied domains with straight-forward decay through the valley lines into their neighbours). Under these conditions, the domain area calculation can hardly fail to give us $D$ correctly, the mean distance separating tunnelling-inducing valley lines. On the other hand, the decay constant from one domain to another, $\bar{\rho}_E$, is a different matter entirely. As will be discussed in section \ref{MultiDimTun}, the level of approximation involved is very high, and there is no \textit{a priori} assurance that our method yields numbers which faithfully capture the decay of the eigenstates. Therefore, a direct test is in order. This can be done as follows: for the same noise realisation, we perform the full LLT calculation, as well as find the low energy eigenstates by exact diagonalisation. Now, we know that within each domain, the wavefunction remains roughly constant (same order of magnitude). Therefore, we integrate $\left|\psi\right|$ over the domains, and divide by the domain areas to get the average of the wavefunction amplitude on each domain.

Then, by visual inspection of the eigenstates, we find examples of eigenstates and domain pairs where it is clear that the wavefunction tunnels from one domain to the other, as opposed to an independent occupation of the two domains (or any of the more complex behaviour described in section \ref{HigherEs} which is encountered at higher energies). We also avoid higher local modes than the fundamental. Having identified suitable candidates, we take the ratio of the mean amplitudes on the two domains and compute the logarithm. The resulting number is equivalent to $\bar{\rho}_E$ from LLT, the exponential cost of going specifically between these two domains (in this noise realisation), at an energy equal to the eigenvalue corresponding to the eigenstate examined.

\begin{figure}[htbp]
\includegraphics[width=6in]{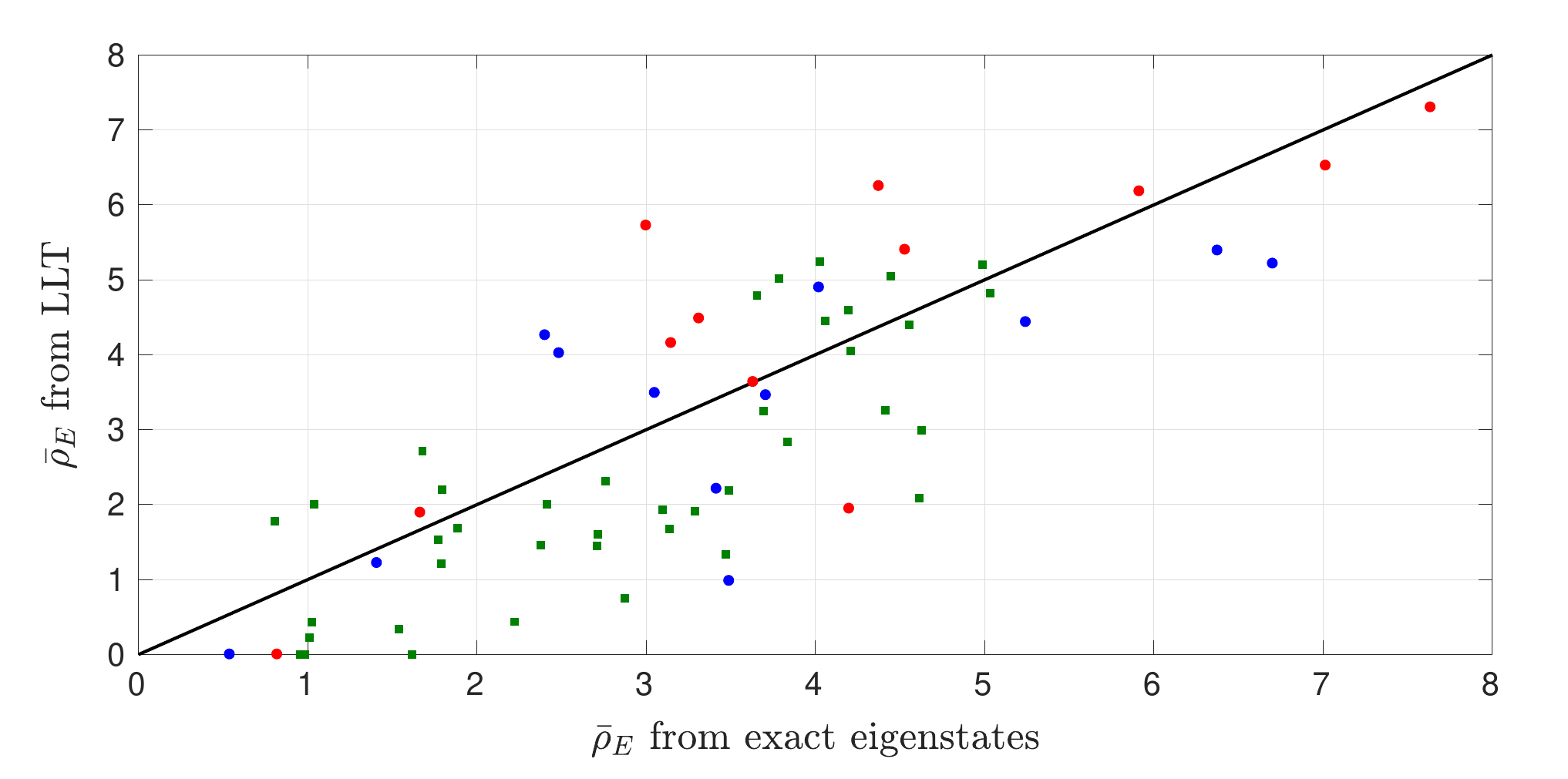}
\includegraphics[width=6in]{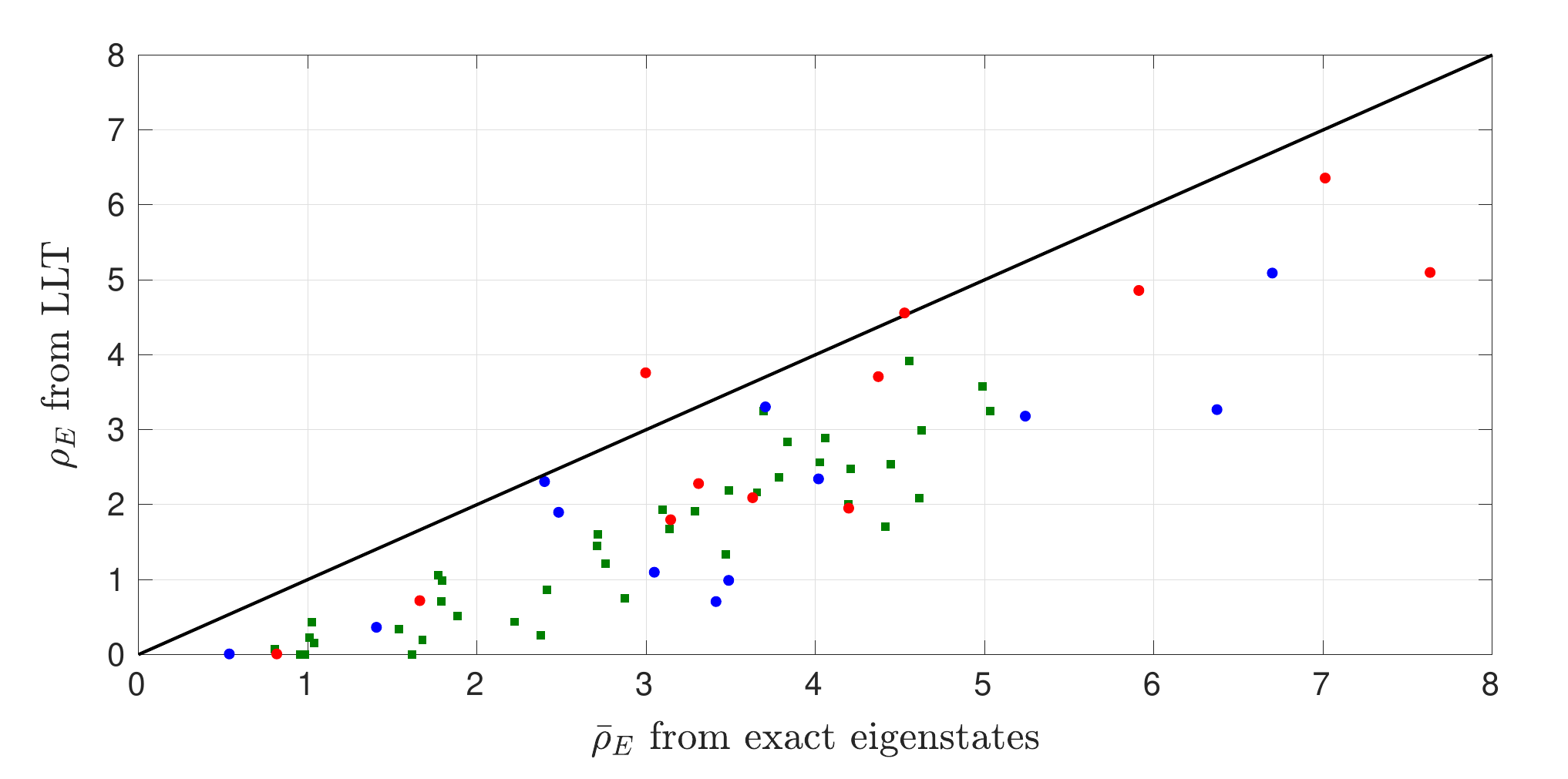}
\caption{\label{rhoComp} Exponential decay cost linking two neighbouring domains, plotting the values measured from exact eigenstates and LLT against each other. Top: the LLT calculation shown uses the average over all linking paths between domain pairs, and there is a very clear correlation to the eigenstate decay coefficient: the data points fall nicely around the identity map, shown as a black solid line. Bottom: data for the exact same test-cases is now plotted so that the LLT calculation shows the minimal path only over all linking paths between domain pairs, and it is obvious that this method systematically underestimates the true decay rate. All data points presented were obtained for a system with $L=W=25\ell$, $V_0=21.33E_0$, $\sigma=0.48\ell$. Blue and red circles have $f=0.02$, with blue  coming from diagonalising $H$ with $W_E$ and red with $V$, while green squares used the real potential $V$ and $f=0.1$.}
\end{figure}

We have performed this test, and the results are shown in the top panel of Fig.~\ref{rhoComp}. A clear correlation is seen, whether the predictions of LLT are compared to the eigenstates of $H$ with potential $V$ or $W_E$. The performance of the LLT method is equally good for arbitrary strengths of localisation (compare sparse and dense scatterer results), simply because the only numbers included in the test are those for which the eigenstates and domains chosen are sensible (sufficiently low energy, correct local modes, decay as opposed to independent occupation, etc.). Of course there is scatter about the identity function, but since much averaging is performed during the calculation of $\xi_E$, this scatter will disappear in the mean. This gives us confidence in the validity of our novel computational method for very low energies.

In contrast, as mentioned earlier, the Agmon distance itself, $\rho_E$, systematically falls short of the true decay coefficient (being a formal lower bound), as depicted in the bottom panel of Fig.~\ref{rhoComp}.

We emphasize that there is no other available method to compare our calculation of the localisation length to. The only reliable approach is to run time-dependent simulations, integrating the Schr\"odinger equation. The simplest test would be to initiate a translating Gaussian wavepacket with a fairly narrow energy distribution outside the disorder, allow it to propagate, and observe the resulting exponential decay set in with time. The (unnormalised) energy distribution for our translating 1D Gaussian initial condition is simply
\begin{equation}
\label{Edistro}
g(E)\ dE = \exp\left[-2\bar{\sigma}^2\left(\sqrt{\frac{2mE}{\hbar^2}}-k_0\right)^2\right]\frac{\sqrt{m}}{\hbar\sqrt{2E}}\ dE.
\end{equation}
One would have to average 20 to 30 realisations to get accurate results, measure the decay length scale seen in the density, and compare to that obtained from the energy-resolved $\xi_E$ obtained from LLT by reconstructing the expected density profile for the given energy distribution according to, e.g., equation (63) of Ref.~\cite{DelandeLectures}. However, taking into account the energy distribution of the wavepacket would in this case only provide a fine-tuning of $\xi_E$ taken at the mean energy of the narrow Gaussian wavepacket, which would provide a very good estimate already.

We have attempted precisely such testing of our LLT $\xi_E$ (in parameter regimes and at low enough energies where the curve $\xi_E(E)$ is smooth and monotonically increasing), to find that the LLT prediction greatly \textit{underestimates} the real localisation length, by up to as much as an order of magnitude. For example, using a system geometry given by $L=50\ell, W = 25\ell, R = 30\ell$, noise with $f=0.1, \sigma = 0.48\ell, V_0=21.33E_0$, initial condition specified by $\bar{\sigma}=5\ell, k_0=1/\ell$, and evolving the state for a total time of $100t_0$, we find that quasi-steady state in the density profiles is achieved at $\sim 40t_0$, after which the exponential profile changes slowly and can be meaningfully fitted. We extract a time-dependent localisation length from the density profiles $\xi(t)$ which increases from $10\ell$ to $13\ell$ over the fitted time interval ($60t_0$) of the simulation, and would only increase further with time before eventually equilibrating to a constant. Meanwhile, the LLT calculation, combined with equation (63) of Ref.~\cite{DelandeLectures} and the energy distribution (\ref{Edistro}), together with an exponential fit to the overall predicted density profile, yields a value of $\langle \xi \rangle \approx 2.746\ell$, which is considerably smaller.

The reason for this discrepancy is that the simple decay model that we have been assuming is only valid at very low energies, after which more complex mechanisms of how the eigenstates can spread out spatially come into effect. These are beyond quantum tunnelling and the semiclassical theory thereof, and are described in detail in section \ref{HigherEs}. In the example above, the energy distribution lay fully outside of the applicability regime of the LLT calculation. Comparison to time-dependent simulations in a regime where the simple decay model applies are further discussed in section \ref{HigherEs}.
\subsection{Effect of parameters}
Let us consider -- and when possible, examine -- the effect of the different parameters in the model on the localisation length obtained via the prescription given in this section. Firstly, the calculation can be performed as a function of energy, and as expected, the computed number increases with energy monotonically until one reaches the regime where the finite extent of the system limits the calculation and artificially reduces $\xi_E$, as well as the mobility edge predicted by LLT but found unphysical in section \ref{ME}, beyond which it is no longer possible to perform the calculation. However, the computation ceases to be valid much earlier than that, because the pure decay model we assumed breaks down, as illustrated in section \ref{HigherEs}. In fact, it is usually only very low energy eigenstates that are captured correctly by our description, and the only method known to us of establishing when the complex decay behaviour (section \ref{HigherEs}) begins is by visual inspection of the exact eigenstates. This ``complex decay'' is beyond quantum tunnelling and semiclassical theory, and is attributed directly to Anderson localisation. We will therefore only show data for $E=0$, where the results have been confirmed as meaningful across the range of parameters shown.

\begin{figure}[htbp]
\includegraphics[width=6in]{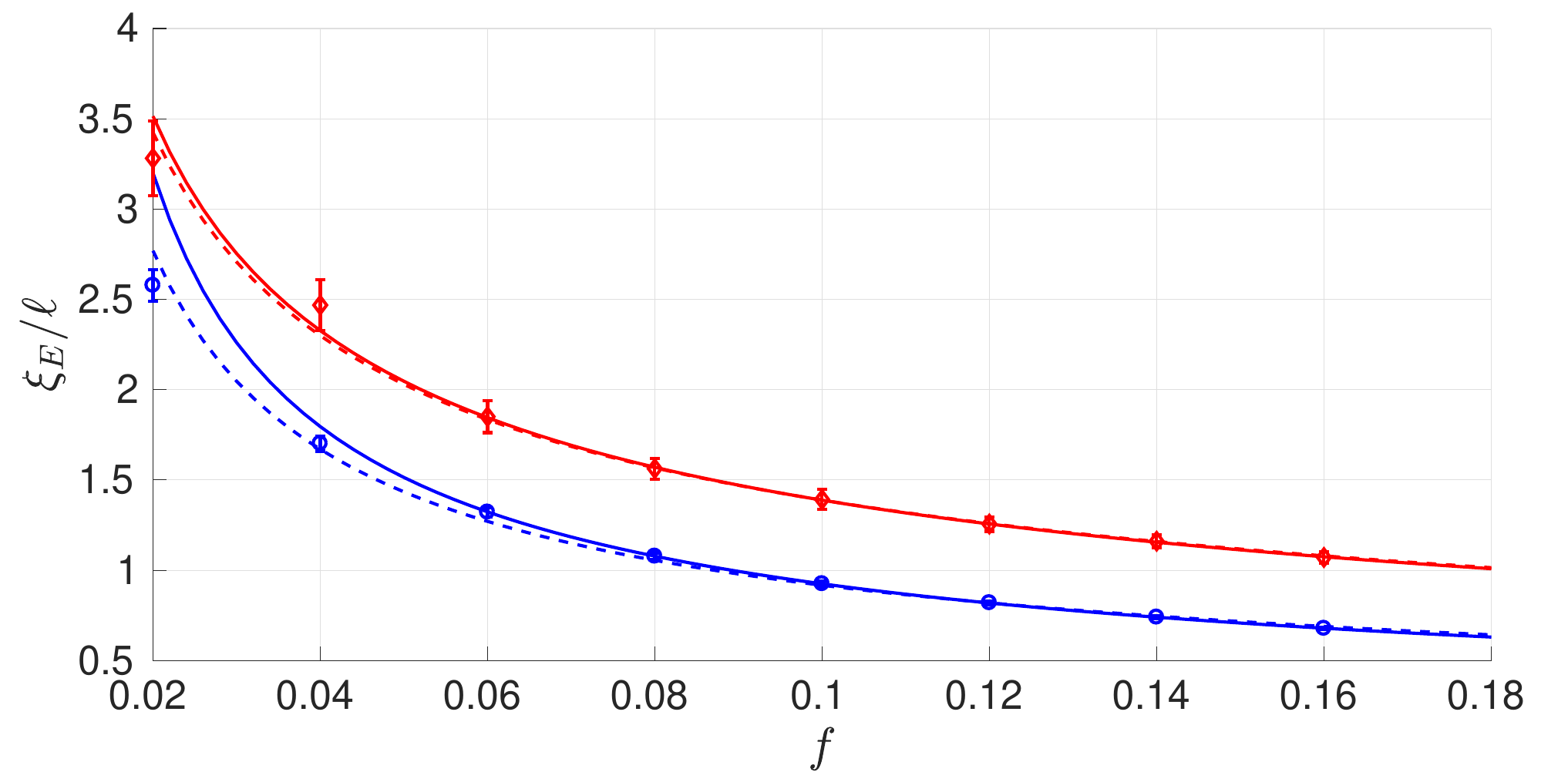}
\caption{\label{xi_E} The eigenstate localisation length $\xi_E$ at zero energy as a function of fill factor for two scatterer heights, blue circles: $V_0=21.33E_0$, red diamonds: $V_0=5E_0$. For all simulations $W=25\ell$, while $L=125\ell$, $\sigma=0.48\ell$ for the data shown in blue and $L=25\ell$, $\sigma=\ell/2$ for the data plotted in red. The lines are fits according to equation (\ref{fit_eqn}), with the colour matching the data set being fitted. Dashed lines fit all the data points, while solid lines exclude the first two (as it increases the fit quality and it is possible that the lower fill factor points are not very accurate). Error bars show the standard error. The effect on $\xi_E$ of using different system sizes in the two cases is about $0.03\ell$ (using $L=25\ell$ for the $V_0=21.33E_0$ data set increases $\xi_E$ by about $0.03\ell$), which is an order of magnitude smaller than the effect seen in the figure. Thus, increasing the fill factor or scatterer height decreases $\xi_E$, as expected.}
\end{figure}

Figure \ref{xi_E} demonstrates that the localisation length is reduced by strengthening the disorder by either increasing the scatterer height or the fill factor. Increasing the width of the scatterers also decreases the localisation length, but we do not simulate this directly in this paper. System size only influences the results weakly due to finite size effects studied thoroughly in section \ref{WidthDep}.

Since we have the opportunity, we compare our results to the analytical formula for the localisation length in 2D
\begin{equation}
\label{Xi2D}
\xi \sim \ell_B \exp\left(\frac{\pi}{2}k\ell_B\right),
\end{equation}
where $\ell_B$ is the Boltzmann mean free path and $k$ the wavenumber associated with the energy at which the localisation length is evaluated. The Boltzmann mean free path (the distance over which the wave loses memory of its initial direction) is related to the scattering mean free path $\ell_s$ (the mean distance between scatterers) through the scattering cross section of a single scatterer, which includes information about the scatterer height and shape, as well as the energy of the wave. We recall that while this formula is quite freely used in the literature (e.g.~\cite{BS}), it is not expected to be correct, as it is derived (for a classical wave) by first assuming weak localisation and then forcing the diffusion coefficient to zero \cite{Sheng, Vollhardt} (in addition, we do not have white noise or an infinite system).

One may relate the mean free path to the fill factor rather trivially by simple geometrical arguments, yielding $\ell_s \propto 1/\sqrt{f}$, and then fit the numerically-obtained $\xi_E$ as a function of fill factor to
\begin{equation}
\label{fit_eqn}
\xi \sim \frac{a}{\sqrt{f}} \exp\left(\frac{b}{\sqrt{f}}\right).
\end{equation}
This has been done in Fig.~\ref{xi_E}, and the fits are of reasonable quality. However, this does not prove the validity of equation (\ref{Xi2D}), as one would have to check the energy dependence of the fit coefficients for consistency with the formula, an impossible task in our case since the LLT calculation is limited to such low energies.
\section{Breakdown at higher energies}
\label{HigherEs}
\begin{figure}[htbp]
\begin{center}
\includegraphics[width=3in]{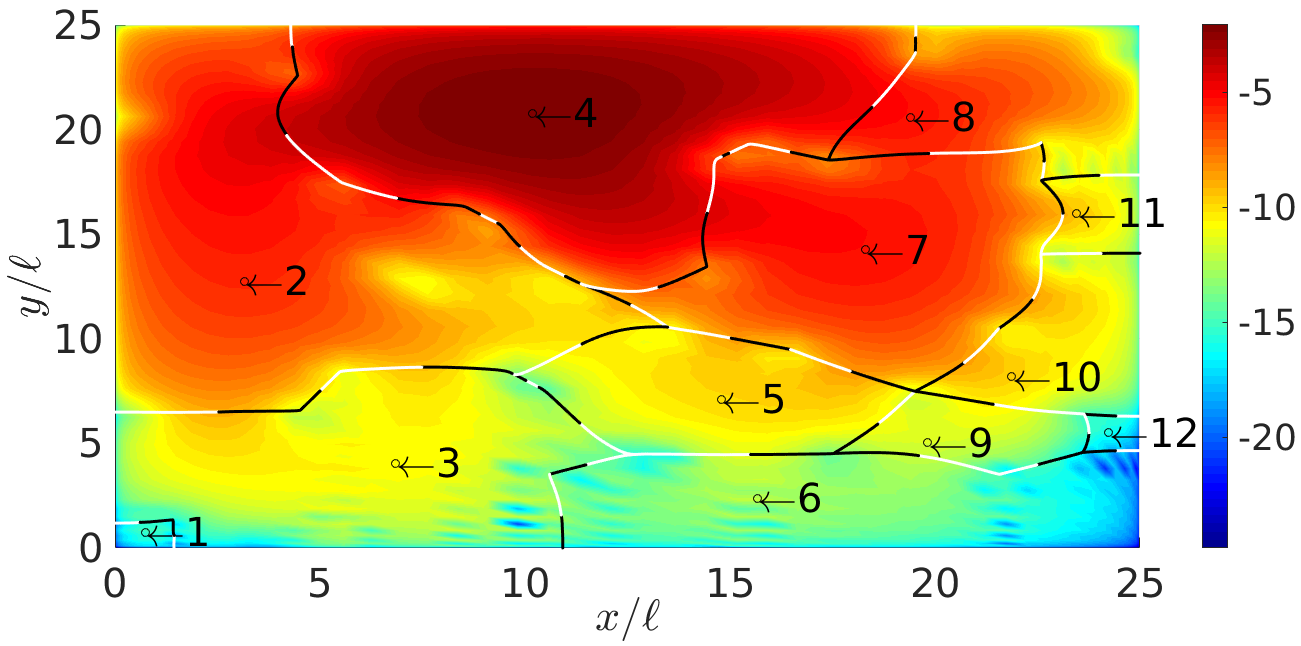}
\includegraphics[width=3in]{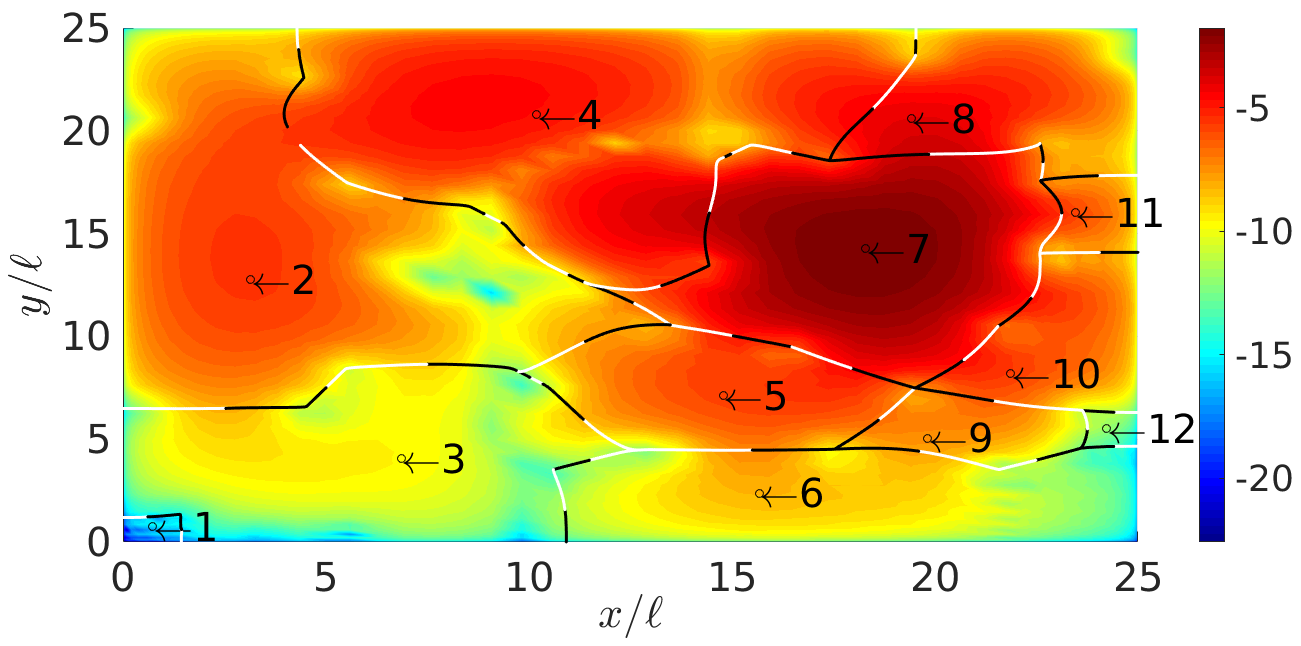}
\includegraphics[width=3in]{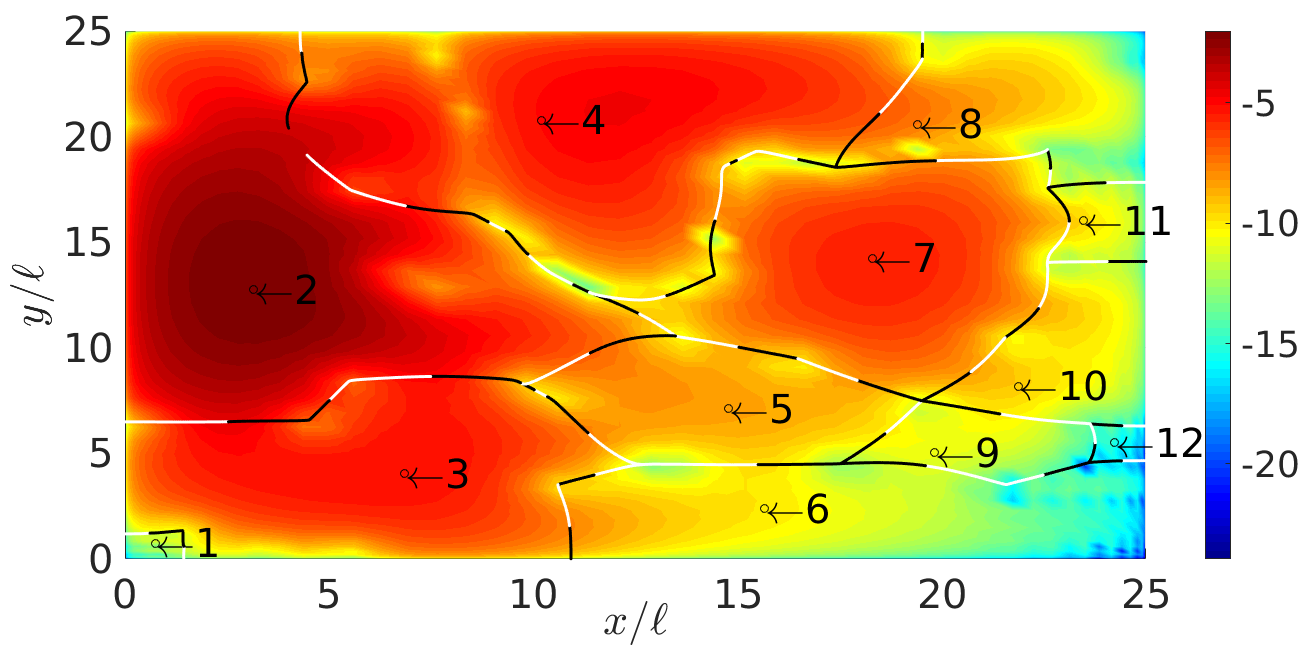}
\includegraphics[width=3in]{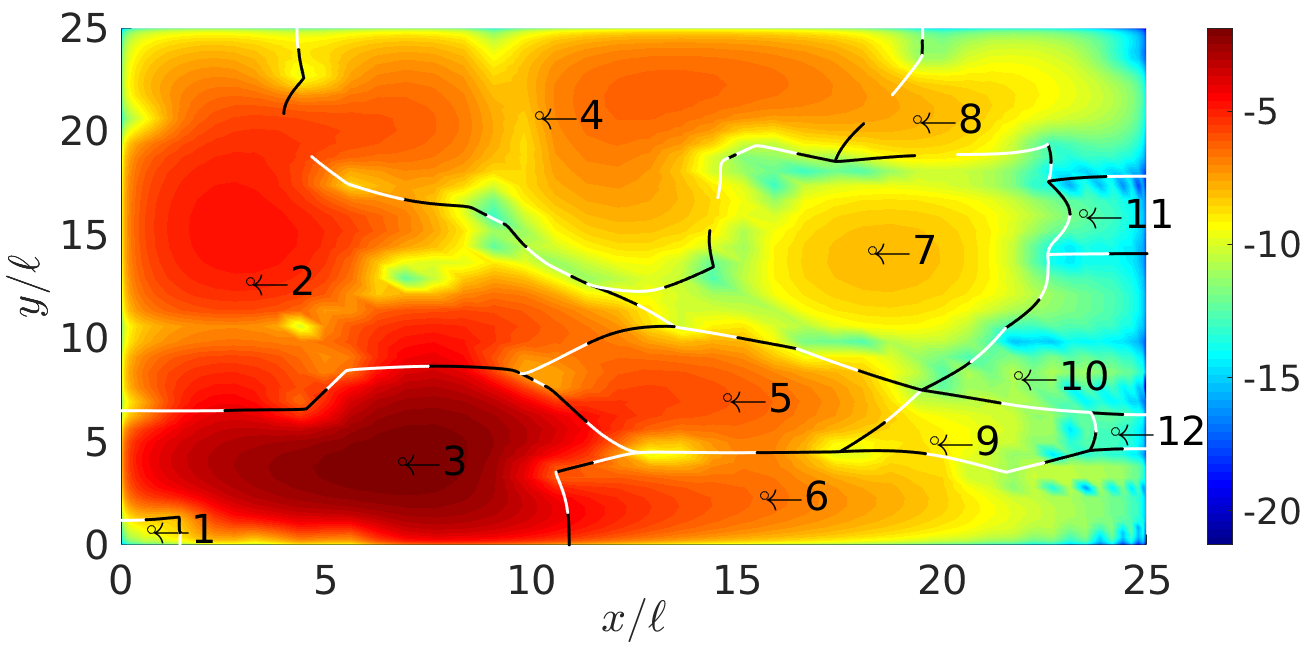}
\includegraphics[width=3in]{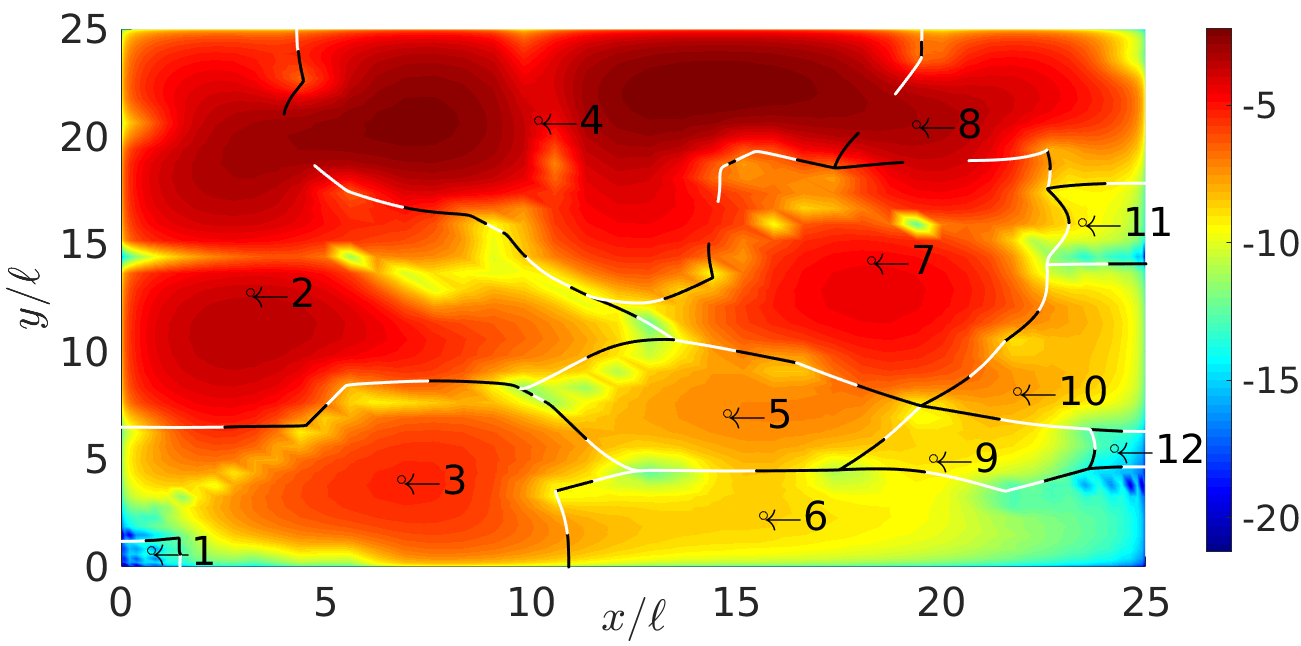}
\includegraphics[width=3in]{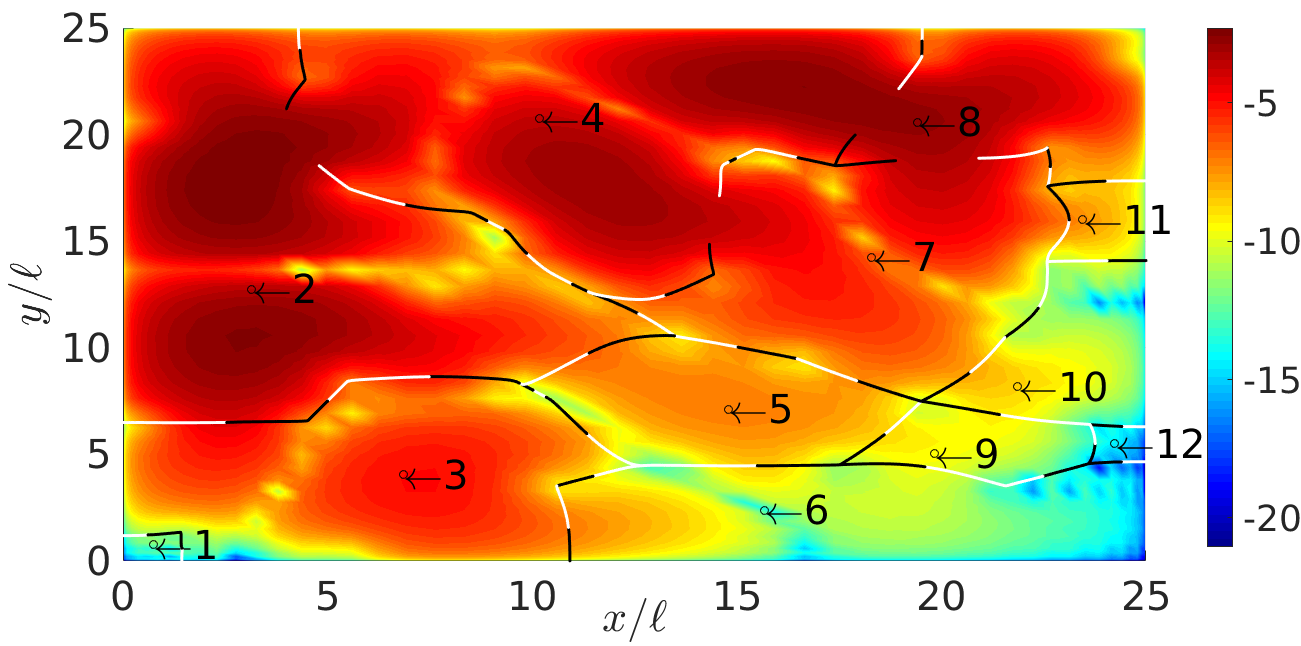}
\end{center}
\caption{\label{decay_mechanisms} Six of the lowest energy eigenstates (going across the panels and then down) for a single noise realisation in a system with $L=W=25\ell$, $V_0=21.33E_0$, $\sigma=0.48\ell$, $f=0.06$. The colour map depicts the logarithm of the amplitude of the wavefunction. Black and white lines (the different colours are used simply to make the structure of the valley network clearer) show the effective domain walls (where $E<1/u$) at the given eigenstate energies, having removed any open valley lines first. Maxima of $u$ are marked with open black circles and the corresponding domains are numbered for ease of reference. The eigenstates demonstrate the various possible mechanisms by which the wavefunction can spread across the system, other than by pure decay (see discussion in the text). Once these mechanisms come into effect, our calculation of the localisation length loses meaning; this constrains the regime of its applicability to very low energies.}
\end{figure}

As we have briefly mentioned in the previous section, the localisation length extracted from time-dependent simulations disagrees with our LLT prediction, even when we limit ourselves to sufficiently low energies where $\xi_E$ is smooth and monotonically increasing. In fact, the localisation length from time-dependent simulations is considerably larger, by up to as much as an order of magnitude. In this section we explain how and why this occurs, based on an analysis of the structure of the eigenstates, using Fig.~\ref{decay_mechanisms} for illustration. In particular, we find that the pure decay model (applicable for example to the first eigenstate in Fig.~\ref{decay_mechanisms}) we have assumed thus far ceases to be relevant beyond very low energies, and describe the mechanisms by which the wavefunction spreads out across the system that come into play at higher energies. These effects are beyond quantum tunnelling and its semiclassical approximation, violating the Agmon inequality (\ref{decay}), and are best ascribed to Anderson localisation directly.

We have already explained that as the energy increases, the valley lines of LLT cease to be effective and domain walls break open, as segments of the potential barriers between them are ``submerged''. When the breaks in the domain walls are small, one still sees some exponential decay through such walls (e.g.~third eigenstate in Fig.~\ref{decay_mechanisms}, decay from second to fourth domain), even though semiclassically (according to the formal Agmon distance), it is now possible to go across the barrier at no cost at all. In this low-energy regime, our use of $\bar{\rho}_E$ to capture the tunnelling \textit{and} base the domain area merging on its vanishing is sensible. However, as the gaps in the domain walls grow, it becomes common to have single-amplitude bumps extending between domains through these gaps (e.g.~sixth eigenstate in Fig.~\ref{decay_mechanisms}, between the fourth and eighth domains), and one can no longer talk of decay. In this regime, it would be better to use $\rho_E$ proper (which indicates that no tunnelling occurs) together with the criterion $\rho_E=0$ to merge domains. This is one mechanism that causes the true localisation length to be greater than the one we compute. Since there are many others (see below) that make a sensible calculation of $\xi_E$ at higher energies impossible anyway, we choose to persist with $\bar{\rho}_E$, which is the correct number to use at low energies.

A prominent, strongly dominant mechanism going beyond the pure tunnelling picture is what we shall term the ``seeded excitation'' scenario. Here, ordinary tunnelling from a strongly occupied domain into its neighbour excites a local mode inside that domain (usually manifesting as a separate bump), of an amplitude set by the decayed wavefunction in the ``receiving'' domain. Many examples of this can be seen in Fig.~\ref{decay_mechanisms}, with the lowest-energy case occurring in the second eigenstate, going from the seventh and eighth domains into the fourth, as well as the fourth to second (although here the local excitation and the original decayed amplitude are merged and it is the amplitude maximum in the second domain that is the tell-tale sign of seeding). The effect of seeded excitation is to strongly increase the weight of the eigenstate on the ``receiving'' domain (that is, increase the average value of the wavefunction on this domain), and as a result, decrease the decay coefficient between the domain pair in question.

Another mechanism that comes into play at higher energies is ``resonant excitation''. Occasionally, we find domains excited without any significant decay into them from other, strongly occupied domains (e.g.~seventh domain in the third and fourth eigenstates of Fig.~\ref{decay_mechanisms}). In such cases, the excitation is caused by ``resonance'' with a mode in a near-by occupied domain (in the examples provided, it is probably the fourth domain which is responsible). Note that such resonances can happen even between domains that are of considerably different areas, as long as higher modes are involved, so that the \textit{mode} energy is close. This scenario allows the eigenstates to cover a larger area without undergoing a decay. As energy increases and higher mode excitations become more prevalent, more and more resonances are possible as the range of available energies to match grows.

An interesting observation regarding resonant excitations is that the distance between the two domains in question is never very large (perhaps a gap of two or three domains at most), so that overall, the occupied domains are still clustered and the states are localised. A possible explanation may be that intermediate detuned domains reduce the coupling between the resonant domains, which only allows fairly local resonant excitations.

Clearly, higher-order modes (e.g.~Fig.~\ref{decay_mechanisms}, the fourth domain in the fifth eigenstate has a prominent node) are not accounted for in our description of section \ref{XiSaddles}, but this is not a serious problem, as usually, all the bumps within a single domain have similar amplitudes and the nodes between them do not reduce the mean value of the amplitude on the domain by much.

The final complicating factor is that even simple decay can occur from several nearest neighbours (e.g.~in the sixth eigenstate, the seventh domain gets a contribution from both the fourth and eighth domains), which implies that the mean value of the wavefunction on that domain will be greater than it would have been if only one such decay contributed to its population.

\begin{figure}[htbp]
\includegraphics[width=6in]{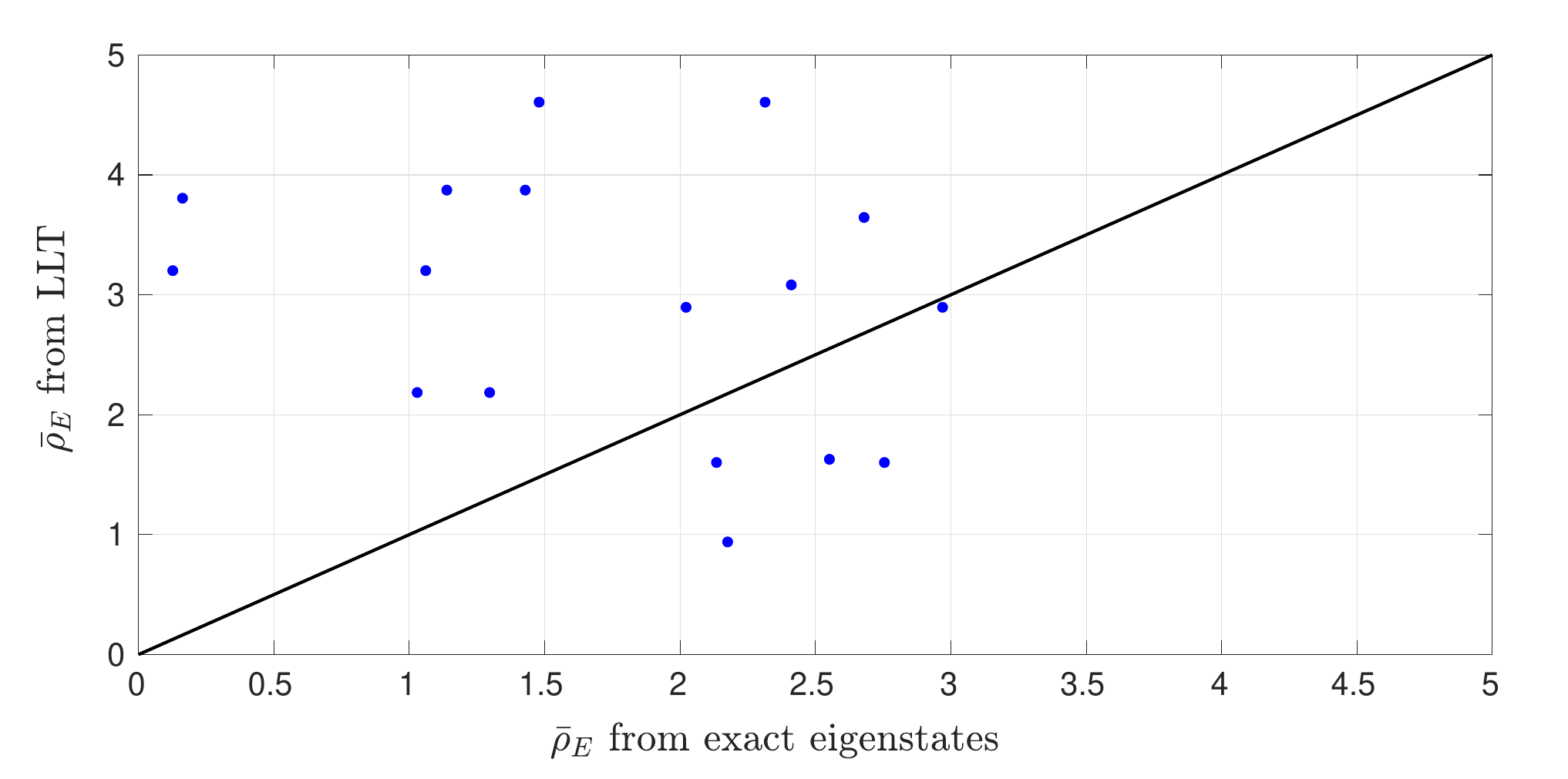}
\caption{\label{rhoCompBad} Exponential decay cost linking two neighbouring domains, plotting the values measured from exact eigenstates and LLT against each other. Data points were obtained for a system with $L=W=25\ell$, $V_0=21.33E_0$, $\sigma=0.48\ell$, $f=0.06$. Only cases that do \textit{not} fit the clean decay model tested in Fig.~\ref{rhoComp} were included in this test, and we see that overall, the LLT method yields a larger decay coefficient than the true value due to all the additional complex mechanisms, effective at higher energies, discussed in section \ref{HigherEs}.}
\end{figure}

All (but the last) of the factors outlined so far are beyond quantum tunnelling, violate the Agmon inequality (\ref{decay}), and should be thought of directly as quantum interference effects. They serve to increase the localisation length beyond the value calculated according to our LLT method, which is therefore only valid at very low energies, for maximally-localised states. This happens due to both larger effective distances separating decay events, which is fairly straight-forward to both understand and visualise, and weaker decay when such events do occur. The latter has been quantitatively confirmed by comparison to exact eigenstates (Fig.~\ref{rhoCompBad}), this time choosing domain pairs that do \textit{not} fit the pure decay model, but involve one or several of the more complex mechanisms discussed in this section. It is clear that LLT overestimates the decay coefficient, and the data shown can certainly accommodate the observed difference between LLT and time-dependent simulations in a regime where these mechanisms are prevalent. Considering the contribution from the larger effective area (compared to that assumed by the pure decay model of LLT) which also serves to increase the localisation length, these results are consistent with and explain why density profiles from time-dependent Schr\"odinger simulations indicate a larger localisation length than that predicted by LLT.

We note that it is possible to set the disorder strength so high that the pure decay model is applicable up to sufficiently high energies (confirmed by examining eigenstates) that a slowly translating Gaussian can fit in to the range of energies where our LLT calculation should show agreement with time-dependent simulations. We have done this test, but found that once again LLT underestimates the localisation length extracted from time-dependent simulations. For example, in a system with $L=25\ell, W = 25\ell, R = 30\ell$, noise parameters $f=0.2, \sigma = 0.48\ell, V_0=21.33E_0$, a translating Gaussian with $\bar{\sigma}=5\ell, k_0=0.5/\ell$, and total evolution time of $100t_0$, quasi-steady state is reached at $\sim 50t_0$, after which the fitted localisation length $\xi(t)$ stays roughly constant at the average value of $8.3\ell$. On the other hand, the localisation length predicted by LLT is $\langle \xi \rangle \approx 0.62\ell$, indicating much stronger localisation. In this case, the LLT calculation is valid over the entire range spanned by the energy distribution of the wavepacket used.

This can be explained by ``edge effects'': for the direct time-integration of the Schr\"odinger equation, one needs empty, noise free ``reservoirs'' on either side of the disordered system to initiate the wavefunction in and to collect any transmitted atoms. The coupling of the noisy region to the empty reservoirs modifies the valley network in the vicinity of those edges of the system, as the wavefunction is not forced to zero by boundary conditions, but is allowed to take on arbitrarily high values. In particular, localisation is weakened through the fact that some of the valley lines disappear upon addition of the reservoirs, and the barriers in the effective potential along the remaining valley lines are lower as $u$ near the ``coupling'' edges is higher. When the disorder is so strong that the (``internal system'') localisation length, as computed from LLT, is smaller than the typical domain size, these edge effects become important. Strong decay happens over one or two domains, and because in the time-dependent simulations, it is precisely the affected regions that the wavefunction passes through as the exponential decay sets in, we observe a larger localisation length than is found away from the edges. This explains why we still see a discrepancy in the regime where the LLT method should in principle work well.

Now, an interesting observation is that it is not only the lowest energy eigenstates in $H$ with $W_E$ conforming to the pure decay model which are similar to those in $H$ with $V$, but also a few modes beyond this regime, when quantum tunnelling in $W_E$ is already insufficient to capture the structure of the eigenstates. This can be seen in Fig.~\ref{CompSpec}, for the second to fifth eigenstates. The reason that a few low energy modes are similar even in the regime where the tunnelling model is no longer applicable is simply the similarity of the two potentials, $V$ and $W_E$, the latter being a smoothed version of the former. While the correlation between the eigenstates in the two landscapes is lost at higher energies, all the mechanisms discussed in this section can also be seen in the eigenstates of $H$ with $W_E$ beyond that point.

The understanding that the pure tunnelling picture in LLT is only applicable at very low energies is novel, and establishes when and how one can use LLT in a useful manner. Two very recent papers \cite{LLT_Dirac, LLT_highE} have developed generalisations of LLT to allow treatment of systems with internal degrees of freedom, with \cite{LLT_highE} explicitly extending their technique to arbitrary energies, while the method in \cite{LLT_Dirac} is amenable to such an extension \cite{LLT_highE}. These generalisations of course come at the price of added complexity, but have additional advantages as well: for example, the method of \cite{LLT_highE} removes the constraint that the physical potential cannot be negative on any part of the system domain, and yields some helpful features arising from the different normalisation of the eigenstates chosen therein. On the other hand, it should be noted that the paper \cite{LLT_highE} has only presented examples of their calculation performed at zero energy.
\section{Multidimensional tunnelling}
\label{MultiDimTun}
The Agmon distance of LLT (\ref{Agmon}), including minimisation over all paths connecting the two points in space, gives a prescription to predict the minimal decay of eigenstates through the barriers of $W_E$ as they tunnel out of each domain -- a local potential well -- and spread across the system. In section \ref{XiSaddles} we have heuristically outlined and tested a method to quantitatively estimate $\rho_E$ between neighbouring domain minima of $W_E$, avoiding the path minimisation stage, but using the usual expression for the integrand along the path.

Multidimensional tunnelling is in fact an old and thoroughly-investigated problem. Of course, brute force quantum mechanical calculations are possible, but physicists have been striving to obtain \textit{insight} into the process by generalising the WKB approximation to dimensions higher than one to describe it. In 1D, WKB is a straight-forward and methodical approach (see, e.g., \cite{Knudson}) -- a controlled approximation that is fully understood. The generalisation to several dimensions is a different matter entirely: there is a large body of literature developing and discussing different methods, their limitations, suggesting improvements, and utilising these techniques to solve practical problems. In this section, we will provide an overview of this topic, to place our method of section \ref{XiSaddles} in perspective.

Let us see where the result (\ref{Agmon}) comes from. The starting point of the derivation is usually the Feynman propagator, none other than the Green's function of the system. One has to go through a series of approximations, listed below, in order to arrive at this semiclassical formalism:
\begin{enumerate}
\item The propagator is expanded in powers of $\hbar$, and only the zeroth order term is retained\footnote{An equivalent approach is to write the wavefunction in polar form and expand the phase similarly.} \cite{Kazes1990,DasMahanty}.
\item Next, one usually assumes that Hamilton's principle function is pure imaginary \cite{Kazes1990,Gregory1991,Takada1994}.
\item In principle, if we want to use the Feynman propagator to describe tunnelling from one region of space where the wavefunction is initially contained to another, we must consider all source points, all target points, and all possible paths to arrive from each source to each target point. In the simplest approximation, one uses the fact that the contribution of the classical path is the largest, and as we move away from it in configuration space, the contribution of the other paths is exponentially suppressed. Therefore, one usually only examines the classical path, or at most a ``tube'' of paths around the classical one. Moreover, it is common to only consider one source point (at which the wavefunction is maximal) and one target point (say the minimum in the potential on the other side of the barrier). The classical trajectory method was developed and used in many papers, e.g.~\cite{BBW1,BBW2,Coleman1977,DasMahanty}, and relies on minimising the action via the Euler-Lagrange equations.
\end{enumerate}

Assumption 1 is already a strong limitation, and to the best of our knowledge, first order solutions were only ever obtained in the classically allowed region \cite{Kazes1990}. However, taking $\hbar\rightarrow0$ is the essence of the semiclassical nature of the method, and not much can be practically done to overcome this approximation.

Assumption 2 is certainly not generally justified \cite{Kazes1990,Gregory1991,Takada1994}. These three references have superbly dealt with the case of a general complex action, and demonstrated that a geometrical ray construction, following two surfaces (equi-phase and equi-amplitude) along two orthogonal paths, is necessary to solve the problem in earnest. They have proven that the imaginary action approximation breaks down if one considers a general incoming wavefunction, incident on a barrier such that its $k$-vector is arbitrarily predetermined. It has also been argued that this approximation can even fail for tunnelling out of a potential well \cite{Takada1994}. The geometrical construction proposed in these papers is extremely involved, and is completely impractical for our purposes.

While in principle, accuracy could be improved by including more than one source and target point, as well as considering multiple paths as in \cite{DasMahanty}, all three simplifications of the third assumption are essential for our case: we cannot afford (computationally) to calculate many paths or to describe each domain by anything more than the point at which $W_E$ attains its minimum. This is because the calculation needs to be done so \textit{many} times that it is simply impractical.

The usual final form of the semiclassical approximation in the forbidden region involves solving the classical equations of motion with negative the potential and the energy, or equivalently, in imaginary time. The differential equations are based on Newton's laws, imposing energy conservation as a constraint, and seek out the path of minimal action. In 2D, they take the form
\begin{eqnarray}
\frac{d^2x}{ds^2} &=& \frac{ \frac{\partial V}{\partial x} - \frac{dx}{ds}\left( \frac{\partial V}{\partial x}\frac{dx}{ds} + \frac{\partial V}{\partial y}\frac{dy}{ds} \right) }{2(V-E)},\nonumber\\
\frac{d^2y}{ds^2} &=& \frac{ \frac{\partial V}{\partial y} - \frac{dy}{ds}\left( \frac{\partial V}{\partial x}\frac{dx}{ds} + \frac{\partial V}{\partial y}\frac{dy}{ds} \right) }{2(V-E)}.
\end{eqnarray}
Here, $s$ is the arc length along the path defined by the coordinates $(x,y)$, and $V$ is the potential the particle with energy $E$ moves in. In this parametric form, the equations for the allowed and forbidden regions are identical.

In the context of tunnelling out of a potential well, the trajectory is usually required to pass through the turning surface (where kinetic energy vanishes) normally, so that it can connect smoothly to a classical trajectory in the allowed region. On the turning surface, the velocity is aligned along the gradient of the potential \cite{Coleman1977,DasMahanty}. An alternative constraint was used in \cite{BBW1}: the authors required their escape paths to pass through the saddles of the potential and be aligned along the correct axis of the saddle at those points (which is closer in spirit to our approach, but is less rigorous). Essentially, if the direction of the incoming wave is predetermined and it impinges on the turning surface at any angle other than normally, the action must be taken as complex and the classical equations are insufficient. This is the chief difference between tunnelling out of a local well and the transmission of an incoming wave through a barrier.

We highlight that in the final form of the semiclassical approximation (\ref{Agmon}), the minimal path is energy-dependent: one must solve the set of ordinary differential equations defining the minimal path for each energy separately. If we wish to find the classical path that connects two specific points, knowledge of the energy gives us the magnitude of the velocity vector, but its direction is unknown. Trial and error is called for to discover the latter: one needs to try different initial directions of motion until a path that arrives at the desired end point is found. Furthermore, if the two points of interest are separated by one or more turning surfaces (which cannot be crossed classically), one must begin at each of the two points, and try different directions until a trajectory that hits the turning surface normally and is reflected back on to itself is found. In order to connect points lying on turning surfaces, in principle, the initial direction of the velocity can be found from the gradient of the potential, but in practice, fine-tuning is still necessary. Thus, in general, finding the true classical path is a piece-wise process and takes multiple rounds of guessing the direction of the velocity. This makes the traditional (and formally correct) solution of the semiclassical problem (\ref{Agmon}) impractical for our purposes.

Our method of section \ref{XiSaddles} overcomes these difficulties: no differential equations need to be solved at all (one only needs to know the localisation landscape $u$), one path is computed for all energies, and there is no need to guess the initial condition. As we have seen in Table \ref{AgmonTest}, it performs well, which justifies its use despite the many approximations in deriving the semiclassical formulation, as well as our heuristic way of computing the escape paths. In either case, no other level of approximation is practical for our purposes, as we need to compute the Agmon distance between every two neighbouring domains at all energies for many noise realisations (twenty are used in practice), at each set of parameters investigated.

The additional discovery that by averaging over all candidate paths of LLT, the mean Agmon distance $\bar{\rho}_E$ in fact yields the average decay rate (rather than the minimal) is a further point of merit to our approach. The other candidate paths cannot be obtained from the rigorous semiclassical formalism, which so far has only been able to provide a lower bound on the decay rate.

A few final notes are in order, without which any review of multidimensional tunnelling would be incomplete. References \cite{Auerbach1984,Auerbach1985} have developed the path decomposition expansion method, which allows one to divide space into separate regions, minimise the action in each one using whatever method happens to be optimal in that vicinity (chosen based on physical considerations), and then collate the solutions using global consistency equations. Reference \cite{DasMahanty} deserves special attention, as an exceptional effort was made to consider many classical paths from many source points, deriving the tunnelling current and transmission coefficient through the potential barrier.

For a more comprehensive review of the topic, the reader is referred to \cite{GoodReview}, as well as the original literature cited above.
\section{The question of transport}
\label{Transport}
The fact that an Anderson-localised system possesses localised eigenstates has profound implications on transport properties, constituting an experimentally-accessible handle to probe the nature of the system. It is therefore important to understand this aspect of the physics, a goal we address in the next section, with the relevant literature reviewed here as a form of introduction to the topic.

Recall that in section \ref{Intro}, we have glimpsed the extensive literature on the experimental detection of Anderson localisation. When it comes to classical systems involving, e.g., light and sound waves, it is natural to transmit the wave through the disordered sample and detect the outgoing signal \cite{Wiersma, Scheffold, Storzer2006, Hu}. However, a common problem with this approach is that absorption of the wave by the medium reduces the output intensity, and because nothing \textit{but} the outgoing intensity can be measured, it is difficult to separate the effect of absorption from that of localisation \cite{Wiersma, Scheffold}. One proposed solution to this is to instead examine the transverse spreading of the ``beam'' at the sample output \cite{deVries}, a method that was successfully employed in \cite{Sperling, Lahini2008, Segev2007}. Now, in the setting of ultracold atoms, one can measure the density \textit{everywhere}, including inside the ``sample'' (in fact a disordered potential), and while there is certainly some loss of atoms with time, one can easily differentiate whether the atoms have been lost from the system (and are no longer detectable by the imaging procedure) or are ``stuck'' inside the disorder where they can be measured and shown to accumulate. The goal is then to prove that this accumulation occurs due to interference effects, as opposed to classical trapping in the potential which may have local wells deeper than the atomic energy.

The question of transport through a disordered sample in the localised regime has been studied thoroughly in the literature. It is widely accepted that the conductance (see \cite{russian_guys, Heinrichs, Sheng, Greek_review} for a general discussion) through a sample is proportional to $\exp(-2L/\xi)$, where $L$ is the length of the system and $\xi$ the localisation length \cite{Kunz, Kirkman, Heinrichs, Fan, LeeFisher, finite_scaling, Greek_review, Sheng, Hilke2}. The conductance can be directly expressed through the quantum-mechanical transmission coefficient \cite{Kirkman, finite_scaling, Greek_review, Sheng, Hilke2}, or through the transmission matrix in the case of multiple conduction channels \cite{Heinrichs, LeeFisher, Chinese, Greek_review, deSterke, Sheng}. The conductivity can also be expressed through the Kubo-Greenwood formula \cite{Thouless2, Fan, ScalingTheory, LeeFisher}, and has been evaluated explicitly in \cite{Fan, LeeFisher, Chinese}. The closely-related dimensionless conductivity (``Thouless number'') has been numerically computed in \cite{finite_scaling, Sheng, deSterke, Kaiser, Chinese}, often by the transfer matrix method, as well as the conductance \cite{Hilke2, Aspect2017, topological} and the transmission coefficient \cite{russian_guys, Hilke, Aspect2017, topological}. Explicit finite-size effects on the conductivity have been numerically demonstrated in \cite{Fan, Chinese, topological, 2D_corr, deSterke}, in support of the famous scaling arguments of Anderson localisation, specifically predicting the behaviour of the dimensionless conductivity \cite{ScalingTheory, Sheng, Vollhardt, Greek_review, LeeFisher, finite_scaling, deSterke}.

A diffusive picture of transport for Anderson localisation has been thoroughly developed, from the classical description, to renormalised diffusion in the weakly-localised regime \cite{Sheng, Vollhardt}, to a self-consistent formulation of diffusion \cite{Sheng, Vollhardt, Greek_review, Ono}, which has been shown to perform well also with stronger scattering. A less complicated concept, the semiclassical conductivity, has occasionally been of interest \cite{Fan, Thouless2}, as well as a many-body generalisation of the quantum-mechanical conductivity in an interacting system \cite{Mirlin}.

Thus, overall, the idea of transport \textit{through} a disordered system is not new, neither from the theoretical nor the experimental points of view. However, for the cold-atom setting, it was quite a novel approach, first proposed in \cite{Aspect2017} for a 1D system, tested experimentally in \cite{Berthet}, and in parallel, experimentally realised in 2D \cite{BS}.

Traditionally, e.g.~in electronic or light systems, a transmission experiment involves examining the steady-state current at the output of the noisy potential. This is made possible by the fact that such systems are macroscopic, essentially coupled to infinite reservoirs on either side of the noise, and equilibrium is attained fairly quickly on experimental time-scales. When it comes to cold atoms, the typical system size is considerably smaller, the number of atoms in the system is finite and not overly large, and total experimental times are quite short compared to the time-scales of the dynamics. In the macroscopic systems, measuring the output transmission as a function of system size has long been believed to yield an exponentially decaying curve, with the length scale given by the localisation length, $\xi_E$. With ultra-cold atoms, it is not always possible to reach a steady-state transmission (at least in the specific scenario we will consider), but the initial particle current exiting the disordered potential is readily accessible.

Using translating Gaussian wavepackets, we will show that the initial flow rate out of the disordered potential also decays exponentially with system size, but its governing length scale is only correlated to the localisation length, not equal to it, being in general smaller. For our system of randomly-positioned Gaussian scatterers in 2D, the true localisation length at the energies of interest can only be (accurately) obtained through time-dependent simulations, initiating fairly narrow Gaussian superpositions outside the disorder, allowing them to transmit through a very long section of the noisy potential, and extracting the decay rate of the evolving density profile as a function of time. The time taken for the apparent localisation length to stop changing is usually very long. Therefore, while the true localisation length can be accessed in such experiments, it requires very large systems and very long experimental times. The initial flow rate, however, requires short segments of the potential (in comparison), and short evolution times, as we are interested in the initial dynamics, when the atoms first exit the disorder. 

Furthermore, the decay length-scale of the flow rate (which we shall call $\bar{\xi}$) is unambiguously defined and easily extracted from simulations (or measurements), in contrast to the localisation length, which depends on which segment of the density profile is chosen for the exponential fit -- a time-dependent question -- and how one extrapolates the extracted localisation length to the infinite-time limit. The resulting number, $\bar{\xi}$, quantifies the strength of localisation in the sense that it is qualitatively correlated with the true localisation length, but is usually smaller than $\xi_E$.

The use of reasonably narrow (in momentum space) Gaussian superpositions approximates a monochromatic plane wave, and allows us to neglect averaging over the energy distribution when making predictions for the entire wavepacket, as the refinement to the results evaluated at the central energy is very minor. It also allows one to easily tune the energy of the wavepacket by changing the central momentum value, thus enabling an energy-resolved measurement of localisation properties. It is, however, imperative to initially place the wavepacket in a region free from potential scatterers in order for the free-system energy distribution to apply. If the same Gaussian wavefunction is placed inside the disorder, the extra energy arising from the disordered potential shifts the entire energy distribution to higher energies and thereby completely changes (weakens) the observed localisation properties. This implies that the noise-filled region must have empty ``reservoirs'' on either side of it, first to hold the atoms before they go through the disorder, and then to collect them once they emerge. The effect of adding such reservoirs is to weaken localisation in a fairly narrow region along the edges of the system that open on to the reservoirs (as already mentioned in section \ref{XiSaddles}). If the localisation length is very short or the system length is very small, these can be noticeable or even dominant, but if the localisation length and the system length are considerably larger than the regions affected by these edge effects, then they are negligible.
\section{Transmission scenario}
\label{Trans}
In this section we continue investigating what can be learned about Anderson localisation by passing incident wavepackets through the disorder and examining the transmitted signal, using direct time-dependent simulations.

The idea of using transmission to search for localisation in 2D, in the context of ultra-cold atoms, was pioneered by the experimental study \cite{BS}. Here, a Bose-Einstein condensate (BEC) was prepared in one reservoir of a dumbbell-shaped potential, and allowed to expand through a channel filled with randomly placed potential scatterers, eventually arriving at the second reservoir. While the system was set up in a transmissive configuration, the wavefunction did not have CoM translation, and as such was purely expanding. We will examine such a scenario in section \ref{Expansion}. In addition, the 2D plane was tilted so that the atoms experienced acceleration, and interactions were not tuned away, so that a single particle picture would be inapplicable, at least for early times. The effect of these two factors is explored in section \ref{Secondary}.

Here we will consider a much cleaner scenario: a transmissive set up with a translating wavepacket. Plane waves would of course be ideal candidates for probing the disorder because they would allow us to resolve the energy dependence in detail, but this is not realistically possible as it implies the wavefunction must have infinite extent in position space. Fortunately, it is not necessary to realise true plane waves: Gaussian wavepackets that have a fairly narrow momentum distribution are equally useful. To generate these, one must simply have a large Gaussian cloud in real space, which is not difficult. In an experiment, one could use a Feshbach resonance to tune interactions to zero, and create an initial BEC in the ground state of a weak harmonic trap, which would then be transferred into the 2D trap by adiabatic ramping of potentials, as in \cite{BS}. A Bragg pulse could then be used to impart momentum to the cloud, making the entire wavepacket move at a constant velocity through the system. The momentum transferred can be finely tuned by varying the angle between the Bragg beams. In \cite{Berthet}, the authors have instead boosted their non-interacting wavepacket to a finite velocity by allowing it to accelerate in a given linear potential which is switched off when the cloud reaches the noisy potential.

The proposal in the previous paragraph would of course create a 2D Gaussian wavefunction, which would be a very natural object to study experimentally. In our theoretical work here, however, we will use a rectangular system, as described above equation (\ref{1DG}), with empty reservoirs of length $R$ added on to each side of the usual region filled with noise (length $L$ and width $W$), from now on referred to as the ``channel''. Considering the Cartesian symmetry of the system, it is more natural for us to use a 1D Gaussian, given by (\ref{1DG}), to probe the disorder. Experimentally, the scheme outlined above could be easily adapted to create a 1D Gaussian: one must simply (adiabatically) turn off the harmonic trap in the transverse direction after loading the atoms into the 2D trap and turning on the SLM potential (which in this case would confine the atoms to the rectangular domain modelled here), but before the Bragg pulse and before the longitudinal harmonic trap is switched off.
\subsection{Compartment populations and the flow rate}
Thus, a 1D Gaussian wavepacket with some CoM translation is initiated in $R_1$ and transmits through the disordered potential to $R_2$ (see appendix \ref{apptdep} for details on implementation). We now propose an observable that can be easily measured in this scenario, highlight its advantages and disadvantages, and discuss its physical meaning. Define the normalised populations of the three compartments, $r_1$ (first reservoir), $c$ (channel) and $r_2$ (second reservoir), as the atom number in each compartment divided by the total number of atoms in the system. As a function of time, $r_1$ will begin from one, decrease as atoms pass from $R_1$ into the channel, causing $c$ to rise from zero, and eventually, as atoms arrive at the second reservoir, $r_2$ will increase from zero as well at the expense of $c$ diminishing.

\begin{figure}[htbp]
\includegraphics[width=6in]{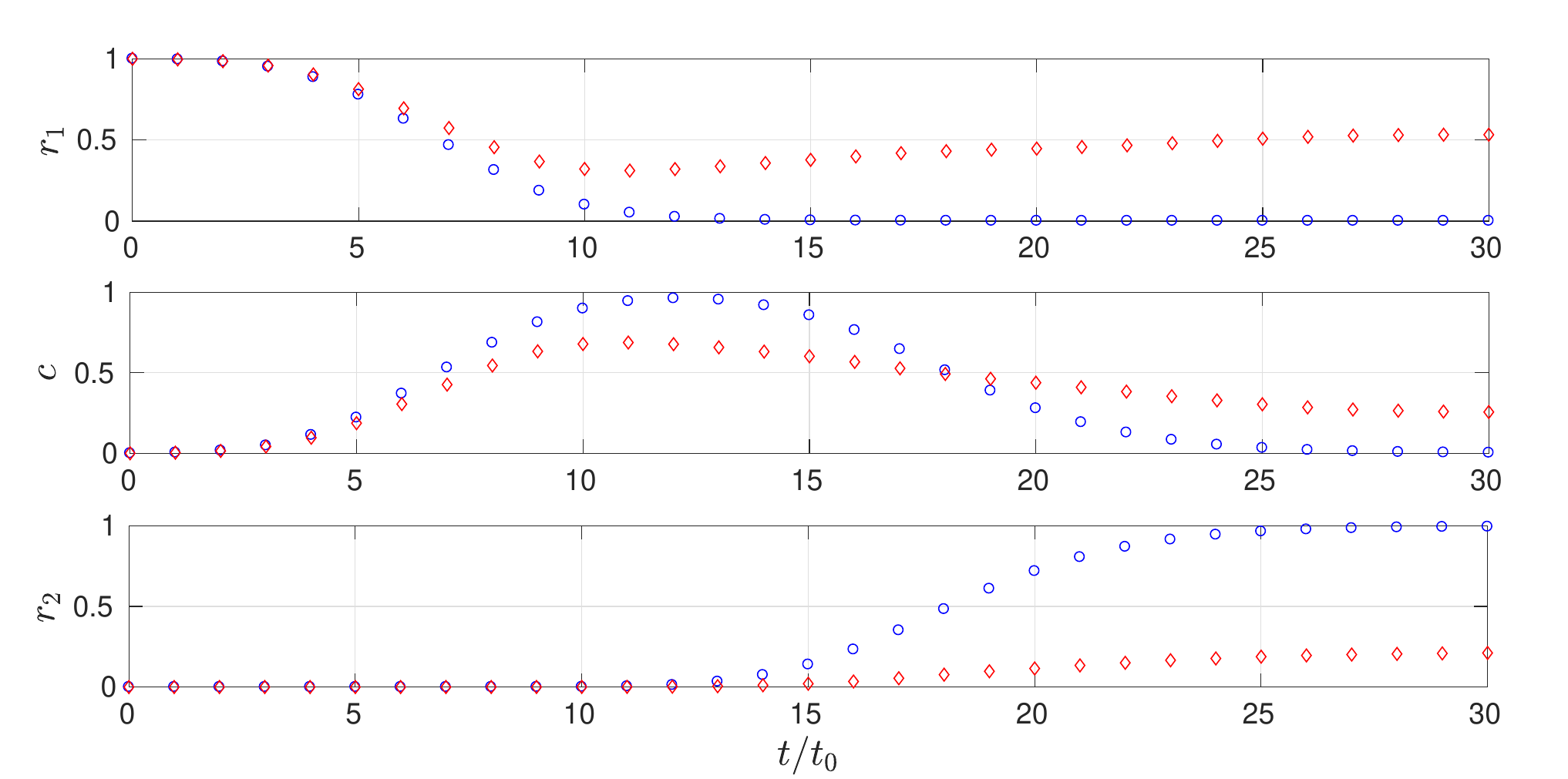}
\caption{\label{eg_pops} Normalised populations of the three compartments -- the two reservoirs and the channel -- when a 1D Gaussian wavepacket transmits from (the centre of) $R_1$ to $R_2$ through an empty channel (blue circles) and in the presence of potential scatterers with $f=0.05$ (red diamonds). Parameters used are $L=W=25\ell$, $R=30\ell$, $\sigma=\ell/2$, $V_0=5 E_0$, $\bar{\sigma}=5\ell$, $k_0=1/\ell$. The maximal rate of growth of $r_2$ is coined the ``flow rate'', denoted by $\rho$, and is our observable of choice.}
\end{figure}

If the channel is empty, all the atoms get through to $R_2$, moving as a slowly spreading lump at constant speed. The population dynamics for this case are shown in Fig.~\ref{eg_pops}, and it is evident that there is a nicely linear segment in $r_2(t)$, as it increases from zero and settles into a temporary equilibrium (the entire cloud is in $R_2$ by $t/t_0=30$). Of course, as time goes on, the wavefunction reflects off the right hand side of $R_2$, reverses direction and travels back into the channel. This latter behaviour is more complicated, usually involving two-way transport to and from $R_2$, and is thus to be avoided for the purpose of making quantitative measurements. Instead, we measure the maximal rate of the initial rise of the $r_2(t)$ curve, and denote this quantity $\rho$.

Figure \ref{eg_pops} also shows an example of the population dynamics in the presence of noise. For this illustration, we have chosen parameters to ensure the incident wave experiences Anderson localisation, but system size and localisation length are such that a visible portion of the atoms arrive at $R_2$. As is immediately evident, the flow rate is strongly reduced. This is due to several factors. First, note that when potential scatterers are present in the channel, the Gaussian wavepacket no longer travels as a single lump, but breaks up and smears out over the system as a result of multiple scattering events. Now, some of the incoming wave is reflected off the scatterers at the entrance to the channel and travels back to $R_1$. Out of the portion of atoms that enter the channel, some get localised and remain ``stuck'' in the channel. Some atoms transmit through to $R_2$, and it is this fraction that is detected by our measurement.

The flow rate out of the channel, $\rho$, is determined by the product two quantities: the linear density (having integrated over the width) of the atoms at the channel-to-second-reservoir interface, and the velocity of the atoms at that point. It is simple to confirm that almost all of the initial energy remains in kinetic form as the atoms move down the channel, which suggests that the presence of the scatterers does not cause a serious change in the atomic speed distribution. The density, on the other hand, is exponentially suppressed under Anderson localisation. A reduction in the flow rate can thus be directly linked to the density of the atoms at the end of the channel.

We measure $\rho$ as the gradient of a fitted straight line to the segment of $r_2(t)$ that sees the atoms flow into the drain reservoir as a one-way process, before flow back into the channel upon reflection at the far end sets in. For short channels, this translates to the maximal slope of the initial rise of $r_2$. For long channels, $r_2$ keeps on growing with time, and the slope may well increase as time goes on. In these cases, one must examine the spatially resolved density in $R_2$ to identify the correct time interval to fit. While this may not be possible in all experimental settings (e.g.~in systems of light or electrons), it is readily doable with cold atoms, and seeing the particle influx in the density is essential: $r_2(t)$ alone does not suffice to correctly decide which time interval to choose.

While the flow rate reflects on the density, a decrease in the flow rate does not necessarily report on localisation -- other factors can contribute to a reduction in the density at the channel output. A translating Gaussian wavepacket in an empty channel may be naively expected to have a channel-length independent flow rate. In practice, a roughly linear and non-negligible decrease is seen in $\rho(L)$ with no potential scatterers (see Fig.~\ref{rho_of_L}). This is due to the spreading of the wavepacket. As time goes on and the cloud moves further down the channel, the atomic density becomes more diffuse. Thus the rate of matter influx into the second reservoir is reduced (it takes longer for the entire cloud to move in to $R_2$). This effect arises from the density, and is ``real''.

One might consider normalising this dependence out to isolate the effect of Anderson localisation in the presence of a noisy potential, but that makes little sense: in the latter case, the wavefunction breaks up and moves through the system in a completely different way, so this empty-channel behaviour is not embedded in the noisy results and thus cannot be taken out. We just need to compromise on the fact that $\rho(L)$ falls roughly linearly even without noise -- this is the price we pay for not being able to use plane waves directly. This empty-channel effect is much more dramatic for a purely expanding wavefunction, such as that studied in section \ref{Expansion}, for understandable reasons. Therefore, we must always compare the flow rate to that in an empty channel to detect localisation effects. Meanwhile, classical trapping should be excluded by comparison to a channel filled with ordered scatterers, as discussed in section \ref{Ordered}.

Some of the advantages of using $\rho$ to quantify transport are that it is easy to measure, both from direct theoretical simulations and experimental results, and it has a high signal-to-noise ratio (because early $r_2$ dynamics provide a clean, isolated signal on a null background, essentially). Only a linear fit is required to extract $\rho$, which is numerically robust and keeps data processing artefacts to a minimum.

On the other hand, the flow rate is only a useful observable when $\xi_E$ is considerably greater than the scatterer size (constraining ourselves to point-like disorder). This is because we need to use system lengths such that an observable transmission into $R_2$ takes place: one should ``sample'' the density profile that would result in a very large system using several finite system lengths, $L$. Each finite system of length $L$ used should contain a reasonable number of scatterers to ensure that different noise realisations are significantly different, and that we do not need to average over too many realisations.

In addition, a complicating factor in the interpretation of the flow rate is that in the transmission scenario, some of the wavefunction is reflected from the noisy potential at the entrance to the channel. The value of the reflection coefficient depends on the strength of the noise (both the scatterer height and the fill factor), so whenever examining $\rho$ as a function of $f$ or $V_0$ it is important to normalise $\rho$ by the flow rate into the channel to isolate the effects of localisation.

The physical meaning of the flow rate is also clear: $\rho$ is proportional to the transmission coefficient of the channel. If we had infinitely long reservoirs and could run the experiment or simulation for a very long time, we could measure the transmission coefficient directly, by waiting until the population of $R_2$ stops rising. The value to which it would equilibrate would then be the transmission coefficient. But since time is limited and the reservoirs are very finite, this cannot be done: the wave reflects off the end of $R_2$ and comes back into the channel, after which it is no longer a one-way process. However, the initial slope of $r_2$ is clearly proportional to the final value to which it would rise, and a reasonably short segment of $r_2(t)$ is sufficient to obtain the slope.

To summarise, the flow rate $\rho$ carries information about the atoms that go through the disorder, and by inference, those that do not -- those that are ``stuck'' in the channel due to Anderson localisation. The other important quantity that we will examine is the density profile of the wavefunction in the channel, which is readily accessible both in theory and in cold atom experiments.
\subsection{Flow rate length scale}
The flow rate observable can be used to quantify the strength of localisation by extracting a length scale, $\bar{\xi}$, associated with its exponential fall-off with system length $L$. We will show that this number is qualitatively correlated with the localisation length, $\xi_E$, which can be obtained by extrapolating the decay rate of the density profile in a very large system to the infinite-time limit. However, we should explain from the outset that these two numbers cannot be expected to be equal. While the flow rate is closely connected to the density profile, it is measured at the initial surge of the atoms into the drain reservoir, while the localisation length is determined by allowing the system to relax to steady state, so the times at which the two measurements are made are different. Furthermore, the flow rate does not sample the density profile at any one time, as by varying $L$, we change the time taken for the atoms to reach $R_2$. This implies that $\bar{\xi}$ does not have to be equal to the observed localisation length in the density $\xi(t)$ at \textit{any} time $t$.

Thus, for disordered systems governed by Anderson localisation, we hypothesise that the flow rate decays as
\begin{equation}
\label{rho}
\rho \propto \exp(-2L/\bar{\xi}),
\end{equation}
for a quasi-monochromatic probing wave, but we will find that reasonably narrow Gaussian superpositions also obey this equation. The length scale $\bar{\xi}$ would depend on all the properties of the disorder, and will have a weak dependence on the channel width due to finite size effects, as discussed in section \ref{WidthDep}.

Note that we are assuming that the flow \textit{into} the channel is constant (up to reflection off the noise), as is indeed the case in our rectangular system. In a dumbbell geometry as was used in \cite{BS}, the flow in would also depend on $W$, as the reservoir and the atomic cloud are larger than the entrance to the channel. In order to make our discussion here applicable to the dumbbell system, one needs to normalise the flow out of the dumbbell channel by the rate of flow in, which will remove the explicit dependence on $W$.

Now, usually, the wavefunction probing the disorder is not monochromatic, but has an energy distribution, as in \cite{BS}. Our 1D Gaussian wavepackets also have a non-vanishing width in momentum space which ideally should be accounted for, but we will choose a fairly narrow superposition so that a single-energy description is not a bad approximation.

Recall that it is very important to average $\rho$ from time-dependent simulations over noise realisations. The shot-to-shot fluctuations in the strongly-localised regime are very high, so that we use 20 realisations to get accurate results. Typically, the standard deviation of the flow rate measurements is very much of the same order of magnitude as the mean. Note that high fluctuations in the transmission from a strongly-localised system have been independently found by other researchers \cite{Chabanov, Hu, Kaiser, DelandeLectures} and even put forward as a ``smoking-gun'' of Anderson localisation.

\begin{figure}[htbp]
\begin{center}
{\includegraphics[width=4in]{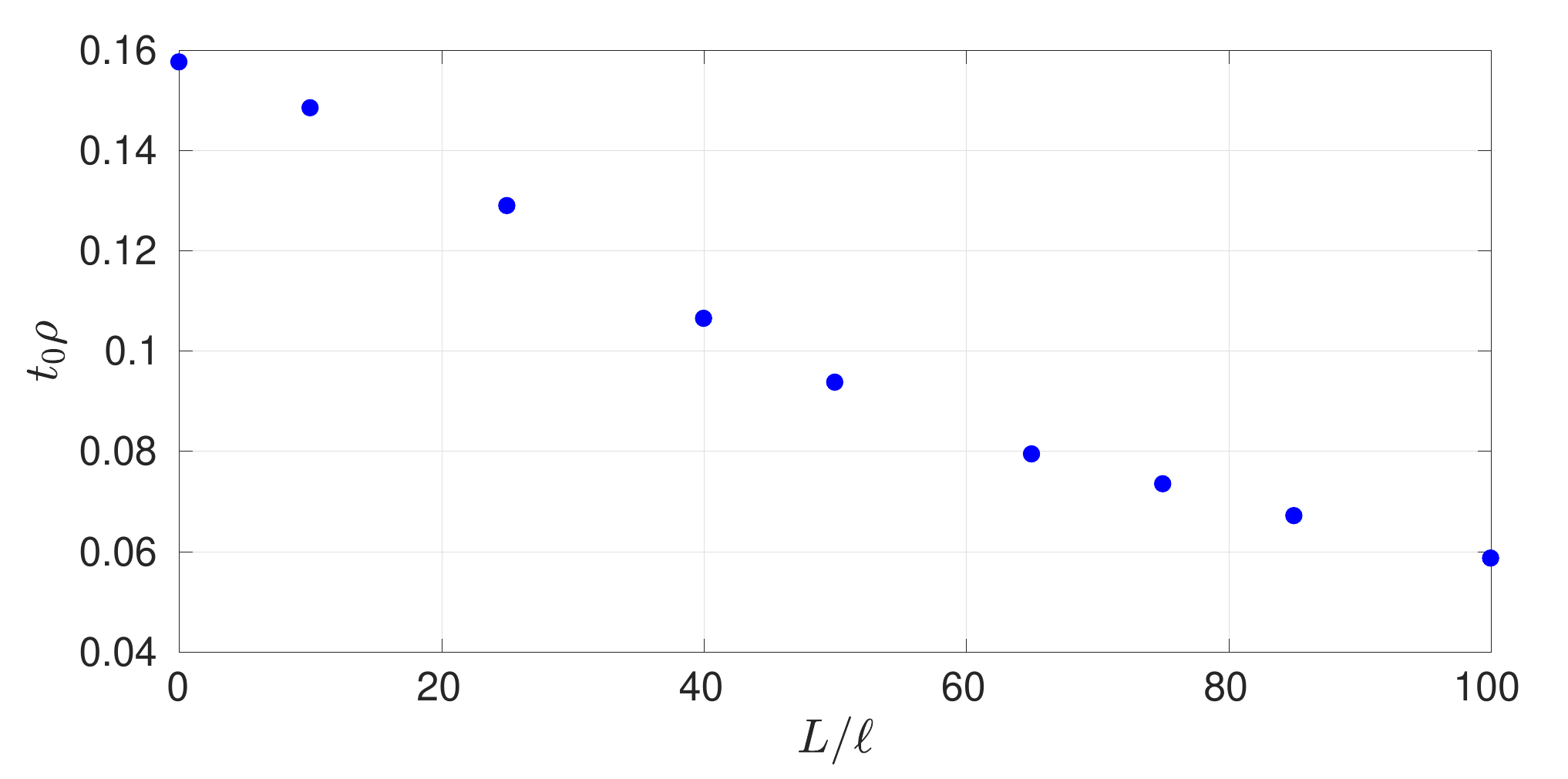}}
{\includegraphics[width=3in]{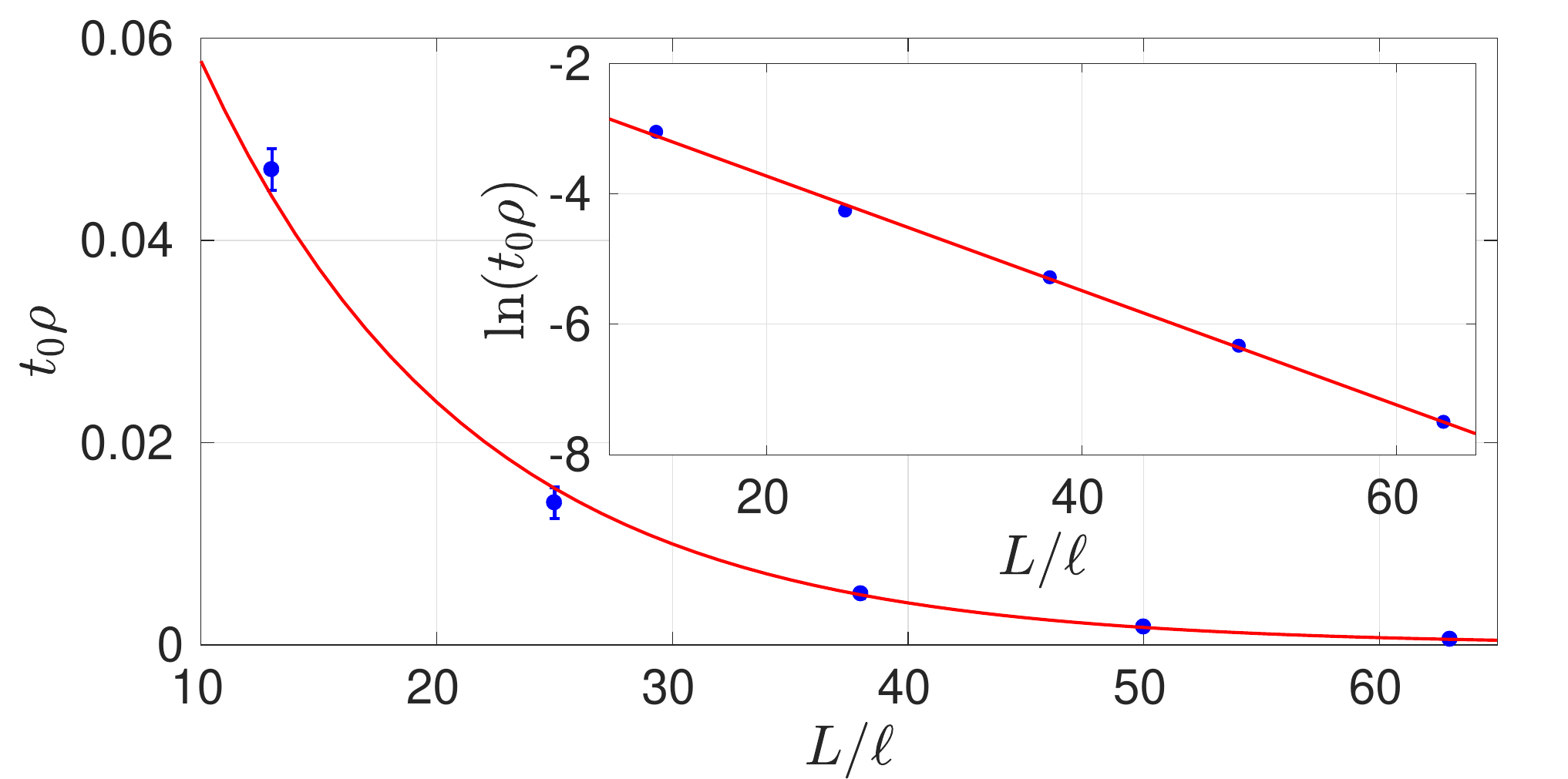}}
{\includegraphics[width=3in]{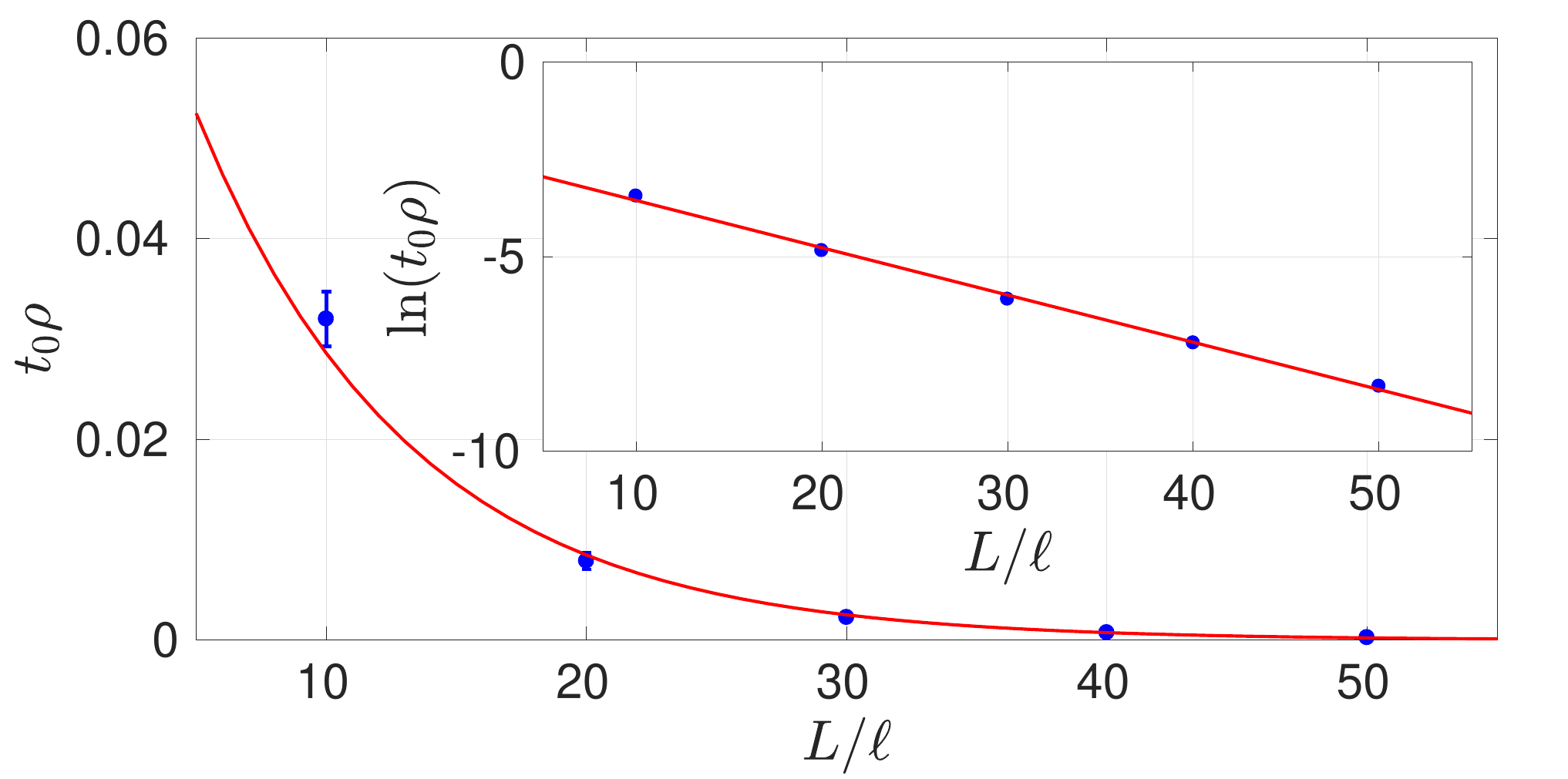}}
{\includegraphics[width=3in]{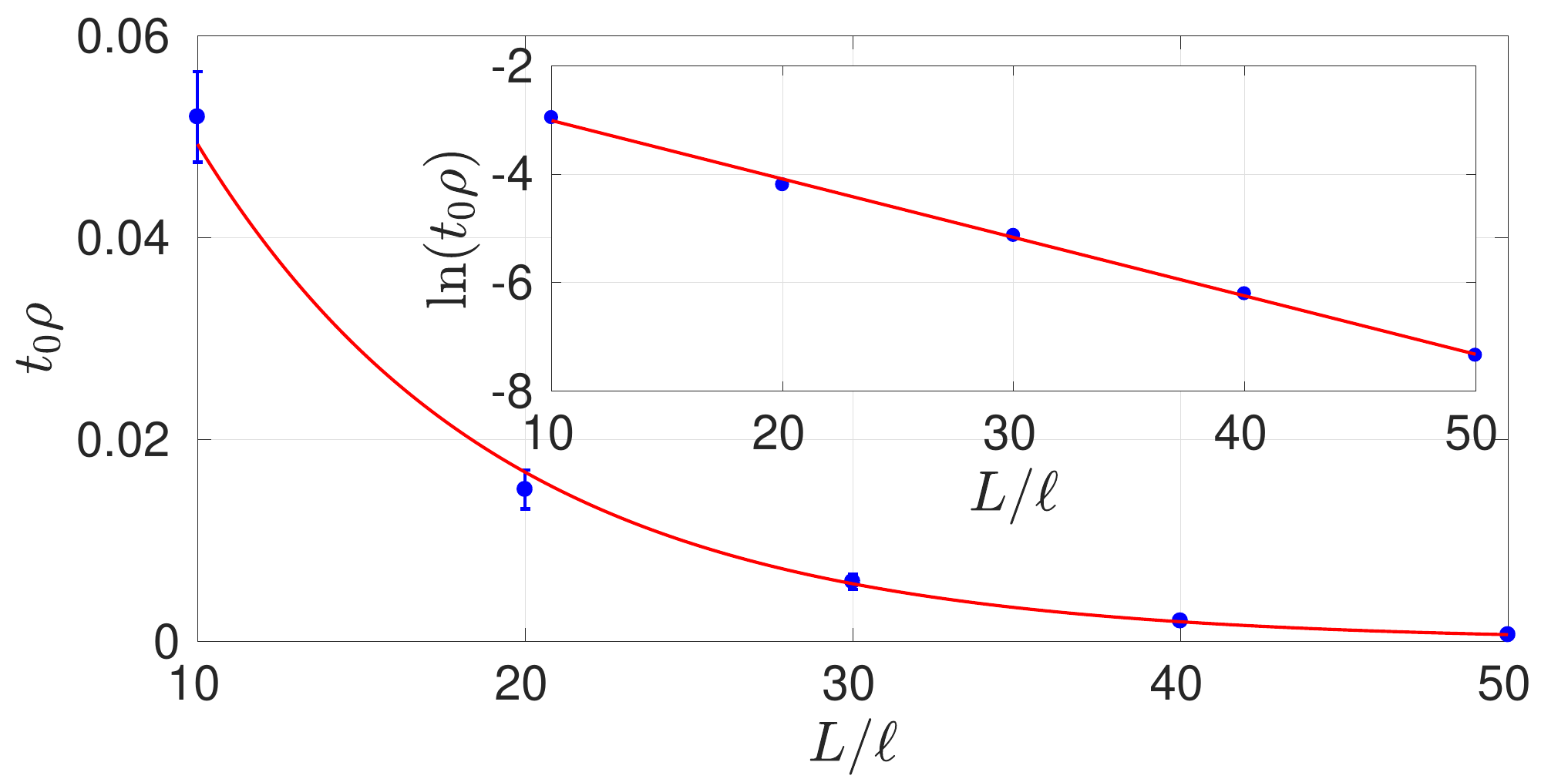}}
{\includegraphics[width=3in]{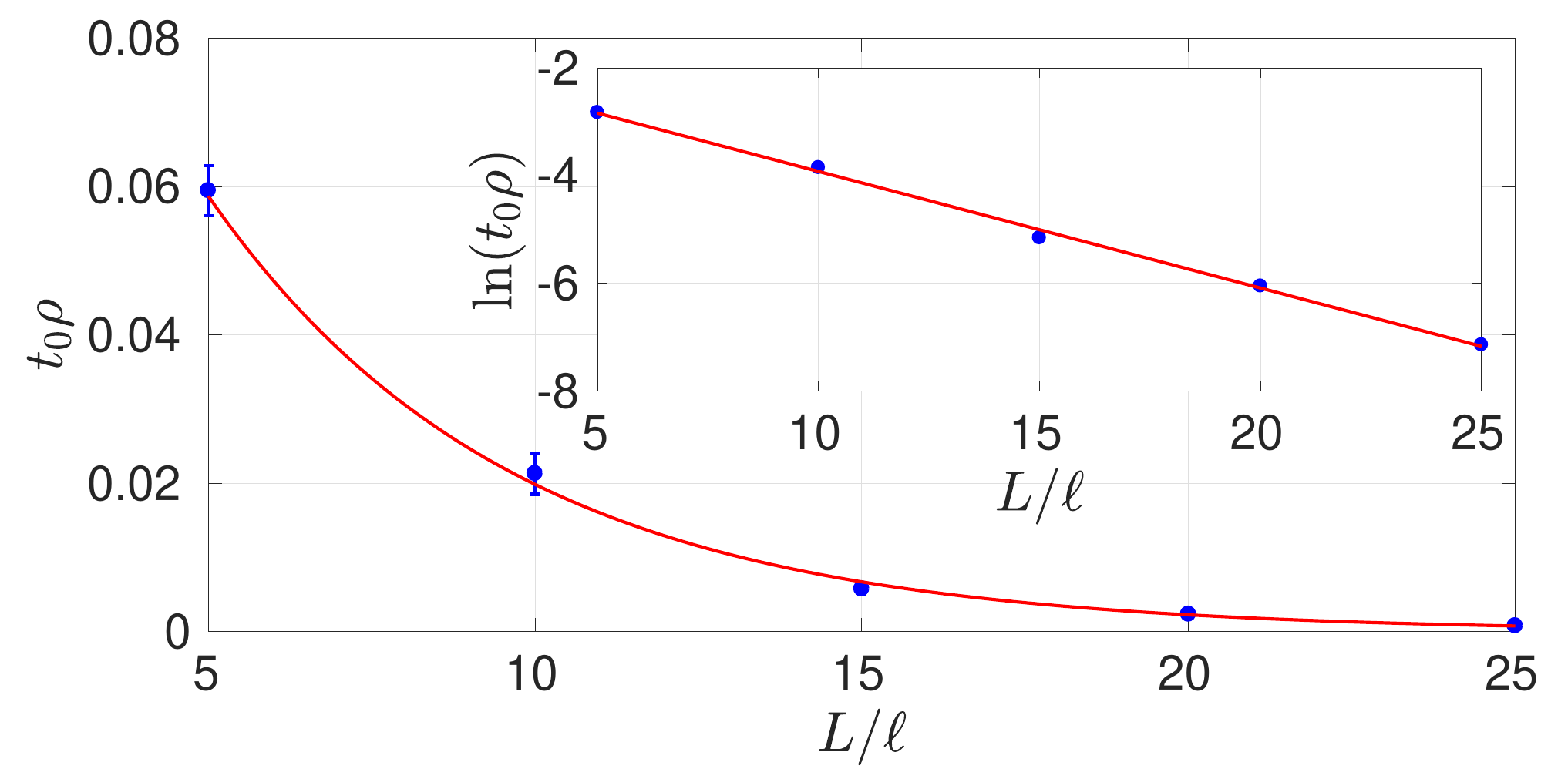}}
\end{center}
\caption{\label{rho_of_L} The flow rate $\rho$ from time dependent simulations, for an empty channel (top) and disordered configurations (the other four panels). Common parameters to all panels are $W=25\ell$, $\sigma=\ell/2$, $R=30\ell$, and the initial 1D Gaussian is placed in $R_1$ with $\bar{\sigma}=5\ell$, $k_0=1/\ell$. For the second row, $V_0=5E_0$, and for the third row, $V_0=10E_0$. Left panels of the second and third rows have $f=0.06$, and the right panels have $f=0.1$. Blue circles show the data points obtained by averaging over 20 noise realisations (in one case, we used 40 to increase the precision). The red lines are obtained from a linear fit to the logarithm of the flow rate, as shown in the insets of the last four panels. The extracted length scale of exponential decay is as follows: $V_0=5E_0, f=0.06$: $\bar{\xi} = 22.8\ell$; $V_0=5E_0, f=0.1$: $\bar{\xi} = 16.45\ell$; $V_0=10E_0, f=0.06$: $\bar{\xi} = 18.55\ell$; $V_0=10E_0, f=0.1$: $\bar{\xi} = 9.22\ell$. Error bars show the standard error. Thus, the flow rate is exponential in the channel length and a length-scale of this decay can be extracted.}
\end{figure}

We are now in a position to examine the results. Figure \ref{rho_of_L} shows the flow rate extracted from time dependent simulations, transmitting a 1D Gaussian through the disordered potential $V$, as a function of channel length for two different values of the scatterer height and density, as well as for an empty channel for comparison. The empty channel data shows a strong decrease arising from the dispersion of the Gaussian wavepacket with time, but it is by no means exponential. A strong and clearly exponential decrease is visible when disordered scatterers are present in the channel, as one would expect from our reasoning above. In section \ref{Ordered}, we show that the flow rate is considerably greater with ordered scatterers, confirming that Anderson localisation is indeed at play. We can readily extract the length-scale of this decay, as quoted in the caption of Fig.~\ref{rho_of_L} for the specific examples shown. In the next section, we will show that it is correlated to the localisation length extracted from the density profiles.
\subsection{Comparison to density profiles}
We now compare the localisation length extracted by exponentially fitting $\rho(L)$ at constant $f$ and $V_0$ to that extracted from the density profiles. We wish to obtain the localisation length from its direct definition through the density profiles. As always, we must average twenty simulations where the same wavepacket transmits through a channel long enough to allow the wavefunction to decay essentially to zero within its limits, using different noise realisations. The 1D density profiles at each point in time are averaged, and the resultant is visually examined. At early times, the density is changing rapidly as the atoms penetrate into the channel and the density gradually takes on an exponential profile. At later times, the profile changes fairly slowly, so that its logarithm can be linearly fitted and a time-dependent, observed localisation length $\xi(t)$ can be reported. The true localisation length $\xi_E$ would be obtained in the $t\rightarrow\infty$ limit of this data, but convergence is very slow, interpolation is not straight-forward, and we do not actually need the value of $\xi_E$: $\xi(t)$ is sufficient for our purposes.

\begin{figure}[htbp]
\includegraphics[width=6in]{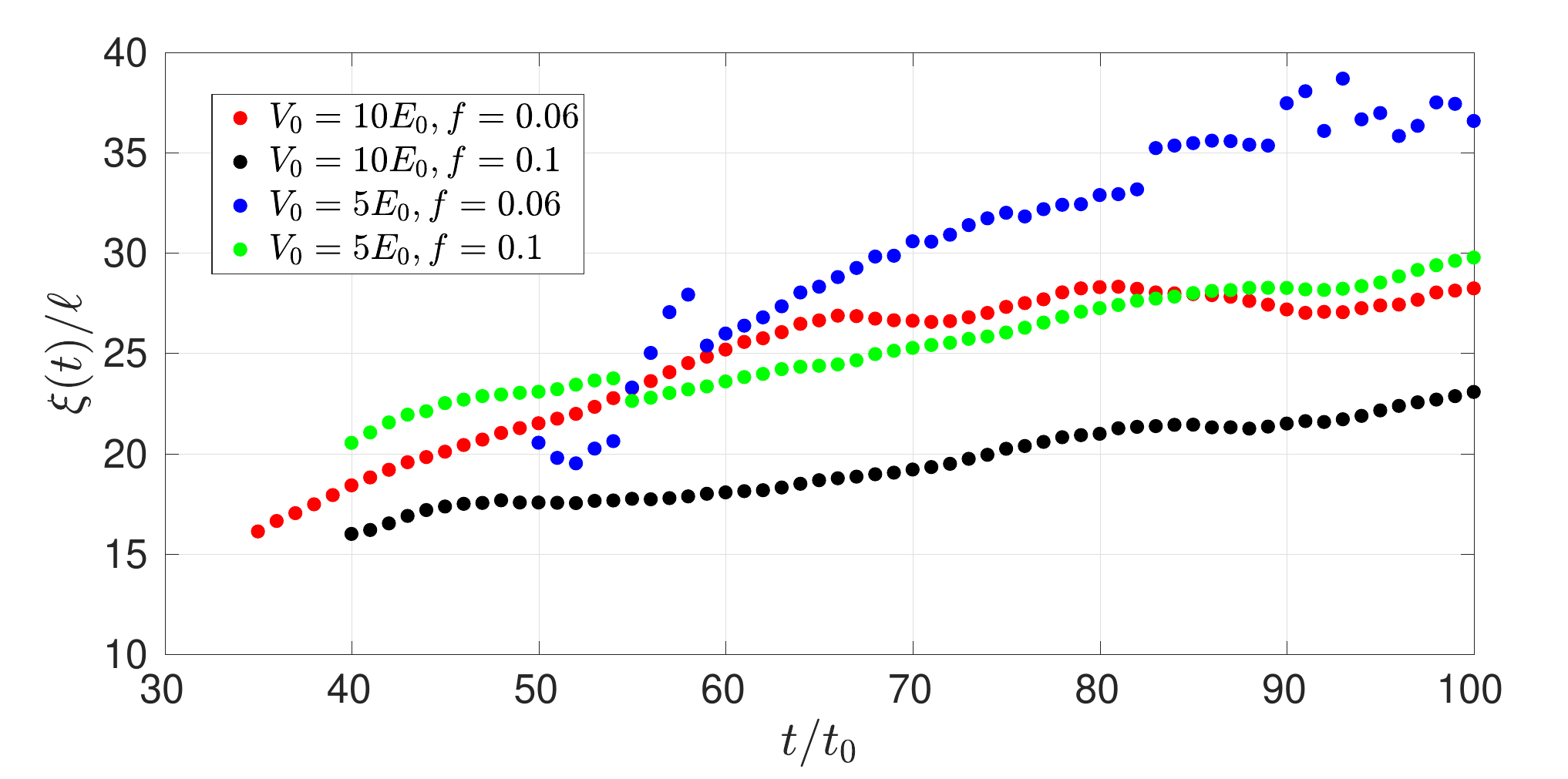}
\caption{\label{xioft} Observed localisation length as a function of time, extracted from linear fits to the logarithm of the density, as a 1D Gaussian transmits through a long channel and experiences Anderson localisation. Parameters common to all four simulations were $W=25\ell$, $\sigma=\ell/2$, $R=30\ell$, $\bar{\sigma}=5\ell$, $k_0=1/\ell$. The fill factor and scatterer height are specified in the legend, and the corresponding channel lengths were taken as: $V_0=5E_0$, $f=0.06$: $L=150\ell$; $V_0=10E_0$, $f=0.06$: $L=75\ell$; $V_0=5E_0$, $f=0.1$: $L=100\ell$; $V_0=10E_0$, $f=0.1$: $L=60\ell$. The length scale $\bar{\xi}$ extracted from the flow rate is qualitatively correlated with the long-time values of $\xi(t)$.}
\end{figure}

Looking at the curves $\xi(t)$ shown in Fig.~\ref{xioft}, obtained for the same noise parameters and probing wave as the flow rate measurements in Fig.~\ref{rho_of_L}, we see that the long-time limit of $\xi(t)$ is qualitatively correlated with $\bar{\xi}$. This is evident from the data shown as the curves $\xi(t)$ are monotonically increasing, barring minor oscillations. The difference between the two intermediate localisation strength cases is clearly not resolved  (as the two are quite close to each other), but it is apparent that $\bar{\xi}$ can be used to gauge the strength of localisation, and is in fact of the order of $\xi(t)$ at early times, when the profile first begins settling down to its eventual exponential shape.

We have thus shown that the decay of the flow rate with channel length happens on a length scale correlated to that of the decay in the density profiles in a very long channel where the wavepacket essentially decays to zero within the noisy system. The advantages of this new approach are that it permits $\bar{\xi}$, correlated to $\xi_E$, to be extracted where transmission through a sample can be measured but the density profiles are inaccessible (e.g.~electronic or photonic systems), or where system size can be varied, but is limited and cannot be pushed to a sufficiently large $L\gg\xi_E$ that would enable a direct observation of strong localisation and a measurement of $\xi_E$ from the density.

We should remark that the ``observed'' localisation length $\xi(t)$, as well as the length scale $\bar{\xi}$, heavily depend on the energy distribution of the wavepacket used to probe the disorder. Moreover, we will see in section \ref{Secondary} that it is strongly affected by some secondary factors which may be present in experiments, such as acceleration and interactions. The functional dependence of $\xi_E$ and $\bar{\xi}$ on $f$, $V_0$, and the shape of the scatterers is currently unknown and is a matter for a future investigation. This would require a prior development of an efficient method of extracting $\xi_E$ from $\xi(t)$.
\section{Dimensional crossover}
\label{WidthDep}
In this section we will investigate finite size effects using LLT, and in particular, the effect of system size on intrinsic localisation properties of the system. Because the valley lines have an associated mean distance between them , $\bar{D}$ (see appendix \ref{appLLTnew} for details on evaluation), if any dimension of the system is smaller than or comparable to this distance, the presence of the system boundary affects the structure of the valley network significantly. In fact, we will see that as we start from a width smaller than the spacing of the valley lines and progressively increase it, $\bar{D}$ increases and equilibrates as it approaches its infinite-system value. Precisely the same effect would be seen but in both dimensions if we held $L=W$ and changed the two together, except that the network would always have a 2D structure, first constricted, then allowed to spread out as the system walls move further out. This exercise if left for another study.

In this paper we choose to use our LLT technology specifically to illustrate the dimensional crossover from 1D to 2D as the width of the channel is increased from $W\ll L$ to $W\approx L$. This is quite natural in our ``channel'' system, inherited from the experiment \cite{BS}. First of all, as already pointed out, if the flow into the channel increases with $W$ (as it does, for example, in the dumbbell geometry where the reservoirs are circular), there will be a proportional increase in the flow rate out of the channel for obvious reasons. This has been discussed in the literature: the Landauer conductance, thoroughly studied over the years \cite{Sheng}, is in fact conceptually extremely similar to our transmissive scenario with the flow rate as an observable. This formalism explicitly brings out the dependence on macroscopic quantities, but does not attempt the essential computation of the quantum mechanical transmission coefficient. The width is treated as the number of independent transport channels, the contributions of which to the conductance are added incoherently. This leads to the prediction of a linear dependence of the current on the width (and gives rise to quantised conductance). The same ``extensive'' linear dependence is clearly discussed in \cite{Fan}.

A more subtle dependence on the width arises from finite size effects of the intrinsic localisation properties. There exists a large and thorough body of literature studying these effects, and a clear consensus has been established. Assuming that the length of the system is sufficiently large to have practically settled into the infinite limit, we are left with the interplay between the width and the localisation length, the latter of course being tunable by the strength of the disorder. In the limit when the width is smaller than the localisation length, $\xi_E$ increases linearly with $W$. As $W$ increases further, $\xi_E$ slows down its growth and saturates to the infinite limit. This has a direct effect on the conductance and the conductivity of the disordered sample. The (dimensionless) conductance has been computed across the dimensional crossover in \cite{Fan, Sheng, finite_scaling, Chinese, topological, 2D_corr}, and the conductivity in \cite{Fan, Chinese, topological}.

An elegant result was found in \cite{finite_scaling} and then confirmed by numerous other authors \cite{Fan, Romer, Romer2, Sheng, Su, Benoit}: the strength of the disorder can be scaled out, and the transition from 1D to 2D is completely determined by the ratio of the two length-scales: the localisation length and the width of the system. A direct computation of the localisation length as a function of the system width was performed in the regime of weak scattering \cite{russian_guys, Heinrichs, Ossipov, Stefan}, mostly confirming the linear dependence mentioned above. The exception is \cite{Heinrichs}, who find an inversely-proportional dependence on the number of channels, but considering the overwhelming amount of evidence (see below) supporting an increase of $\xi_E$ with $W$, this result is likely to be incorrect. All disorder strengths in 2D were considered by \cite{Sarma, Fan, Romer, Romer2, Sheng, finite_scaling, Benoit, 2D_corr, Su, Chinese}, and Refs.~\cite{Sarma, finite_scaling, Sheng} have repeated the calculation for 3D bars, as well. The behaviour found in these references fully supports the width dependence described in the previous paragraph. In 3D, there is a metal-insulator transition as a function of disorder strength, which is clearly evident in the results. In fact, \cite{Su, Chinese} treat 2D systems with complicating features that give rise to a mobility edge, and the behaviour of the localisation length with width is then very similar that is seen in 3D.

\begin{figure}[htbp]
{\includegraphics[width=3.1in]{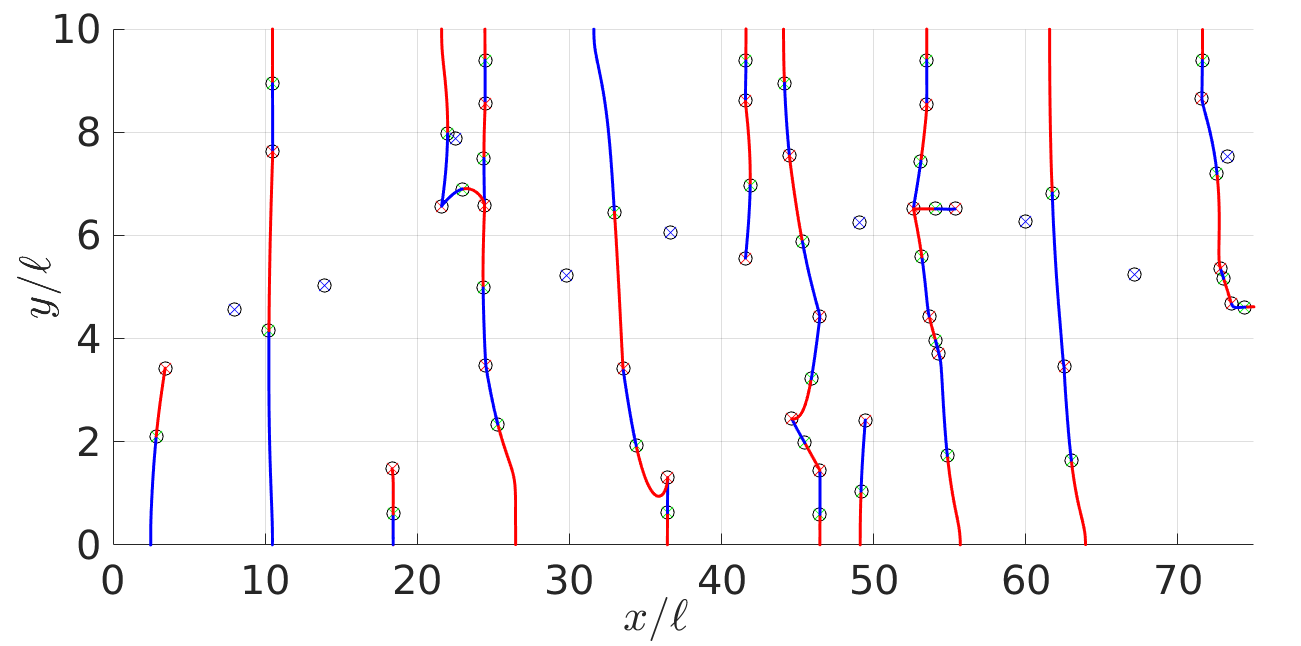}}
{\includegraphics[width=3.1in]{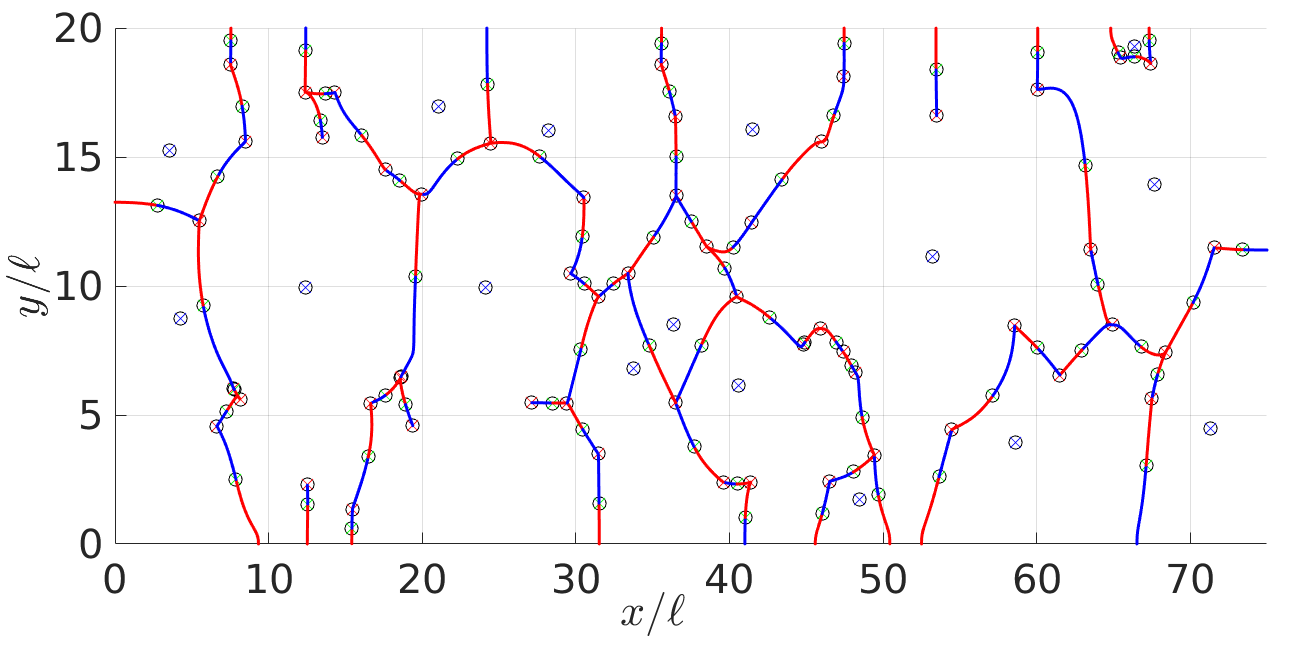}}
{\includegraphics[width=3.1in]{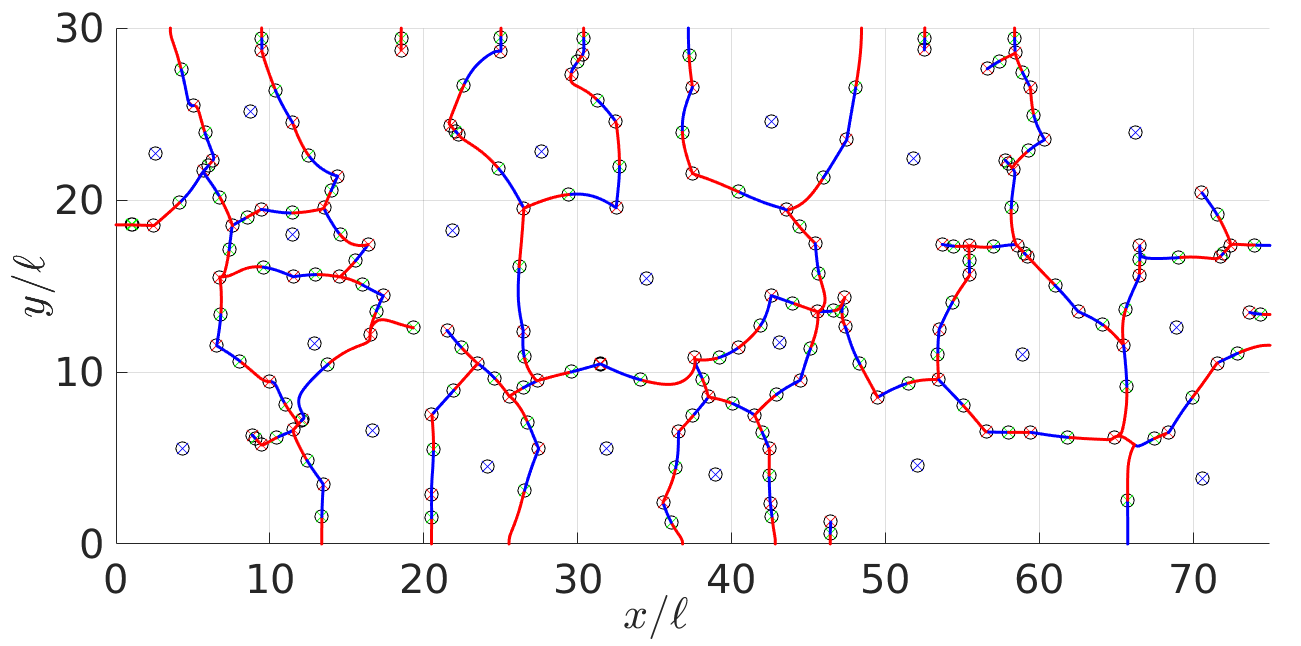}}
{\includegraphics[width=3.1in]{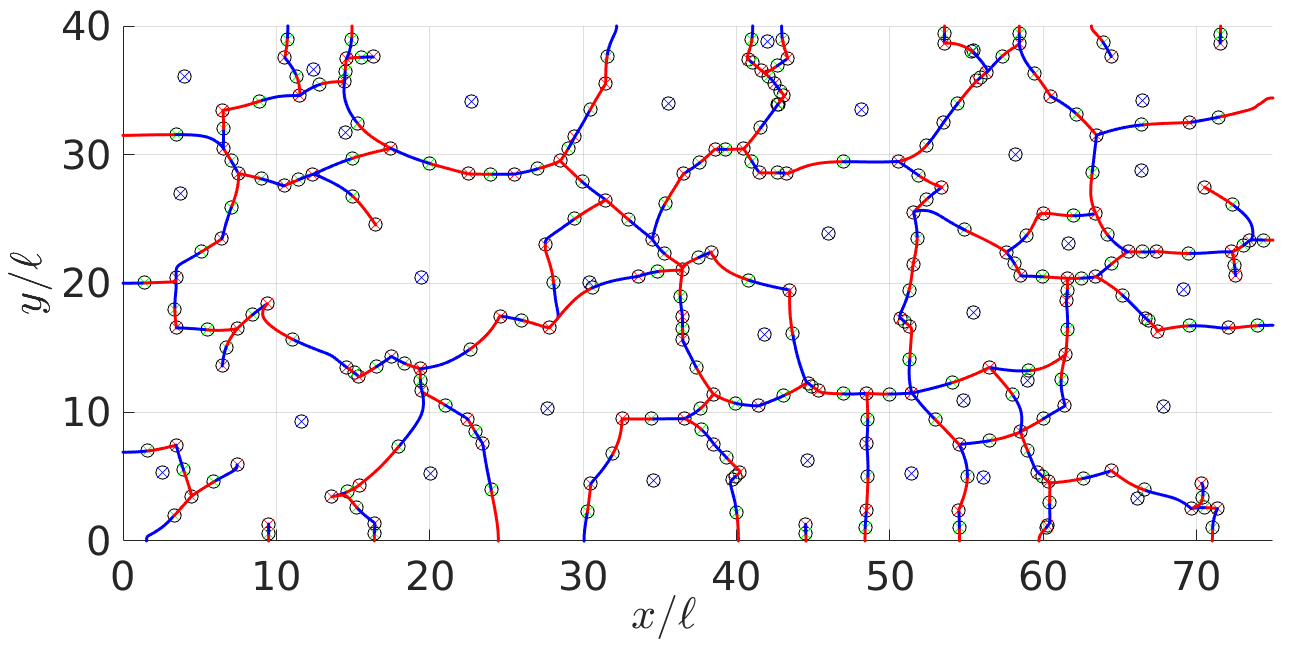}}
{\includegraphics[width=3.1in]{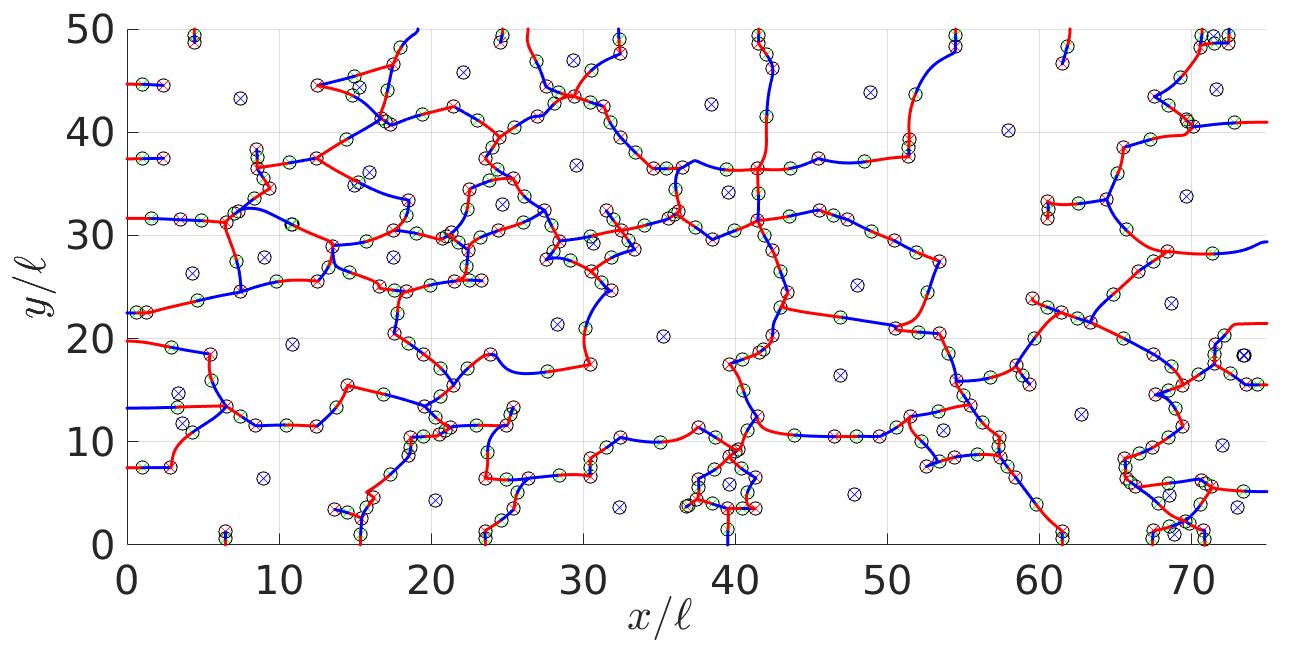}}
{\includegraphics[width=3.1in]{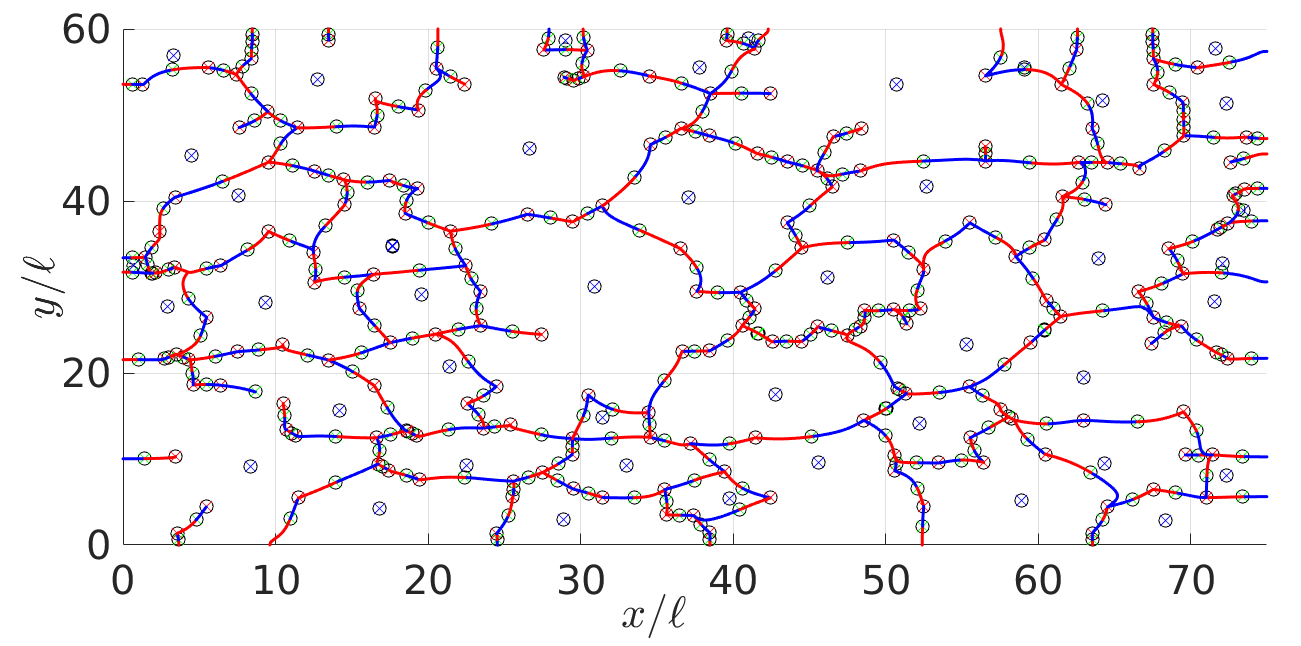}}
\caption{\label{W_nets} Valley networks for different channel widths (with lines and symbols having the same meaning as in the top panel of Fig.~\ref{C2demo}). The length is kept fixed at $L=75\ell$, while the width starts from $W=10\ell$ and increases by $10\ell$ in each panel going across and down. Other parameters are $f=0.05$, $V_0=21.33E_0$, $\sigma=0.48\ell$. For narrow channels, the valley lines almost always run straight across the channel, as the size of an average domain is larger than the width, which corresponds to an effective 1D regime. As $W$ is increased, the network gradually transforms to accommodate many full domains in the transverse direction, and eventually the two dimensions become equivalent, reaching a true 2D regime.}
\end{figure}

Here we choose to keep the flow into the channel constant in order to isolate any small, intrinsic finite-size effects on Anderson localisation. Our results are completely consistent with the literature, but give new insights that elucidate the mechanism behind the transition. Let us begin by visually examining the valley networks as we go through the dimensional crossover. Figure \ref{W_nets} demonstrates that when $W<\bar{D}$, the valley lines simply run across the width of the channel, which corresponds to a 1D regime (in a true 1D system, the valley lines are reduced to points). Gradually, as $W$ increases, structure appears also in the transverse direction, soon reaching the stage where \textit{locally}, without reference to the system boundaries, it is impossible to tell which direction is which. This corresponds to a true 2D regime. Thus LLT directly allows us to visualise the transition from one to two dimensions. We remark that it is $\bar{D}$, and not the localisation length (as is commonly believed), that is the relevant length scale to be compared to system size.

\begin{figure}[htbp]
\includegraphics[width=6in]{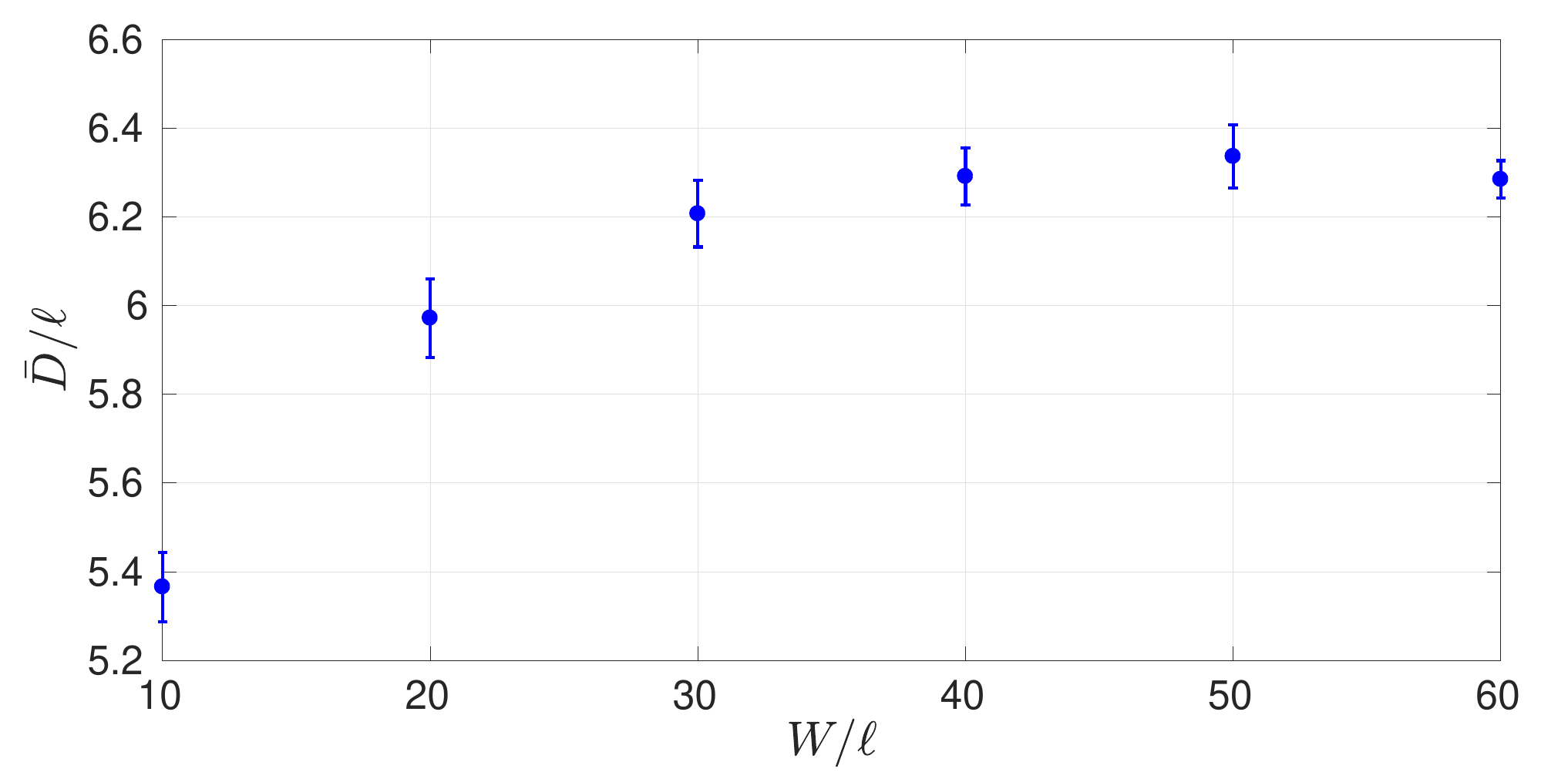}
\caption{\label{Dbar_of_W} The mean distance between valley lines for varying channel widths. Other parameters are $L=75\ell$, $f=0.15$, $V_0=21.33E_0$, $\sigma=0.48\ell$, 20 noise realisations are averaged over and the error bars show the standard error. There is a clear increasing trend in $\bar{D}$ as we go through the dimensional crossover from 1D to 2D, which also reflects on the localisation length.}
\end{figure}

\begin{figure}[htbp]
\includegraphics[width=6in]{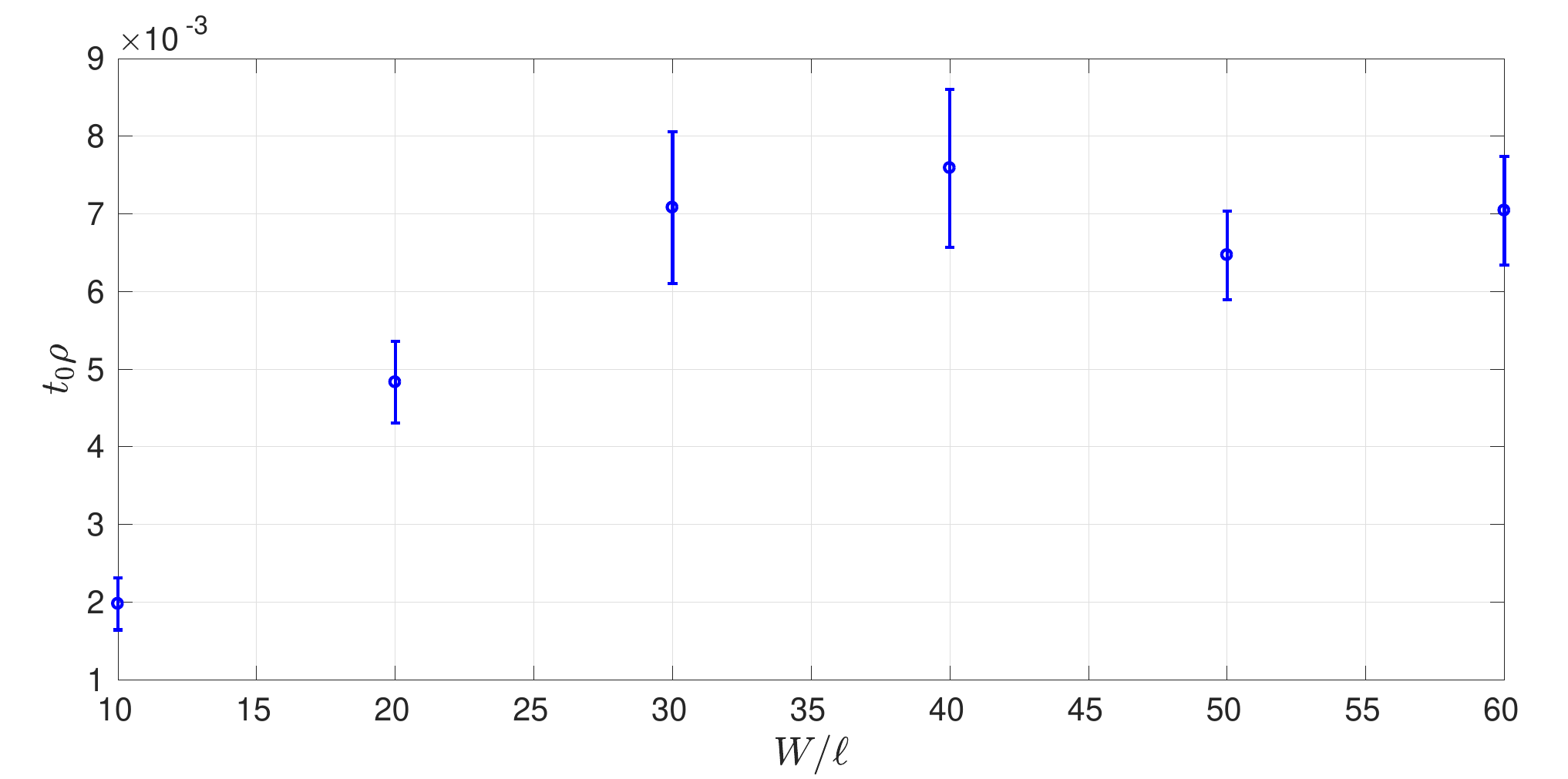}
\caption{\label{Wrho} Flow rate out of the channel for a 1D Gaussian with $\bar{\sigma}=5\ell$, $k_0=1/\ell$, in a system with $R=30\ell$, $L=25\ell$ $f=0.05$, $V_0=21.33E_0$, $\sigma=0.48\ell$. Twenty noise realisations are averaged over and the error bars show the standard error. Transmission increases with width as localisation weakens across the crossover from 1D to 2D.}
\end{figure}

We can quantify this effect by inspecting the mean distance between the valley lines $\bar{D}$ (closely connected to the localisation length) across this transition, as shown in Fig.~\ref{Dbar_of_W}. Indeed, as the valley lines move further apart, localisation weakens, manifesting as a larger distance between the valley lines. Since this is a finite-size effect, it is much weaker than the dependence on the noise parameters, for example, and requires rather strong Anderson localisation to resolve the trend. It also depends on $\bar{D}$ compared to system size, so it is quite easy to find parameters where the trend is drowned out in the fluctuations; it is, however, undoubtedly real (we have seen the same pattern emerge for many parameter values). This can be confirmed via time-dependent simulations, using a translating 1D Gaussian in the transmissive scenario, measuring the flow rate out of the channel. This data is displayed in Fig.~\ref{Wrho}, and reveals the same trend: localisation weakens with increasing $W$, resulting in a higher transmission through the channel. Note that this finite size effect would not be visible in the flow rate as a function of $L$ (even though it is present if $L<\bar{D}$) because of the explicit dependence of $\rho$ on $L$ which is exponential and completely eclipses this weak trend. The reason it is clearly resolvable as a function of the width is because $\rho$ has no explicit dependence on $W$ -- only through $\bar{\xi}$ (the length-scale of exponential decay of $\rho(L)$).

To conclude, while finite-size effects of localisation properties have been extensively studied previously (and are consistent with our findings), our work here constitutes the first clear explanation of \textit{why} they occur. This is exclusively due to the conceptual and computational power of LLT to (a) elucidate the significance of the valley network to localisation and (b) access the mean distance between the valley lines, which is the relevant length scale to be compared to system size. 
\section{Mobility edge}
\label{ME}
Localisation landscape theory guarantees that the eigenstates are suppressed at the valley lines of the localisation landscape $u$, as long as $E<1/u$, where $E$ is the corresponding eigenvalue of the Hamiltonian and $u$ is evaluated along the valley lines \cite{Marcel2012}. This can be viewed as a tunnelling process through a barrier in the effective potential $W_E=1/u$ \cite{FnM2016b}. In the course of our LLT-based calculation of the localisation length at low energies in section \ref{XiSaddles}, we have found that quantitatively, the tunnelling picture breaks down at quite low energies. However, the eigenstates certainly continue respecting the suppressing effect of the valley lines, and so one can still link the Anderson localisation exhibited by these to the valley network of LLT. On the other hand, the peaks of $W_E$ have a finite height (which depends on the strength of the disorder), and when $E$ surpasses all the peaks of $W_E$, there is nothing more to cause any confinement or decay in the eigenstates, at least from the point of view of LLT. Logically, this does not imply that there \textit{cannot} be any more decay -- it just means that there does not \textit{have} to be any, based on mathematically rigorous inequalities established in \cite{Marcel2012}. This raises the question of whether there is a true mobility edge in our system.

At first, this may be surprising, since the results of scaling theory are so strongly engraved in our understanding of Anderson localisation. According to \cite{ScalingTheory}, no mobility edge exists in 1D and 2D, but is present in 3D. This result holds under ``normal'' conditions (to be specified below), and has been confirmed repeatedly by many authors for 1D \cite{ScalingTheory, Kunz, HerbertJones, Thouless, Kirkman, Sarma}, 2D \cite{ScalingTheory, Kunz, Sarma, Romer, Romer2, Fan, finite_scaling, Benoit}, and 3D \cite{ScalingTheory, HerbertJones, Sarma, finite_scaling} systems. Sheng \cite{Sheng} provides many different arguments and a collection of evidence to support these claims, \cite{Ossipov} re-establishes that two is the marginal dimension (i.e.~all states are localised in 1D, and a mobility edge exists for all dimensions higher than two), while \cite{Piraud2012} discusses the situation in all dimensions.

Several physical mechanisms are known that can give rise to a mobility edge in lower dimensions. One of the more thoroughly explored ones is correlations: if the parameters in the Anderson model or the distribution of scatterers (e.g.~delta-function or Gaussian bumps) in a continuous system are not completely random (white noise), or if the Fourier transform of the continuous potential spans a finite frequency range, then the disorder is said to be correlated and a mobility edge in lower dimensions is possible \cite{Piraud2012, Aspect2017}. This has been shown in 1D for discrete \cite{Jan, 1D_corr, Makarov} and continuous \cite{Hilke, Hilke2, Makarov, Aspect2018} models, as well as in 2D \cite{2D_corr}, while in 3D, correlations allow one to tune the mobility edge out of existence \cite{Pilati}. Another commonly investigated mechanism is the introduction of a magnetic field which breaks time-reversal symmetry, thus weakening and eventually destroying localisation: demonstrations in 2D systems include \cite{LeeFisher, Su, itscomplicated, Ulrich, Imry, Ono, Vollhardt, Sheng}, with 1D studies also available \cite{Imry, Vollhardt, Sheng}.

Interparticle interactions have a detrimental effect on localisation and cause a metal-insulator transition in lower dimensions, as has been found experimentally \cite{Greek_review, openquestions} and theoretically \cite{Pun, Apel}. Chiral symmetry \cite{Chinese, Chiral}, spin-orbit coupling \cite{Su, itscomplicated, SOC, SOC2, Imry, Sheng}, topology and spin \cite{itscomplicated, topological}, and symplectic symmetry \cite{deSterke} can also create a mobility edge, as can the presence of acceleration in the system \cite{Berthet, Vollhardt}. We examine the effect of interactions and acceleration in section \ref{Secondary}. Often, the transition is studied through a finite-size scaling analysis \cite{Su, Chinese, itscomplicated}. Curiously, it is sometimes possible to have delocalised states in lower dimensions without adding any of the above complicating ingredients \cite{Wolfy, Gong, doped, Thouless2}, while on the other hand, not all studies including a magnetic field find a transition \cite{Stefan} and not all frequency components outside the spectral window of correlated disorder are necessarily delocalised \cite{Segev}.

For us, the relevant factor to consider is correlations. To reiterate, the results of scaling theory only hold for white noise, where the spatial Fourier transform of the disordered potential $V$ is uniform in frequency space\footnote{S.S.S. gratefully acknowledges Donald H.~White and David A.W.~Hutchinson for initially bringing this fact to her attention.}. If it is not, there is some spatial correlation to the noise: in our case, the mean spacing between the scatterers is tuned by the fill factor, and the size and shape of the scatterers determine the smallest-scale features, thus setting an upper frequency cut-off in the spectrum. As a result, the spectrum has a finite extent and a completely non-trivial shape.

\begin{figure}[htbp]
\includegraphics[width=6in]{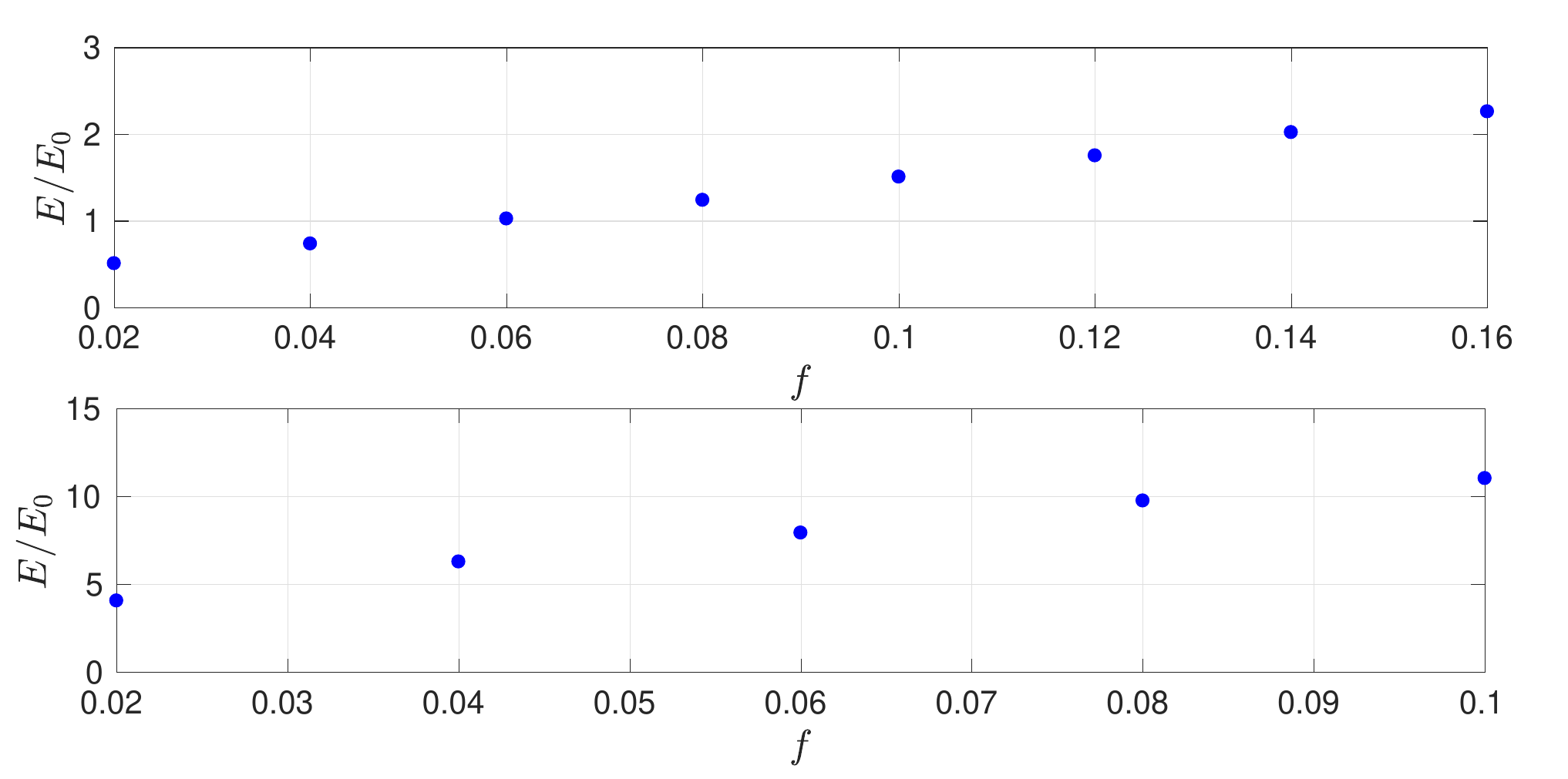}
\caption{\label{MEdep0} The mean energy of the maxima of $W_E$ averaged over 20 noise realisations as a function of fill factor. Top panel has $V_0=5E_0$, $L=W=25\ell$ and bottom panel $V_0=20E_0$, $L=50\ell$, $W=25\ell$. Both panels used $\sigma=\ell/2$. The mean energy at the maxima of $W_E$ can be viewed as a rough measurement of the mobility edge predicted by LLT. It is clearly strongly increased by both $f$ and $V_0$.}
\end{figure}

As we have repeatedly demonstrated throughout the article, the height and density of the scatterers strongly influence localisation properties, which usually implies the localisation length (the scatterer shape has a weaker effect, not studied in this paper). The idea we discuss here is that the same parameters may also shift the mobility edge -- if it exists -- just as strongly. This can be intuitively understood by thinking about the effective potential of LLT, $W_E$. We know that the exponential decay associated with quantum interference effects of Anderson localisation at low energies can be viewed as tunnelling through this potential landscape. The higher the surface of $W_E$, the stronger the localisation. As soon as the energy of the eigenstates $E$ exceeds the peaks of $W_E$, no further localisation is guaranteed by LLT. As $f$ and $V_0$ increase, the entire surface of $W_E$ moves up, including the extrema, i.e.~the (apparent) mobility edge moves, as illustrated in Fig.~\ref{MEdep0}. To summarise, LLT predicts the existence of a true mobility edge that is affected by the properties of the disorder.

At this point, it is important to mention that the presence of peaks in the effective potential is insufficient to guarantee that there is Anderson localisation in the system -- for this, the on-domain, fundamental eigen-energies must be randomised, which occurs when the Hamiltonian features a disordered potential (see sections \ref{LLTold} and \ref{BHM}). We proceed with the understanding that this is the case (otherwise speaking of mobility edges is meaningless).

We can further ensure that the LLT prediction of the mobility edge in itself is not a finite-size effect. It could in principle be possible that as $L,W\rightarrow\infty$, $W_E$ stretches up to infinity and the mobility edge vanishes. This can be easily ruled out by checking for the dependence of the mobility edge on system size. Modest changes in $L,W$ are then sufficient to reassure one that the mobility edge does not shift with system size (for small $L,W<\bar{D}$, the mobility edge decreases as the system grows and eventually settles down as finite-size effects cease to be important\footnote{Recall that due to Dirichlet boundary conditions on $u$, $W_E$ diverges to infinity at the edges of the system. Thus, the mean of all peak energies on the surface of $W_E$ is artificially shifted higher than the ``true value'' if the edges of the system -- a strip of a given width running along the perimeter of the system domain where the boundary conditions pull $W_E$ up higher than the average values in the interior -- constitutes a significant portion of the total area. As the system area increases, the contribution of the strip to the mean diminishes, and the mobility edge falls lower.}).

Let us now elaborate on the nature of the LLT-predicted mobility edge. To date, in the conventional understanding of Anderson localisation, it has always been viewed as a sharp phase transition -- a single energy below which one has localisation and above which one does not. However, according to LLT, the mobility edge is determined by the peaks of $W_E$. While one could define such a sharp mobility edge by simply averaging over all maxima (and over many noise realisations, of course), it does not give us the full picture. Naturally, the values of $W_E$ at the set of extrema form a \textit{distribution}. Energies within the range over which this distribution extends still experience some localisation, because there are still parts of $W_E$ above their energy that suppress the wavefunction. As $E$ sweeps through the range covered by this distribution, fewer and fewer maxima remain that affect this energy component, and finally, when we reach the high-energy end of the mobility edge distribution, localisation truly vanishes. If a phase transition can be claimed at any point, then it is here: when no constraining extrema remain.

\begin{figure}[htbp]
{\includegraphics[width=3.1in]{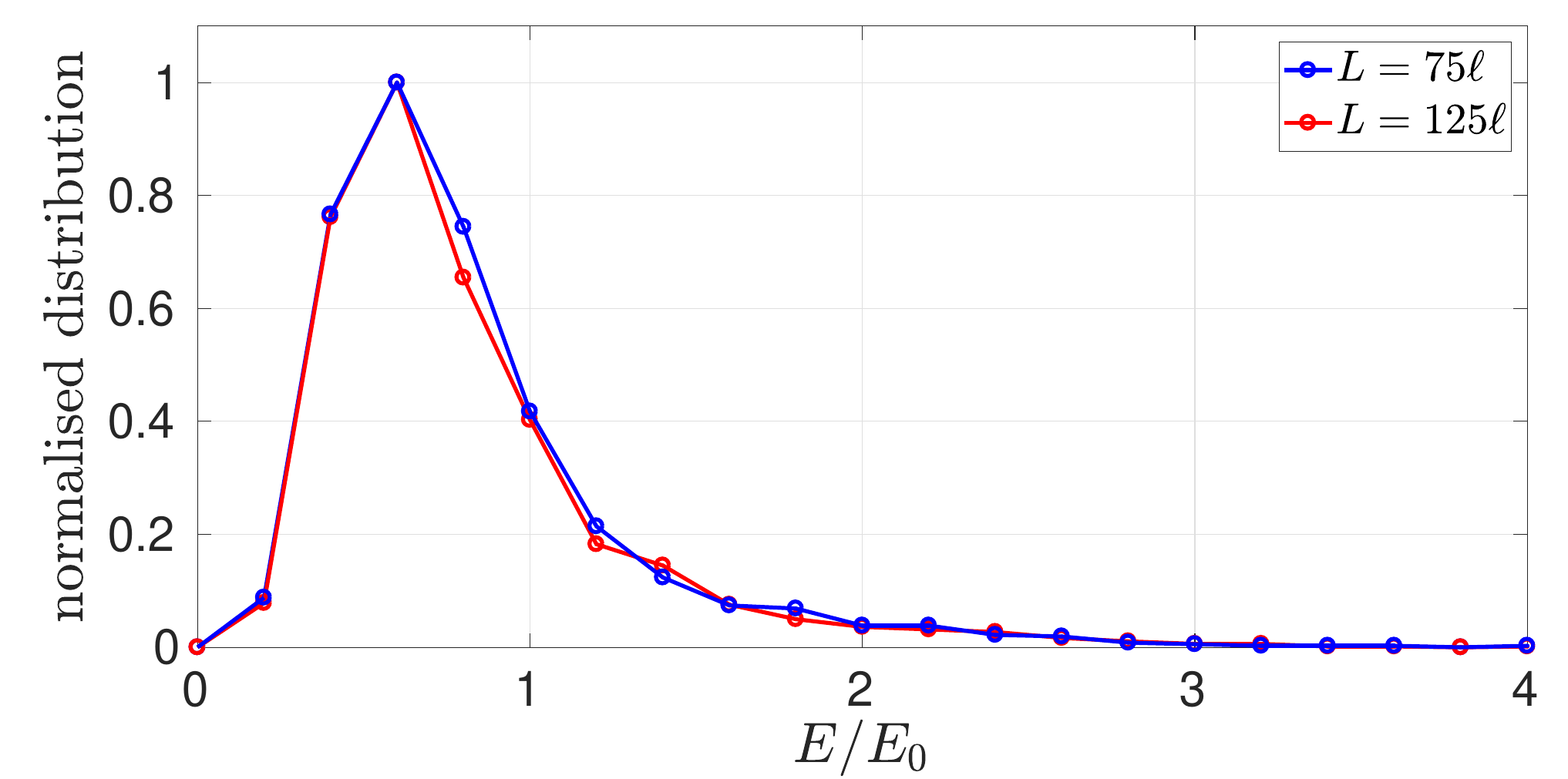}}
{\includegraphics[width=3.1in]{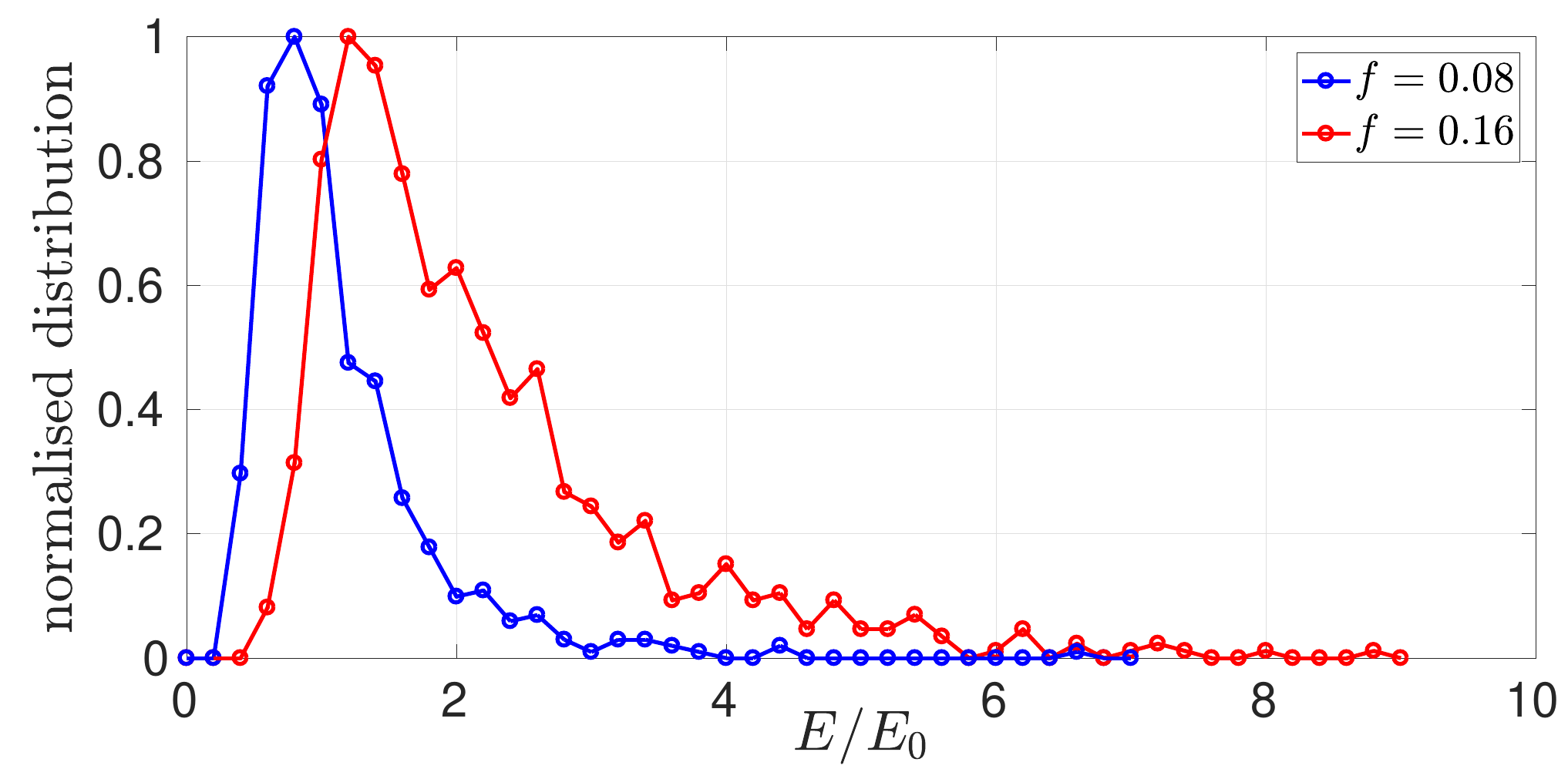}}
{\includegraphics[width=3.1in]{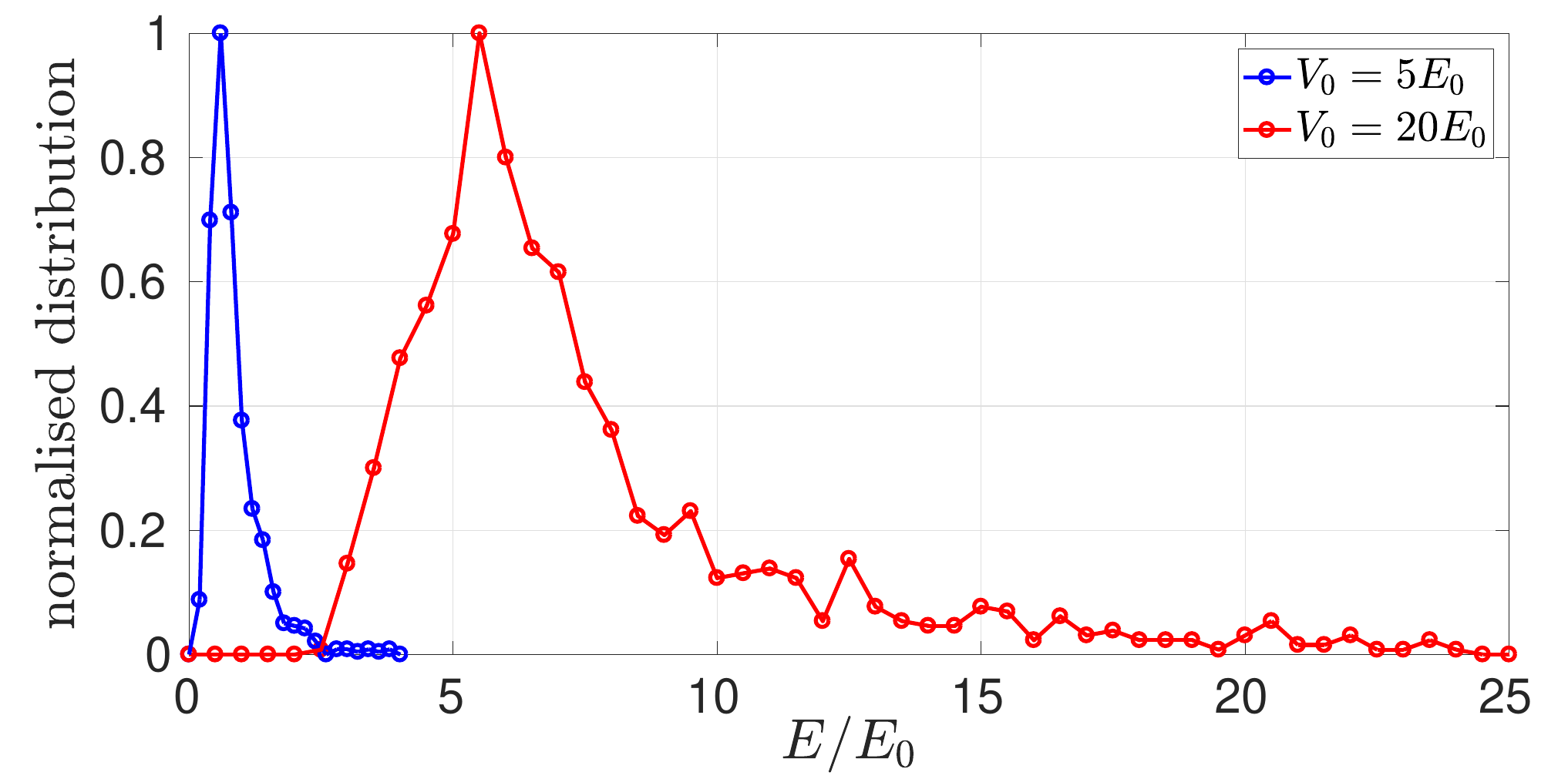}}
{\includegraphics[width=3.1in]{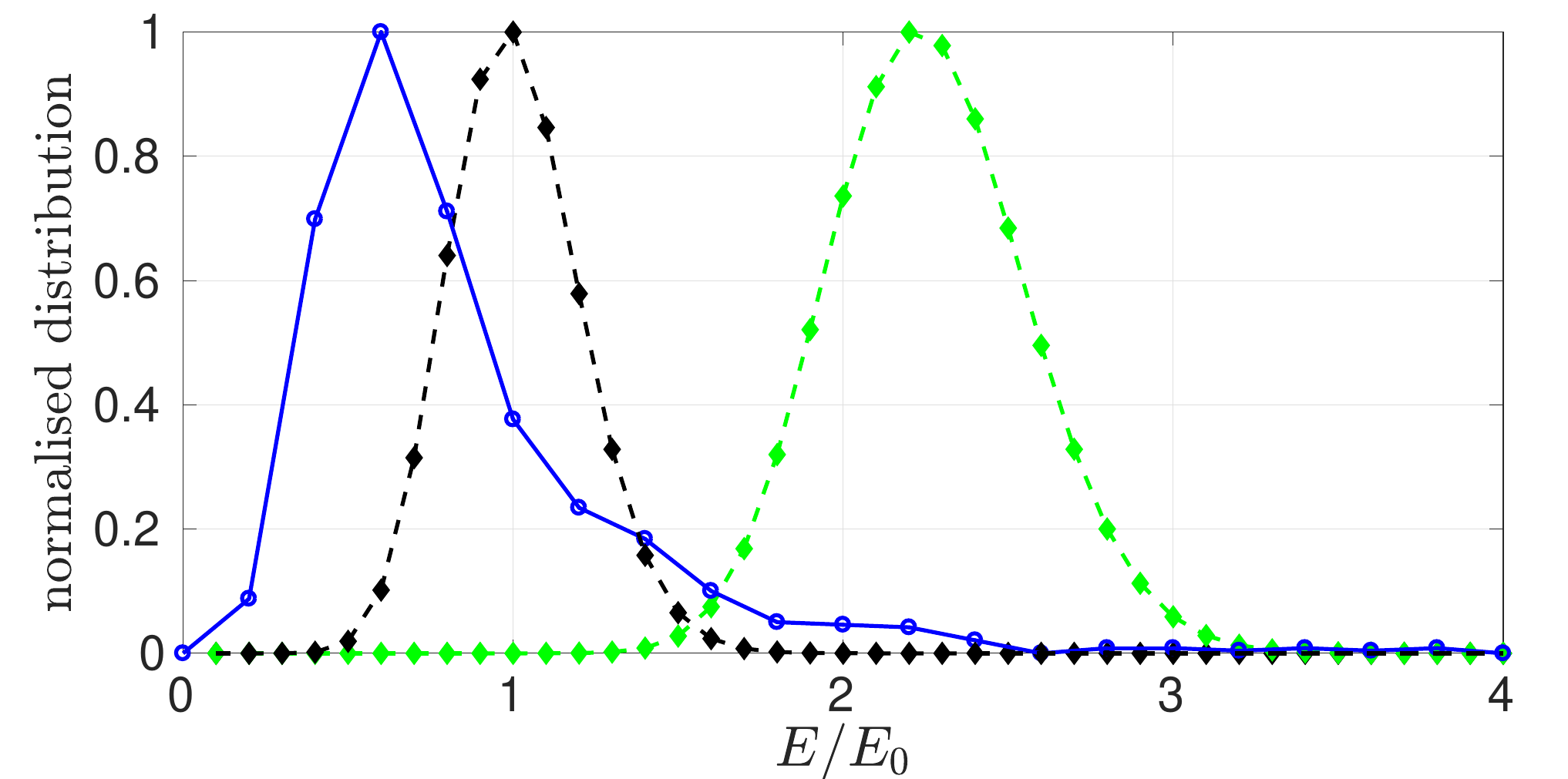}}
\caption{\label{MEdep} First three panels: Mobility edge distributions. The extrema are binned in intervals of $0.2E_0$ for $V_0=5E_0$ and $0.5E_0$ for $V_0=20E_0$, using 20 noise realisations, counting the number of extremum points in each interval. The distributions are normalised such that the maximal value is one. Top left panel: $W=25\ell$, $f=0.06$, $V_0=5E_0$, $\sigma=\ell/2$, two values of $L$ are used (see legend) and the results confirm that the mobility edge is independent of system size as long as $L,W>\bar{D}$. Top right panel: $L=W=25\ell$, $V_0=5E_0$, $\sigma=\ell/2$, showing two values of $f$ (see legend). Bottom left panel: $L=50\ell$, $W=25\ell$, $f=0.06$, $\sigma=\ell/2$, comparing the distributions at low and high $V_0$, as indicated in the legend. Both the fill factor and the height of the scatterers strongly move the mobility edge. Bottom right panel: mobility edge distribution for $L=50\ell$, $W=25\ell$, $f=0.06$, $\sigma=\ell/2$, $V_0=5E_0$ (blue circles, solid line) and the energy distribution (\ref{Edistro}) for a 1D Gaussian with $\bar{\sigma}=5\ell$, $k_0=1/\ell$ (black diamonds, dashed line) and $\bar{\sigma}=5\ell$, $k_0=1.5/\ell$ (green diamonds, dashed line) normalised such that the maximal value is one. The flow rate $\rho(L)$ shown in Figs.~\ref{rho_of_L} and Fig.~\ref{MEtest} was obtained with these noise parameters and these two initial conditions, respectively.}
\end{figure}

Next, we demonstrate the effect of parameters on the mobility edge distribution. As shown in Fig.~\ref{MEdep}, the mobility edge is independent of system size as long as $L,W$ exceed the mean distance between the valley lines so that finite-size effects are no longer important (for smaller $L,W$, the mobility edge falls with increasing system size). In contrast, it is strongly shifted by both the density and height of the scatterers. The effect of the scatterer shape will be investigated in a future study; we can expect interesting results because the shape and size of the scatterers determine the momentum cut-off of the Fourier transform of $V(x,y)$, and thus should affect the mobility edge, if it truly exists.

Clearly, our understanding of the mobility edge so far fully relies on the interpretation provided by LLT. However, there is already ample evidence (see section \ref{HigherEs}) that LLT ceases to be useful at fairly low energies, so it is quite possible that at the energies where the peaks of $W_E$ are attained, LLT is simply no longer relevant and the prediction of a mobility edge may well be simply wrong. In principle, this could be tested by performing exact diagonalisation and direct time-dependent simulations above and below the mobility edge predicted by LLT, increasing system size, and showing that below the mobility edge one \textit{eventually} sees signs of localisation, while above it, one does not. On the other hand, such a test could well be inconclusive because the localisation length grows (roughly) exponentially with the wavenumber \cite{Sheng} (possibly diverging at the mobility edge, if it truly exists). As we increase system size, we can never be sure that it is large enough and that going a little further will not reveal signs of localisation. Moreover, these numerical methods are not easily scaled up to large systems, and one soon hits the computational wall.

Fortunately, we can circumvent this problem by making use of our flow-rate formalism. If we examine the flow rate out of the channel in the transmissive scenario with a translating wavepacket as a function of channel length $L$, we will find that it obeys the functional form
\begin{equation}
\label{rhoc2}
\rho = c_1\exp(-2L/\bar{\xi}) + c_2.
\end{equation}
We have introduced the off-shift $c_2$ to account for the possibility that some energy components may lie above the mobility edge (if it is real), and thus would always contribute to a non-zero transmission through the noisy potential, regardless of $L$.

If all energy components present in the probing wave were localised -- with different localisation lengths, but nonetheless \textit{localised} -- then taking $L\rightarrow\infty$, should cause the flow rate to vanish. If there exists a mobility edge and a portion of the energy distribution of the atoms lies above it, then these energy components will never be localised and $c_2$ will be non-zero. Such an off-shift to the exponential decay of the flow rate allows one not only to detect the mobility edge, but also observe its dependence on parameters. Changing either the energy distribution of the initial condition or the noise parameters will shift the two distributions with respect to each other and change the value of $c_2$.

Thus, a non-zero off-shift $c_2$ to the rate of flow out of the channel is a practical approach to test the prediction of a true mobility edge arising from LLT. It does not require increasing system size beyond what is accessible via standard techniques, relying instead on taking a limit as the system size increases. We look for changes in the transmission with system size, in other words. A complementary option -- if computational resources allow it -- is to observe the density profiles as a function of time for a system the channel of which is taken progressively longer, checking whether the density decays exponentially, and if it is possible to reach a regime where no significant portion of the atoms transmit to the second reservoir.

In Fig.~\ref{rho_of_L}, we have shown several examples of $\rho(L)$ for a given low energy wavepacket, and there was no visible off-shift $c_2$ as $L\rightarrow\infty$ for any of the disorder strengths considered. Comparing the energy distribution (\ref{Edistro}) of the translating Gaussian utilised to the mobility edge distribution for the weakest disorder case in the last panel of Fig.~\ref{MEdep}, we see that while the atomic distribution peaks at a higher energy than the mobility edge distribution, there are still some maxima of $W_E$ across the entire range of atomic energies. Since all energy components within the mobility edge distribution are still localised (albeit to varying degrees), by forcing the wavefunction to pass through a sufficiently long channel, it is possible to fully attenuate the signal detected in the second reservoir. Thus, the outcome of exact time-dependent simulations is not inconsistent with LLT in this case.

\begin{figure}[htbp]
\includegraphics[width=6in]{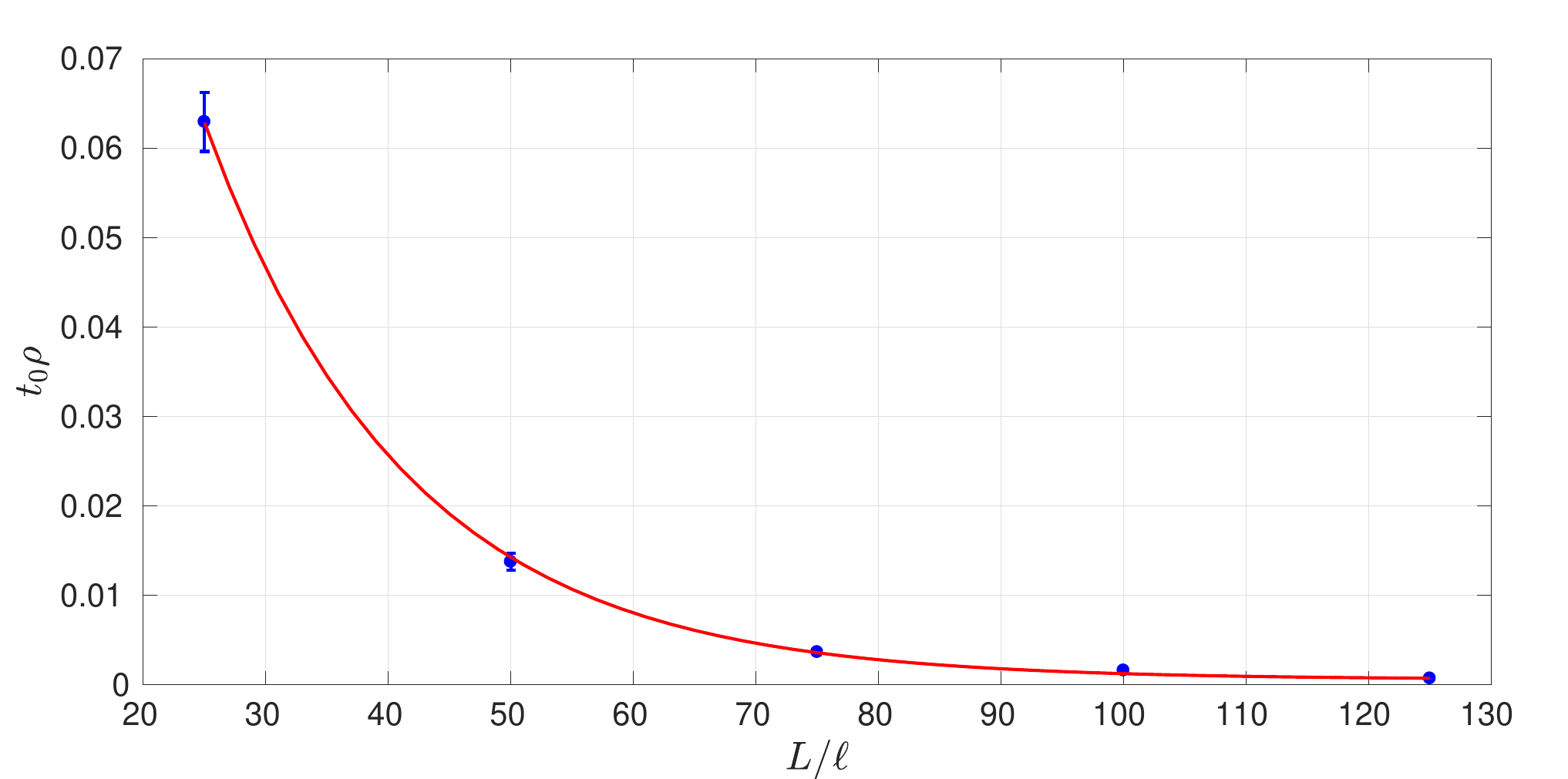}
\caption{\label{MEtest} The flow rate out of the channel as a function of channel length, $L$. Error bars indicate the standard error and the red line is an exponential fit. Parameters were $f=0.06$, $W=25\ell$, $\sigma=\ell/2, V_0=5E_0$, $R=30\ell$, the 1D Gaussian was placed in $R_1$ and $\bar{\sigma}=5\ell$, $k_0=1.5/\ell$. No significant off-shift $c_2$ to the exponential decay is observed, indicating that all the energy components are localised. This is confirmed by running simulations with $L=250\ell$ and ensuring the wavefunction decays almost fully within the channel, with practically no population arriving at $R_2$.}
\end{figure}

Now let us increase the translational momentum of the 1D Gaussian wavepacket to $k_0=1.5/\ell$ such that most of its energy distribution lies beyond the mobility edge (see Fig.~\ref{MEdep}, bottom right panel) and repeat the test. The corresponding flow rate measurements are shown in Fig.~\ref{MEtest} and reveal no significant off-shift, $c_2$. This suggests that all the energy components are localised, in complete contradiction to the predictions of LLT. We have confirmed this by using a very long channel ($L=250\ell$), where we observed the density eventually decay almost completely within the disordered potential, with no significant fraction of the atoms reaching the final reservoir. We have also run single-shot simulations for a 1D Gaussian with $k_0=2/\ell$ and $\bar{\sigma}=5\ell$ for quite long channels -- just about reaching the limit of what our computational resources can accomplish at the moment -- and still found no evidence of a delocalised atom fraction.

Thus, while we have documented the prediction of LLT regarding the mobility edge, it is not supported by direct Schr\"{o}dinger evolution, confirming that LLT is restricted to the low-energy, well-localised end of the spectrum, as already suggested by the applicability range of our LLT-based calculation of the localisation length in section \ref{XiSaddles}.

The original article on LLT suggested that the mobility edge was physical \cite{Marcel2012}, but in a later paper \cite{FnM2016b}, the authors realised that above the peaks of $W_E$, localisation was still present, but the tunnelling picture is no longer applicable (in fact, this description breaks down at much lower energies; see section \ref{HigherEs}). Furthermore, since $W_E$ inherits the random nature of $V$, it also supports Anderson localisation, hence enabling its eigenstates to be localised beyond the range of energies dominated by quantum tunnelling (this was also pointed out in \cite{FnM2016b}). Therefore, the energy at which the peaks of $W_E$ are attained does not represent a cut-off to the energy components that can be localised, neither in $V$ nor in $W_E$.
%
%
\section{Expanding wavepackets}
\label{Expansion}
So far throughout the paper, we have considered the localisation properties of a translating 1D Gaussian wavepacket which has the advantage of a fairly compact energy distribution. Furthermore, the translational momentum provides a direct means of tuning the average energy of this distribution and thus allows one to scan the energy of the wave probing the disorder. The other advantage of the translating Gaussian wavepacket is that since all momentum components are fairly close together in energy, they all behave more or less similarly and thus the quantitative interpretation of the results is straight-forward.

On the other hand, experiments with cold atoms to date have almost exclusively utilised expanding wavefunctions with no CoM translation. While imparting momentum to the atoms via a Bragg pulse is not impossible (and the machinery for it is in place at the laboratory where \cite{BS} has been performed \cite{therm}), it does require an extra laser and quite a lot of care to ensure clean operation. The method employed in \cite{Berthet} -- boosting a non-interacting wavepacket to a finite velocity by allowing it to accelerate in a given linear potential -- requires fast control over the linear potential through magnetic fields, which could be difficult. The question we address in this section is whether an expanding Gaussian without CoM motion can be equally well used to study Anderson localisation. Apart from providing a guide for future research, this investigation is useful also for the interpretation of all the experiments and theoretical studies performed to date that used purely expanding wavefunctions (of course, different studies used different specific wavepackets, but the common underlining principle should still be helpful).

\begin{figure}[htbp]
\includegraphics[width=6in]{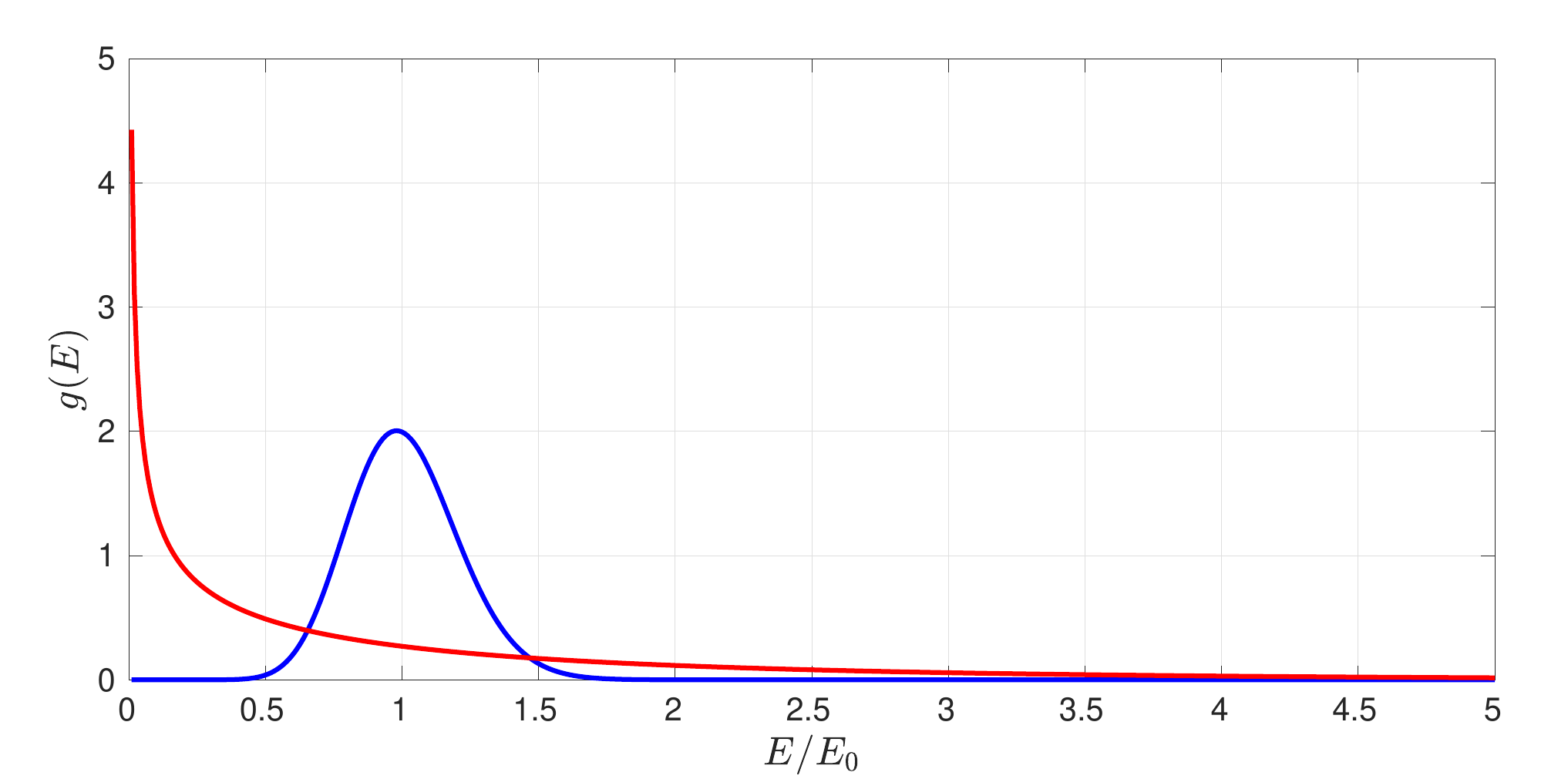}
\caption{\label{EDcomp} Normalised energy distributions for a moving 1D Gaussian wavepacket (blue) with $\bar{\sigma}=5\ell$, $k_0=1/\ell$, $E\approx1.17E_0$ and an expanding one (red) with $\bar{\sigma}=0.5\ell$, $k_0=0$, $E\approx1.05E_0$. While for the translating case the energy distribution is fairly compact, centred on the mean energy, the expanding Gaussian distribution has most of its weight around zero energy and possesses a long, weakly populated tail extending to high energies.}
\end{figure}

To begin with, let us inspect the energy distributions of a translating and expanding Gaussian of approximately the same mean energy. Figure \ref{EDcomp} shows the energy distribution of the translating wavepacket mostly used so far with $\bar{\sigma}=5\ell$ and $k_0=1/\ell$, giving a mean energy of $E\approx1.17E_0$. If we insist on a stationary CoM ($k_0=0$) for our expanding wavefunction, then the only way to increase the energy is through the width of the momentum distribution. A comparable mean energy can be attained with $\bar{\sigma}=\ell/2$, yielding $E\approx1.05E_0$. The energy distribution for this initial condition is also illustrated in Fig.~\ref{EDcomp}, and the difference is quite striking: the translating Gaussian is centred on the mean energy and is fairly compact, involving only energies close to the mean. In contrast, the expanding wavefunction spans a large range of energies, with most of the weight concentrated at very low energies and a long, thin tail extending to very high energies.

Now, most experiments with cold atoms to date have initiated an expanding wavefunction inside the disordered potential, allowed it to evolve for a long time, and observed the exponential decay in the wings of the wavefunction. Is such a technique useful for studying Anderson localisation? First of all, the energy distribution of a wavepacket in free space is altered when it is initiated within a potential. The additional potential energy shifts the entire distribution upwards, and one can never truly know what the resulting distribution is, seeing as the potential is of a random nature. As a result, the same very wavepacket initiated inside the disorder will localise more weakly (with a longer localisation length) than if it was initiated outside the noise and allowed to transmit into it. A further problem is that if we change the strength of the disorder (either $f$ or $V_0$), keeping the probing wave the same, the energy distribution of this wave (initially placed inside the potential) changes together with the noise parameters, as the energy distribution shifts by more or less compared to its free-space distribution, depending on how much energy is added by the potential. Thus, we would not see strong changes in the observed localisation properties if we change the noise strength, as the probing energy components inevitably change with the noise. This makes wavepackets initiated inside the disorder less than ideal for studying Anderson localisation. Note that all the statements made in this paragraph have been confirmed by direct numerical simulations.

This reasoning leads us to focus on the case when the atoms are initiated outside the disorder, allowed to transmit through it, and collected on the other side. Thus, the set up remains identical to that used in section \ref{Trans}, except here we will use purely expanding 1D Gaussians rather than translating ones. We note that the addition of empty reservoirs on either side of the noisy potential weakens localisation in fairly thin strips along the channel edges opening onto the reservoirs, but this effect is only noticeable (or even dominant) when the localisation length $\xi_E$ is similar to or smaller than the mean distance between valley lines $\bar{D}$ (we shall work away from this limit).

\begin{figure}[htbp]
\includegraphics[width=6in]{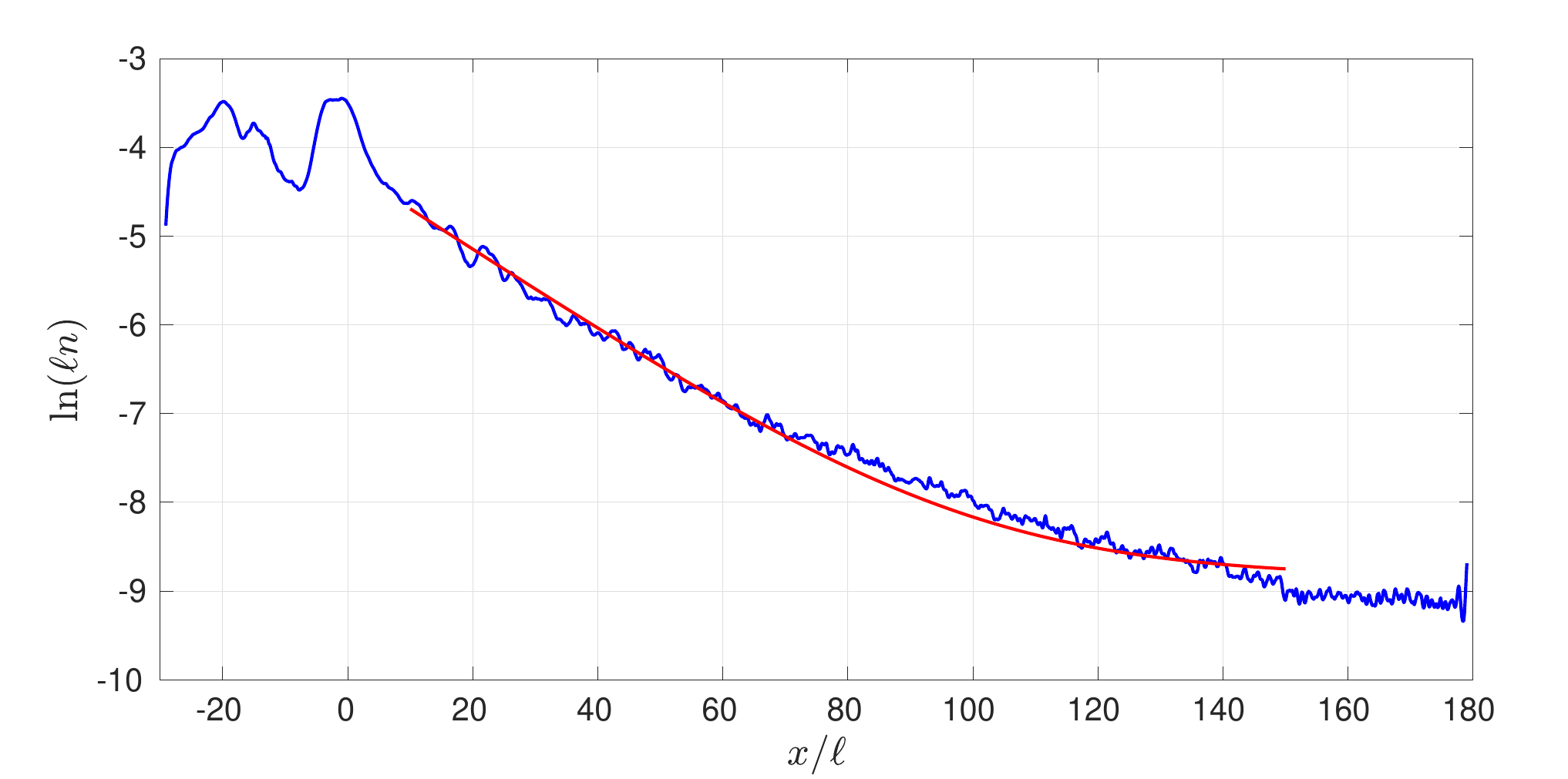}
\caption{\label{TransDen} Logarithm of the density profile at $t=50t_0$ (by which time a quasi steady state has been reached) upon transmission of a stationary 1D Gaussian with $\bar{\sigma}=0.5\ell$, $k_0=0$ initiated in the centre of $R_1$. Parameters were $L=150\ell$, $W=25\ell$, $R=30\ell$, $f=0.06$, $V_0=5E_0$, $\sigma=\ell/2$. The profile shown is averaged over 20 noise realisations, and the red line is an exponential fit.}
\end{figure}

\begin{figure}[htbp]
{\includegraphics[width=3in]{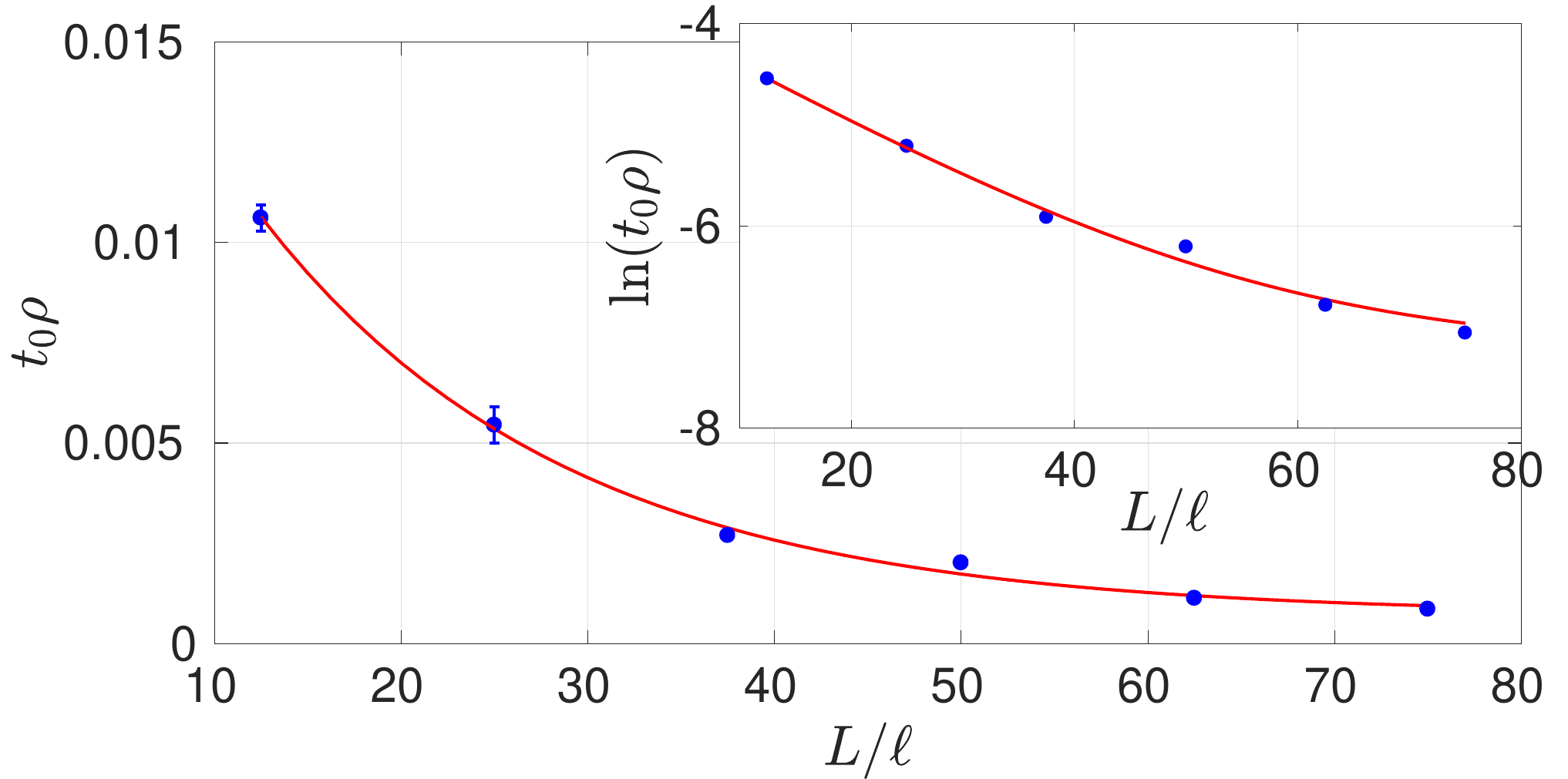}}
{\includegraphics[width=3in]{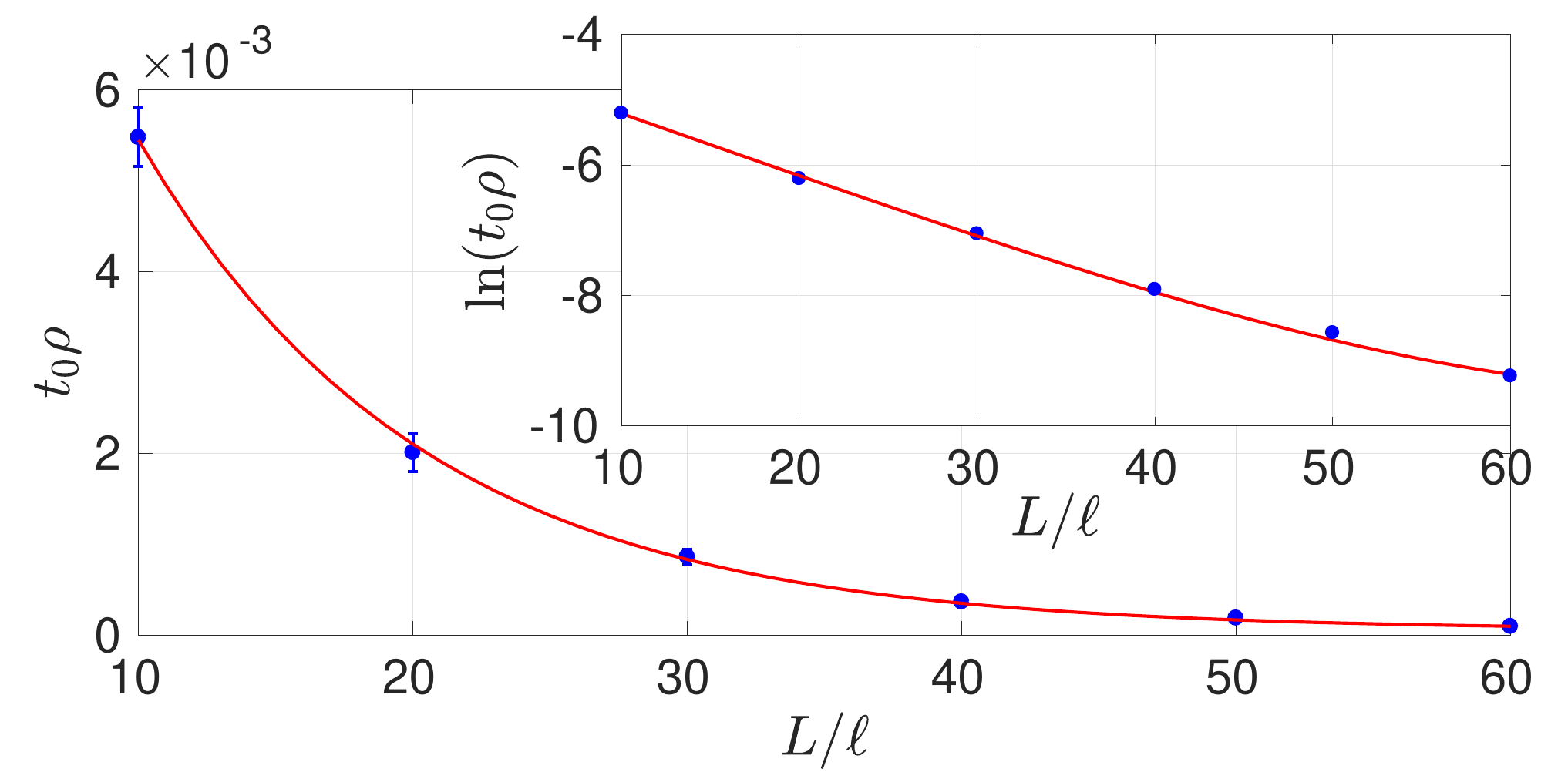}}
\caption{\label{EvsT} Flow rate of a purely expanding Gaussian with no CoM motion in the transmission set-up. Parameters common to both panels were $W=25\ell$, $R=30\ell$, $\sigma=\ell/2$, $\bar{\sigma}=\ell/2$, $k_0=0$. Left panel: $f=0.06$, $V_0=5E_0$, right panel: $f=0.1$, $V_0=10E_0$. Every data point was obtained by averaging over 10 noise realisations. The error bars show the standard error. The solid line is an exponential fit, allowing for a small off-shift, resulting in the following length-scales: $\bar{\xi} = 32.825\ell$ (left panel), and $\bar{\xi} = 20.636\ell$ (right panel).}
\end{figure}

Next, because in an expanding wavefunction there are so many widely different energy components (propagating at different speeds) and since all of them localise on different length scales, there is no \textit{a priori} guarantee that the overall density profile will have an exponential functional form. Using a very long channel in an attempt to localise as many of the energy components as possible, we still see a small fraction of the atoms -- the ones at the high end of the distribution -- arrive at $R_2$. The overall density profile reaches quasi-steady state fairly quickly, however, and if any deviation from an exponential curve (with a small off-shift) exists, it is not obvious (see Fig.~\ref{TransDen}). The same is true of the flow rate, $\rho(L)$ (see Fig.~\ref{EvsT}). Both the density profiles and the flow rate depend strongly on the properties of the disorder, as we now demonstrate.

Extracting the flow rate $\rho(L)$ in the transmission set-up with an expanding wavefunction for two different disorder strengths yields the results shown in Fig.~\ref{EvsT}. We note that the normalised population of the drain reservoir, $r_2$, shows a clear and clean initial linear increase (even for long channels where one must be somewhat careful when using translating Gaussians; see section \ref{Trans}) which is easily fitted. The resulting curves $\rho(L)$ are indeed exponential, but it seems that a small off-shift should be allowed for in the fitting function to capture the observed dependence. This does not indicate a mobility edge: we have tested this by increasing system size and measuring the density in the drain reservoir, to confirm that the transmitted population keeps decreasing, so if we extended our $\rho(L)$ data to higher $L$ we would see it decay essentially to zero. The length scale of this decay, $\bar{\xi}$, is readily extracted via such fitting.

\begin{figure}[htbp]
\includegraphics[width=6in]{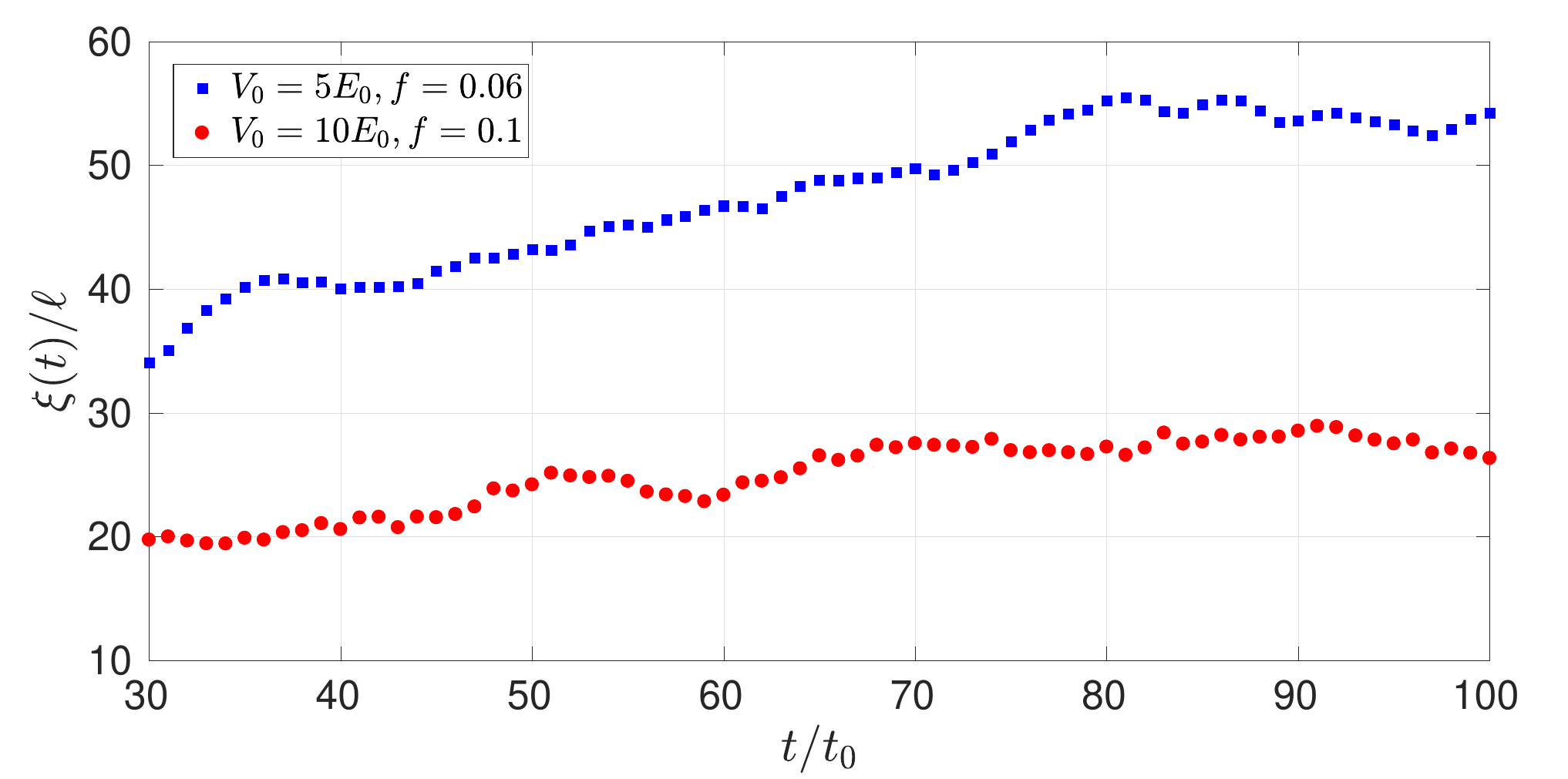}
\caption{\label{xioft_EinT} Observed localisation length as a function of time, extracted from non-linear fits to the density (allowing for a small off-shift), as a 1D Gaussian expands through a long channel and experiences Anderson localisation. Parameters common to both simulations were $W=25\ell$, $R=30\ell$, $\sigma=\ell/2$, $\bar{\sigma}=\ell/2$, $k_0=0$. The fill factor and scatterer height are specified in the legend, and the corresponding channel lengths were taken as: $V_0=5E_0$, $f=0.06$: $L=150\ell$; $V_0=10E_0$, $f=0.1$: $L=75\ell$. The length scale $\bar{\xi}$ extracted from the flow rate is qualitatively correlated with the long-time values of $\xi(t)$ and is very close to its early quasi-steady-state values.}
\end{figure}

Furthermore, in section \ref{Trans}, we have shown that for translating Gaussians, the length scale $\bar{\xi}$ was correlated with the decay rate seen in the density profiles as a function of time, $\xi(t)$. The same is true for expanding wavefunctions, as demonstrated in Fig.~\ref{xioft_EinT}. With expanding wavefunctions -- in contrast to translating ones -- true steady state is attained much more rapidly, and the similarity of $\bar{\xi}$ to $\xi(t)$ in early quasi-steady-state (the earliest $t$ values for which $\xi(t)$ is shown) is even stronger.

Thus, the flow rate analysis is equally helpful for studying the overall, observed decay of the density profiles when using purely expanding wavefunctions in the transmissive set-up. The only disadvantage of such probing waves is that it is not simple to resolve the energy dependence of the localisation length. For arbitrary wavepackets, one should reconstruct the expected density profile from the energy-resolved localisation length $\xi_E(E)$ according to, e.g., equation (63) of Ref.~\cite{DelandeLectures}. For energy distributions that are fairly localised -- such as translating Gaussians of fairly narrow width in momentum -- this can be neglected to a reasonable level of approximation, and by tuning the central energy of the wavepacket, one could sample $\xi_E(E)$. This is, however, completely impossible with purely expanding wavepackets due to the shape of the energy distribution (see Fig.~\ref{EDcomp}).
\section{Connection to the Anderson model}
\label{BHM}
So far, from the point of view of LLT, we have seen that Anderson localisation arises due to tunnelling of the wavefunction through the peaks of the effective potential $W_E$. But how does the disorder -- the random component -- come into the picture? We address this question using two different approaches in this and the next section. The first approach relies on establishing a connection between our continuous system and the Anderson model (also commonly referred to as a tight-binding model), the work-horse of localisation studies, e.g.~\cite{Sheng, HerbertJones, Thouless, Kirkman, Sarma, russian_guys, Fan, Heinrichs, Greek, Romer, Romer2, He, finite_scaling, Benoit, Aoki, Chinese, 2D_corr, LeeFisher, Malyshev, Weaire, Greek2, 1D_corr, Makarov}. In its simplest form, the Anderson model consists of a lattice of ``sites'', described by a single quantum-mechanical state each, that are coupled to their nearest neighbours by tunnelling. In the canonical form of the model, the on-site energies are drawn out of a random distribution, a case which is known as ``diagonal disorder''. On the other hand, off-diagonal disorder occurs when the hopping strengths between sites are randomised. The latter case has been studied in \cite{Romer, Romer2, LeeFisher, Chinese, Greek2, 1D_corr, Makarov, Weaire}, with several of these papers \cite{Romer, Chinese, Greek2, 1D_corr, Weaire} concluding that the two ways of randomising the model are not equivalent. In particular, off-diagonal disorder was found to be not as efficient at inducing localisation. Interestingly, our results in this section point in the same direction, providing a new vantage point to an old problem.

To begin with, we recall that the valley lines of $u$, corresponding to the peak ranges of $W_E$, divide the system into a collection of domains, which in the effective potential look like local wells, i.e.~local oscillators. These are not completely decoupled, of course: we know that it is possible to tunnel out of each domain into its nearest neighbours. If we consider a single, isolated domain, then LLT allows us to construct its fundamental eigenmode and calculate the corresponding eigen-energy trivially through equations (\ref{local_mode}) and (\ref{local_energy}). These equations are simple to verify in practice, and an example was shown earlier in Fig.~\ref{TestLocal}. If we now bring the rest of the system -- all the other domains -- back into the picture, the local eigenstate will ``spill out'' into its nearest-neighbour domains and the result will be very close to a true low-energy eigenstate of the entire system. We will refer to these as ``nearest-neighbour coupled'' (NNC) states. In fact, our work in section \ref{XiSaddles} provides a method for computing the amplitude of the wavefunction on each of the neighbouring domains at very low energy: their occupation is linked to the amplitude of the strongly populated domain through the ``mean'' Agmon distance, and we have shown that the latter can be reasonably calculated by following the paths connecting the maxima of $u$ through the saddles, the approximate candidate paths of least cost.

Since we know that the wavefunction remains more or less constant within each domain, we could approximate the amplitude on the secondary, nearest-neighbour domains as constant. Furthermore, due to the exponential suppression of the amplitude each time a valley line is crossed, only nearest-neighbour coupling needs to be considered. Let us calculate the average value of the wavefunction amplitude in the fundamental mode on the main domain: 
\begin{equation}
\bar{u}_j = \frac{1}{A_j}\int\limits_{\Omega_j} u\ d\mathbf{r},
\end{equation}
where the area of the domain is
\begin{equation}
A_j = \int\limits_{\Omega_j}1 \ d\mathbf{r}
\end{equation}
and we denote the region occupied by domain $j$ by $\Omega_j$. In the NNC states, the nearest neighbours have a constant amplitude of
\begin{equation}
\bar{u}_j\exp(-\bar{\rho}_E(j,i)),
\end{equation}
where the indices $j,i$ label the main and neighbouring domains (respectively) in question. The mean Agmon distance is of course energy-dependent, as always, and should be evaluated at the energy of the on-site fundamental mode $E_j$ for approximating NNC states.

We can build up the entire set of NNC states, where each one has a single strongly occupied domain in its fundamental mode and a small, constant amplitude on its nearest neighbours. These states must be normalised to unity, as usual, taking into account their population on all occupied domains. The normalisation constant becomes
\begin{equation}
n_j = \sqrt{\int\limits_{\Omega_j}u^2\ d\mathbf{r} + \sum\limits_{i\in \mathrm{nns}} A_i \left[\bar{u}_j e^{-\bar{\rho}_E(j,i)}\right]^2},
\end{equation}
where ``nns'' stands for the set of nearest neighbours of domain $j$.

Our goal now becomes to use this picture to develop a simple model of dynamics. Note that it is not easy to account for the fact that as energy increases domains merge. It is much simpler to keep this description a low energy one, such that the number of domains remains energy independent. Furthermore, beyond the energy at which the mean Agmon distance stops correctly predicting the decay of eigenstates, we can no longer compute the coupling coefficients between the domains. Therefore, since imposing the extremely low-energy restriction does not come at the cost of losing insight, we adopt it.

We remark, in addition, that a very strong approximation is implicit in only considering NNC states: \textit{most} states in the exact spectrum of the Hamiltonian are dropped, and not all of the ``retained'' NNC states necessarily exist as part of the complete spectrum. In addition, there is no clean energy cut off below which all states are NNC and above which reside all the other possible complicated states.

Now, Hamiltonian matrix elements can be easily computed between any two states where the fundamental domains are nearest neighbours. In this picture, the system is reduced to a discrete model of several coupled linear oscillators, equivalent to Anderson's original model for localisation in 2D. The difference is that in our case, all parameters come out of LLT and all have a random component, each site has a different number of nearest neighbours and each noise realisation is different. The Hamiltonian takes the form
\begin{equation}
H_A = \sum\limits_j E_j a^{\dagger}_ja_j + \sum\limits_{<i,j>}t_{i,j}(a^{\dagger}_ja_i+a^{\dagger}_ia_j),
\end{equation}
where $a_j^{\dagger},\ a_j$ are bosonic creation and annihilation operators, the coupling is restricted to nearest neighbours, $E_j$ is given by (\ref{local_energy}), the overlap integrals are
\begin{equation}
t_{i,j} = \int d\mathbf{r}\ \psi_i(\mathbf{r}) H \psi_j(\mathbf{r}),
\end{equation}
and $\psi_i(\mathbf{r})$ are the NNC states. The matrix elements $t_{i,j}$ can be evaluated by making use of the defining property of the localisation landscape, $Hu=1$, leading to
\begin{equation}
t_{i,j} = \frac{1}{n_in_j}(\bar{u}_jA_ie^{-\bar{\rho}_E(j,i)} + \bar{u}_iA_je^{-\bar{\rho}_E(i,j)}).
\end{equation}
The Heisenberg equations of motion for this system of coupled harmonic oscillators are
\begin{equation}
i\hbar\frac{da_k}{dt}=E_ka_k+\sum\limits_{j}t_{j,k}a_j,
\end{equation}
but we will make a classical field approximation and treat the amplitudes $a_k$ as complex numbers.

With this simple, discrete model at our disposal, we can understand the role played by disorder in transport suppression. First, recall that LLT tells us that eigenstates of the full (continuous) Hamiltonian only spread across two adjacent domains if the energy is high enough for the domain wall between them to break down, or if the two domains have very similar local energies. It is the second case that is of interest to us now. The same physics is captured by the realisation that if two coupled oscillators have very different frequencies, energy transfer between them is suppressed (due to energy conservation): the detuning limits the transfer of excitations, just like for a pair of coupled pendula.

If we use an ordered lattice of potential scatterers (see the next section), the landscape $u$ and the associated valley network are completely regular. Every domain is identical and all domains have the same on-site energy. While there are still barriers in $W_E$ (since $u$ still has valleys), the eigenstates are all completely extended and the transfer of excitations from one site to the next in our discrete, dynamical model is complete. Thus eigenstate localisation results from the combination of two factors: tunnelling through the potential barriers of the effective potential $W_E$, and the energy mismatch between the domains arising from the randomness. Equivalently, transport suppression originates from the fact that the coupling strengths $t_{i,j}$ are small (slowing down the transfer of excitations) and the set $\{E_k\}$ has a random component, limiting the amount of population transfer between detuned nearest neighbours. This is why randomising the two sets of parameters has a different effect on inducing localisation, as was found in previous studies, with only diagonal disorder leading to true strong localisation.

We highlight the fact that the LLT-Anderson model we have presented is very limited by several strong approximations. In particular, only nearest-neighbour coupling is accounted for, only states where a single domain is strongly occupied in its fundamental mode (and tunnels into nearest neighbours), the coupling coefficients between domains can only be computed at very low energies, and we do not attempt to push the model up to and beyond the point where domains begin to merge. Furthermore, note that the spatial extent of domains is completely neglected -- they are taken as discrete points -- which eliminates travel time across domains. In principle, this could be remedied by incorporating time delays into the coupling terms in the Hamiltonian, proportional to the distances between domains. In practice, differential equations with time delays behave very unnaturally, with populations diverging to infinity due to feedback from the past, so we opt out of this approach.

The limitations listed in the paragraph above are ``acceptable'' in the sense that the model can still be implemented and studied, with the results providing useful qualitative insight. There is another, much more severe problem with reducing the continuous 2D system to the discrete Anderson model, which cripples the entire enterprise.

One of the main problems with using the LLT-Anderson model in practice lies in the fact that the coupling strengths $t_{ij}$ have non-trivial \textit{signs}. Inspecting the exact eigenstates of $H$, we find that they can be expressed as purely real functions, but the signs of the wavefunction on the primary occupied domain and all its nearest neighbours certainly need not be the same. Localisation landscape theory gives us a way to compute the amplitude of the wavefunction on the nearest-neighbour domains at low energy, but tells us nothing about the signs. Since the LLT-Anderson model is a coherent one, phase information such as relative signs cannot be dropped at any cost. Thus for this model to be at all useful, one needs a method to deduce the signs of the wavefunction on the different domains in the NNC states.

One idea that could be developed and used in the future is requiring the set of all NNC states to be approximately orthogonal. In a valley network with $N_D$ domains, we could represent all the NNC states as discrete vectors in $\mathbb{R}^{N_D}$ where the basis states are the domains and the amplitude is the average of the wavefunction on the domains in each state. The signs of all these entries are initially undetermined. One could pose a minimisation problem that requires the inner product of all states with an overlap (that is, all NNC state pairs where the strongly occupied domains are up to twice removed neighbours) to be as small as possible. We only need to solve for binary variables ($\pm1$), but the problem can easily be under-determined. This is a possible avenue to pursue in the future but we have not yet attempted to implement this idea.


If the sign problem is solved at some stage, then the first test of the LLT-Anderson model would be as follows. Imagine initiating the system with only one internal domain excited in its fundamental mode at $t=0$. It is straight-forward to derive an expression for the population of its nearest-neighbours as a function of time, following similar logic and methodology to that presented in this section, based on LLT. The evolution in the discrete model can then be compared to exact Schr\"{o}dinger evolution in the continuous model where the same domain is initiated in its fundamental local mode, and the evolving wavefunction is integrated over the area of each domain to produce an output that can be directly compared to the discrete model. At best, we could hope for qualitative agreement in the low energy limit due to the large number of states missing from the LLT-Anderson model representation and the fact that the coupling coefficients between domains can only be computed at very low energies.

\begin{figure}[htbp]
{\includegraphics[width=3.1in]{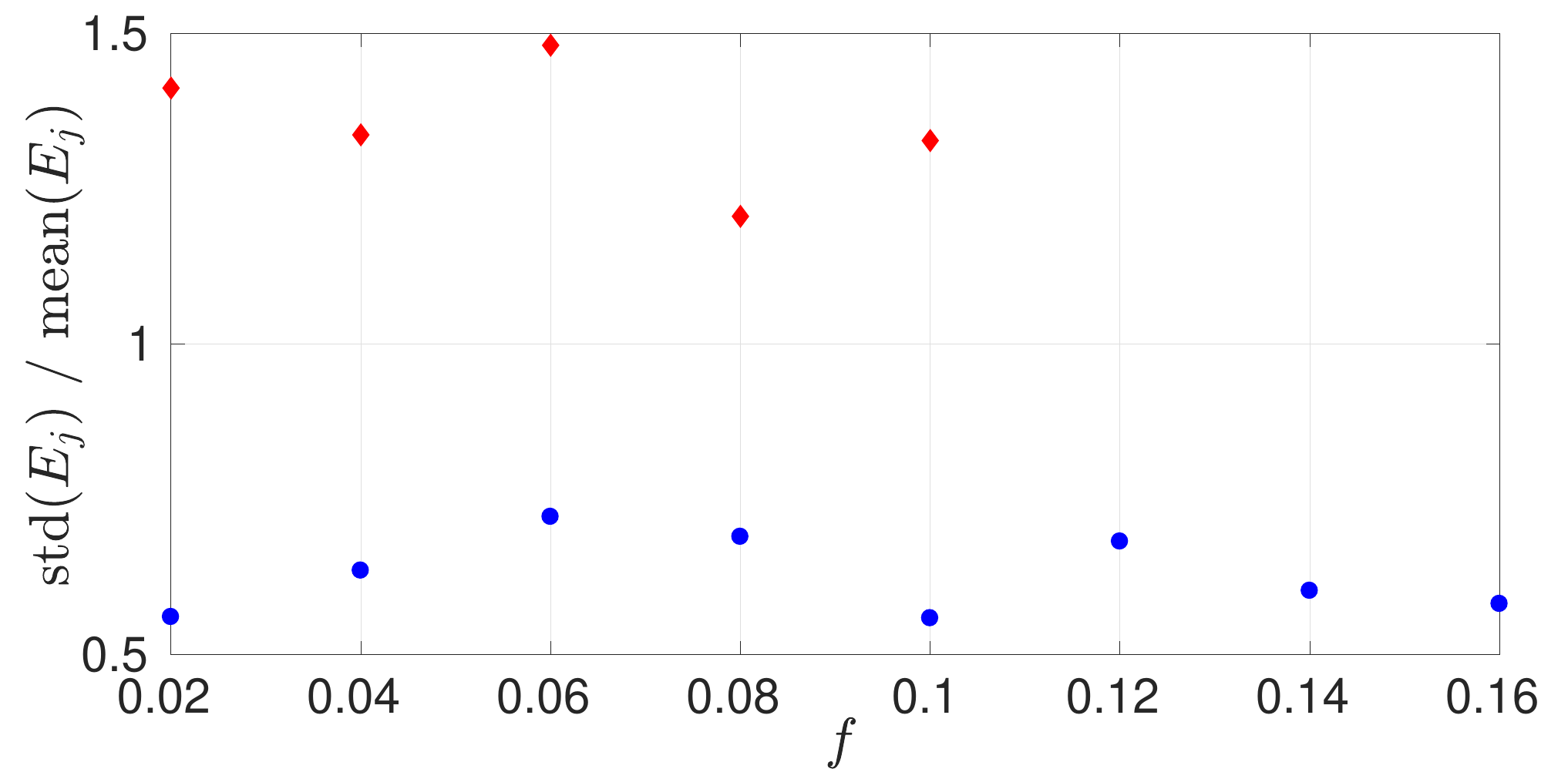}}
{\includegraphics[width=3.1in]{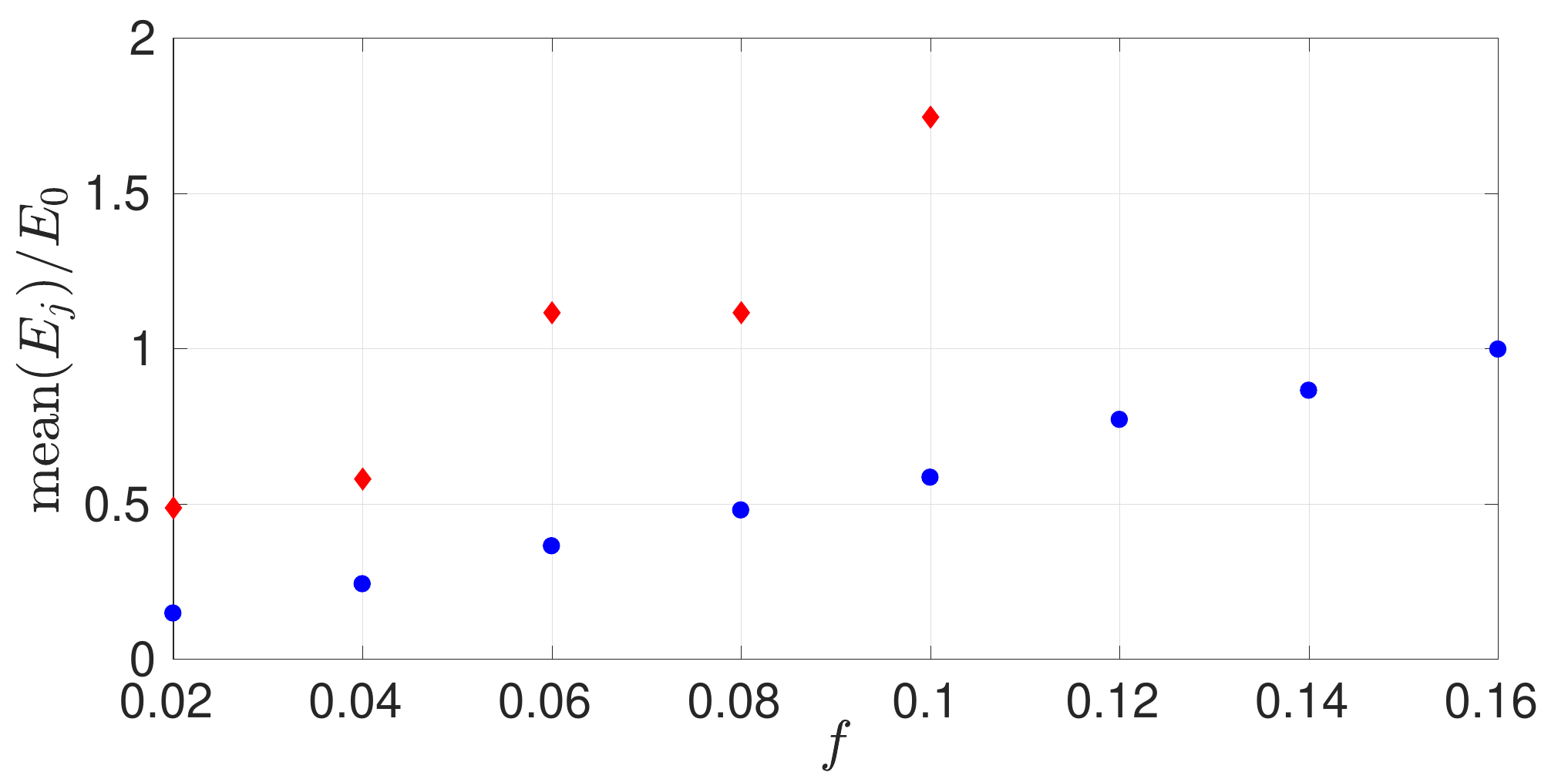}}
{\includegraphics[width=3.1in]{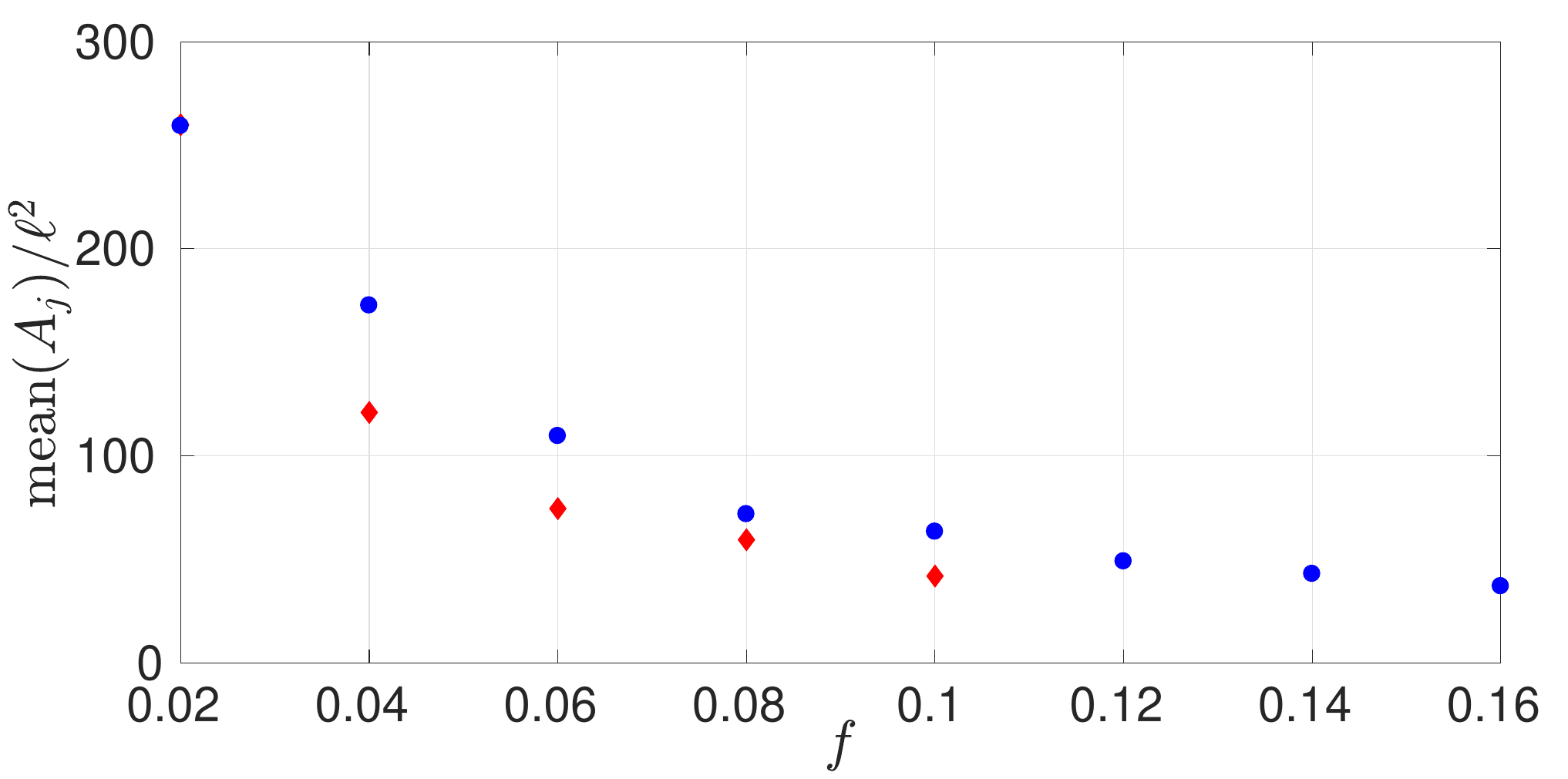}}
{\includegraphics[width=3.1in]{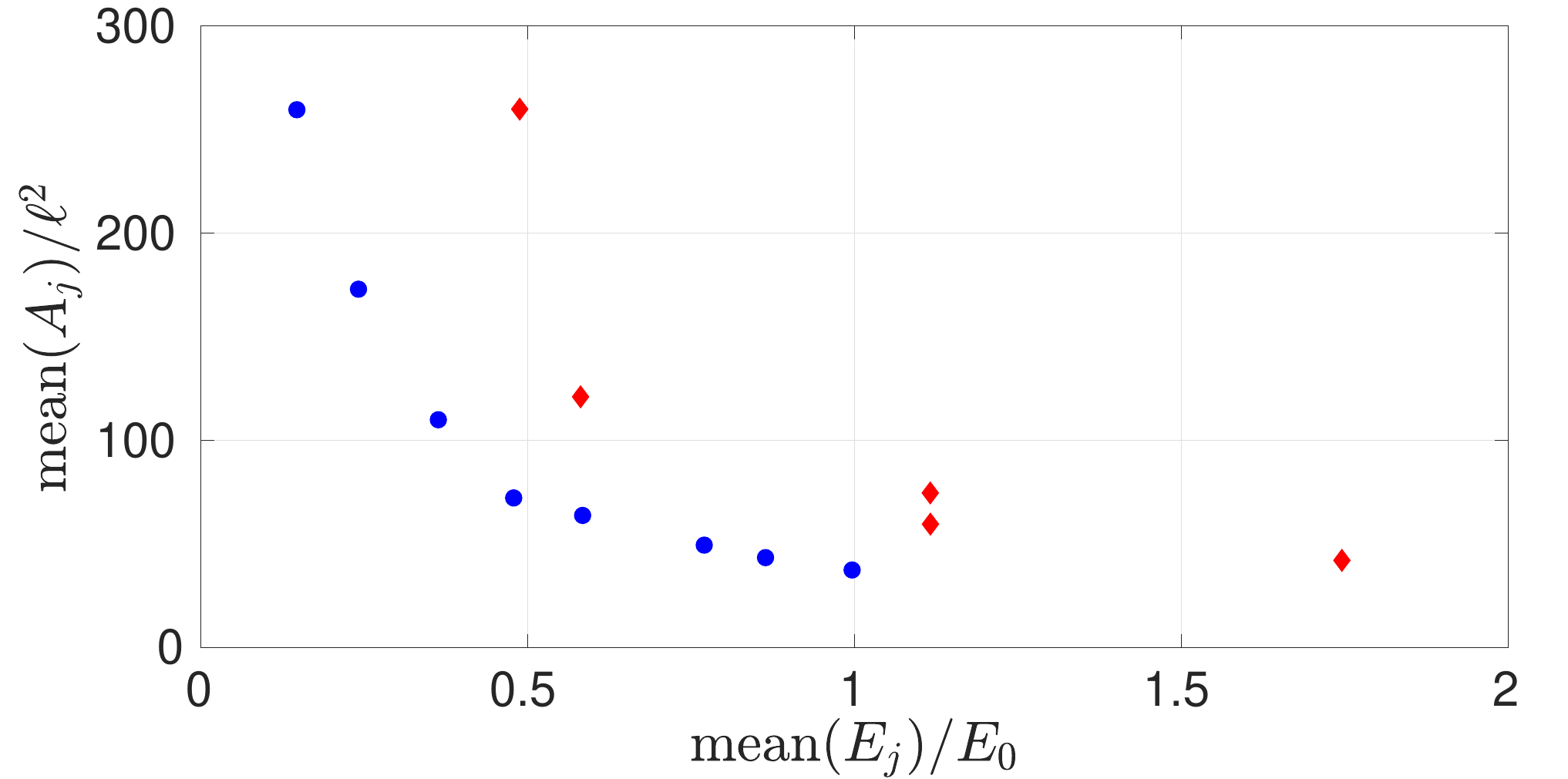}}
\begin{center}
{\includegraphics[width=3.5in]{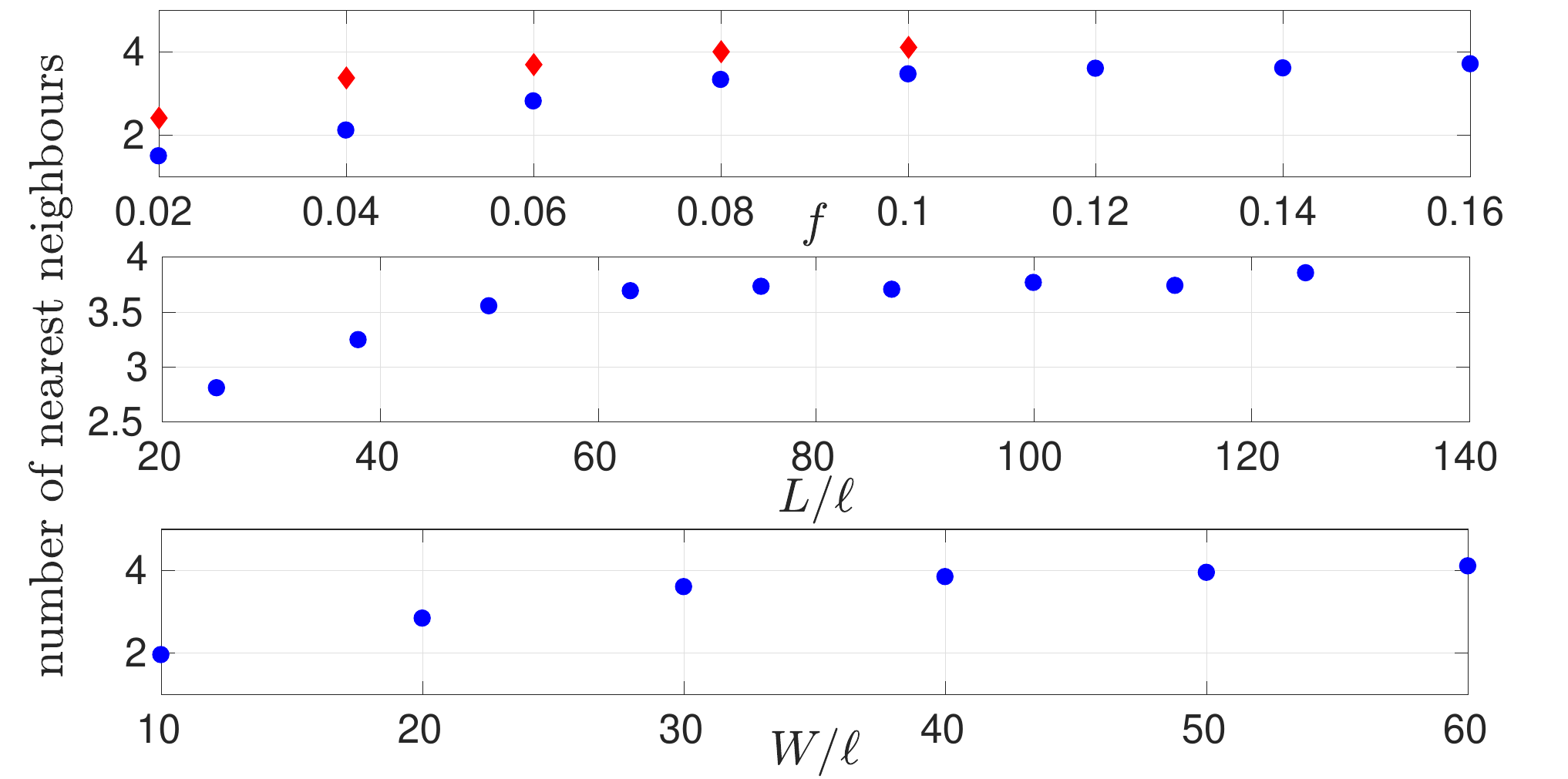}}
\end{center}
\caption{\label{Stats} Dependence of various quantities featuring in the LLT-Anderson model, averaged over domains and over 20 noise realisations. Blue circles: $V_0=5E_0$, $\sigma=\ell/2$, red diamonds: $L=50\ell$, $W=25\ell$, $V_0=20E_0$, $\sigma=\ell/2$. First four panels: blue circles correspond to $L=W=25\ell$. Bottom figure, first subplot: same parameters and colour code as the first four panels. Second subplot: $f=0.06$, $W=25\ell$, third subplot: $f=0.05$, $L=50\ell$. Top left: the relative fluctuations of the local domain energies display no dependence on fill factor but increase with $V_0$. Note that we have explicitly confirmed that this increase is not attributed to the larger $L$ used for the red data points. Top right: the local energies increase with both fill factor and scatterer height (the dependence on $V_0$ seen is not due to $L$, as a higher $L$ decreases mean $E_j$). Middle left: the domain area decreases with both fill factor and scatterer height (the dependence on $V_0$ seen is not due to $L$, as no consistent change is seen in $A_j$ with increasing $L$). Middle right: There is a clear inverse correlation between the energy and the area of a domain, much like for a simple harmonic oscillator. The shift of the curve for higher $V_0$ is not due to higher $L$, as neither the mean of $A_j$ or of $E_j$ shows a consistent dependence on $L$. Bottom panel: the mean number of nearest neighbours each domain possesses grows with increasing disorder strength as the domains shrink and the valley network becomes more compactly packed. The observed growth with $L$ and $W$ is a finite-size effect that is eliminated in the large system limit. The width dependence is a quantification of the dimensional crossover from 1D to 2D, and the qualitative restructuring of the valley network seen in Fig.~\ref{W_nets}.}
\end{figure}

While the LLT-Anderson model cannot be used to accurately model dynamics (see above), we can learn about our system by inspecting statistical properties of quantities that enter the simplified model as a function of parameters (the sign problem is irrelevant for this exercise). First of all, we might wonder if the relative detuning between domains, on average, depends on the properties of the disorder. We may compute it as the standard deviation of the local energies over their mean, and average over 20 noise realisations, as usual. In Fig.~\ref{Stats}, we show that surprisingly, the relative fluctuations do not depend on the fill factor, but do increase with scatterer height. The mean on-site energy increases with fill factor and scatterer height, while the mean domain area decreases. The energy and area are inversely correlated, as might be expected from the relation for a harmonic oscillator. We note that there is no consistent change in the average area, local energy or the variability of the latter with system size, except at very small $L,W<\bar{D}$.

Our computational scheme (see appendix \ref{appLLTnew}) allows for the identification of nearest neighbour domains; in Fig.~\ref{Stats} we also show the mean number of neighbours each domain has. It increases with fill factor and scatterer height because the domains become smaller and more compactly fitted, but it also grows with $L$ and $W$, a finite size effect which vanishes for sufficiently large systems. The increase with $W$ is more pronounced as the explored values of $W$ are smaller -- this is a manifestation of the dimensional crossover from 1D to 2D discussed in section \ref{WidthDep}, illustrated in Fig.~\ref{W_nets}. 

In summary, this section demonstrated how our continuous 2D system may be reduced to a discrete lattice model, equivalent to Anderson's model for localisation. The entire enterprise is highly approximate and can only be used to derive qualitative insights. It served to highlight the importance of disorder in detuning the domains -- thought of as local oscillators -- from each other, thus limiting the efficiency of excitation transfer.
\section{Distilling the effect of disorder}
\label{Ordered}
Anderson localisation is usually identified by its trade-mark property: an exponentially decaying density profile for a wave travelling in a disordered potential. However, there are other mechanisms at play which can often create similar effects and lead to the misinterpretation of experimental data and simulation results. One such mechanism is classical trapping: if the potential $V$ is sufficiently dense and the scatterers are considerably higher than the atomic energy, the wavefunction may become trapped in a local minimum of $V$, tunnelling out, causing exponential decay, but for reasons other than Anderson localisation. Even when there are no trapping regions in $V$, if it is sufficiently dense and high, waves passing through will feel a degree of attenuation. An excellent way of determining whether there is any observable effect from the randomness of the noisy potential is to compare it directly to a regular lattice of scatterers of the same height and density as used in the disordered case\footnote{This idea was developed during work towards the research presented in \cite{BS}, benefiting from formidable contributions from Donald H.~White, to whom we are grateful for his input.}, as was also done in \cite{BS, Malyshev, russian_guys}. In this section we do just that.

\begin{figure}[htbp]
{\includegraphics[width=3.1in]{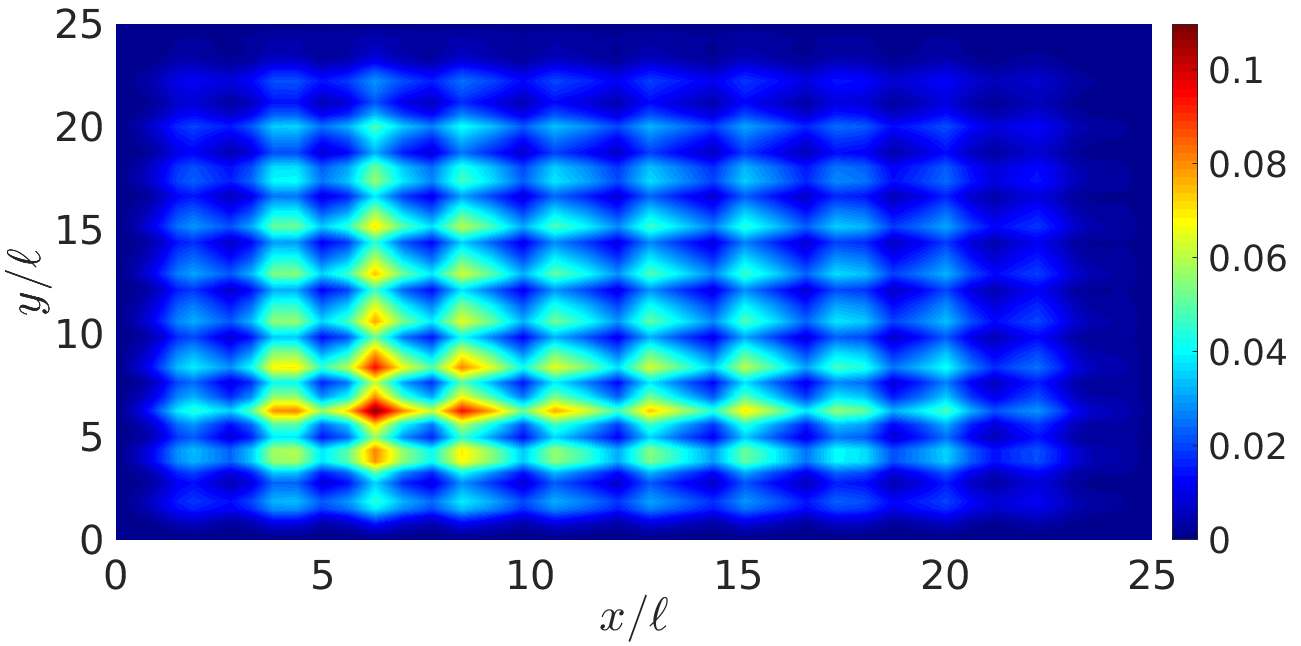}}
{\includegraphics[width=3.1in]{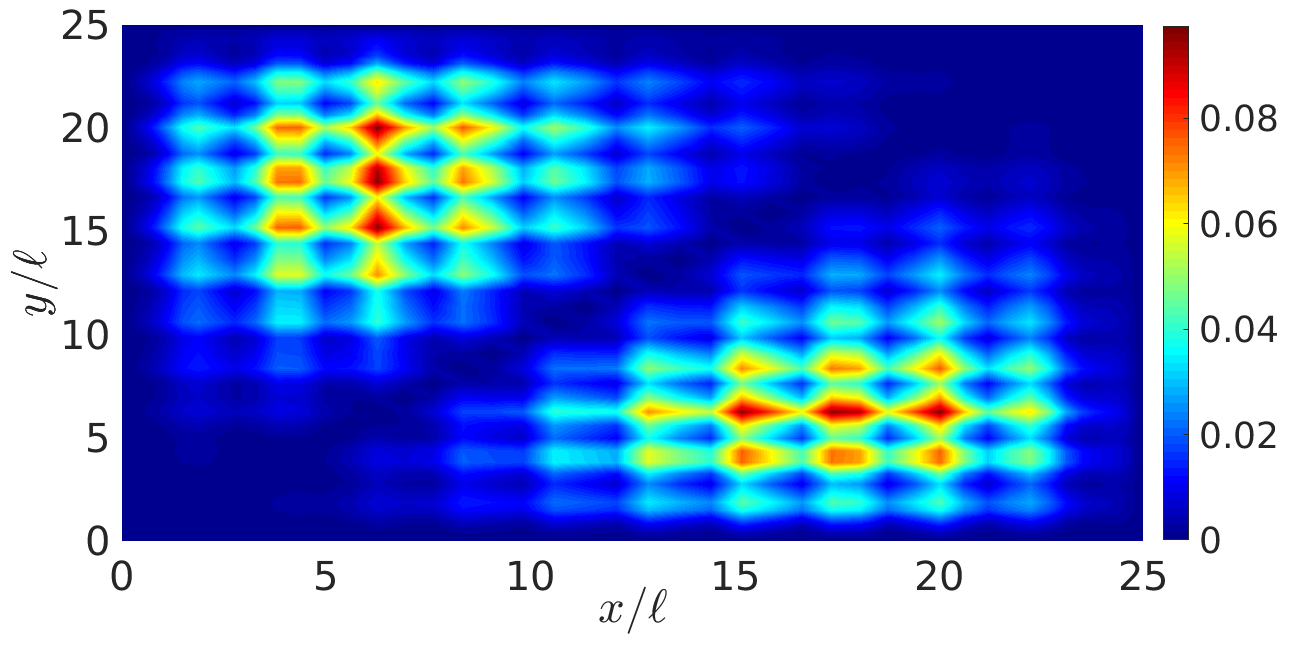}}
\caption{\label{OSpec} The lowest two eigenstates of the Hamiltonian (\ref{Ham}) for ordered scatterers with $L=W=25\ell$, $f=0.2$, $V_0=20E_0$, $\sigma=\ell/2$, plotting $\ell\left|\psi\right|$ as a colour-map. All of the eigenstates are completely delocalised and have much higher energies than their localised counterparts with a random potential.}
\end{figure}

As always, the eigenspectrum of $H$ is a good place to start our investigation. Figure \ref{OSpec} shows two of the lowest energy eigenstates using high and dense ordered scatterers, where for the same parameters with random scatterers the eigenstates are extremely localised. Clearly they are completely delocalised (as are higher states), as expected, which confirms that the localisation seen in the eigenstates is caused by the randomness. The numerous nodes visible in these states come at an energy cost: the lowest eigenvalues are much higher than the equivalent typical numbers with a noisy potential.

\begin{figure}[htbp]
{\includegraphics[width=3.1in]{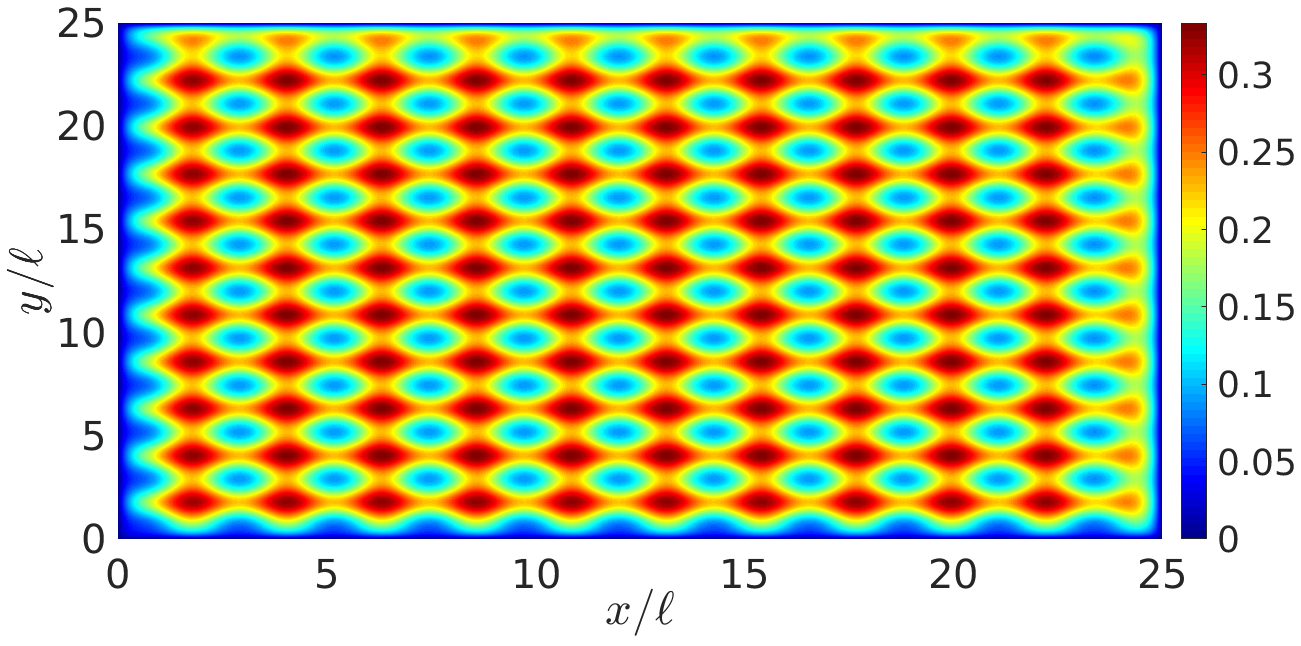}}
{\includegraphics[width=3.1in]{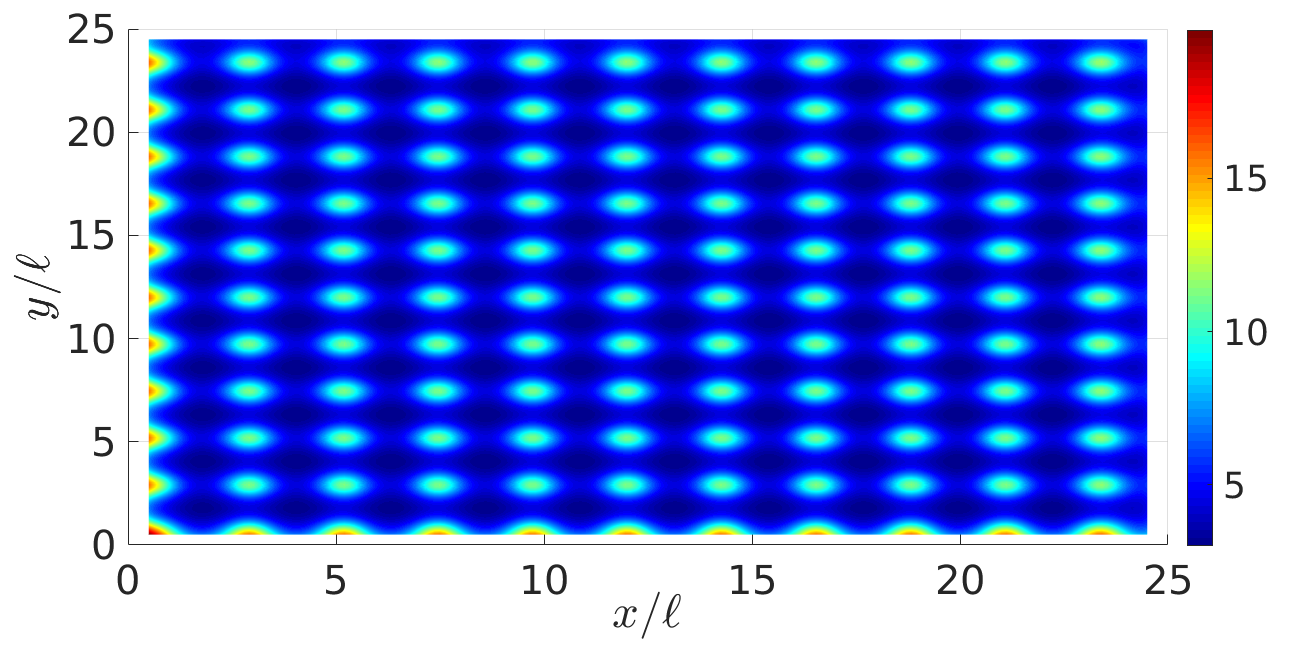}}
\begin{center}
{\includegraphics[width=3.1in]{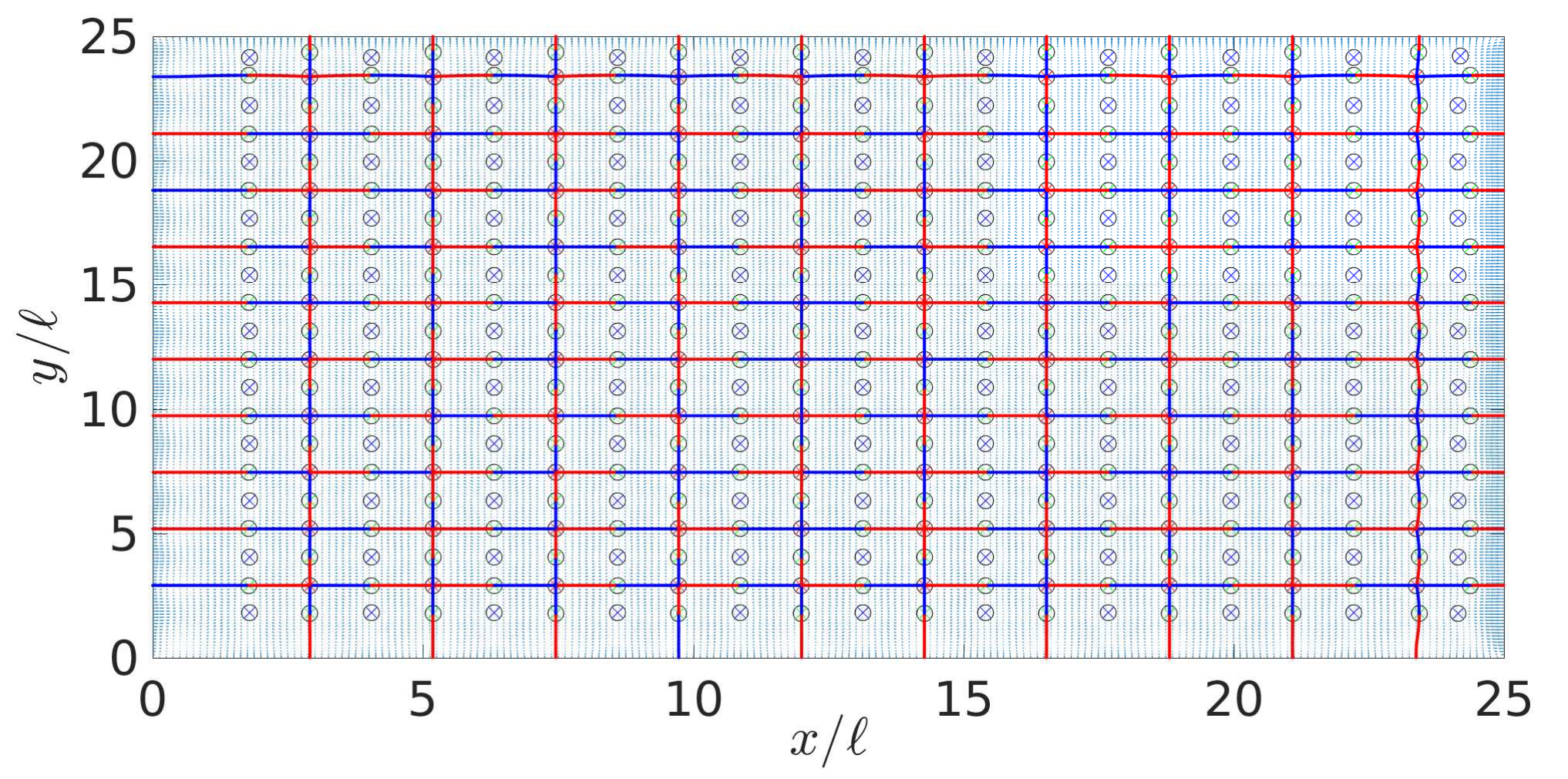}}
\end{center}
\caption{\label{Or_LLT} The localisation landscape $u$ (top left) and effective potential $W_E$ (top right) shown as a colour-map for ordered scatterers with $L=W=25\ell$, $f=0.2$, $V_0=20E_0$, $\sigma=\ell/2$. The associated valley network is depicted in the bottom panel. The lines, symbols and vector field plot have the same meaning as in Fig.~\ref{Neteg}. All three objects are completely regular, with all domains identical.}
\end{figure}

Next, we inspect the key objects of LLT: Fig.~\ref{Or_LLT} shows the localisation landscape, effective potential and valley network for an ordered lattice of high and dense scatterers. While there are many domains and the peaks in $W_E$ are high (due to the strength and density of the scatterers), all domains are identical and would have the same exact local eigen-energies. As discussed in the previous section, according to LLT, in this case the eigenstates can have many domains occupied at once, and transport is unhindered (a full transfer of excitations) in the LLT-Anderson model as all domains are resonant with each other. 

Evidently, both the eigenspectrum and LLT reveal a stark contrast between the case of ordered and disordered scatterers. However, one usually does not have access to either in realistic experiments, so let us test if measurable quantities show the same strong difference.

If we transmit a 1D Gaussian through an ordered array of scatterers\footnote{Note that the reflection at the entrance to the channel can be very strong for high fill factors and strong scatterers, to the point where almost no atoms propagate into the channel.}, we see the atoms transmit freely and fill up the entire channel. On the other hand, with a noisy potential, the atoms propagate some distance into the channel and then freeze out exponentially. Thus, completely different dynamics are seen in transmission of wavepackets in the presence of ordered scatterers compared to a random potential. Such a comparison should be possible in experiments where the atomic density can be measured and the potential controlled (e.g.~\cite{BS}), allowing one to differentiate the effects of Anderson localisation from other mechanisms.

\begin{figure}[htbp]
\includegraphics[width=6in]{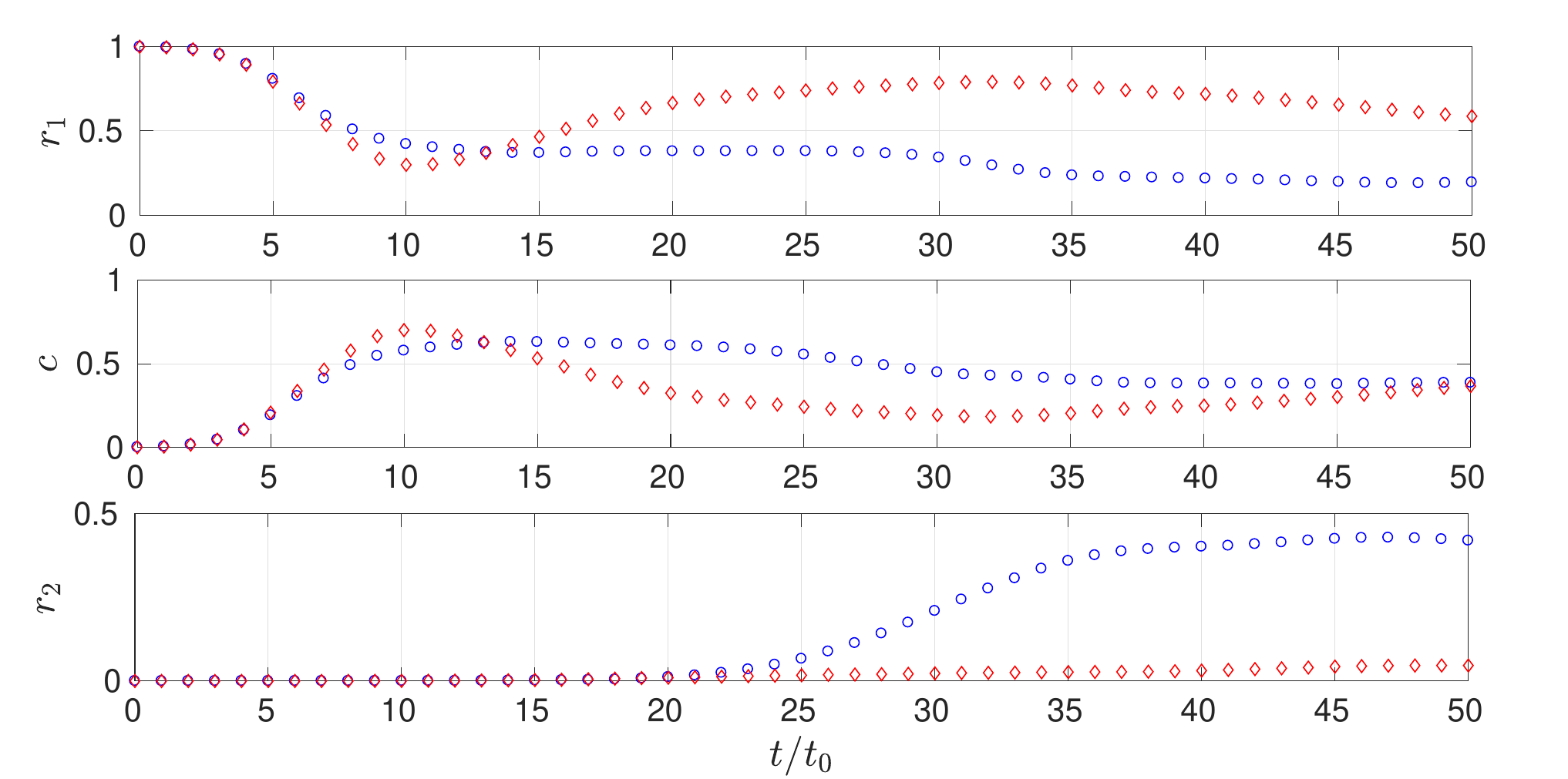}
\caption{\label{Or_Pops} Normalised populations of the three compartments -- the two reservoirs and the channel -- when a 1D Gaussian wavepacket transmits from (the centre of) $R_1$ to $R_2$ through an ordered lattice of scatterers (blue circles) and a disordered one (red diamonds). Parameters used are $L=W=25\ell$, $R=30\ell$, $f=0.1$, $\sigma=\ell/2$, $V_0=5 E_0$, $\bar{\sigma}=5\ell$, $k_0=1/\ell$. It is obvious that the flow rate out of the channel is much smaller in the presence of noise.}
\end{figure}

If we compute compartment populations for the transmitting 1D Gaussian and compare to a typical disordered run with the same parameters, as shown in Fig.~\ref{Or_Pops}, we see that the flow rate $\rho$ and the final population of $R_2$ are much smaller in the presence of noise, validating the fact that the disordered runs display strong evidence of Anderson localisation. Needless to say, the 1D density profiles in the ordered lattice case display no signs of localisation whatsoever -- there is no exponential decay involved (we highlight that the flow rate can be used to quantify transport regardless).

\begin{figure}[htbp]
{\includegraphics[width=6in]{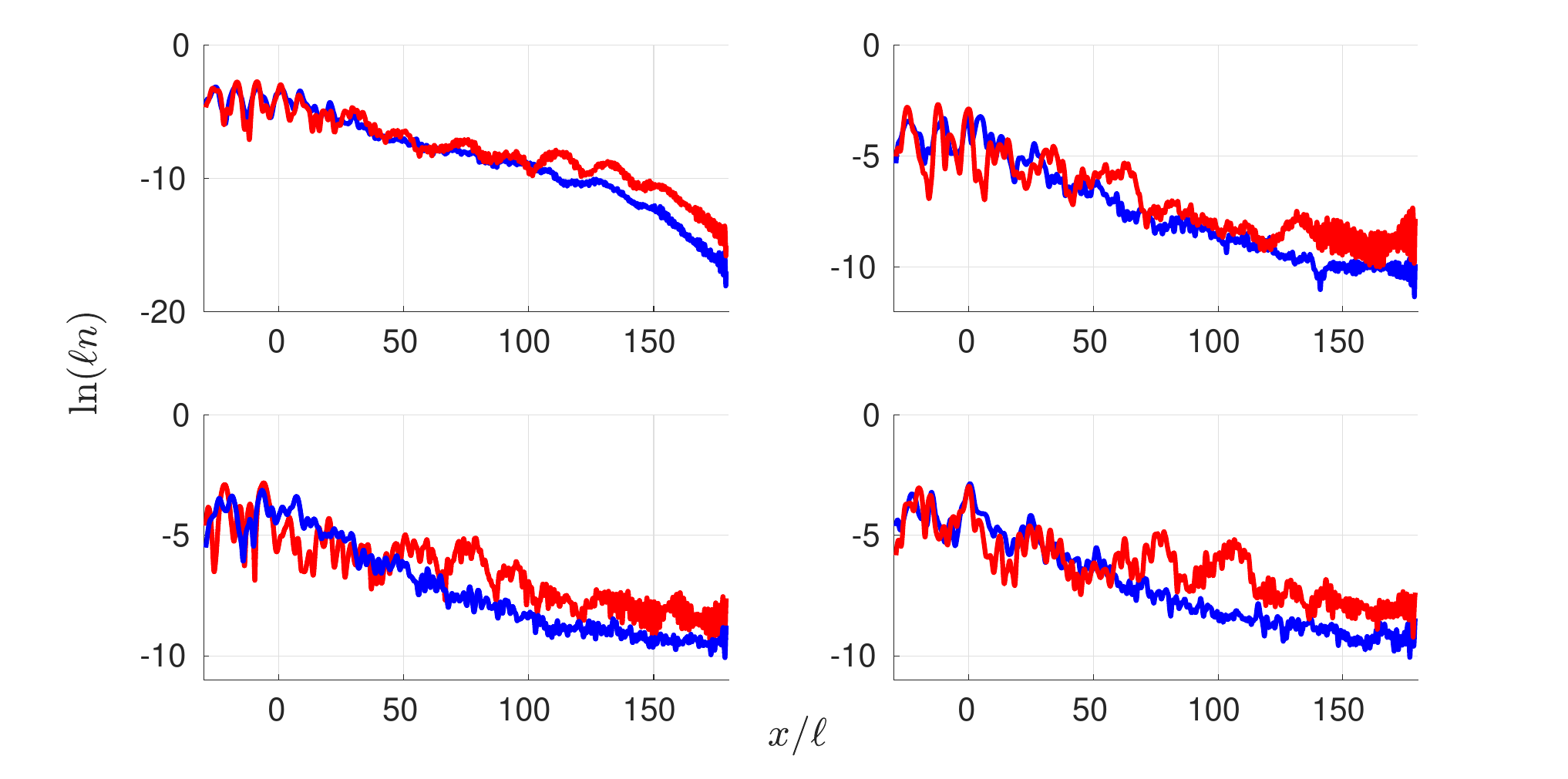}}
{\includegraphics[width=6in]{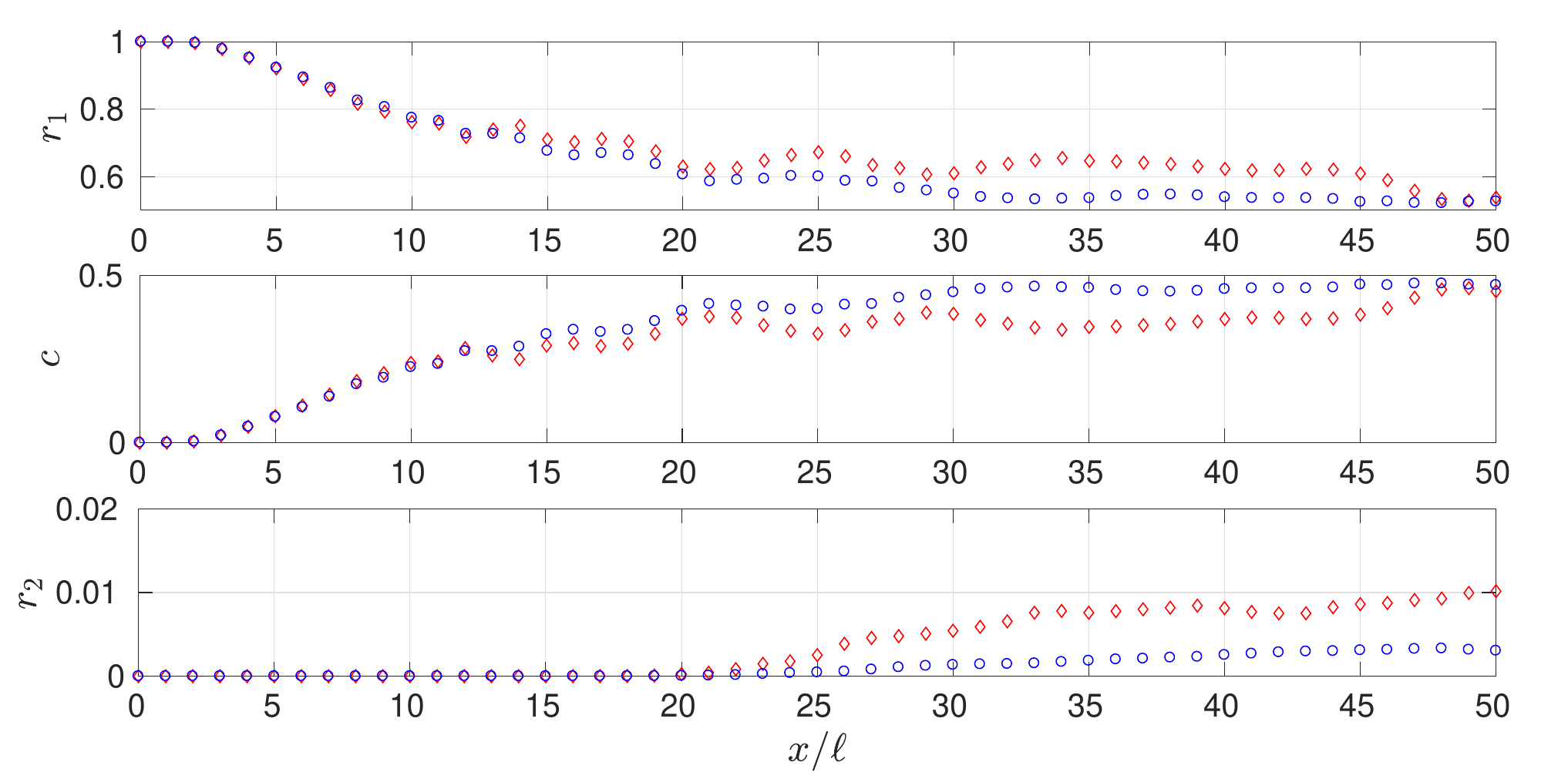}}
\caption{\label{EinT_OrDis} Top: logarithm of density profiles at $t/t_0=20, 30, 40, 50$ (going across and down) upon transmission of a stationary 1D Gaussian with $\bar{\sigma}=0.5\ell$, $k_0=0$ initiated in the centre of $R_1$. Blue lines correspond to disordered scatterers and red to a regular lattice. Parameters were $L=150\ell$, $W=25\ell$, $R=30\ell$, $f=0.06$, $V_0=5E_0$, $\sigma=\ell/2$. Bottom: compartment population curves for the same simulations; the same colour code is used. The difference between ordered and disordered scatterers is much easier to see in the density profiles on a logarithmic scale (compared to linear) and is evident in the growth rate of the population of $R_2$.}
\end{figure}

As for expanding wavepackets (see Fig.~\ref{EinT_OrDis}) in the transmissive set up, the difference in the density profiles on a linear scale is a little difficult to see, but is revealed on a logarithmic scale, as well as through the flow rate.

In conclusion, we have shown that in the regime where Anderson localisation dominates the physics in a noisy potential, the contrast between an ordered and a disordered array of scatterers can be used to clearly distinguish localisation effects. Whenever a comparison of these two scenarios is not significantly different, one cannot claim Anderson localisation with any degree of confidence. We reiterate, however, that reflection at the entrance of the channel can be very severe for an ordered lattice, and so in the transmission set up, it is crucial to examine influx into the channel as well as the output current; quantitatively, the flow rate out should be normalised by the flow rate in to the channel.

Another method of identifying when Anderson localisation is at play is by inspecting the shot-to-shot fluctuations. Throughout our investigation, we have found that whenever Anderson localisation dominated, fluctuations between different noise realisations were very large indeed, requiring averaging over 20 runs. In fact, we have confirmed that the relative error in the flow rate out of the channel (quantified by the standard error divided by the mean) clearly and strongly increases with both fill factor and scatterer height. This was done with several different initial conditions for the time evolution, in the transmission set-up, indicating that it is a fundamental localisation effect, rather than being caused by the specific details of the numerical simulation. If we weaken localisation by, for example, considerably reducing scatterer height or the fill factor, then fluctuations also fall significantly (as they must, because the empty channel case is of course deterministic). Alternatively, tuning atomic energy to higher values where the scatterers are weakly felt also reduces fluctuations as the localisation length increases (or diverges all together, in cases when a mobility edge exists). Recall that high fluctuations in the transmission from a strongly-localised system have been independently found by other researchers \cite{Chabanov, Hu, Kaiser, DelandeLectures} and even put forward as a ``smoking-gun'' of Anderson localisation. We will see in the next section that adding interactions or acceleration also weakens localisation and reduces variability between realisations.
\section{Effect of realistic experimental features}
\label{Secondary}
\subsection{Specifics of the experiment \cite{BS}}
\label{criticism}
In this section we will discuss the effect of several ``secondary'' features present in the experiment \cite{BS}. We begin from the geometry of the system: a dumbbell was used in \cite{BS}, involving large circular reservoirs connected by a rectangular channel. In our study here, we used rectangular reservoirs of the same width as the channel itself. The difference is of course that with circular reservoirs, the atoms quickly expand in $R_1$ to a cloud diameter that exceeds the channel width, so the flow rate into the channel is strongly increased as the channel is widened. With rectangular reservoirs of width $W$, this effect is completely absent. However, the increased influx with the width would need to be scaled out in any case in order to expose the more interesting finite-size effects that we have examined in section \ref{WidthDep} -- we are able to simply skip this step.

Next, it is important to realise that the initial condition used to probe the disorder influences the outcome of the experiment. Whether the wavepacket is radially expanding (i.e.~is subject to angular dispersion) or along $x$ only may make a difference to the measured results. Clearly interparticle interactions would completely change the picture (this scenario is considered separately below). Assuming the cloud is allowed to expand sufficiently and become so dilute that interactions are negligible before entering the disordered channel, the effect can be still captured with the linear Schr\"{o}dinger equation, with the initial condition taken as the asymptotic limit of a Thomas-Fermi (TF) cloud in 2D \cite{Kamchatnov}. This wavefunction will have a completely different energy distribution to, for example, the Gaussian wavepackets used in this paper. In other words, it is crucial to use the correct initial condition if one wishes to model/theoretically reproduce the experiment.

In our modelling so far, we have taken the system to have Dirichlet boundary conditions. This can be achieved experimentally by having a very high repulsive potential that covers a very large area with a ``hole'' in it which comprises the system. This is precisely what is done in \cite{BS} through the use of an SLM, and in this case the ``hole'' is dumbbell-shaped, with the atoms confined inside. However, what if the repulsive potential is not so high so as to prevent the atoms from leaking out over the edges of the dumbbell/rectangular system? We have implemented such a model and ensured that even if the potential is quite low and ``spill-over'' the sides of the system is quite noticeable, nothing important changes in the dynamics or observations. The fact that we normalise the three compartment populations by the total number of atoms in the system practically renders this issue unnoticeable.

Furthermore, the confining SLM potential discussed in the previous paragraph is produced by a very wide and powerful laser beam, which is nonetheless Gaussian in profile. We have checked that for the parameters used in \cite{BS}, the variation in intensity over the size of the system is negligible. It would be possible to simulate the case of a position-dependent confining potential (corresponding to a less well-expanded laser beam), but we have not modelled this directly yet. The interesting aspect of this idea is that the potential scatterers are created via the SLM from the same repulsive beam, and in this scenario, their height would also vary throughout the system. While this variation would not be random, it is quite likely that it would have a noticeable effect on the dynamics.

Next, to enter the 2D regime, one usually loads the atoms into a 2D trap from a 3D one where the BEC is initially prepared, which is precisely what was done in \cite{BS}. The aspect ratio of the 2D trap in this experiment is superb: 800-to-1 in the horizontal-to-vertical directions, suggesting that the trap is very shallow in the plane and very deep vertically. Under these circumstances, it would seem reasonable to reduce the description to a 2D one, and leave out the harmonic confinement in the plane all together. However, computing the 2D trap harmonic potential and comparing it to the atomic energies used in the experiment, we see that the 2D trap cannot be safely neglected. Its depth, while small, is comparable to atomic energies, and the length-scale on which it varies is similar to the size of the dumbbell. It influences the motion of the atoms, and placing its centre in the correct position in simulations is certainly desirable. As far as we are able to determine, the vertical confinement is indeed so tight that one may safely reduce the dimensionality of the system to 2D and leave the vertical direction out.

This completes our discussion of the minor features relevant to the experiment \cite{BS}. We now move on to consider two more-general, important physical mechanisms: acceleration and interactions are believed to weaken or even destroy localisation, but concrete, direct tests and understanding of the observed effects are an on-going effort in the literature. With the infrastructure built up so far in this article, we can try to fill this knowledge gap.
\subsection{Acceleration}
\label{Accel}
The question of acceleration, resulting from a linearly varying background potential, is an interesting matter to consider. It is known that a system must posses time-reversal symmetry in order for full Anderson localisation to be possible, as the probability amplitude for closed Feynman paths that are traversed clockwise and anti-clockwise must be able to fully cancel \cite{Sheng}. The most common way to break time-reversal symmetry in the context of Anderson localisation is the introduction of a magnetic field, but a time-dependent potential will also serve the same purpose. A magnetic field has been shown to weaken localisation in \cite{LeeFisher, Pichard, Imry}, as was spin-orbit coupling \cite{Pichard, Imry}. A common result of several studies is that in 1D (or quasi-1D) systems, in the limit of strong symmetry breaking, the localisation length is multiplied by a constant factor \cite{LeeFisher, Pichard, Imry}, while in higher dimensions it diverges as localisation is fully destroyed \cite{Imry}. Other systems investigated include the kicked rotor \cite{Blumel} where a general anti-unitary symmetry is broken, and a continuous superfluid system \cite{Cohen} where quantised persistent currents break time-reversal symmetry.

Now, when acceleration is included in the system, the atoms are more likely to move downstream than upstream, of course, but this is not true time-reversal symmetry breaking: reversing the direction of time and conjugating the wavefunction (to reverse momenta) leaves the Schr\"{o}dinger equation unchanged. Nevertheless, the effect on the amplitude cancellation is similar, reducing localisation and even creating a mobility edge in the lower dimensions \cite{Vollhardt}. This does not mean that Anderson localisation cannot be effectively studied in the presence of acceleration -- in fact, essentially all experiments performed in the solid state setting involved a voltage applied across the system, explicitly included in the theory of Landauer conductance \cite{Sheng}. More recently, a pair of companion studies \cite{Aspect2017, Aspect2018} have considered cold atoms transmitting through and expanding into (respectively) a disordered potential, using both white and correlated noise. Both papers included acceleration as a key feature (also see references therein for other examples), relevant for the two experiments \cite{Berthet, BS}, both of which included an acceleration to help their atoms transmit through the noise. In fact, \cite{Berthet} can almost be considered a direct test of \cite{Aspect2017}, confirming the single-parameter scaling of \cite{Aspect2017}, the algebraic decay of the density profiles, and finding a delocalisation transition with correlated noise.

As the problem currently stands, it is not entirely clear whether acceleration makes localisation fundamentally weaker (or even impossible), requiring the addition of a new element to the theoretical description, or simply increases the energy of the atoms, thereby weakening localisation via the usual energy-dependence of the localisation length. We will address this matter directly. This formulation of the problem raises another important question: for atoms moving in a changing background potential landscape, the total energy is given by the kinetic plus the potential energies. Assuming the disorder itself is placed on a flat background section, intuitively, it would make sense if only the kinetic energy was relevant for determining localisation properties, not the total, but this needs to be demonstrated directly -- this is our second goal.

The gravitational potential takes the form $-max$, where $a$ is the acceleration, $m$ the mass and $x$ the longitudinal spatial coordinate. Since this term is negative, adding it on to our disordered potential will allow $V(x,y)$ to become negative in parts of the system domain. This invalidates the use of LLT \cite{Marcel2012, part1}: if $V<0$ in some region, the localisation landscape $u$ also takes on negative values and all the structure and logic of the theory fail. In principle, one could simply add an absolute energy shift to $V$ to ensure that the total, including acceleration, is positive everywhere. However, unlike conventional quantum mechanics, LLT is not invariant with respect to absolute energy shifts -- these actually change the physical predictions \cite{part1}. As such, this is not a satisfactory solution. If we turn to exact diagonalisation, include the acceleration term in the Hamiltonian, slowly increase $a$ for the same noise realisation and observe the eigenstates, we do see that they become progressively more spread out, but it is difficult to quantify. This leaves time-dependent simulations as the best method of approaching the question of acceleration.

\begin{figure}[htbp]
{\includegraphics[width=6in]{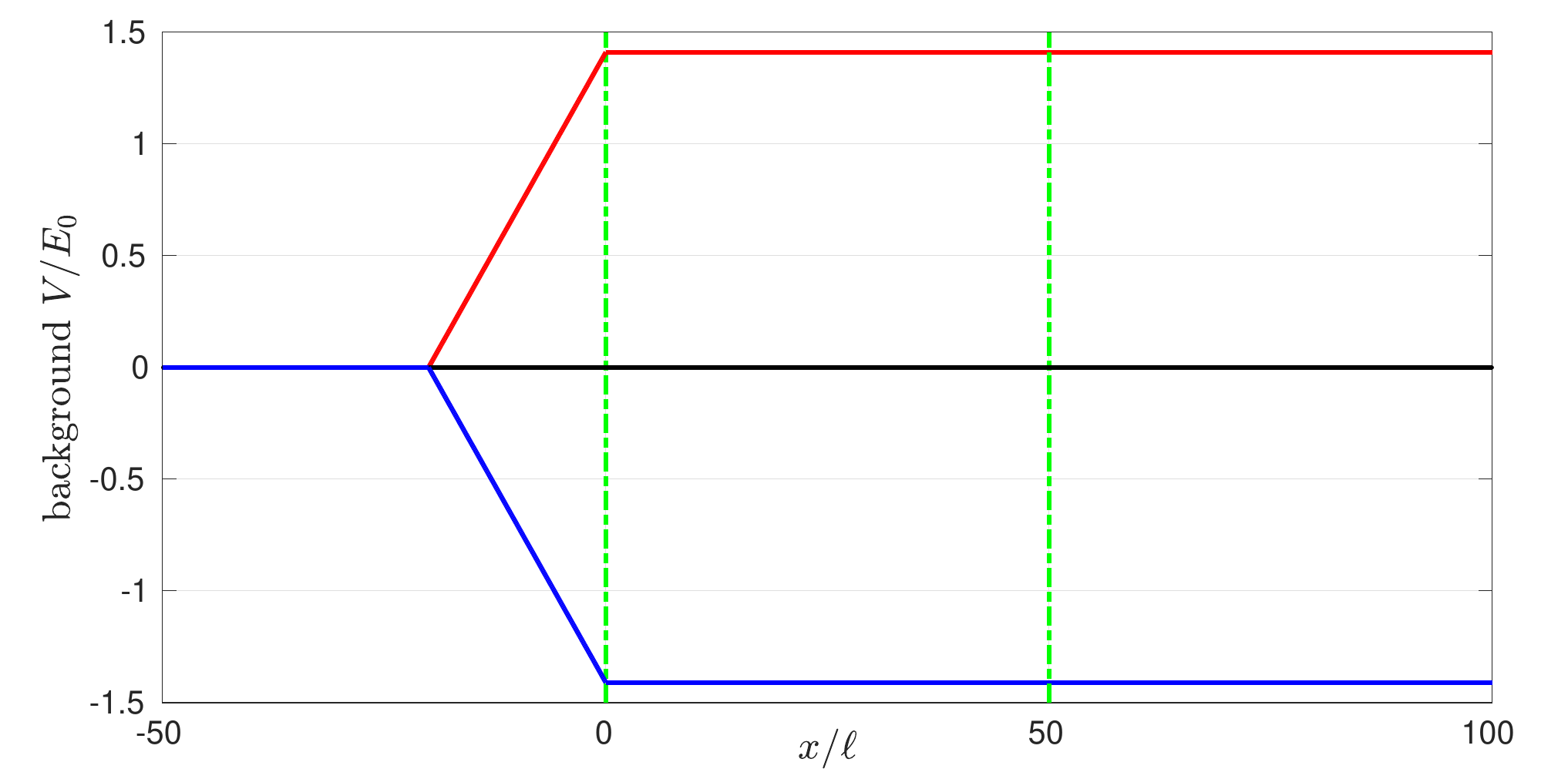}}
{\includegraphics[width=6in]{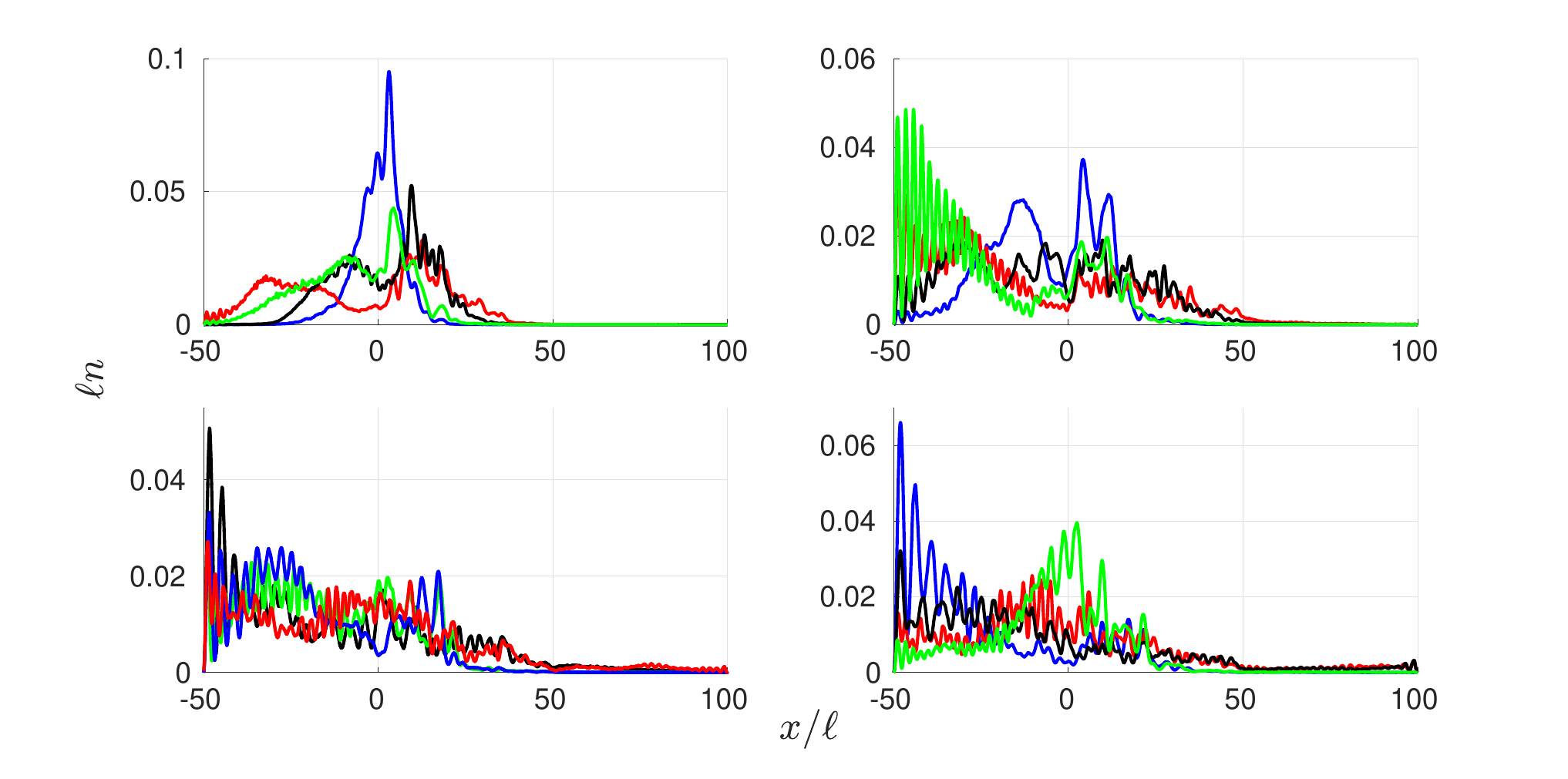}}
\caption{\label{ramp} Top: background potentials used for the ramp test. We wish to compare two wavepackets, one with $\bar{\sigma}=5\ell$, $k_0=1/\ell$ ($E=1.17E_0$) and one with $\bar{\sigma}=5\ell$, $k_0=1.5/\ell$ ($E=2.58E_0$). We choose system geometry $W=25\ell$, $L=R=50\ell$, and the green dash-dotted lines depict the reservoir-channel boundaries. The black solid line shows a flat background potential, used normally. The initial wavepacket is placed at $x=-35\ell$ and fits in to the interval $[-50,-20]\ell$. Over the section $[-20,0]\ell$, we ramp the background potential up or down, resulting in two alternative background potentials that change the potential energy by $1.41E_0$, the difference between the mean energies of the two wavepackets of interest. Bottom: four simulations are shown. Panels correspond to  $t/t_0 = 20, 30, 40, 50$ going across and down. The same noise realisation is used for all four runs, with $f=0.1$, $\sigma=\ell/2$, $V_0=5 E_0$. The blue (red) lines show the 1D density in a flat background with $k_0=1/\ell$ ($k_0=1.5/\ell$), and the black (green) correspond to a run with the potential ramped down (up) and using the $k_0=1/\ell$ ($k_0=1.5/\ell$) wavepacket. While reflection at the entrance to the channel depends somewhat on the presence and nature of the ramp potential, the similarity between the blue and green (red and black) lines proves that it is the kinetic energy part only that governs localisation properties.}
\end{figure}

To determine whether it is the total energy or only the kinetic energy part that sets localisation properties, we perform a ramp test, as illustrated in Fig.~\ref{ramp}. The basic idea is to use a potential ramp before the atoms enter the disorder to change their kinetic energy while keeping the total fixed, and observe their transit through the channel. We wish to compare two wavepackets, both translating 1D Gaussians, but with different energies. We use larger reservoirs than usual to accommodate the ramp and perform all tests with this geometry. First, we propagate both initial conditions with a flat background for comparison. We then create two ramps (before the noise begins) that change the potential energy of the two wavepackets by the difference between them. The high energy wavepacket travels through the ramp-up and the low energy one through a ramp-down potential. The same noise realisation is used for all four simulations. While reflection at the entrance to the channel depends somewhat on the presence and nature of the ramp potential, the results are unmistakable: wavepackets with the same kinetic energy at the point of entering the noisy section of the potential behave the same way, while those with different kinetic energies but the same total behave differently.

This result is perfectly sensible: it states that as long as the noisy potential ``lives'' on a flat background, this background energy can be set as the reference point for measuring energy (as absolute energy shifts are irrelevant). Then the only factor determining localisation physics is the energy of the probing wave, initiated outside the disorder, at the point when it is about to enter the random potential, relative to this background zero-energy mark, that is, the kinetic energy alone. In other words, because of our choice of the reference energy, when the atoms arrive at the noisy potential, their potential energy is nil by definition and all energy is in kinetic form. Therefore, a non-uniform potential landscape outside of the noisy region is not fundamentally interesting: all one needs to know is how much kinetic energy the wavefunction has at the point of entering the disorder.

We now wish to find out whether adding an acceleration to the disordered system proper makes it impossible to localise the atoms completely, or if by changing noise parameters, it is still possible to achieve practically full attenuation of the density in the channel, which would indicate that acceleration does not fundamentally destroy localisation but rather just increases the kinetic energy of the atoms and thus weakens it.

\begin{figure}[htbp]
\begin{center}
{\includegraphics[width=5in]{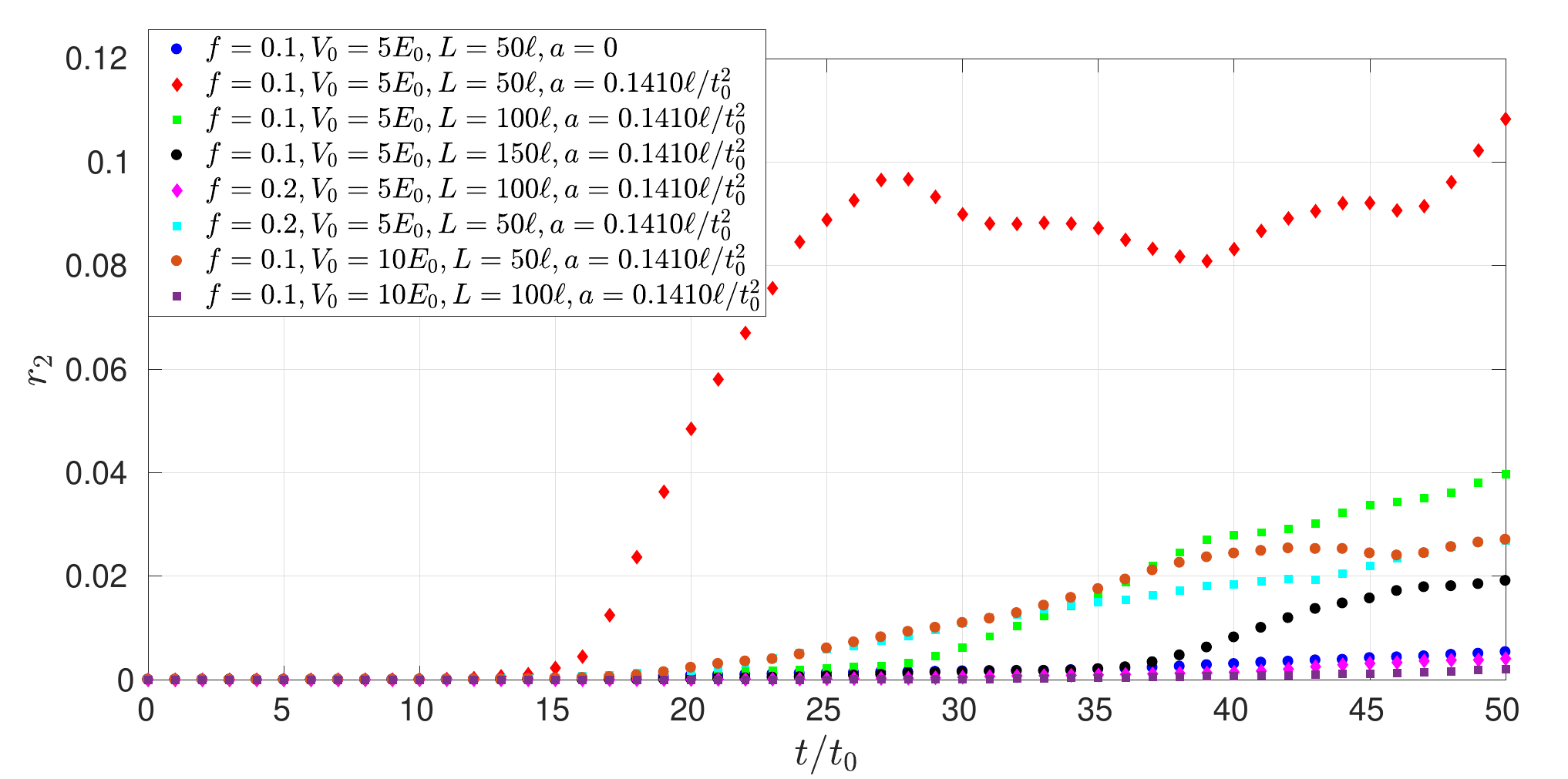}}\\
{\includegraphics[width=5in]{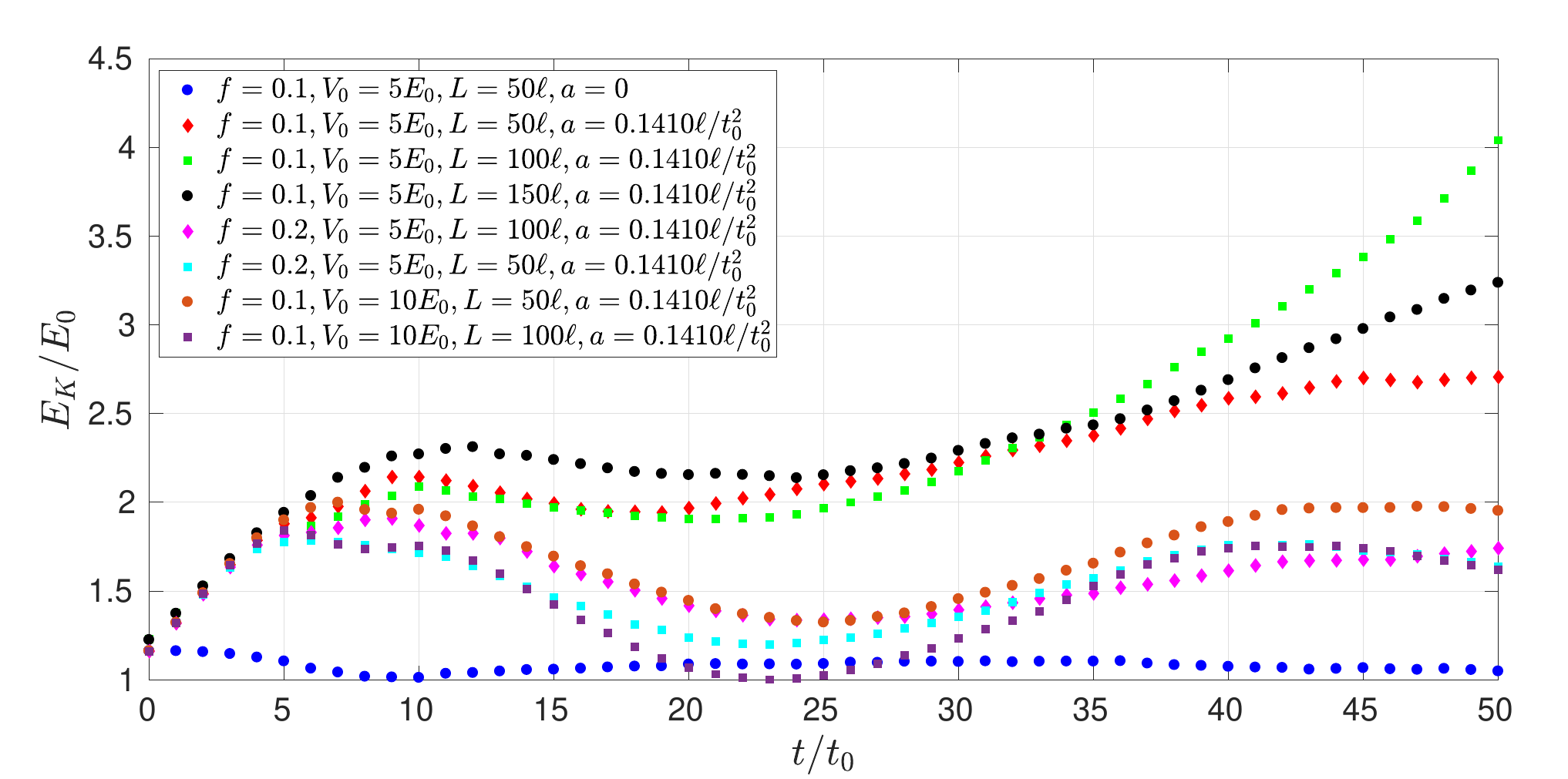}}
\end{center}
\caption{\label{Atest} Top: normalised population of the second reservoir, bottom: kinetic energy as a function of time, computed for single noise realisations. In all cases we use a 1D Gaussian initiated in $R_1$ with $\bar{\sigma}=5\ell$, $k_0=1/\ell$. Parameters common to all simulations are $W=25\ell$, $R=30\ell$, $\sigma=\ell/2$. Blue circles depict the case of no acceleration with $L=50\ell$, $f=0.1$, $V_0=5 E_0$, sufficient to achieve strong localisation in the channel, with very little population arriving to $R_2$. Kinetic energy stays mostly constant, oscillating between $E_0$ and $1.16 E_0$. Red diamonds show the result of adding an acceleration of $a=0.1410\ell/t_0^2$ (arbitrarily taken equal to the value used for the ramp test of Fig.~\ref{ramp}), which is present in all the other runs as well. Transmission into the second reservoir increases very strongly, and the kinetic energy $E_K$ climbs to much higher values than before. Crucially, it keeps climbing as time goes on. If we try to restore localisation by increasing channel length ($L=100\ell$ -- green squares, $L=150\ell$ -- black circles), the transmission into $R_2$ progressively decreases but from the density profiles, it does not appear that full localisation would be possible if $L$ was increased further. Again, the kinetic energy continuously grows with time. On the other hand, if we double the fill factor ($L=50\ell$ -- cyan squares, $L=100\ell$ -- magenta diamonds), or double the scatterer height ($L=50\ell$ -- brown circles, $L=100\ell$ -- purple squares) it is easily possible to reach strong localisation in the channel. This is because the kinetic energy does not grow unbounded, but rather oscillates with time, remaining between $E_0$ and $2E_0$, very reasonable energies to be localised at these noise parameters (judging from experience gained without acceleration).}
\end{figure}

We perform the following test, illustrated in Fig.~\ref{Atest}. We use a low-energy 1D Gaussian wavepacket and begin from a given set of system parameters where the density essentially decays to zero within the length of the channel with a flat background potential. Then we add on an acceleration, and indeed a large fraction of the atoms now transmits to the second reservoir. Progressively increasing $L$ causes a smaller fraction of the atoms to transmit, but does not induce strong localisation inside the channel. Observing the dynamics, it is not obvious that increasing $L$ further would lead to strong localisation. However, doubling the fill factor or the scatterer height (with the original channel length) immediately causes strong localisation in the channel, with little arriving in $R_2$. Increasing $L$ in both cases confirms that only a short further channel length was needed to attain an essentially full decay of the density.

This tell us that acceleration 
\begin{itemize}
\item weakens localisation compared to a flat background when the same initial condition and system parameters are used,
\item for a given set of noise parameters, may make it impossible to localise certain energy components,
\item does not fundamentally render localisation impossible, as increasing the strength of the disorder leads to strong localisation even in the presence of a large acceleration.
\end{itemize}

Since we now know that the kinetic energy $E_K$ determines localisation properties, we inspect it in order to gain insight into our observations. As demonstrated in Fig.~\ref{Atest}, without acceleration, $E_K$ changes very little over time, oscillating slowly. With acceleration and without strengthening the disorder -- in the cases when we saw that one could not readily localise all the atoms in the channel -- the kinetic energy increases with time. This explains our inability to localise the wavefunction: we may make the channel longer, but as the atoms travel further down, they gain energy, and localisation length increases even more, leading to a ``vicious circle''. In contrast, when either the density or the height of the scatterers is increased, the kinetic energy remains bounded from above; in fact, it oscillates between $E_0$ and $2E_0$, energies that are readily localised at these noise parameters. This last remark is based on observations made regarding the behaviour of the system \textit{without} acceleration. As such, by inspecting the kinetic energy we can predict if localisation is possible for a given wavepacket, noise regime, and acceleration value.

Finally we remark that the increase in kinetic energy of the atoms as they travel down the channel in the presence of the potential scatterers is much smaller than in an empty channel. This is sensible: filling up the channel with (positive) Gaussian scatterers raises the total potential landscape, leaving less room for change in the kinetic energy. No doubt precisely the same observation is responsible for the fact that by increasing the strength of the disorder, we were able to qualitatively change the behaviour of $E_K(t)$ from increasing to bounded and oscillatory. All this insight into the effect of acceleration was, to the best of our knowledge, thus far unknown.
\subsection{Interactions}
\label{Ints}
Now that we understand the effect of acceleration on Anderson localisation, we may wonder how interparticle interactions change localisation properties. We know that Anderson localisation causes particles to bunch together, while repulsive interaction try to push them apart. In this sense, it is likely that interactions will be detrimental to localisation -- in fact, this is the generally accepted picture.

Experimental evidence, including specifically in 2D, has shown that interacting systems with disorder can possess a mobility edge even in low dimensions \cite{openquestions, Pollak, Greek_review}. Previous theoretical work that included interactions involves studies in 2D \cite{Aoki2, Pun}, 1D (or quasi-1D) \cite{Apel, Mirlin}, and at low but non-zero temperature \cite{Apel, Mirlin}, with several studies finding an induced mobility edge \cite{Pun, Apel}. In addition, \cite{Kimball} has demonstrated that an Anderson insulator can be distinguished from a Mott insulator via the spin configuration (a glass or anti-ferromagnetic phase, respectively), and \cite{Pikovsky} solved the nonlinear Schr\"odinger equation in 1D on a lattice, confirming that interactions oppose localisation.

The attempt to study Anderson localisation in the presence of interactions has given rise to a new concept -- many-body localisation -- which has grown into a research field in its own right over the last decade \cite{Abanin, Huse, AbaninReview, Bloch2015, Bloch2016, MBL}. The 1D case is well-understood, and while in 2D the situation is more complicated, conceptually similar ideas to the methods used in 1D are being pursued. However, many-body localisation has diverged away from the question of what happens to Anderson localisation in the presence of interactions and focused on ergodicity, thermalisation and localisation in Hilbert space instead. Incidentally, these ideas appeared in the literature very early on \cite{Pollak}, and it is the remarkable advance in computational techniques and resources that has allowed for the explosion of research to take place only recently.

Several studies have approached the question of interactions in the presence of disorder in cold-atom systems via the Gross-Pitaevskii equation, e.g., \cite{BS, Garcia} in 2D and \cite{Zhen, Min, Donsa} in 1D. Reference \cite{Donsa} is particularly interesting, predicting that 1D experiments would not be able to clearly detect Anderson localisation due to the presence of interactions. On the other hand, we will show that with realistic experimental parameters, interactions visibly weaken localisation, but the latter remains sufficiently strong to be detected.

Overall, there is little direct knowledge on how interactions would modify the picture we have built up in this article so far. This question is important because not all experiments with cold atoms tune interactions to zero. For example, like many of the earlier localisation experiments, the 2D study \cite{BS} has not attempted to eliminate interactions, relying on the fact that the atoms will spread out over a large area in the course of the experiment and interactions should become negligible (that was the intention), leaving bare Anderson localisation to be observed. We are in a prime position to test the validity of this assumption. The rest of the section is dedicated to precisely this cause.

First of all, we remark that the initial condition can be a complicated, unknown function, but we will restrict our exploration to a TF profile. The excellent paper by Kamchatnov \cite{Kamchatnov} contains analytical approximations of the order parameter of a BEC in a 2D TF profile after the confining harmonic potential is abruptly removed, assuming the atoms are evolving in an infinite 2D plane. These can be used to obtain an order of magnitude idea of what we might expect, but the presence of boundaries in our system means that the free predictions soon lose relevance and one needs explicit simulations to capture the dynamics. We remark that using the results in \cite{Kamchatnov}, it is straight-forward to show that the energy distribution in the long-time limit of the condensate evolving in 2D with no external potential is \text{linear} (falling with energy), which is very different from the shape one might expect naively for an expanding wavepacket (e.g.~a 2D Maxwell-Boltzmann or a Bose-Einstein distribution).

Note that the geometry of the system -- a fully rectangular system studied earlier in this article or a dumbbell potential used in \cite{BS} -- is now very important, as it determines the shape and size the wavefunction can assume, setting interaction energy, which then drives future dynamics. For the rest of the section, we will use a dumbbell geometry in our simulations, firstly to allow the condensate to expand significantly before entering the disordered region (the circular reservoirs are much larger than the rectangular ``channel extension'' reservoirs), and second, to ensure our results retain relevance to the experiment. The dumbbells are constructed by placing two circular reservoirs of radius $R$ on either side of a rectangular channel (dimensions $L\times W$), and bringing the reservoirs in towards the channel (keeping the reservoir centres on the $y=W/2$ axis) until their circumference touches the corners of the channel. The channel is then filled with noisy scatterers, as always. Furthermore, we will use a purely expanding wavefunction, with no added translation or acceleration, in order to make our investigation as simple and transparent as possible.

It is worth pointing out that when interactions are added to the system, clearly exact diagonalisation and LLT are no longer applicable at all, but time-dependent simulations can still be performed. Instead of solving the Schr\"{o}dinger equation, we must use the GP equation, but density profiles and the flow rate observable are still perfectly well defined and fully accessible.

The GP equation for the mean-field order parameter in the grand-canonical ensemble reads
\begin{equation}
i\hbar \partial_t\psi(\mathbf{x},t) =  \left[-\frac{\hbar^2}{2m}\nabla^2 + V_0(\mathbf{x}) - \mu + g_{2D}\left|\psi(\mathbf{x},t)\right|^2\right]\psi(\mathbf{x},t).
\end{equation}
Here the order parameter is normalised to $N$ particles, the total number of atoms in the BEC. In the experiment \cite{BS}, $N\sim15,000$, so we will use numbers of this order of magnitude. A dimensionally-reduced interaction strength is used, $g_{2D}$, obtained as usual from the 3D s-wave scattering length, which for Rb-87 is $95a_0$ ($a_0$ being the Bohr radius).

Since we intend to use a TF profile for the initial condition, we must first compute the ground state of a harmonic trap, the centre of which is positioned in the centre of the circular reservoir $R_1$. If
\begin{equation}
V = \frac{1}{2}m\omega_t^2r^2
\end{equation}
where $r$ is the polar radial coordinate centred on the centre of $R_1$, then for a given chemical potential $\mu$ (which controls $N$), the TF profile is given by an inverse parabola
\begin{equation}
\left|\psi\right|^2 = \frac{\mu-\frac{1}{2}m\omega_t^2r^2}{g_{2D}},
\end{equation}
and the TF radius $R_0$ where the density vanishes is determined by $\mu=\frac{1}{2}m\omega_t^2R_0^2$. By setting $N$ and $R_0$, one can obtain the corresponding trap frequency $\omega_t$ from equation (49) of \cite{Kamchatnov}, and from there, the chemical potential. For the aspect ratio of the 2D trap and its harmonic oscillator length-scales we use the experimental values from \cite{BS}.

Before presenting results, let us outline the strategy we will adopt to elucidate the effect of interactions. Since the theory is now nonlinear, we can no longer speak of an ``energy distribution'', as the superposition principle does not hold. However, we can still compute the kinetic energy of an interacting cloud, and by Fourier analysis of the order parameter, access its distribution. This is to be done with the understanding that this is only part of the energy (in an empty channel, the rest is in interactions) and that this distribution changes with time. We would like to compare two cases, both with the same kinetic energy distribution, with and without interactions present. This could be approximately accomplished by evolving a TF profile with the GP equation, waiting until the cloud is close to entering the channel, and then performing two runs: one where interactions are switched off at that point and one where they are left on. If we monitor the kinetic, interaction, and potential energy components and see that the interaction energy stays more or less constant in the GP simulation as the atoms transit through the channel, we will have achieved our goal. If we observe a serious difference between the two simulations, this will be due to the effect of interactions, and would imply that they directly influence localisation properties.

\begin{figure}[htbp]
{\includegraphics[width=3.1in]{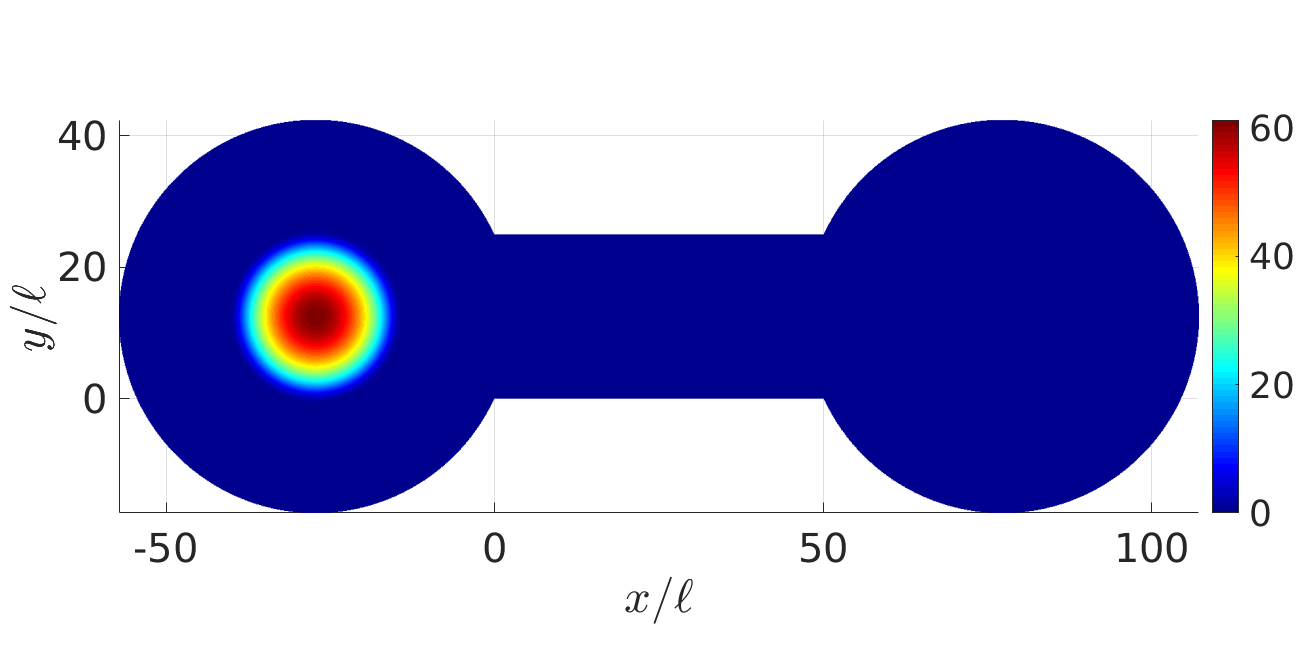}}
{\includegraphics[width=3.1in]{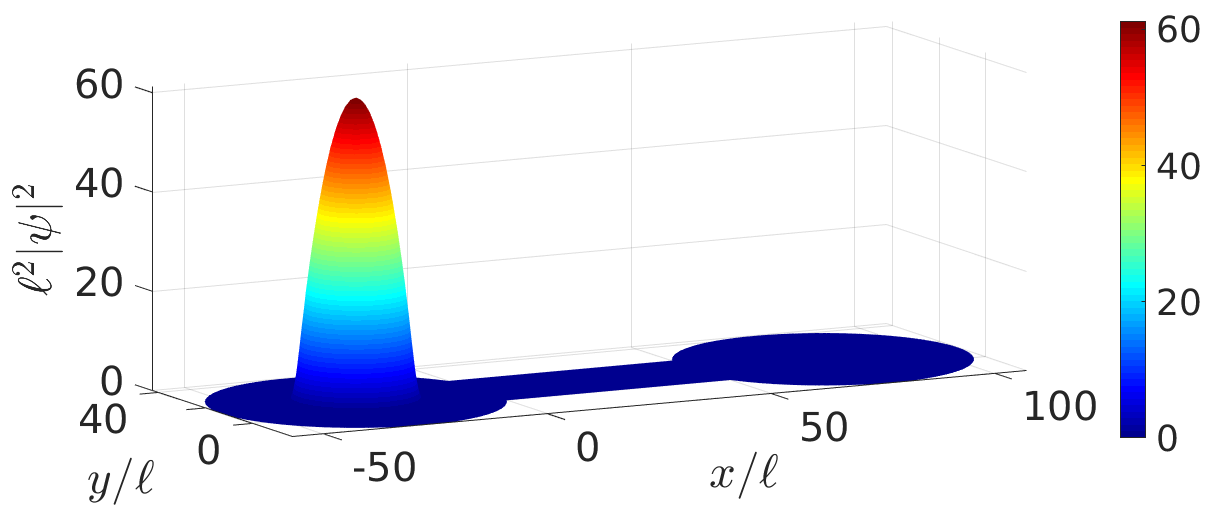}}
\caption{\label{egDB} View from the top and from the side of the dumbbell system, illustrating the geometry, depicting the density of the BEC in the ground state of a harmonic trap corresponding to $R_0=12.5\ell$, with $N=15,000$ atoms. The dumbbell dimensions are $L=50\ell$, $W=25\ell$, $R=30\ell$ and no potential scatterers are present in the channel. Gross-Pitaevskii parameters were set to $\mu=8.05E_0$, $\omega_t=0.4536/t_0$, and $g_{2D}=0.1315 E_0\ell^2$.}
\end{figure}

\begin{figure}[htbp]
\begin{center}
\includegraphics[width=6in]{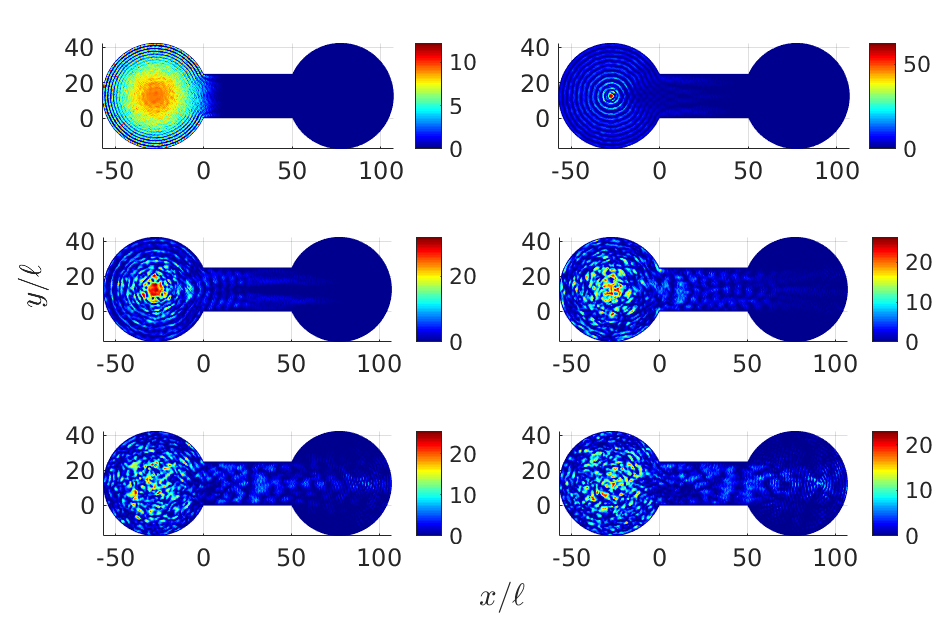}
\end{center}
\caption{\label{GP2Dpics} Time evolution of the initial density profile shown in Fig.~\ref{egDB} with the harmonic trap removed at $t=0$. Panels going across and down correspond to $t/t_0=5, 10, 15, 20, 25, 30$. The wavefunction expands and transmits through the empty channel (with no CoM motion), arriving at the second reservoir.}
\end{figure}

\begin{figure}[htbp]
\begin{center}
\includegraphics[width=6in]{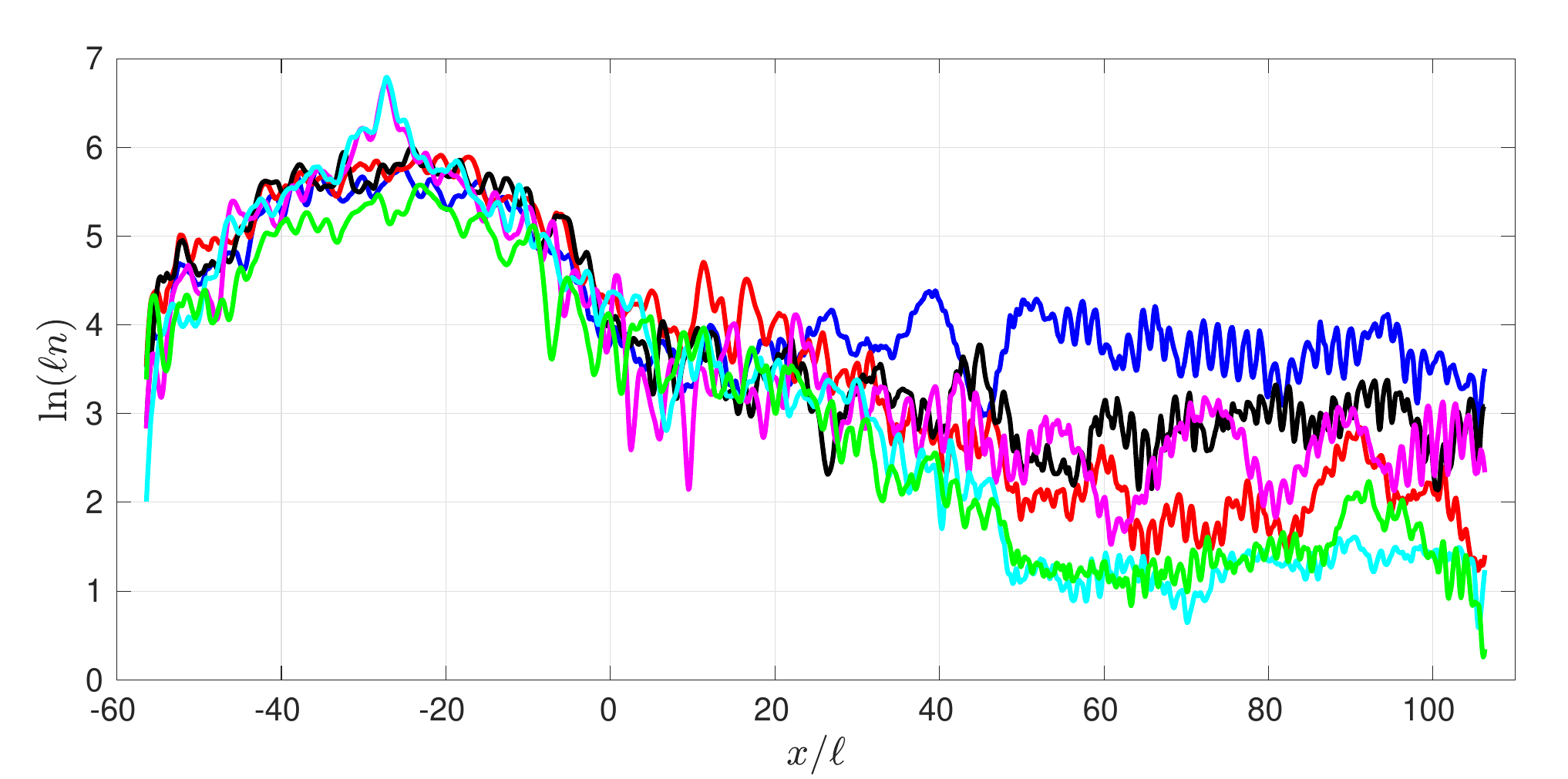}
\end{center}
\caption{\label{GP1Dden} Long-time ($t=30t_0$) density profiles resulting from evolving the initial condition shown in Fig.~\ref{egDB} with the harmonic trap removed at $t=0$. Blue: empty channel. Red: a single noise realisation with $f=0.1$, $V_0=5E_0$, $\sigma=\ell/2$. Transmission into the second reservoir is clearly suppressed, and the density in the channel now looks exponential. Black: an ordered lattice of scatterers with the same density and height, showing a much slower fall off of the density in the channel. Magenta: ordered lattice with interactions turned off at $t=5t_0$. Cyan: disordered scatterers (using the same noise realisation as for the red curve) with interactions turned off at $t=5t_0$. There is no significant difference between the linear and nonlinear ordered density profiles, while the disordered profiles run parallel to each other (on a logarithmic scale), with the linear case lying visibly and consistently lower. This suggests that Anderson localisation is weakened by interactions, but still survives with approximately the same localisation length, except that a fraction of the atoms are effectively delocalised for the given system size. Green: including (linear-in-time) atom loss at a rate of 200 atoms per $t_0$ and using the same noise realisation as for the red curve. The entire BEC is smaller and as a result, the density profile is lower than in the conservative case. On the other hand, the two still run parallel to each other, which implies no significant changes to the localisation length experienced by the particles.}
\end{figure}

\begin{figure}[htbp]
\begin{center}
\includegraphics[width=6in]{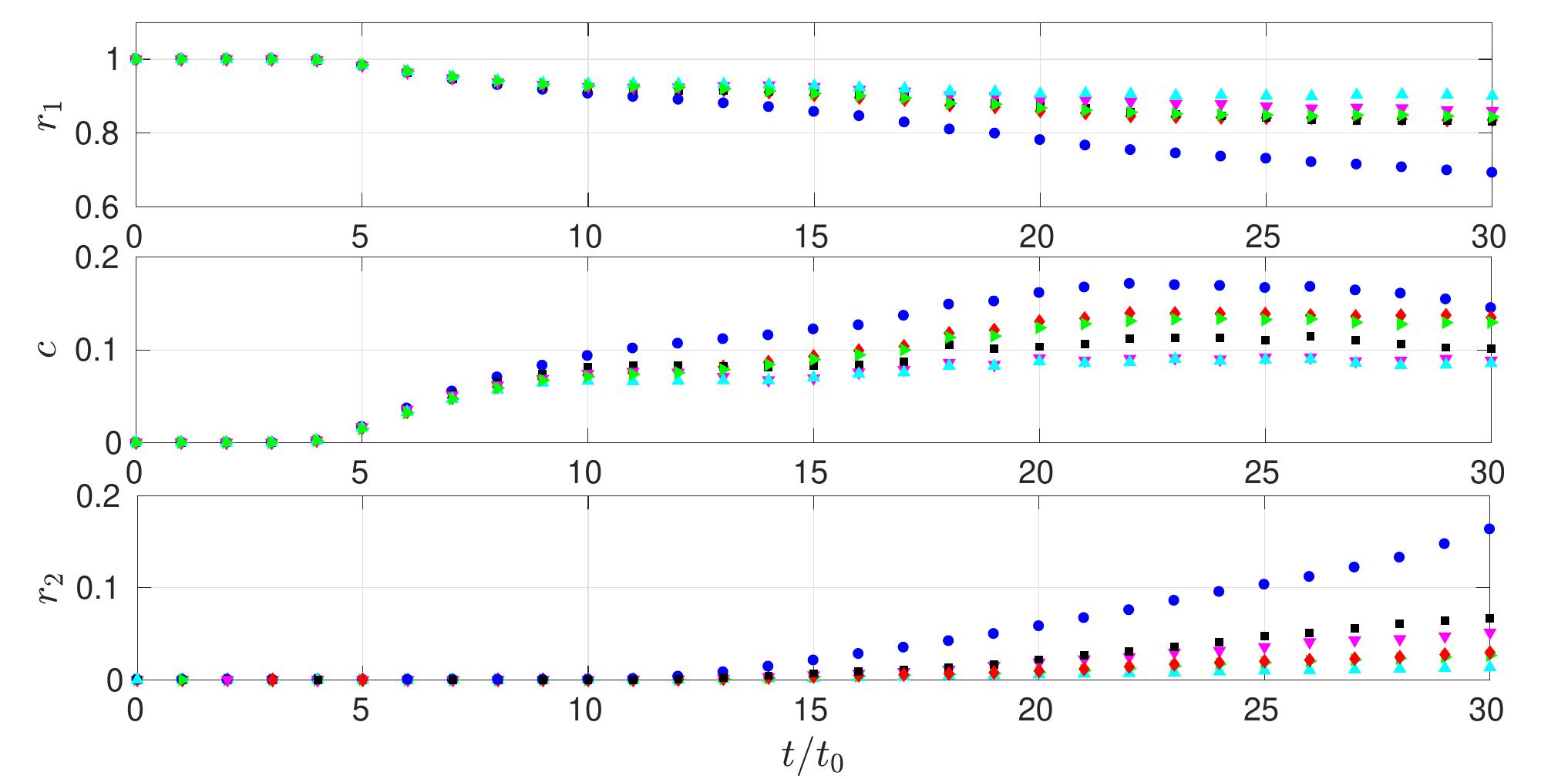}
\end{center}
\caption{\label{GPpops} Normalised compartment population curves for the initial condition shown in Fig.~\ref{egDB} with the harmonic trap removed at $t=0$. Blue: evolution in an empty channel. Red: a single noise realisation with $f=0.1$, $V_0=5E_0$, $\sigma=\ell/2$. Transmission into the second reservoir is clearly suppressed. Black: an ordered lattice of scatterers with the same density and height. The flow rate is indeed higher for the ordered lattice, indicating some effect of the disorder is certainly present. Magenta: ordered lattice with interactions turned off at $t=5t_0$. Cyan: disordered scatterers (same realisation as was used for the red curve) with interactions turned off at $t=5t_0$. The curve $r_2(t)$ is much more linear without interactions and is clearly reduced by the removal of interactions. Normalising the flow rate out of the channel by the flow rate into it reveals that indeed localisation is stronger in the linear case, but is present with interactions also (see text). Green: including (linear-in-time) atom loss at a rate of 200 atoms per $t_0$ and using the same noise realisation as for the red points. The populations are only very slightly below their conservative counterparts, and the flow rates both in and out of the channel are very similar.}
\end{figure}

To begin with, we examine the empty channel behaviour to gain intuition into the dynamics. Figure \ref{egDB} shows the dumbbell system as well as the initial condition, the ground state solution in a harmonic trap corresponding to the chosen TF profile. We will use a TF radius comparable to the size of the initial cloud in the experiment \cite{BS}, $R_0=12.5\ell$. The time evolution of this order parameter is depicted in Fig.~\ref{GP2Dpics}, allowing a direct visualisation of the transport. Note that the wavefunction is purely expanding, with no CoM motion. The long-time 1D density profile is shown in Fig.~\ref{GP1Dden}. Observe in particular that at long times, the profile in the channel is constant and $R_2$ is generously filled with atoms. This is also reflected in the corresponding compartment population curves, which are plotted in Fig.~\ref{GPpops}.

\begin{figure}[htbp]
\begin{center}
\includegraphics[width=6in]{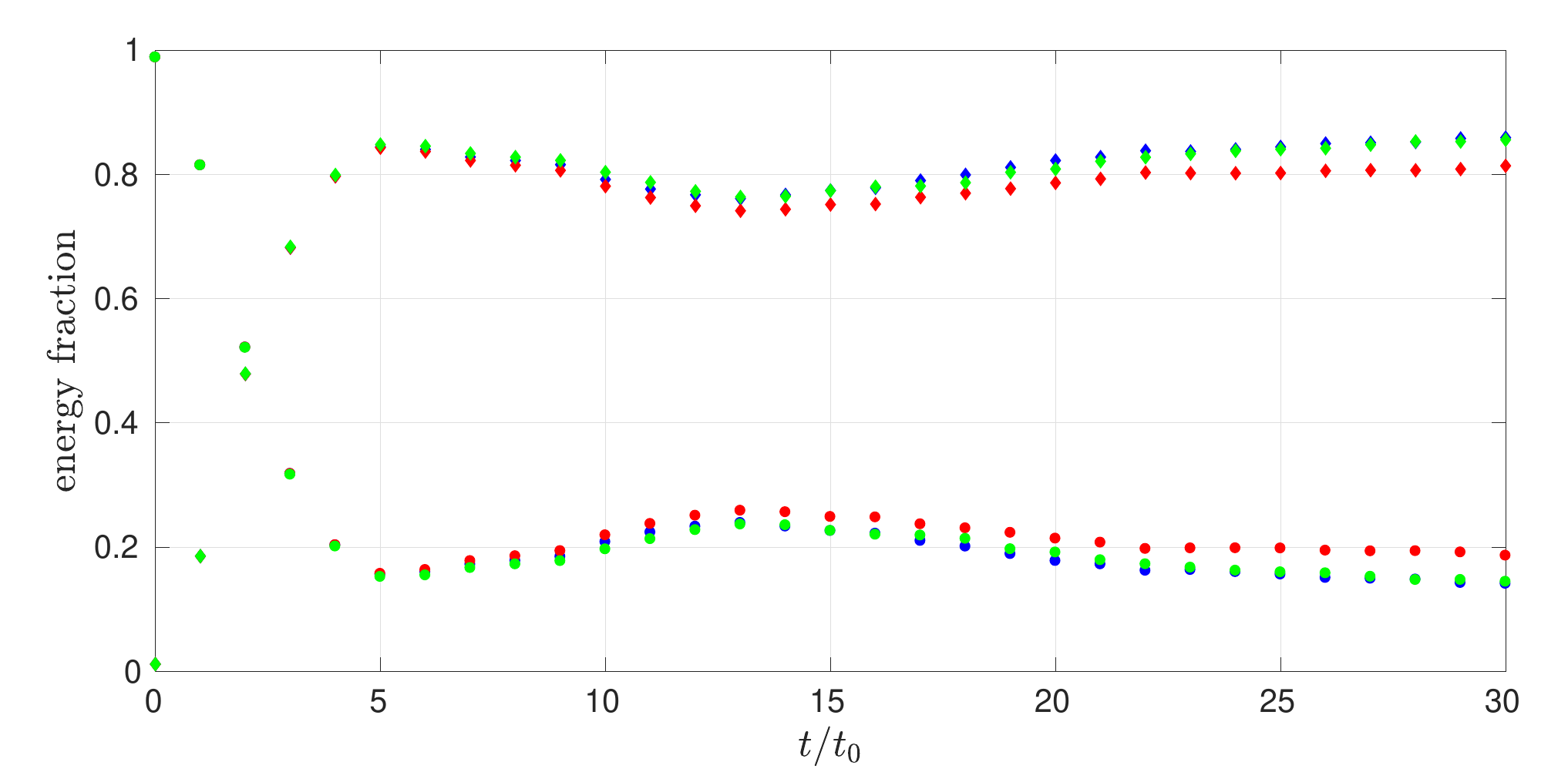}
\end{center}
\caption{\label{Efrac} Kinetic (diamonds) and interaction (circles) energy divided by their sum as a function of time for the initial condition shown in Fig.~\ref{egDB} with the harmonic trap removed at $t=0$. The total energy in this example is $E_T\approx2.7E_0$. Blue: empty channel. As soon as the cloud expands to the size of $R_1$ and enters the channel (in this case, at around $t=5t_0$), the energy fractions stabilise to roughly a 20\%-to-80\% interaction-to-kinetic energy splitting, which indicates that the effect of interactions should be readily observable. Red: a single noise realisation with $f=0.1$, $V_0=5E_0$, $\sigma=\ell/2$. The addition of scatterers has a weak effect on the energy fractions, with most of the conversion still happening during the initial expansion in $R_1$. Green: including (linear-in-time) atom loss at a rate of 200 atoms per $t_0$ and using the same noise realisation as for the red points. Interaction energy becomes less important as a result of loss, but not significantly.}
\end{figure}

Next, we would like to find out how significant interactions are and how quickly their importance decreases with time as the atomic cloud expands. An obvious way to accomplish this is to inspect the fractional contributions from kinetic and interaction energy as a function of time in the empty channel case, as shown in Fig.~\ref{Efrac}. We see that at the point when the cloud expands to the size of $R_1$, which in this case happens at around $t=5t_0$, the kinetic and interaction energy fractions stop changing rapidly, and settle in to a roughly constant ratio of 20\%-to-80\% interaction-to-kinetic energy splitting. One fifth of the energy typically stays in interaction form -- this is certainly not negligible and the effect should be easily observable. We have confirmed that changing the particle number or TF radius by 20\% does not drastically change the empty channel results shown thus far, and so we will focus on this set of parameters for further investigation.

The empty channel results are to be contrasted to the case when Gaussian scatterers are present in the channel. We add disordered scatterers of moderate strength to the system, and examine 1D density profiles, the compartment populations and the energy fractions for the effect (see Figs.~\ref{GP1Dden}, \ref{GPpops} and \ref{Efrac}). The 1D density profiles now reveal what appears to be exponential decay in the long-time limit, the flow rate out of the channel is significantly reduced, and the energy fractions are mostly unchanged, such that it is still valid to approximate them as constant after the initial expansion in $R_1$. To confirm that the transport suppression is indeed due to Anderson localisation, we compare also to an ordered lattice of the same scatterer density and height (see Figs.~\ref{GP1Dden} \& \ref{GPpops}). The density profile at long times decays considerably slower than in the noisy case, and the flow rate reveals that ordered scatterers allow more transmission through the channel, so Anderson localisation survives the presence of interactions, at least to some degree.

We are finally in a position to test directly if interparticle repulsion indeed weakens localisation. Turning off interactions at $t=5t_0$ and repeating the ordered and disordered scatterer runs, we find that the long-time density profile in the channel for ordered scatterers overlaps that obtained with interactions, while for random scatterers, it runs parallel to and below it\footnote{This observation was confirmed for a different noise realisation with the same parameters.} (on a logarithmic scale; see Fig.~\ref{GP1Dden}). This suggests that adding interactions (keeping the kinetic energy distribution fixed) weakens localisation (much stronger suppression is seen in the linear case), but the average localisation length of the particles that are still localised for this system size is unaltered. In both the linear and nonlinear cases there is a strong difference between ordered and disordered scatterers which proves that Anderson localisation is at play despite being weakened by interactions.

Before proceeding further, notice the different shape of the atomic cloud in $R_1$ with and without interactions, visible in Fig.~\ref{GP1Dden}. As may be expected, upon quenching interactions to zero, the density in the source reservoir (slightly) collapses (or ``refocuses'') on the centre of the reservoir. However, we have directly checked that the density in the channel after the quench does not withdraw back into the reservoir, and in fact, differences between the density profiles with and without interactions develop from the far end of the channel. This is confirmed by the fact that the flow rate into the channel in the two cases is not significantly different, while the flow out is (see later). Finally, the fact that for ordered scatterers the profiles with and without interactions overlap means that the effects we see in the disordered case are not due to the specific dynamics after the quench (e.g.~the minor collapse of the condensate), but due to Anderson localisation in the channel. With this established, we may continue to analyse the results.

The key difference induced by the presence of the nonlinearity seems to be that a fraction of the atoms is effectively delocalised for the given system size, which is not the case in the linear regime. Doubling the channel length with interactions included reduced the flow rate strongly, as well as the fraction of the atoms accumulated in $R_2$. It is quite possible that by increasing the channel length sufficiently, we would be able to fully localise the entire cloud. In other words, the localisation length for the higher energy components may still be finite, but considerably increased compared to the linear system. At least, this is the observation we make for the given set of parameters -- it is no doubt possible to easily find a regime where adding interactions will destroy localisation (the kinetic energy distribution and the proportion of energy in interactions would be the key handles to enter this regime).

To confirm our density-based observations, we examine the population curves and the flow rate observable in the relevant simulations (Fig.~\ref{GPpops}). As there is a small variability in the flow rate into the channel between the cases studied, as a precaution, we normalise the flow rate out by it. Note that the flow rate out is strongly different between the five relevant simulations depicted in Fig.~\ref{GPpops} (while the flow rate in is very similar), and causes the majority of the difference observed in the normalised flow rate. The normalised flow rate for the five cases studied is as follows: empty channel $0.46634$, nonlinear with disorder $0.12089$, nonlinear with regular lattice $0.27933$, linear with disorder $0.048194$, and linear with regular lattice $0.20111$. These measurements support the conclusions reached based on the long-time density profiles. As a final note, we remark that shot-to-shot fluctuations are significantly reduced when nonlinearity is included, which further evidences the fact that interactions weaken Anderson localisation.

Thus, we were able to definitively expose the effect of interactions on Anderson localisation using the formalism developed in the earlier parts of the paper for the linear system. It would be highly desirable to apply the procedure followed in this section to more cases -- different parameters, more noise realisations, etc.~-- to confirm that our conclusions are indeed correct. This is left for future work.
\subsubsection{Atom loss: condensate depletion}
In any real experiment, and certainly also in \cite{BS}, the BEC is subject to loss mechanisms, such that the total particle number and energy are no longer conserved. A rather strong depletion of the condensate is observed in \cite{BS} on a long time scale, arising from collisions of the condensate atoms with background thermal atoms in the vacuum chamber. This can be phenomenologically modelled, to lowest order of approximation, as a linear decrease of the particle number. How would such loss change the effect of interactions as a function of time? Intuitively, as $N$ drops, we approach the linear regime, so we would expect localisation to be strengthened by the presence of loss. Let us test this.

We model the process by introducing a linear loss of 200 atoms every $t_0$ unit of time, which over $30t_0$ means $N$ falls from 15 to 9 thousand particles. Rerunning the GP simulation with disordered scatterers studied above with losses present, we find that the total energy drops monotonically and more or less linearly from $2.7E_0$ to $2.3E_0$. Meantime, the population curves lie only a little below the no-loss case (Fig.~\ref{GPpops}), yielding a (normalised by the input current) flow rate of 0.1, while the energy fractions are restored practically to the empty channel results (Fig.~\ref{Efrac}). The 1D density profile at long times lies significantly below its conservative version (Fig.~\ref{GP1Dden}), but that is because the entire atomic cloud is smaller -- the density is lower throughout, including in $R_1$ -- this is not to be taken as a sign of stronger localisation. The gradient on a logarithmic scale is largely unchanged, which suggests no major strengthening of localisation takes place.

Our results imply that realistic condensate depletion would not cause a serious improvement in localisation properties, while it would certainly deteriorate the experimental signal-to-noise ratio. Thus, minimising loss is advisable, as always, and tuning interactions to zero is a better solution for removing the effect of the nonlinearity.
\section{Conclusions and future work}
\label{Conc}
In this paper we have carried out a thorough study of Anderson localisation in 2D with point-like Gaussian scatterers, of particular relevance to the recent experiment \cite{BS}. We used both the height and the density of the scatterers to control the strength of the disorder, and varied system size to demonstrate the strong and direct effect of the number of scattering events on the degree of localisation, as well as much more intricate finite size effects where the localisation length itself depended on system size.

We used three complementary methods to tackle the problem: exact diagonalisation, LLT, and time-dependent simulations (solving the Schr\"{o}dinger and the GP equations). We found that exact diagonalisation was quite limited by system size, and highlighted the difficulty in extracting useful numbers out of the calculations. We then presented a complete review of LLT to date, going on to extend it to new frontiers. We also showed that the effective potential $W_E$ can replace the real potential $V$ in the Hamiltonian in terms of reproducing the low-energy eigenspectrum.

Then we used LLT to calculate the eigenstate localisation length at very low energies, quantifying the decay length scale of the eigenstates. This required us to develop a practical approximation to multidimensional tunnelling and a formidable extension of LLT techniques and machinery. It also involved considerable conceptual progress, linking together domain size and the decay exponent (the ``cost'') of tunnelling through the peak ranges of $W_E$ separating domains through the saddle points. We accounted for the effect of increasing energy by merging domains as the domain walls separating them broke down. Crucially, we explicitly tested the decay coefficients computed from LLT against exact eigenstates, validating our computational method and the many approximations involved. We improved on the direct use of the Agmon distance, which gives a lower bound for the true decay coefficient, and found a way of computing the latter in a way that avoids a one-sided bias. We gave a thorough discussion of how the eigenstates spread out over larger areas at higher energies, beyond the regime where quantum tunnelling in $W_E$ and its semiclassical description is applicable, and explained why the mechanisms involved are not captured by our method for computing the localisation length, thus necessarily limiting it to very low energies. We also reviewed multidimensional tunnelling to set our method in context.

We explored the transmission scenario, introduced in the experiment \cite{BS}, the non-equilibrium cold-atom analogue of steady-state current measurements in macroscopic systems. For our model, direct integration of the Schr\"odinger equation is the only currently known method of calculating the localisation length, but even then one must use very large channels, evolve to long times, and extrapolate the results to the infinite time limit (which is far from straight-forward). We proposed an experimentally feasible set-up, using translating Gaussian wavepackets initiated outside the disorder to better resolve the energy dependence, obtaining cleaner results that are easier to interpret. We introduced the flow rate out of the channel as an excellent physical observable, which enables the extraction of useful information using small to moderately sized systems and short to intermediate evolution times. We proposed a phenomenological equation to capture the dependence of the flow rate on the channel length, a very useful tool for quantifying localisation in systems which are too small to allow full localisation within their range, but where system size can be varied. This length scale is unambiguously determined from the flow rate data.

We found strong shot-to-shot fluctuations in the regime where Anderson localisation governed the physics, which is to be expected since the mechanism relies on the randomness of the potential. We measured the flow rate as a function of system length for several noise parameters, and showed that the length scale of exponential decay of $\rho(L)$ was correlated to that seen in density profiles for sufficiently long channels and times. It was also quantitatively similar to the observed localisation length in the density at early steady-state.

Next, we studied finite size effects in the 1D-to-2D dimensional crossover. Our work took a distinctly different approach from all previous exploration of the width dependence in the literature, but is fully consistent with known results, showing that the localisation length increases monotonically as the width grows and settles into its infinite-system value as the width becomes larger than the mean distance between valley lines. Our approach helped highlight the physical mechanisms that give rise to this dependence, an achievement fully ascribed to LLT.

We examined the predictions of LLT regarding the existence of a true mobility edge, introducing the idea that (according to LLT) the mobility edge was not a sharp phase transition but a distribution of finite width, with localisation vanishing gradually, until finally none is left. We motivated the possible existence of a mobility edge in our system by noting that our continuous random potential had a minimal length scale (which can be thought of as a form of correlations) and documented the LLT prediction for its dependence on the noise parameters.

We overcame the challenge of searching for evidence of the mobility edge by using the limiting behaviour of $\rho(L)$ as $L\rightarrow\infty$, a method that can be equally well applied in experiments. We tested the LLT prediction using two progressively higher energy translating Gaussian wavepackets in the transmission scenario, but found no direct evidence to support it. This confirmed that LLT is only valid at low energies, where states are well localised.

From here, we turned our attention to expanding wavepackets with no CoM translation, and explained the disadvantages associated with starting the wavepacket inside the disorder (as in commonly done with expanding wavepackets in cold-atom experiments) for efficiently studying localisation. We then analysed the behaviour of such purely expanding wavepackets if they are initiated outside and are allowed to transmit through the disorder, finding that they behave in a very sensible way, yielding quantitatively meaningful results, but the energy dependence of the localisation length cannot be probed this way.

Next, we demonstrated how a discrete 2D Anderson model emerges from the continuous LLT description, highlighting the importance of detuning between sites arising from the disorder. The resulting ``toy'' model is very approximate and can only be used for gaining qualitative insights. We approximately reconstructed some of the lowest energy eigenstates of the Hamiltonian: those which have a single strongly occupied domain that has tunnelled across valley lines to its nearest neighbour domains (NNC states). This was possible to achieve due to the new LLT technology we have developed as part of this work. Such a simple, discrete model would allow one to predict low-energy dynamics: for example, the transfer of population from one domain to its nearest neighbours. Unfortunately, the severe approximations made \textit{en route} render the model unhelpful in practice, with one of the major problems uncovered being the lack of knowledge of the signs of the eigenstates as they cross domain boundaries. We also analysed statistical properties of the quantities that enter the LLT-Anderson model. 

Then we used an ordered lattice of scatterers to isolate the effect of disorder, and found that in the regime where Anderson localisation dominated, a clear, strong difference could be seen, present in all the different kinds of simulations and calculations presented throughout the paper. Large fluctuations between different noise realisations is another way of identifying when the disorder is governing the physics.

For completeness, we discussed the effect of several secondary features that affected the experiment \cite{BS}, with specific emphasis on two key physical mechanisms that are believed to weaken localisation. The first important aspect we investigated was acceleration. We began by clearly demonstrating that as long as the disordered potential resides on a flat background, localisation properties are set purely by the kinetic energy of the probing wave as it impinges on the disorder. We further found that it is possible to chose parameters such that with acceleration, a wavepacket that is easily localised in the flat background potential cannot be fully localised by increasing system size alone. However, this can be remedied by increasing the strength of the disorder. Second, we explored how Anderson localisation is affected by interparticle interactions. Our results suggested that the nonlinearity allows part of the atoms to become effectively delocalised at the given system size, but does not change the localisation length itself (averaged over the energy distribution). Interactions do not necessarily completely destroy Anderson localisation -- the latter can still be seen if the nonlinearity is not very strong. Depletion of the condensate did not have a major effect on the results.
\subsection{Future work}
While we have attempted to perform a complete and self-consistent study, there are many still open questions that need to be answered. Several extensions of the work have also been mentioned throughout the article. These have all been left for future research at this point, as we think it may be more beneficial to share the insight obtained thus far with the scientific community to allow others to make use of our results if they find them helpful. Here we list the ideas for future work that were generated in the course of our research to date.
\begin{enumerate}
\item It may be possible to perform the LLT calculation of $\xi_E$ for a system where the Green's functions approach would be applicable, even if it would only provide approximate results, and compare the two.
\item It would be excellent to generalise LLT to 3D, where the logic and conceptual picture are largely unchanged, but the practical framework and the technology are not yet in place (everything beyond obtaining $u$ and performing simple mathematical operations on it). This would open the door to a large number of possible studies in 3D.
\item While the LLT prediction of a mobility edge has been found unphysical, the finite size and given shape of the potential scatterers used imply that there \textit{may} still exist a mobility edge due to the presence of a form of spatial correlations. If so, this energy cut off must lie quite high. Can it be reached and investigated, with time-dependent simulations or otherwise?
\item One should investigate the functional dependence of $\xi_E$ (obtained from $\xi(t)$) and $\bar{\xi}$ on the fill factor, $V_0$, and the width of the scatterers. At the moment, this can only be done by running large numbers of simulations at different parameters and examining the dependence explicitly, hoping to discover the functional form by inspection.
\item What effect does the shape of the scatterers have? We have limited ourselves to 2D Gaussian peaks (of more or less constant width) for the entire paper. What would happen if we changed the width, or even made the scatterers, say, square?
\item It would be interesting to study finite size effects on the localisation length in a similar fashion as we have done here, but holding $L=W$ and increasing the size of the system progressively.
\item Can the sign problem of the LLT-Anderson model be solved, possibly through requiring approximate orthogonality of the NNC states? How useful is the model if it can be?
\item One should test the effect of interactions more thoroughly. There are many possibilities here -- so far we have only done the simplest, most basic tests.
\end{enumerate}
Thus, the stage is set for quite an extensive and active future research in this field.
\section*{Acknowledgements}
S.S.S. warmly thanks the following researchers for extremely helpful discussions on the topics indicated in parentheses after each name: Daniel V.~Shamailov (the entire project), Antonio Mateo-Mun\~{o}z (spectral methods in exact diagonalisation), Xiaoquan Yu (importance of the density of states for Anderson localisation), Jan Major (mobility edges in lower dimensions), Mojdeh Shikhali Najafabadi (time-dependent Gross-Pitaevskii simulations of the system), Marcel Filoche and Svitlana Mayboroda (the Agmon distance). Jan Major is further gratefully acknowledged for reading the manuscript and providing useful comments.
\vspace{10pt}
\noindent\rule{\textwidth}{1pt}
\begin{appendix}
The following appendices describe the numerical implementation of various computational methods discussed in the main text, providing details at the level needed to reproduce our work. All simulations are performed in Matlab on a standard laptop, without parallelisation. We also describe the testing performed for each method/code, in terms of convergence, agreement with other solvers, and comparison to analytical results. Thus, we feel completely confident that the numerics are fully under control and that our results are reporting real physics rather than computational artefacts.
\section{Numerical implementation of exact diagonalisation}
\label{appDiag}
We use the position basis to represent the Hamiltonian operator. The scatterer potential $V$ is diagonal in this basis, and so is trivial to evaluate. As for the Laplacian, the simplest approach is to employ stencils to construct it. This can be easily achieved, and while such an algorithm can diagonalise extremely large systems easily, the eigen-energies of the discretized Laplacian converge to those of the continuous operator extremely slowly. This is a common issue in numerical analysis and the solution is to turn to spectral methods.

When Dirichlet or Neumann boundary conditions are needed, the optimal solution is to employ Chebyshev spectral differentiation (see, e.g.,~\cite{Denys}). An extremely powerful implementation of these methods is the ``Chebfun'' toolbox \cite{chebfun}, which is heavily used in our work. In order to obtain a representation of the Laplacian the eigen-spectrum of which converges rapidly, we adapt the new algorithm developed in \cite{ChebDiag} and which is implemented in the Chebfun toolbox for 1D operators by extending it to 2D. Note that as a result we must solve a \textit{generalised} eigenvalue problem, the consequences being that the eigenvectors are not orthogonal.

The spectral algorithm is more trust-worthy and robust than the stencil one, but is very limited in terms of the system size it can handle and suffers from small numerical artefacts to some degree (e.g.~see Fig.~\ref{est_ffg}).
\subsection{Testing}
The diagonalisation code was tested in several ways. First of all, clearly, the number of Chebyshev points in each dimension was increased until satisfactory convergence was achieved. Moreover, the eigenvalues and eigenfunctions of the free Hamiltonian (Laplacian term only, no external potential) are well known. We ensured that both the spectrum and eigenstates of the \textit{continuous} Laplace operator are reproduced correctly by our code (see, e.g., \cite{WikiSub}).

In order to test the Chebyshev-based code in the presence of a disordered potential, a stencil Laplacian operator was constructed making use of the function available at \cite{Laplacian} (which in turn was tested against the Matlab function \texttt{del2.m}). Since the eigenstates converge quickly for a stencil-based code (it is only the eigenvalues which are problematic), we ensured the eigenstates of the Hamiltonian including a random potential compared well to those returned by the Chebyshev code. Meantime, the eigenvalues of the stencil Laplacian (i.e.~the free Hamiltonian) are in excellent agreement with analytical results for the \textit{discrete} Laplacian \cite{exact} (also see \cite{WikiSub}).

Finally, we have also implemented and tested the ``traditional'' square matrix Chebyshev representation of the Laplacian (included in the Chebfun toolbox), and confirmed that its spectrum converges much slower than that of the rectangular representation discovered in \cite{ChebDiag}.
\section{Numerical implementation of known LLT}
\label{appLLTold}
Localisation landscape theory to date deals with two main objects: the localisation landscape $u$ (and its inverse, $W_E$), and the valley network of $u$. Here we describe how both of these can be computed, even for truly large system sizes (where other approaches are completely impractical) and in very reasonable computational time. The first step is of course to solve the stationary PDE for $u$, (\ref{uPDE}). The optimal method we are aware of is the domain decomposition method, an implementation of which is available making use of the legacy solver of the PDE toolbox in Matlab \cite{DD_ML}. We find excellent performance with partitions about $25\ell \times 25\ell$ in size, together covering the entire area of the system (with no overlap).

We employ spectral differentiation, using the square Chebyshev derivative matrices \cite{ChebDiag, chebfun}, to accurately obtain the first and second derivatives of $u$. Then the functional approximation capability of the Chebfun toolbox (see appendix \ref{appDiag}) is utilised to expand $u$ and its derivatives as analytical series of Chebyshev polynomials. The root-finding routines of the Chebfun toolbox are then able to identify all the extrema of $u$. Note that spurious extrema are often picked up very close to the edge of the system and need to be discarded.

Since the root-finding algorithm is also limited by system size, it is necessary to use partitioning not only for obtaining $u$, but also for identifying its extrema. The same main partitions are used in both cases for simplicity. It is quite likely that the analytical approximations of the solution and its derivatives will not be (sufficiently) accurate on the ``joint lines'' separating the partitions. A region within a thin frame running around the edge of the partitions cannot be trusted as it may suffer from spurious features, and therefore, we place additional partitions centred around all the joint lines, referred to as ``patches''. Note that this is only done when scanning for exterma, not when solving for $u$ with the domain decomposition method. All the partitions and patches are taken larger than necessary, and the solution on the extra ``padding'' frame is not used after the Chebyshev approximations are obtained. The trust regions of all the partitions and patches can be arranged such that they do not overlap and cover the entire system domain.

It is possible that extrema which are on the joint lines of the \textit{trust regions} would be missed in the root-finding step in both partitions. Therefore, we employ a ``safety net'' around each trust region: we look for extrema on the trust regions proper as well as within a thin frame around them (overlapping with other trust regions). Then extrema in the safety nets are compared to all those picked up in the main trust regions, and if one has indeed been missed, it is added to the list. All extrema in the safety nets which are also identified in the trust regions are discarded.

At this point, we have a complete solution for $u$ on the entire system and have found and classified all its extrema. The valley network is constructed in the following way. We begin from each saddle and follow the gradient of $u$ (forward and backward) until we arrive at a minimum or the system boundary, at which point the valley line is terminated. The valley lines are paths of steepest descent connecting saddles to minima of $u$.

A prudent remark is in order. The valley network construction, as outlined here, is not fault-proof: imperfections do occur. If one is interested in overall, average properties of $u$, $W_E$ or the network, then these imperfections are not important. However, for our purposes in sections \ref{XiSaddles} and \ref{BHM}, the valley network really has to be immaculate. This requires fixing any accuracy issues that can result after directly following the prescription in this appendix. The problems encountered and their solutions will be described in appendix \ref{appLLTnew}.
\subsection{Testing}
Several other methods have been employed to solve for $u$ and the performance of the solvers compared. The valley network has also been constructed based on the localisation landscape from all these alternative approaches to ensure agreement at this more refined level as well. The first method is a Chebyshev spectral solver included in the Chebfun toolbox \cite{chebfun}, developed in \cite{Townsend}. It is extremely accurate, but (like all alternative methods described here) limited by system size. Next, we tested the modern PDE solver in Matlab's PDE toolbox, which, like all the Matlab solvers, is based on the finite element method. The legacy solver has similar performance, it is only the interface that is different. A multi-grid implementation \cite{MultiGrid} has also been trialled, using the Matlab legacy solver underneath. Two other implementations of the domain decomposition method have further been evaluated for performance \cite{DD_Bernd}.

We confirmed that all methods agreed with each other, and convergence has been tested with respect to all available precision/resolution parameters in the different codes. This statement also holds for the step size used for constructing the valley network.
\section{Numerical implementation of time-dependent simulations}
\label{apptdep}
All time-dependent PDEs are solved using Matlab's modern PDE solver, \texttt{solvepde.m}, with Dirichlet boundary conditions. The algorithm uses the finite element method for the spatial dependence, and an adaptive-time-step, variable-order stiff ordinary differential equation solver for the time-evolution. Because both the spatial and temporal problems consist of many coupled equations that need to be solved simultaneously, this implementation cannot be easily parallelised, and the domain decomposition method (see appendix \ref{appLLTold}) could not be successfully adapted to solve this issue. Other codes, however, have taken a different approach and overcame this obstacle, e.g.~\cite{Dylan}. As such, our computational capabilities are limited by system size and resolution (a high resolution is required to capture a state with high kinetic energy). Needless to say, all reported results have been checked for convergence.

Whenever we solve the GP equation, starting from a TF profile in a harmonic trap that is removed at $t=0$, we first use the analytical expression for the TF cloud \cite{Kamchatnov} as a guess, and solve for the true ground state in the harmonic trap. The resulting wavefunction is then used as an initial condition for time evolution.
\subsection{Testing}
The performance of the code was tested against two analytical results: the expansion of a 2D Gaussian wavepacket (standard textbook material) and the expansion of an initial TF cloud in 2D after the harmonic trap is removed \cite{Kamchatnov}. Both were reproduced faithfully, thereby inspiring confidence in the numerics.
\section{Numerical implementation of new LLT}
\label{appLLTnew}
In this appendix we give details on how all the extensions of LLT put forward in this article are implemented. First of all, one must solve for $u$ and compute the valley network, as described in appendix \ref{appLLTold}. As mentioned at the end of that appendix, the resulting valley network can have imperfections that need to be ``cleaned up'' before proceeding any further. Below we give a list of known possible issues:
\begin{itemize}
\item Rarely, due to inaccuracies in the gradient of $u$, valley lines may start at a saddle and arrive at a maximum of $u$ (rather than a minimum), or get ``stuck'' at some point (not near an extremum) in the 2D plane.
\item Spurious saddle-minimum pairs (or even entire chains) may be identified during the extremum search.
\item Any type of extremum -- a maximum, minimum or a saddle -- may be accidentally picked up twice during the root-finding stage.
\end{itemize}
All these are reasonably simple to correct, and most of the process can be automated, with only occasional need for human judgement. Once the valley network and the list of extrema are perfected, we may move on to implementing the new features of LLT discussed in the main text.

The first step is the removal of any valley lines that do not constitute part of a closed domain. This can be accomplished by counting the number of valley lines terminating at each minimum and searching for minima that only have one. For each such minimum, we then remove the minimum, the valley line that links it to a saddle, the saddle, and the second valley line originating from that saddle. The process is repeated until there are no more minima with only one valley line connecting to them.

Next, we trace candidate paths of (\textit{approximately}) least-cost with respect to the Agmon distance $\rho_E$ from every saddle point to two maxima of $u$ by following the gradient of $u$, much in the same way as the valley lines are obtained. Thus, two valley lines emanate from each saddle and connect to two minima (or exit the system), following paths of steepest descent. Two lowest-lying paths originate from each saddle and connect to two maxima, following paths of steepest \textit{ascent}.

For any domain that is fully internal to the system, the collection of all valley lines associated with saddles that connect to the (unique) maximum of this domain constitute the domain walls. For domains that lie on the edge of the system, the valley lines are not enough -- we must complete the collection of the domain walls by including the relevant segments of the system boundary. This is not a trivial task, but in brief, one creates a collection of all the exit points of valley lines that terminate by exiting the system and breaks up the perimeter into segments separated by these exit points. Then, we assign a domain to each segment using the knowledge of which saddle the exit valley lines ``belong'' to and which maxima are connected to these saddles. The relative position of the valley line exit points and maxima is also instrumental in correctly assigning system boundary segments to the right domains. The process can be quite intricate if localisation is very strong: it is not uncommon to have several valley lines that practically merge as they exit the system, which complicates matters further. Mostly, we have successfully automated this process, but occasionally human judgement is needed to correctly complete the task.

Now, once we know which saddles connect to which maxima, it is simple to identify domains which are nearest-neighbours, as they will share at least one common saddle. A list of potential problems which we have encountered in the process described so far can be summarised as follows:
\begin{itemize}
\item The minimal-cost paths may be (incorrectly) computed to run on top of the valley lines originating from that saddle.
\item The least-cost paths may get ``stuck'' at some point (not near an extremum) in the 2D plane.
\item The maximum to which a minimal path should connect may not be identified if the path terminates a little too far away from it.
\item The two minimal-cost paths connecting a saddle point to two different maxima of $u$ may (erroneously) overlap, leading to both maxima associated with this saddle being identified as the same one.
\end{itemize}
All of these problems have been solved, largely automating the problem identification and clean-up process, but not fully -- some issues require human judgement, and a selection of the appropriate solution via trialling. All results presented in the main text have of course been fully tested and corrected, whenever necessary.

Once we know the collection of paths (valley lines and potentially segments of system boundary) that make up the domain walls, we can perform integrals of various functions, restricted to the domains. A sufficiently fine rectangular grid is set up such that it just covers the area of the domain (i.e.~we draw a rectangular box that just fits the arbitrarily-shaped domain inside it and set up a regular grid on it). The domain walls are rounded to this grid, and for each $y$-value, it is then simple to find the smallest and largest $x$ values in the set of the domain wall points. The integrand is evaluated on the rectangular grid, with all values outside the domain walls replaced by zeros. A 2D integral is then trivial to perform.
\subsection{Average distance between valley lines}
Here we provide a method to compute the average distance between the valley lines $\bar{D}(E)$ in a network that is not fully closed (i.e.~it includes some ``open'' valley lines). A good solution is to essentially sample the \textit{density} of the valley lines. We begin by cutting down the network to a given energy $E$. Then a large number ($10^3$ seems to work well) of random points in the interior of the system is chosen, and each point is assigned a random direction. Then one measures the distance from each random point in its associated random direction to the nearest valley line. If no valley line is encountered and the ``ray'' exits the system domain, it is discarded. The remaining distances are averaged. This gives a number which is \textit{proportional} to the actual mean distance between the valley lines. The proportionality constant can be determined by running the calculation on a series of closed networks where a domain-area based computation can be performed as well (of course, configurational averaging is needed to obtain meaningful numbers). Comparing the true distance to $\bar{D}(E)$, the two measurements are indeed found to be out by just a constant scaling factor of about 1.84. This scaling factor can then be used to convert the ``proportional'' measure of the distance between valley lines to a real physical length also for open networks.
\end{appendix}
\bibliography{MyRefs.bib}
%
%
\end{document}